\documentclass[a4paper,11pt]{article}

\pdfoutput=1
\usepackage{amsmath}
\usepackage{amssymb}
\usepackage{cite}
\usepackage[breaklinks=true,colorlinks=true,linkcolor=blue,urlcolor=blue,citecolor=blue, linktocpage]{hyperref}
\usepackage{comment}
\usepackage{mathrsfs}
\usepackage{stmaryrd}
\usepackage{graphicx}
\usepackage{color}
\usepackage[dvipsnames]{xcolor}
\usepackage{xspace}
\usepackage{multirow}
\usepackage{cleveref}		
\usepackage[normalem]{ulem}
\usepackage{makecell}
\usepackage{setspace}
\usepackage{geometry}
\usepackage{enumerate}
\usepackage{anyfontsize}
\usepackage{float}
\usepackage{footmisc}

\hypersetup{
	pdfcreator={Hello},
	pdfproducer={World}
}

\newcommand{\fulleqref}[1]{Eq.~\eqref{#1}}

\newcommand{\matteo}[1]{\textcolor{black}{#1}}

\providecommand{\der}[1]{\text{d} #1}
\providecommand{\lcdm}{$\Lambda$CDM\xspace}

\DeclareMathAlphabet\mathbfcal{OMS}{cmsy}{b}{n}
\definecolor{ULBblue}{RGB}{26,90,155}

\newcommand\id{{\rm d}}       

\author{Matteo Lucca}
\title{The future of cosmology? A case for CMB spectral distortions}
\date{}

\begin{document}
\pagenumbering{gobble}
\thispagestyle{empty}

\begin{figure}
	\vspace{-2.5 cm}\hspace*{-3.2 cm}
	\includegraphics[height=30 cm]{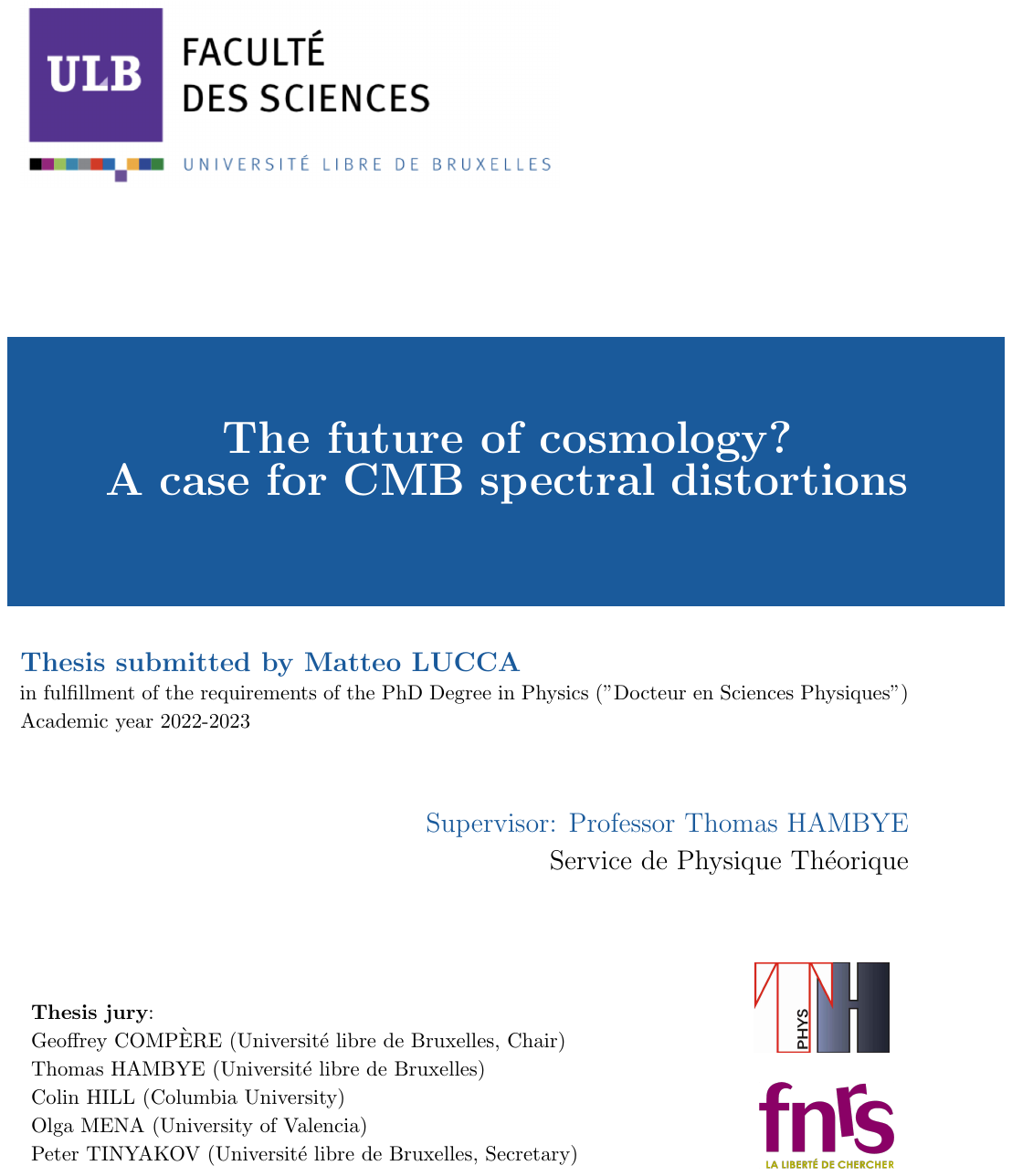}
\end{figure}
\clearpage

\newpage

\begin{center}
	\Huge \bf Abstract
\end{center}

This thesis treats the topic of Cosmic Microwave Background (CMB) Spectral Distortions (SDs). In brief, CMB SDs are any deviation from a pure black body shape of the CMB energy spectrum. They can be generated by any heating, number changing and photon mixing process taking place in the history of the universe. As such, CMB SDs can probe the inflationary, expansion and thermal evolution of the universe. In particular, they are expected to be produced within the \lcdm model as well by the presence of beyond-\lcdm physics like in the case of e.g., features in the primordial power spectrum, primordial gravitational waves, non-standard dark matter properties, primordial black holes, primordial magnetic fields and many models attempting to solve the Hubble tension, among others.

The currently missing observation of this rich probe of the universe makes of it an ideal target for future observational campaigns, as recently highlighted by ESA's long-term science program in the context of the Voyage 2050 initiative. In fact, the \lcdm signal \textit{guarantees} a discovery and the sensitivity to such a wide variety of new physics opens the door to an enormous uncharted territory. With up-coming 
ground-based and balloon experiments expected to start collecting data in the next $1-3$ years, this is an exciting time to do physics with CMB SDs.

In light of these considerations, the thesis opens by reviewing the topic of CMB SDs. The theory of SDs, the many sources, the experimental status and its future perspectives are overviewed in the manuscript in a pedagogical and illustrative fashion, aimed at waking the interest of the broader community in regard to this fascinating and promising topic.

This introductory premise sets the stage for the presentation of the first main contribution of the thesis to the field of CMB SDs: their implementation in the cosmological Boltzmann solver \texttt{CLASS} and the parameter inference code \texttt{MontePython}. The \texttt{CLASS}+ \texttt{MontePython} pipeline is publicly available, fast, it includes all sources of SDs within \lcdm and many others beyond that, and allows to consistently account for any observing CMB SD experimental setup. As such, it is unique and fully opens the door to the cosmological exploration of CMB SDs.

By means of these newly developed numerical tools, the second main contribution of the thesis consists in showcasing the versatility and competitiveness (in terms of constraining power) of CMB SDs for all aforementioned cosmological models as well as for a number of different mission designs and data set combinations. Among others, the results presented in the thesis highlight the fact that CMB SDs could impose the by far most stringent constraints (in the relevant regions of parameter space) on features in the primordial power spectrum as well as on many dark matter and primordial black holes properties. Also, they would be able to play a significant role even within the \lcdm model.

Finally, the manuscript is disseminated with (20) follow-up ideas that naturally extend the work carried out so far, also highlighting how rich of unexplored possibilities the field of CMB SDs still is. The hope is that these suggestions will become a propeller for further interesting developments.  

\newpage

\begin{center}
	\Huge \bf R\'esum\'e
\end{center}
\enlargethispage{\baselineskip}

Cette th\`ese traite du sujet des distorsions spectrales (SDs) du fond diffus cosmologique (CMB). En bref, les CMB SDs sont tout \'ecart par rapport \`a une forme de corps noir pur du spectre \'energ\'etique du CMB. Ils peuvent être g\'en\'er\'es par n'importe quel processus de chauffage, de changement de nombre et de m\'elange de photons ayant lieu dans l'histoire de l'univers. En tant que tels, les CMB SDs peuvent sonder l'inflation, l'expansion et l'\'evolution thermique de l'univers. En particulier, on s'attend à ce qu'ils soient \'egalement produits dans le mod\`ele \lcdm par la pr\'esence de physique au-del\`a de \lcdm comme dans le cas, par exemple, des caract\'eristiques du spectre de puissance primordial, des ondes gravitationnelles primordiales, des propri\'et\'es non-standard de la mati\`ere noire, les trous noirs primordiaux et les champs magn\'etiques primordiaux,
entre autres.

L'observation actuellement manquante de cette riche sonde de l'univers en fait une cible id\'eale pour de futures campagnes d'observation, comme l'a r\'ecemment soulign\'e le programme scientifique \`a long terme de l'ESA dans le cadre de l'initiative Voyage 2050. En fait, le signal \lcdm \textit{garantit} une d\'ecouverte et la sensibilit\'e \`a une si grande vari\'et\'e de nouvelles physiques ouvre la porte \`a un \'enorme territoire inexplor\'e. Avec exp\'eriences qui devraient commencer \`a collecter des donn\'ees dans les prochaines $1-3$  ann\'ees, c'est une p\'eriode passionnante pour faire de la physique avec les CMB SDs.

\`A la lumi\'ere de ces consid\'erations, la th\`se s'ouvre en examinant le sujet des CMB SDs. La th\'eorie des SDs, les nombreuses sources, le statut exp\'erimental et ses perspectives futures sont pr\'esent\'es dans le manuscrit de mani\`ere p\'edagogique et illustrative, visant \`a \'eveiller l'int\'erêt de la communauté \'elargie à l'\'egard de ce sujet fascinant et prometteur.

Ces pr\'emisses introductives pr\'eparent le terrain pour la présentation de la première contribution principale de la thèse au domaine des CMB SDs : l'implémentation des CMB SDs dans le solveur cosmologique de Boltzmann \texttt{CLASS} et le code d'inférence de paramètres \texttt{MontePython}. Le pipeline \texttt{CLASS}+\texttt{MontePython} est accessible au public, rapide, il inclut toutes les sources de SD dans \lcdm et bien d'autres au-delà, et permet de tenir compte de manière cohérente de toute configuration expérimentale d'observation. A ce titre, il est unique et ouvre pleinement la porte à l'exploration cosmologique des CMB SDs.

Au moyen de ces outils numériques nouvellement développés, la deuxième contribution principale de la thèse consiste à montrer la polyvalence et la compétitivité (en termes de pouvoir contraignant) des CMB SDs pour tous les modèles cosmologiques susmentionnés ainsi que pour un nombre de conceptions et de données de mission différentes. Entre autres, les résultats présentés dans la thèse mettent en évidence le fait que les CMB SDs pourraient imposer les contraintes de loin les plus strictes (dans les régions pertinentes de l'espace des paramètres) sur les caractéristiques du spectre de puissance primordial ainsi que sur de nombreuses interactions de matière noire et trous noirs primordiaux. En outre, ils pourraient jouer un rôle important même au sein du modèle \lcdm.

Enfin, le manuscrit est diffusé avec (20) idées de suivi qui prolongent naturellement le travail effectué jusqu'à présent, soulignant également à quel point le domaine des CMB SDs est encore riche en possibilités inexplorées. L'espoir est que ces suggestions deviendront une hélice pour d'autres développements intéressants.

\newpage

\begin{center}
	\Huge \bf Acknowledgments
\end{center}

First of all, I would like to thank Thomas Hambye, my supervisor, for giving me the opportunity to pursue the PhD degree and for helping me discover myself as a scientist. I have truly appreciated the freedom and guidance offered to me, which greatly helped me develop my independence, confidence and professional maturity. I would also like to thank Thomas for the thorough feedback on this manuscript, which has significantly improved its quality.

I am very grateful  to the members of the jury, Geoffrey Comp\`ere, Colin Hill,  Olga Mena and Peter Tinyakov,  for taking the time to review the thesis and to participate to the private and public defenses, which ultimately led to very stimulating discussions.

For the financial support during the course of the thesis I thank the F.R.S.-FNRS, the “Probing dark matter with neutrinos” ULB-ARC convention and the IISN convention 4.4503.15. When necessary, the computational resources have been provided by the C\'ECI, funded by the F.R.S.-FNRS under Grant No. 2.5020.11 and by the Walloon Region.

Of the many collaborators I have had the honor to work with, I would like to especially thank Jens Chluba, Silvia Galli, Deanna Hooper and Julien Lesgourgues for their invaluable mentorship and for being such inspiring scientific and personal role models. Their humility, enthusiasm and impressively deep knowledge of physics have greatly impacted me as a scientist. I would also like to express my gratitude to Lennart Balkenhol, S\'ebastien Clesse, Ga\'etan Facchinetti, Guillermo Franco Abell\'an, Marco Hufnagel, Riccardo Murgia, Lorenzo Piga, Vivian Poulin, Laura Sagunski, Nils Sch\"oneberg, Joeseph Silk, Tristan Smith and Laurent Vandereyden for the very enjoyable and stimulating collaborations.

Of course, I would like to thank all the (current and former) members of the Service de Physique Th\'eorique at ULB, in particular Eleni Bagui, Iason Baldes, Rupert Coy, Quentin Decartes, Nicolas Grimbaum-Yamamoto, Aritra Gupta, Michael Kuznetsov, Laura Lopez-Honorez and Michel Tytgat as well as the other members of the group mentioned above, for creating such a pleasant environment and for the numerous lively conversations.

For the many good memories at conferences, visits and schools I would also like to thank \'Etienne Camphuis, Bryce Cyr, Eleonora Di Valentino, Thomas~Kite, Anna Kormu, Roi Kugel, Elizabeth Lee, Francisco Maion, Lurdes Ondaro-Maella, Th\'eo Simon and Denis Werth on top of several of the aforementioned colleagues.

On a personal side, I would like to thank Ila, Till and Nino for their incredible support. We made it through some very painful times and I am at a loss for words to express how lucky I feel for having had you by my side. Of course, much of it would have been impossible without the refreshing smiles and joy of Ceci, Leo e Sofi. I would also like to thank Elena, Francesca and Mauro for welcoming me so warmly in their family. The precious friendship of Thorben and the splendid group of friends in D\"usseldorf have also been a constant source of positivity and happiness for which I am very grateful.

Finally and most importantly, I would like to thank my soon-to-be wife Giuli. I love you.

\newpage

\hfill\begin{minipage}{0.55\linewidth}
	Cara mami, \\
	ce l'abbiamo fatta! Anche questa \`e andata \\
	e in fondo ce la siamo cavata discretamente dai.. \\
	Vorrei dedicatrti la tesi, alla mia pi\`u grande fan. \\
	Mi manchi tanto. \\
	Ti voglio bene. \\
	Grazie di tutto, \\
	Il tuo Matty \\ \\
	P.S. Qui ti salutano tutti!
\end{minipage}

\newpage

\begin{center}
	\Huge \bf Table of contents
\end{center}

{\fontsize{10}{10}
\tableofcontents
}

\newpage

\begin{center}
	\Huge \bf List of abbreviations
\end{center}
\enlargethispage{\baselineskip}

\noindent AME: Anomalous Microwave Emission \\
BAO: Baryon Acoustic Oscillations \\
BB: Black Body \\
BBN: Big Bang Nucleosynthesis \\
BR: Branching Ratio \\
CIB: Cosmic Infrared Background \\
CL: Confidence Level \\
\texttt{CLASS}: Cosmic Linear Anisotropy Solving System \\
CMB: Cosmic Microwave Background \\
CN: Cyanogen \\
CNC: Cluster Number Count \\
CRR: Cosmological Recombination Radiation \\
CS: Compton Scattering \\
DE: Dark Energy \\
DM: Dark Matter \\
EDE: Early Dark Energy \\
EM: Electromagnetic \\
$EM$: Emission Measure \\
EOS: Equation Of State \\
FLRW: Friedmann-Lemaitre-Robertson-Walker \\
GW: Gravitational Wave \\
GUT: Grand Unified Theory \\
kSZ: kinetic Sunyaev-Zeldovich \\
$\Lambda$D: $\Lambda$ Domination \\
LOS: Line Of Sight \\
MCMC: Markov chain Monte Carlo \\
MD: Matter Domination \\
PBH: Primordial Black Hole \\
PCA: Principal Component Analysis \\
PMF: Primordial Magnetic Field \\
PPS: Primordial Power Spectrum \\
PSR: Potential Slow-Roll \\
PTA: Pulsar Timing Arrays \\
qPBH: quasi-extremal Primordial Black Hole \\
RD: Radiation Domination \\
SD: Spectral Distortion \\
SM: Standard Model \\
SN: Supernova \\
SNR: Signal-to-Noise Ratio \\
SZ: Sunyaev-Zeldovich \\
tSZ: thermal Sunyaev-Zeldovich

\vspace{0.3 cm}
\noindent\textbf{Experiments}

\noindent ACT: Atacama Cosmology Telescope \\
APSERa: Array of Precision Spectrometers for the Epoch of Recombination \\
ARCADE: Absolute Radiometer for Cosmology, Astrophysics and Diffuse Emission \\
COBE: Cosmic Background Explorer \\
DMR: Differential Microwave Radiometer \\
EDGES: Experiment to Detect the Global Epoch of Reionization Signature \\
FIRAS: Far-InfraRed Absolute Spectrophotometer \\
LBL: Lawrence Berkeley Laboratory \\
LIGO: Laser Interferometer Gravitational-Wave Observatory \\
LVK: LIGO-Virgo-KAGRA \\
PIXIE: Primordial Inflation Explorer \\
PRISM: Polarized Radiation Imaging and Spectroscopy Mission \\
SDSS: Sloan Digital Sky Survey \\
SIMBAD: Spectroscopic Interferometer for Microwave Background Distortions \\
SPT: South Pole Telescope \\
TMS: Tenerife Microwave Spectrometer

\newpage

\pagenumbering{arabic}

\section{Introduction}\label{sec: intro}

As a theorist, the current status of the research regarding the Spectral Distortions (SDs) of the Cosmic Microwave Background (CMB) is at the same time a goldmine of opportunities and maddening. There is no other way to put it.

CMB SDs are any type of deviation from a pure Black Body (BB) shape of the CMB energy spectrum and can be induced by any heating, number changing and photon mixing process happening in the history of the universe (see e.g., \cite[\hyperlink{I}{I},\,\hyperlink{XX}{XIX}]{Chluba2016Which, Chluba2019Voyage, Chluba2019Spectral} for recent reviews). In a loose way, they are the spectral analogous of the spatial (temperature and polarization) anisotropies of the CMB. Since their postulation in the late '60s \cite{Zeldovich1969Interaction, Sunyaev1970Interaction, Zeldovich1972Influence}, the effort of a number of scientists made of them an effectively exact science (see e.g.,~\cite{Danese1982Double, Burigana1991Formation, Hu1993ThermalizationI, Hu1993thermalizationII, Hu1995Wandering, Chluba2005Spectral, Chluba2011Evolution, Chluba2012CMB, Chluba2013Green, Chluba2013Distinguishing, Chluba2014Teasing, Chluba2016Cosmospec}). We now believe we know basically everything about their theory: what can source them (within the $\Lambda$CDM model -- ensuring a guaranteed signal -- and beyond), how they evolve across the whole thermal history of the universe and what kind of information can be extracted from their observation. For instance, CMB SDs of primordial origin, i.e., sourced in the pre-recombination era, carry invaluable information on physics to which complementary probes such as the CMB anisotropies are not sensitive, while the late-time production of SDs can teach us a lot on the epoch of reionization and structure formation. Yet, while the numerical advances have been able to keep the pace, the observational status has been dramatically lagging behind: we have no direct observation of the SD signal, only loose upper bounds inferred from data gathered by the FIRAS mission in the early '90s \cite{Fixsen1996Cosmic}. 

A metaphor might help the unaware reader understand the insider perspective. Imagine a beautiful day you decided to go hike in the mountains, exploring a new path. You have your equipment all set up and your hopes high, your experience tells you the view from the top is going to be breathtaking. But after a while some clouds start to appear in the sky and the more you hike the closer and thicker they get. By the time you are midway -- a long way from the start, you can barely see the tip of your nose. As said: maddening. But worthless? Absolutely not, because as always in the mountains (and in life) the clouds come and go, one must always remember that. 
And so you persevere, keep going up the mountain and eventually the clouds start to clear. By the time you reach the top the sky is as clean as it gets, the view as spectacular as advertised.

And so is the situation for CMB SDs. Our current understanding promises some incredibly rich physics to test and discover, we have all the theoretical and numerical tools we need to extract it, but we have not seen anything so far. Yet, the rewarding view is somewhere out there. Not only CMB SDs offer a \textit{certainty of discovery} (either we do observe the \lcdm signal or something is fundamentally wrong about our understanding of the early universe -- two equally exciting possibilities), but their observation would also open the door to uncharted territory in the context of the most fundamental questions of modern cosmology such as the properties of inflation, the nature of dark matter, and the Hubble tension among many others.

\newpage 
Far from the ambition of indicating the way forward, the writing of this thesis is rather the opportunity for me to convince \textit{my five-and-twenty readers}\footnote{A. Manzoni, \textit{I promessi sposi}, 1827 -- a very modern book in a lot of ways.} of the wide-ranging perspectives that an observation of the CMB SD signal could offer and of the fact that its reach might not be too far away in the future, as important steps forward are already in motion. And that is because achieving these goals will only be possible with a community effort.~Whether you are a young student or an established professor, a theorist or an experimentalist, only together we will be able to find our way outside of the clouds.

With this in mind, the discussions carried throughout the thesis will mostly be pedagogical and review-like, aimed at waking the interest of the broader community in regard to this fascinating and promising topic. The results developed during the course of the thesis (a full list is provided at the very end of the manuscript\footnote{The works developed during the thesis will be referenced using the roman notation.}) will be placed in the more general landscape of the literature on CMB SDs. Of course, it will not be possible to cover extensively all that there is to say about CMB SDs, since there is much more in the literature about the subject than what will be feasible to present in the manuscript. To amend for this limitation the interest reader will find an extensive collection of references where to find more in-depth discussions with the related technical details. Furthermore, the text is disseminated with (20) follow-up ideas that are in our opinion worth perusing going forward, also highlighting how rich of unexplored possibilities this field is. The hope is that these suggestions will become a propeller for further interesting developments.

Concretely, we will start this thesis in Sec.~\ref{sec: lcdm} by reviewing the standard cosmological model, $\Lambda$CDM. There we will introduce the key concepts and equations ruling the expansion and thermal history of the universe following in chronological order the main epochs of its evolution. This will set the basis for the discussion more strictly related to CMB SDs that will follow in the subsequent sections. In the context of the metaphor above, this will prepare us for the hike. The information contained in this section has been gathered during the development of several references in the course of the thesis, in particular~[\hyperlink{V}{V}] for inflation (Sec. \ref{sec: infl}), [\hyperlink{IX}{IX},\,\hyperlink{XIII}{XIII}] for Big Bang Nucleosynthesis (Sec. \ref{sec: BBN}), [\hyperlink{X}{X}] (among others) for recombination and the CMB (Sec. \ref{sec: CMB}), [\hyperlink{VI}{VI}$-$\hyperlink{VIII}{VIII}] for structure formation (Sec.~\ref{sec: LSS}) and [\hyperlink{II}{II},\,\hyperlink{IV}{IV},\,\hyperlink{VI}{VI}$-$\hyperlink{VIII}{VIII},\,\hyperlink{X}{X}] in relation to cosmological tensions (Sec.~\ref{sec: tensions}).

In Sec.~\ref{sec: theory} we will then dive deeper into the theory of CMB SDs and the numerical tools used to evaluate its predictions. As we will see there, a number of processes will provide access to a mountain of cosmological information, but will also require increasingly high levels of experimental sensitivity to be observed. The goal of this section is to provide a pedagogical overview on the topic of SDs (Sec.~\ref{sec: ped}) and the related sourcing effects (Sec.~\ref{sec: sources}), following the reviewing references [\hyperlink{I}{I},\,\hyperlink{III}{III},\,\hyperlink{XIV}{XIV}] published in the course of the thesis. The various contributions to the final SD signal within \lcdm are graphically summarized in Fig.~\ref{fig: SD_LCDM_summary}. The discussion will then culminate in the presentation of the implementation of CMB SDs in the cosmological Boltzmann solver \texttt{CLASS} (Sec.~\ref{sec: num}), which has been fully set up in the context of the aforementioned works and represents the first main contribution of the thesis. In fact, the newly developed \texttt{CLASS} implementation is publicly-available, fast, it includes all sources of SDs within \lcdm and many others beyond that, and allows to consistently account for any observing experimental setup. As such, the code is unique (no other can do all of the above, especially not at the same time) and fully opens the door to the cosmological exploration of CMB SDs, exploiting the many models, features and observables already included by default in \texttt{CLASS}.

After having introduced CMB SDs as cosmological observable, the current state of the art and perspectives to observe them will be the subject of Sec.~\ref{sec: exp}. There we will discuss the current limited experimental status (Sec. \ref{sec: firas}), i.e., where the clouds start to thicken, together with up-coming efforts (Sec. \ref{sec: current_exp}) and possible long-term avenues (Sec. \ref{sec: future_exp}) to improve the situation, i.e., how to potentially clear the clouds. The various stages of observational development are graphically summarized in Figs.~\ref{fig: SD_exp_spectrum}, \ref{fig: SD_exp_upcoming} and \ref{fig: SD_exp_future}, respectively, the former of which is an original contribution of this manuscript and represents the most complete (to our knowledge) collection of available data points to date. Furthermore, in complementarity with the aforementioned \texttt{CLASS} implementation of SDs, to be able to perform data analyses and forecasts for the many suggested experimental setups a very general class of (mock) likelihoods has been included in the parameter extraction code \texttt{MontePython} (by default tightly interfaced with \texttt{CLASS}) in [\hyperlink{I}{I}]. Its details are discussed in Sec.~\ref{sec: num_2} and it includes, among others, a realistic treatment of galactic and extra-galactic foregrounds, as implemented in [\hyperlink{V}{V}].

With theory, observational status and numerical pipeline outlined in the previous sections, in Sec. \ref{sec: app} we will then overview the role that CMB SDs could play within the \lcdm model (Sec.~\ref{sec: res_lcdm}) as well as a number of possible perspectives for discovery of new physics that might be achieved with the future observation of a CMB SD signal, i.e., the view from the top of the mountain. The considered scenarios stretch across the whole history of the universe all the way from inflation (Sec.~\ref{sec: res_infl}) to structure formation (Sec.~\ref{sec: res_lcdm}), bridging over (among others) the production of primordial Gravitational Waves (GWs) and Primordial Black Holes (PBHs) in the very early universe (Secs.~\ref{sec: res_further} and \ref{sec: res_PBH_form}), Dark Matter (DM) and PBH properties manifesting in the pre-recombination era (Secs.~\ref{sec: res_DM}, \ref{sec: res_PBH_ev} and \ref{sec: res_PBH_acc}), and the impact of matter inhomogeneities (sourced via e.g., primordial magnetic fields) on the recombination process (Sec.~\ref{sec: res_PMFs}). The role that CMB SDs could play in the context of the Hubble tension will also be discussed (Sec.~\ref{sec: res_tens}). This wide spectrum of applications unequivocally highlights the richness of CMB SDs as a probe of cosmology.

In the context of Sec.~\ref{sec: app}, the work carried out in the course of the thesis either represents (now or a the time of publication) the ``state of the art'' (i.e., the most recent and advanced treatment) of a given scenario, such as in the case of \lcdm in the pre-recombination era~[\hyperlink{I}{I},\,\hyperlink{III}{III},\,\hyperlink{XIV}{XIV}], inflation~[\hyperlink{V}{V}], DM decay and PBH evaporation~[\hyperlink{I}{I}], inhomogeneous recombination~[\hyperlink{XIV}{XIV}] and Hubble tension~[\hyperlink{IV}{IV}], or sets the stage for the further study of other effects thanks to the very general \texttt{CLASS}+\texttt{MontePython} pipeline, such as in the case of primordial GWs and PBH accretion.~This is the second main contribution of the thesis:~a broad, state-of-the-art, internally-consistent, reproducible and extendable collection of analyses showcasing the untapped potential of CMB SDs.

We will conclude in Sec.~\ref{sec: conc} with a summary of the thesis and other closing remarks.

\newpage
\noindent\textbf{Units, notation and conventions}

\noindent Unless stated otherwise we will adopt natural units with $\hbar=c=k_B=1$. To facilitate the unit conversion in the following discussion, we recall that 1 K = $8.62\times10^{-5}$ eV, \linebreak 1~GHz = $4.14\times10^{-6}$ eV and 1 Jy/sr = $1\times10^{-26}$ W/(m$^2$\,Hz\,sr). For the metric we will use the $(-,+,+,+)$ sign convention. For 3D and 4D quantities we will use Latin and Greek indices, respectively, and always implicitly assume the Einstein summation convention over repeated indices. For all cosmological quantities we will use the index 0 to identify today's value.

\newpage
\section{Preparing the equipment: the \lcdm model}\label{sec: lcdm}

The \lcdm model of cosmology is a magnificent success. If one stops and thinks about it, it is astonishing that with our limited (as it covers only 5\% of the total energy budget of the universe) knowledge of particle physics and six additional parameters we can describe the history of the whole universe at large scales since shortly after the big bang. Our understanding of cosmology is so deep that not only we can determine its characteristics to the percent level consistently at vastly different (spatial and temporal) scales, but it has also allowed to place some of the most stringent bounds we have on some of the properties of the neutrinos, of the dark matter and of possible associated dark sectors. In the following sections we will cover these successes while we go over the whole history of the universe in the context of the \lcdm model. 

We start in Sec. \ref{sec: cosmo_princ} by introducing basic concepts such as the cosmological principle and the Friedmann equations, which underlie the \lcdm model. In Sec. \ref{sec: infl} we begin the journey across the history of the universe with inflation, in particular slow-roll inflation. In Sec. \ref{sec: neu} we briefly overview the many phase transitions and particle decouplings that happen in the early universe, setting up the stage for the production of light elements such as deuterium and helium, which we discuss in Sec.~\ref{sec: BBN}. In Sec. \ref{sec: CMB} we describe the successive process of cosmic recombination and the richness of the CMB anisotropies it releases. During the period called ``dark ages'', the universe will eventually start to form structures and this epoch is the focus of Sec. \ref{sec: LSS}. We conclude in Sec. \ref{sec: tensions} with a discussion about emerging tensions in the determination of the aforementioned cosmological parameters between different probes.

\subsection{Cosmological principle, expansion and their consequences}\label{sec: cosmo_princ}
One of the most fundamental pillars underlying the \lcdm model is what is known as the cosmological principle, which states that the matter distribution in the universe is homogeneous and isotropic on cosmological (i.e., large enough) scales. The validity of these assumptions has been proven at large and intermediate scales on the basis of CMB, galaxy clusters, radio and X-ray observations among others \cite{Aluri:2022hzs} (although it might be in conflict with measurements of quasars \cite{Secrest:2020has}). 

Another important realization is that the universe is expanding. This had been first theoretically postulated by Friedmann in 1922 and later emphasized in 1927 by Lemaitre and then in 1929 by Hubble on the basis of the period-luminosity relation found by Leavitt in 1908 for Cepheid variables. What they found is that there is a direct proportionality between the (proper) distance $D$ to an object and its speed along the Line Of Sight (LOS) $v$, suggesting that the universe is expanding and the further away an object is the faster it moves away from us. This led to the Hubble(-Lemaitre) law
\begin{align}
	v=H_0 D\,,
\end{align}
where $H_0$ is the proportionality constant between distance and speed, also known as Hubble(-Lemaitre) constant. The Hubble parameter is often parameterized in terms of its dimensionless equivalent
\begin{align}
	h=\frac{H_0}{100~\text{km/(s Mpc)}}\,.
\end{align}
Today we know the value of $H_0$ to be of the order of 70~km/(s Mpc), although we will come back to its (conflicting) measurements in Sec. \ref{sec: tensions}. 

This relation is often recast in terms of the redshift $z$, which is a measure of how a photon's wavelength $\lambda$ (or, equivalently, its frequency or energy) is modified along its path towards us and can be due to e.g., the presence of a gravitational potential or, in fact, an expanding space. It is formally defined as 
\begin{align}
	z=\frac{\Delta \lambda}{\lambda}-1=\sqrt{\frac{1+v}{1-v}}\simeq v\,,
\end{align}
where the last equality holds in the low-velocity limit and therefore implies that
\begin{align}
	z\simeq H_0 D\,.
\end{align}
Explicitly, this equation means that the further away an object is the more its radiation will be redshifted, i.e., the larger the amount energy lost due to the expansion of the universe.\footnote{Intuitively, one could imagine that the wavelength of a photon traveling in an expanding universe will stretch, and hence increase. As a consequence, its frequency will decrease and so will its energy. On the other hand, being proportional to the wavelength shift, the redshift will also increase. The more time the photon has to travel, i.e., the larger the distance, the more this behavior will be pronounced, and hence the direct proportionality to the redshift.} Note that this particular definition of redshift does not account for the proper motion of the given object, which could, for instance, move towards use and hence blueshift, but  only for the so-called Hubble flow, i.e., the expansion of the universe.

Once these assumptions are made, i.e., homogeneity, isotropy and expansion, one can define a simple metric to describe the corresponding spacetime and, together with Einstein field equations, its evolution. This exercise has been conducted for the first time in the `20-30s by Friedmann, Lemaitre, Robertson and Walker, so that the aforementioned metric is referred to as FLRW metric and reads
\begin{align}\label{eq: metric}
	\der{s}^2=-\der{t}^2+a(t)\,\der{l}^2\,,
\end{align}
where
\begin{align}
	\der{l}^2=\frac{\der{r}^2}{1-kr^2}+r^2\der{\Omega}^2
\end{align}
is the spatial distance with
\begin{align}
	\der{\Omega}^2=\der{\theta}^2+\sin^2\theta\der{\phi}^2\,.
\end{align}
Here $a(t)$ is the so-called scale factor which parameterizes the relative expansion of the universe and can be shown to be related to the redshift as
\begin{align}\label{eq: a_z}
	\frac{a}{a_0}=\frac{1}{1+z}\,,
\end{align}
where $a_0$ is the scale factor of the universe today. Following the standard convention we will fix it to $a_0=1$, noting however that all calculations are independent of this choice. Furthermore, in \fulleqref{eq: metric} $k$ represents the curvature of the spacetime. In the context of this thesis the curvature is not a fundamental quantity and therefore, for sake of simplicity, we will always implicitly set it to zero unless stated otherwise (as we will see below this assumption is very well justified).

As previously mentioned, this metric can then be used to solve the Einstein equations 
\begin{align}\label{eq: Einstein_eq}
	R_{\mu\nu}-\frac{R}{2}g_{\mu\nu}-\Lambda g_{\mu\nu} = 8\pi G T_{\mu\nu}\,,
\end{align}
where the metric tensor $g_{\mu\nu}$ is related to the metric as 
\begin{align}
	\der{s}^2=g_{\mu\nu}\der{x^\mu}\der{x^\nu}\,,
\end{align}
$R_{\mu\nu}$ and $R$ are the Ricci tensor and scalar (both related to the metric via the Christoffel symbols $\Gamma^{\alpha}_{\mu\nu}$), respectively, $\Lambda$ is the cosmological constant, $G$ is the gravitational constant and
\begin{align}
	T_{\mu\nu}=T^{\mu\nu}=
	\begin{pmatrix}
		-\rho & 0 & 0 & 0 \\
		0 & p & 0 & 0 \\
		0 & 0 & p & 0 \\
		0 & 0 & 0 &p
	\end{pmatrix}
\end{align}
is the stress-energy (or energy-momentum) tensor in an homogeneous, isotropic universe (with no stress, viscosity and heat conduction), with $\rho$ and $p$ as the energy density and pressure of the cosmological (perfect) fluid, respectively. Conservation of momentum and energy can then be expressed in tensor notation as
\begin{align}\label{eq: energy_cons}
	\nabla_\nu T^{\mu\nu}=0\,,
\end{align} 
from which follows that
\begin{align}\label{eq: continuity}
	\dot{\rho}=-3H(\rho+p)\,,
\end{align}
where we have the defined the Hubble (or expansion) rate $H=\dot{a}/a$ (its value today, $H_0$, is the Hubble constant introduced above). This equation will be referred to as continuity equation. The (00) and ($ii$) components of \fulleqref{eq: Einstein_eq} deliver what are known as the two Friedmann equations, i.e., 
\begin{align}\label{eq: Friedmann_eq}
	\left( \frac{\dot{a}}{a}\right)^2= H^2 = \frac{8\pi G}{3} \rho \quad \text{and} \quad \frac{\ddot{a}}{a} =-\frac{4\pi G}{3}(\rho+3p)\,.
\end{align}
As a subtlety, note that here we have implicitly performed the transformation $\rho\to\rho+\Lambda/(8\pi G)$ and $p\to p-\Lambda/(8\pi G)$ so as to include the cosmological constant as additional source of energy density and pressure of the cosmological fluid. Being $\Lambda$ a constant this changes nothing about the dynamics of the equations.

In the above equations $\rho$ and $p$ describe the cosmological fluid, which is an ensemble of different components: relativistic and non-relativistic matter, and the cosmological constant. For each of these fluids one can derive a relation between pressure and density, known as Equation Of State (EOS), such that
\begin{align}
	p=w\rho\,,
\end{align}
where $w$ is the EOS parameter. In full generality, $w$ can be a function of the pressure (as it is the case, for instance, for EOSs used to describe neutron stars), but in a cosmological context one can make the simplifying assumption (on top of homogeneity and isotropy) that the particles of each fluid are identical and in kinetic equilibrium (i.e., they all share the same temperature or, else, no temperature gradient is present). In this way, $w$ is just a constant depending on the characteristics of the fluid.\footnote{As a remark, this does not mean that the value of $w$ has to be constant over time. In fact, as in the case of neutrinos, it is possible for a cosmological fluid to transition from being relativistic to non-relativistic over the history of the universe and that changes its EOS.} As it can be simply shown, for a relativistic fluid (radiation) one has $w_r=1/3$, for a non-relativistic (collisionless) fluid $w_m=0$ and for a cosmological constant the EOS parameter is chosen such that $\dot{\rho}=0$, which using Eq.~\eqref{eq: continuity} leads to $w_\Lambda=-1$.

The simple form of the aforementioned EOS bears important consequences. First of all, it allows to re-write Eqs. \eqref{eq: continuity}-\eqref{eq: Friedmann_eq} as a function of the energy density only, i.e., as
\begin{align}\label{eq: e_density}
	\dot{\rho}=-3H\rho(1+w)\,, \quad H^2 = \frac{8\pi G}{3} \rho \quad \text{and} \quad \frac{\ddot{a}}{a} =-\frac{4\pi G}{3}\rho(1+3w)\,.
\end{align}
The new form of the continuity equation can then be solved to obtain
\begin{align}
	\rho=\rho_0 \left(\frac{a}{a_0}\right)^{\matteo{-}3(1+w)}
\end{align}
where $\rho_0$ is the value of the cosmological energy density today (often referred to as the critical energy density), which with the first Friedmann equation takes the form
\begin{align}
	\rho_0 = \frac{3 H_0^{\matteo{2}}}{8\pi G}\,.
\end{align}
In turn, this allows to further simplify the first Friedmann equation to become
\begin{align}\label{eq: Friedmann_eq_2}
	H^2=H_0^2\left[\Omega_{r,0}\left(\frac{a}{a_0}\right)^{\matteo{-}4}+\Omega_{m,0}\left(\frac{a}{a_0}\right)^{\matteo{-}3}+\Omega_{\Lambda,0}\right]\,,
\end{align}
where we have decomposed the cosmological fluid into radiation, matter and the cosmological constant (from now on indexed with $r$, $m$ and $\Lambda$, respectively), and introduced the dimensionless energy density $\Omega_{i,0}=\rho_{i,0}/\rho_0$ with $i=r,m,\Lambda$. The $a$-dependent scalings have been computed according to Eq. \eqref{eq: e_density} with the respective EOS parameters. 

A useful relation that can be derived from Eq.~\eqref{eq: Friedmann_eq_2} is
\begin{align}
	\Omega_{r,0}+\Omega_{m,0}+\Omega_{\Lambda,0}=1
\end{align}
and it will often be referred to as budget equation. For the history of the universe roughly after Big Bang Nucleosynthesis (BBN, see Sec. \ref{sec: BBN}) the radiation energy density is made up of photons and neutrinos (henceforth labeled with the index $\gamma$ and $\nu$, respectively) as long as they are considered relativistic (which can be for ever in the massless case -- which we will always assume here unless stated otherwise), while the matter energy density is composed of Standard Model (SM) particles (often referred to as baryonic matter, hereafter referred to with the index $b$) and a (typically) cold non-interacting component called Dark Matter (DM, indexed as $\rm cdm$). The role of the cosmological constant is often interpreted as a unspecified Dark Energy (DE) component that is time-independent and spatially uniform.

Thanks to the precise measurements of various observables, in particular of the CMB anisotropies and Baryon Acoustic Oscillations (BAO), today we know that \cite{Aghanim2018PlanckVI}
\begin{align}
	\Omega_{\Lambda,0}\simeq0.69\,, \quad \Omega_{\rm cdm,0}\simeq0.26\,, \quad \Omega_{b,0}\simeq0.05 \quad \text{and} \quad \Omega_{r,0}\simeq 9.2\times10^{-5}\,,
\end{align}
assuming neutrinos to be massless for the latter. The curvature energy density has been constrained to be $\Omega_{k,0}<7\times10^{-4}$ \cite{Aghanim2018PlanckVI}, justifying the assumption of a flat universe solution. These values imply that only $\sim$5\% of the energy budget of the universe is in form of SM baryons and radiation. The remaining $\sim95$\% is part of a yet to understand dark sector composed of DM and DE. We will discuss their known properties in the following sections.

From these values and the energy density scalings discussed above one can also infer that at least three phases have occurred in the history of the universe, depending on which fluid is dominating the total energy density: Radiation Domination (RD), Matter Domination (MD) and $\Lambda$ Domination ($\Lambda$D). To determine the redshifts of transition between these eras it is enough to equate the respective energy densities, finding that RD took place at redshifts larger than the matter-radiation equality redshift $z_{\rm MR-eq}\simeq 3400$, while MD in the redshift range between $z_{\rm MR-eq}$ and the $\Lambda$-matter equality redshift $z_{\rm \Lambda M-eq}\simeq 1.3$. The whole history of the universe after $z_{\rm \Lambda M-eq}$ is and will be dominated by the DE.

In summary, a homogeneous, isotropic, expanding universe can be described by the two Friedmann equations and the continuity equation. In turn, these are ruled by the evolution of the fluids composing the energy budget of the universe, i.e., radiation, non-relativistic matter and the cosmological constant.

\subsubsection{Time and its many parameterizations}\label{sec: time}
After this introductory premises, it is necessary to linger further on the concept of time, which, although seemingly intuitive, requires some care. 

In the FLRW metric defined in Eq. \eqref{eq: metric} proper (physical) distances $d$ are defined as $d=aD$, where
\begin{align}
	D=\int \der{l}
\end{align} 
represents comoving distances, i.e., distances that are not affected by the expansion of the universe and are comoving with the Hubble flow. They are independent of $a$ or, equivalently, time. However, for photons moving radially in vacuum one has that ${\der{t}=a(t)}\,\der{l}$ and hence $\der{s}^2=0$, from which follows that
\begin{align}
	D=\int\frac{\der{t}}{a(t)} \equiv \tau\,,
\end{align} 
where $\tau$ is the so-called conformal time. It identifies the time taken for a photon to travel from emission to the furthest observation point (its horizon), which in turn corresponds to the comoving distance of the photon. The conformal time is not the same as the proper time and the two definitions are used in different contexts depending on the physical system in consideration. Among other useful applications, this definition of time allows to rewrite\footnote{From this conformal transformation comes the name ``conformal'' time.} the FLRW metric as
\begin{align}
	\der{s}^2=a(\tau)\left(-\der{\tau}^2+\der{l}^2\right)\,.
\end{align} 

Furthermore, in the previous section we have also already introduced two other definitions of time: scale factor and redshift. The more time passes by the more the universe expands and the larger the scale factor $a$ becomes, so that $a$ is an equivalent measure of time. Explicitly, from the first Friedmann equation one obtains that $a\propto t^{1/2}$ during RD, $a\propto t^{2/3}$ during MD and $a\propto e^t$ during $\Lambda$D. On the other hand, however, in other contexts it is more intuitive to express time in terms of redshift. For instance, when observing radiation there is a direct redshift-dependent correspondence between energy at emission and on Earth, so that redshift becomes a measure of how much energy has been lost as function of time. Using proper time would make this relation less transparent. Note here also that the higher the redshift, i.e., the more in the past one looks, the lower the proper time of the expansion is at that redshift, leading to an inverse proportionality between $z$ and time (as also clear from Eq.~\eqref{eq: a_z}).

\newpage
In this way, $t$, $\tau$, $a$ and $z$ are all possible and useful definitions of time. Yet, when discussing particle physics processes the reference to idea of temperature $T$ becomes very helpful as well. In the case of photons with an energy $E$, since $\matteo{1+}z\propto E \propto T$ then $T\propto \matteo{1+}z$, implying a direct correspondence between temperature and redshift, which is normalized to the temperature of the universe today $T_0=2.7255$ K $\simeq 2.348\times10^{-4}$ eV  \cite{Fixsen2009Temperature}. As an order of magnitude estimate, at the epoch of recombination ($z\simeq1100$) the universe had a temperature of $T\simeq 0.2$ eV, and a temperature $T=m_e$ at $z\simeq2\times10^9$.

Finally, as we will discuss in the very next section, although the universe is homogeneous at first order, small energy fluctuations exist within it. Since these perturbations follow a wave-like behavior, they can be broken down as a superposition of their different (Fourier) modes $k$, which are related to their wavelength as $k=2\pi a/\lambda$. Based on the assumption that every mode evolves independently (the separate universe approximation), it is often useful to use $\lambda$, and hence $k$, as proxy for the (time) scales of a given process. For instance, at very early times, when the universe was small, only modes with small wavelengths need to be considered as those with much larger wavelengths are not significantly affected by the processes happening at small scales. Fourier modes thus become a synonym of the size of the universe, which is in turn related to the time elapsed since the beginning of its expansion. Therefore, one can use the concept of modes or scales and imply time. 

\subsection{Slow-roll inflation}\label{sec: infl}
This simple picture, however, presents several puzzling questions. For example, there is no reason \textit{a priori} for the universe to be as homogeneous, isotropic and flat as we observe it to be. In fact, it can be shown that if there was only RD in the early universe there would not have been enough expansion between the big bang and the recombination epoch for the observable universe to be causally connected and there would be no reason, for instance, for the CMB radiation to have the same temperature in every direction of the sky. On top of this, given the fact that the curvature of the universe could in principle take any value, measuring it to be so close to zero presents a fine-tuning issue rather complicated to explain in absence of a deeper reason. Also the observation that the universe is almost perfectly electrically neutral at cosmological scales would be difficult to justify in absence of causal connection and a common mechanism for the generation of particles and anti-particles. Furthermore, many Grand Unified Theories (GUTs) and string theory predict the existence of topological defects such as domain walls and cosmic strings as well as magnetic monopoles, which have however never been observed. 

To address these fine-tuning issues, a possible idea is then to imagine an early phase of the universe with an accelerated expansion able to causally connect the observable universe and to dilute any type of ``unwanted'' (i.e., incompatible with observations) initial condition. This process was labeled inflation~\cite{Brout1978Creation, Starobinsky1980New, Kazanas1980Dynamics, Sato1981First, Linde1982New, Albrecht1982Cosmology, Linde1983Chaotic} (see e.g., \cite{Linde2005Particle, Baumann2009Inflation, Lesgourgues2006Inflationary} for modern reviews) and has proven to be in remarkably good agreement with experimental measurements, in particular at CMB scales~\cite{Ade2015Joint, Aghanim2018PlanckVI}.

\begin{figure}
	\centering
	\includegraphics[width=0.65\textwidth]{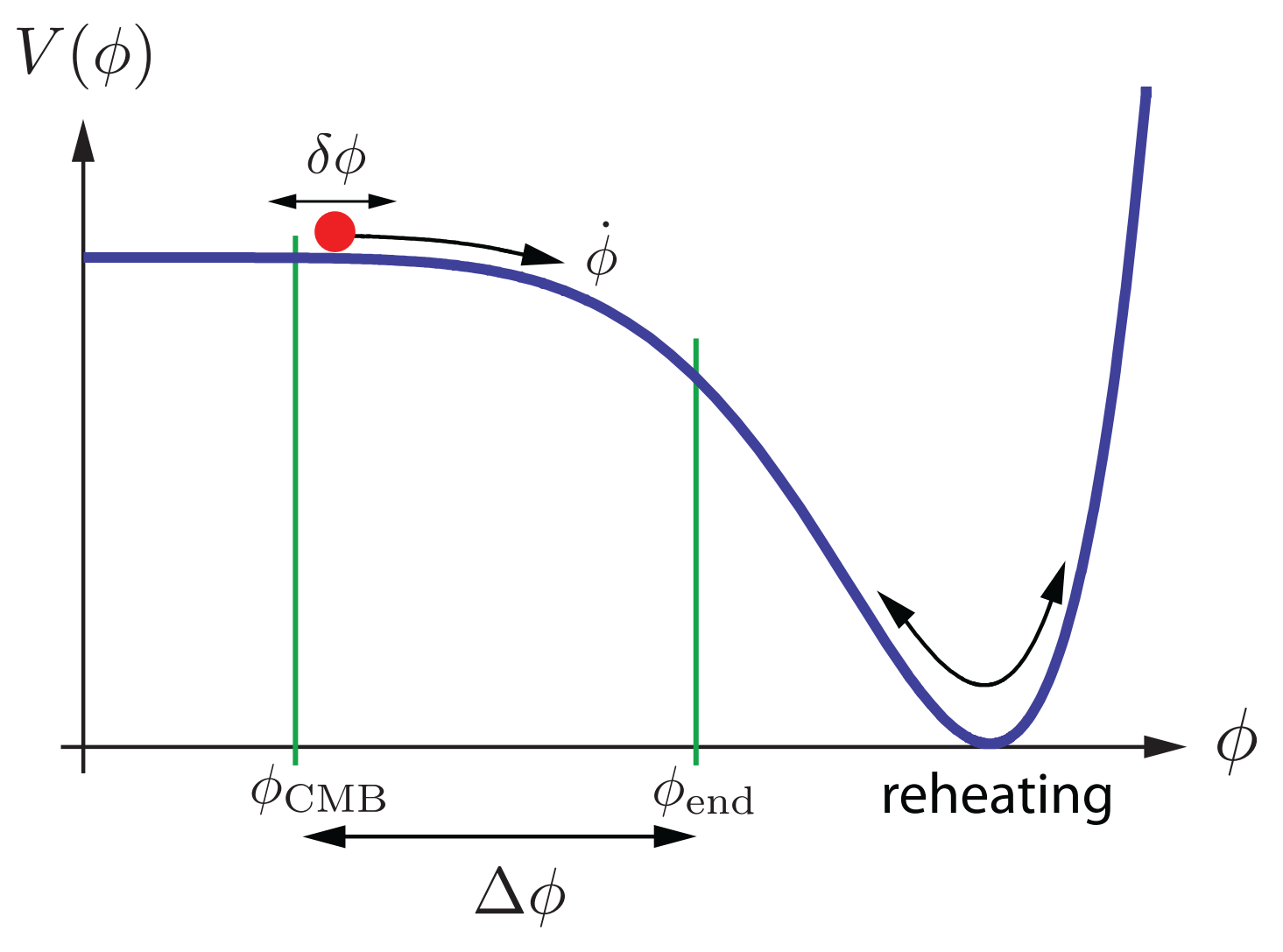}
	\caption{Representative example of inflationary potential. Figure taken from \cite{Baumann2009Inflation}.}
	\label{fig: infl_pot}
\end{figure}

Very naively, based on these arguments inflation has to respect two basic requirements: that it leads to an accelerated expansion and that it naturally stops at some point. A simple possibility to achieve this behavior is to assume the presence of a scalar field $\phi$ with potential $V(\phi)$ whose dynamics is dictated by the Lagrangian
\begin{align}
	\mathcal{L}=\frac{1}{2}\partial_\mu \phi \partial^\mu \phi -V(\phi)\,.
\end{align}
Since the energy density and pressure of such field are given by
\begin{align}
	\rho=\frac{\dot{\phi}^{\matteo{2}}}{2}+V(\phi) \quad \text{and} \quad p=\frac{\dot{\phi}^{\matteo{2}}}{2}-V(\phi)\,,
\end{align}
one can immediately notice that if there is a phase of the field evolution where $\dot{\phi}^{\matteo{2}}\ll V(\phi)$ then $p\simeq-\rho$. Similarly to the case of the DE, this would lead to an exponential expansion rate. Since this inflationary period needs to be sustained for a relatively long period of time, also the derivative of the aforementioned condition needs to be true (i.e., one needs to enforce that the aforementioned condition is not changing over time) and this leads to $|\ddot{\phi}|\ll| \partial_\phi V(\phi)|$, where $\partial_\phi=\partial/\partial\phi$. These two requirements are called slow-roll conditions and define slow-roll inflation. Yet, differently than for the cosmological constant the so-called inflaton field has to decay and this can be ensured by imposing a significant drop in its potential after the slow-roll phase (which formally ends when $p>-\rho/3$). When the field reaches its minimum it will start oscillating around it and decays into all SM particles. This phase of the history of the universe is called reheating and would explain, among others, why the universe has no net electric charge. A graphical depiction of the possible behavior of the inflationary potential as a function of the field $\phi$ is shown in Fig.~\ref{fig: infl_pot}, taken from \cite{Baumann2009Inflation}.

The existence of the inflaton field also has another important consequence on top of accelerating the expansion of the universe. In fact, the evolution of the field will be inevitably populated by quantum fluctuations $\delta\phi$ which perturb its energy density and hence that of the fluids it decays into during reheating. This leads to the formation of over- and under-densities of the energy density of the universe that set the seed for, among others, the adiabatic nature of the cosmological perturbations and the temperature anisotropies of the CMB. 

To understand the implications of these fluctuations, imagine the inflaton field $\phi$ to be a wave with an initial wavelength $\lambda_i$. The presence of quantum fluctuations $\delta\phi$ would make the field roll slower or faster down the potential, thereby changing its energy and hence frequency, which causes corresponding distortions of the initial wave. This effect is graphically represented in Fig. \ref{fig: infl_pot}.   Due to the random nature of the quantum fluctuations this would happen differently in different positions of the universe.\footnote{In terms of the field energy density, this would correspond to a fluid with background energy $\rho$ and to the creation of local fluctuations $\delta\rho$.} The ensemble signal is then a superposition of different waves with different wavelengths. In Fourier space this would correspond to a line at the respective mode $k_i$ centered around a Gaussian distribution of Fourier modes given by the fluctuations. Depending on the characteristics of the field potential, i.e., what is the energy gradient of the field, the fluctuations might imprint a smaller or larger spread of the central mode (since the shifts might be correspondingly smaller or larger). Therefore, the variance of the field in Fourier space $|\langle \phi (k)\rangle^2|$ is informative of the shape of the field potential and it is also the reason why the representation of this process in Fourier space significantly facilitates its mathematical description. In dimensionless units, the variance is commonly expressed as
\begin{align}\label{eq: PPS_phi}
	\mathcal{P}(k)=\frac{k^3}{2\pi^2}|\langle \phi (k)\rangle^2|\,,
\end{align}
defining the dimensionless Primordial Power Spectrum (PPS). 

At the same time, due to the expansion of the universe the initial wavelength $\lambda$ will get stretched more and more as times goes on, implying that the early-time oscillations correspond to increasingly small $k$ modes. This explains why, for instance, for the potential represented in Fig. \ref{fig: infl_pot} small $\phi$ values lead to oscillations as large as the size of the universe at e.g., the time of the CMB ($\phi_{\rm CMB}$), while larger values of the field like e.g., at the end of inflation ($\phi_{\rm end}$) only produce oscillations as large as the size of the universe at that early time.\footnote{Once inflation has ended and the inflaton has decayed, no more quantum fluctuations are generated and one only has a redshifting of the existing ones.} In this way, probing the smallest modes (i.e., the largest scales) means probing the slow rolling of $\phi$ at early times, while the larger the value of $k$ the closer one gets to the end of inflation. 

Similarly, also the size of the quantum fluctuations would become increasingly larger, which would lead to a wider distribution and hence a larger variance $|\langle \phi (k)\rangle^2|$. Being the field slowly rolling during inflation, the scale dependence of the variance at a given initial $k_i$ is expected to be small (since the potential is almost flat) and only increases proportionally to the volume effect due to the expansion. Therefore, one expects that a parameterization of the form $|\langle \phi (k)\rangle^2|/V$ to be almost constant (or, else, scale invariant), where $V$ is the volume enclosing the wavelength corresponding to the given mode $k$. Since dimensionally $V\propto \lambda^3\propto k^{-3}$ this implies that the aforementioned PPS should also be almost scale invariant (as much as $|\langle \phi (k)\rangle^2|$ is), so that it can be parameterized as
\begin{align}\label{eq: PPS_LCDM}
	\mathcal{P}(k)=A_s \left(\frac{k}{k_*}\right)^{1-n_s}
\end{align}
where $A_s$ and $n_s$ are the amplitude and the spectral index of the (scalar\footnote{In general, inflation might be leading to both scalar and tensor perturbations, but we will only focus on the former here.}) PPS and $k_*$ is a given pivot scale.

\begin{figure}[t]
	\centering
	\includegraphics[width=0.8\textwidth]{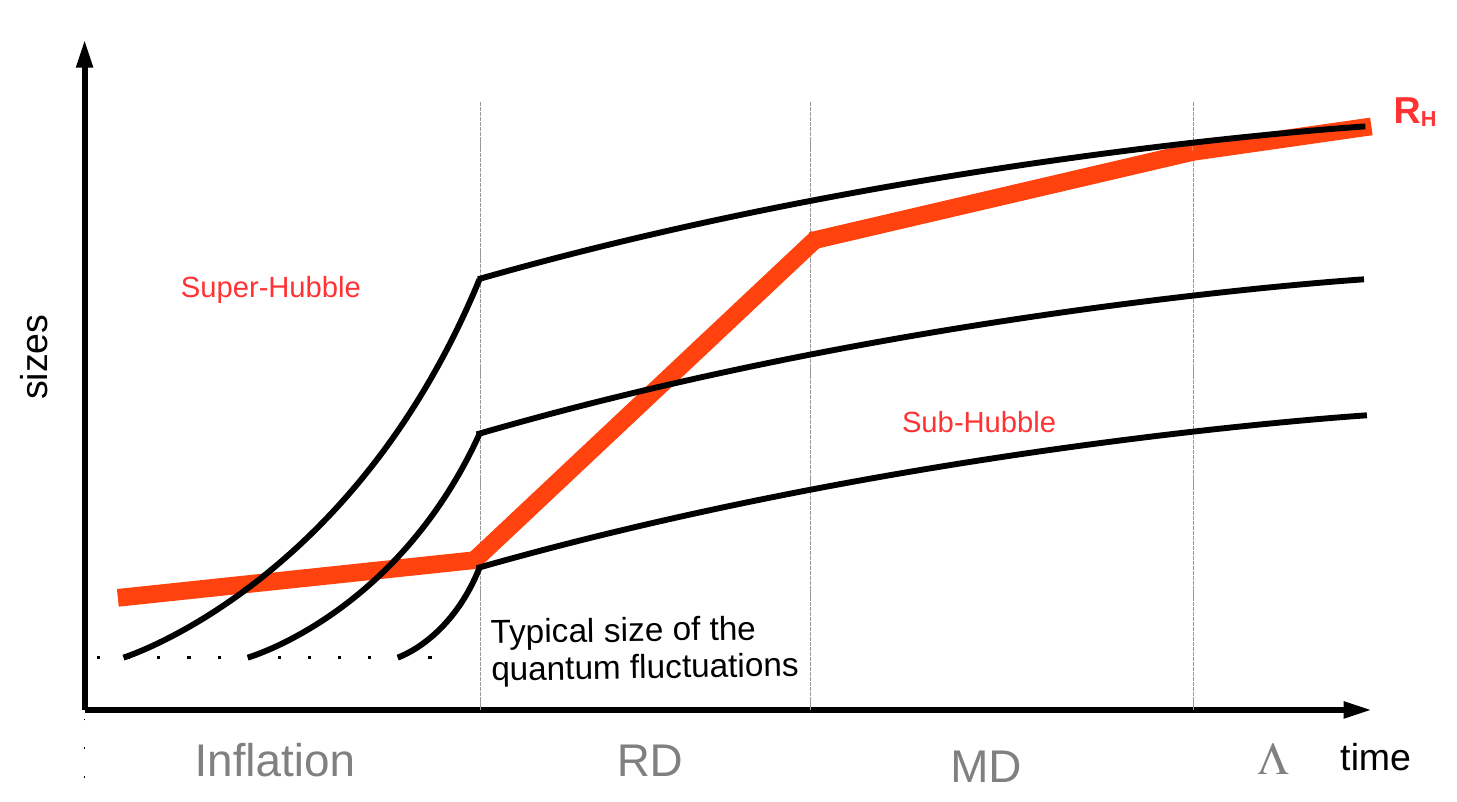}
	\caption{Illustrative comparison between the evolution of the size of the cosmological perturbations (black) and of the Hubble radius (red). Figure adapted from \cite{Hooper2019Extended}.}
	\label{fig: CMB_scales}
\end{figure}

The growth of the fluctuations is graphically displayed in Fig. \ref{fig: CMB_scales}, adapted from \cite{Hooper2019Extended}, as function of time. If one compares the Hubble radius $R_H=1/H$ to the size of the fluctuations $\lambda$ at a given time\footnote{This comparison captures which scales are effectively affected by the physics happening within the Hubble patch.} one obtains
\begin{align}
	\frac{\lambda}{R_H}=\frac{2\pi a}{k}H=\frac{2\pi\dot{a}}{k}\,,
\end{align}
which implies that if the universe is accelerating the fluctuations grow faster than the universe, while if the universe is decelerating the opposite is true. Since during inflation the universe is expanding at an exponential rate, at some point all perturbations grow larger than the size of the universe and enter the so-called super-Hubble regime. On super-Hubble scales, the perturbations are not affected by anything that happens in the history of the (much smaller) universe and remain unchanged. However, once RD onsets the size of the universe starts to increase faster than that of the fluctuations so that, eventually, the perturbations ``re-enter'' the Hubble horizon. Given the scale invariance of the spectrum, once the perturbations enter in this regime they only differ by their size, i.e., they have all just redshifted during inflation. The initial conditions of the perturbations resulting from inflation are hence the same at all scales.

As we will see in Secs. \ref{sec: CMB} and \ref{sec: res_infl}, the shape of the PPS resulting from these fluctuations can be tested by current CMB anisotropy and future CMB SD observations at scales between $10^{-4}-10^4$~Mpc$^{-1}$. For instance, based on Planck 2018 data it has been possible to confirm the validity of the slow-roll scenario to a high precision and even reconstruct the corresponding inflationary potential \cite{Akrami2018PlanckX} (see in particular Secs. 5 and 6 of the reference). Although its origin and exact evolution are still unclear, inflation has thus established itself as the first building block of the standard cosmological model.

\vspace{0.3 cm}
\noindent\textbf{Follow-up idea 1:} To our knowledge, no such analysis as the one performed in \cite{Akrami2018PlanckX} in terms of the inflationary potential and PPS reconstruction has been conducted with the inclusion of ACT, SPT and BICEP2/Keck 2018 data (see however \cite{Forconi:2021que} for a work in this direction). Being sensitive to smaller scales than those probed by Planck (see Fig. \ref{fig: CMB_spectra}), the data gathered by the ground-based experiments should in principle be able to tighten the uncertainties of the Planck results in that direction. The updated B-mode information (BICEP2/Keck 2015 data was used in \cite{Akrami2018PlanckX}, which had twice as large error bars as the 2018 data \cite{BICEP:2021xfz}) might also further constrain the reconstruction at large scales.
\vspace{0.3 cm}

In summary, due to its exponential expansion inflation naturally explains why every patch of observable universe can be causally connected, it predicts a very efficient dilution of curvature and topological defects and gives rise the presence of Gaussian, nearly scale-invariant perturbations.  As we will see in the following sections, all of these predictions and their consequences are in excellent agreement with current data.

\subsection{From reheating to neutrino decoupling}\label{sec: neu}
Inflation can explain many of the necessary initial conditions for the evolution of the universe we live in, but not all of them. In fact, in the simplest inflationary model with only one inflaton field it would be ``natural'' to expect its decay to produce all SM particles and anti-particles in equal measure. To understand exactly why that is concerning, consider for instance a scenario where the primordial number density $n_b$ of a given baryonic particle $b$ with mass $m$ is the same as that of its anti-particle $\bar{b}$. When the temperature $T$ of the universe is higher than~$m$, the radiation bath has enough energy to produce $b\bar{b}$ pairs from the annihilation of, say, two photons, so that there is a perfect (thermal) equilibrium between the number of particles lost due to $b\bar{b}$ annihilation and $b\bar{b}$ pair creation. The number density of the particles is just redshifted. However, when $T$ decreases below $m$ the production rate of $b\bar{b}$ pairs is exponentially suppressed because exponentially (due to the photons following a Bose-Einstein distribution) less and less photons have enough energy to produce $b\bar{b}$ pairs. Since the $b\bar{b}$ and the photon fluid are still in thermal equilibrium, the $b\bar{b}$ annihilation is equally suppressed (due to the increasingly lower number of $b\bar{b}$ particles), and it still takes place as long as the process is efficient.\footnote{Here and henceforth with ``efficient'' we always imply in comparison to the Hubble expansion rate, i.e., if $\Gamma>H$ where $\Gamma$ is the given interaction rate. Qualitatively, one can imagine that if the timescale $\tau=1/\Gamma$ over which a given process is happening becomes much longer than the timescale $t=1/H$ over which the universe expands, i.e., if $\tau>t$ or equivalently $\Gamma<H$, the particles will never be able to induce that process, which therefore becomes inefficient.} This event is commonly referred to as annihilation catastrophe.\footnote{For instance, the annihilation catastrophe of baryons happens at around $T\sim  \mathcal{O}(m_p$/few), where $m_p$ is the mass of the proton.}~Once the particles~$b$ have completed their (kinetic) decoupling from the thermal bath and their annihilation has stopped (due to the much reduced number density) their number density will not change any more aside from the redshifting, it will be frozen out.

Since photons, which are massless, never go through this process, their number density $n_\gamma$ is a good proxy for the ``original'' abundance of a particle, so that the ratio $\eta=n_b/n_\gamma$ after the annihilation catastrophe is indicative of how large the suppression has been. For the case of baryons one can calculate this number to be of the order of $10^{-18}$, which would be the same for particles and anti-particles. There are two problems with this picture: \textit{i)} we do not see the same amount of particles and anti-particles (macroscopically, we do not see any anti-particles at all), and \textit{ii)} observations suggest that $\eta$ in the case of baryons should be of the order of 
\begin{align}\label{eq: eta}
	\eta\simeq2.75\times10^{-8}\,\Omega_{b,0}h^2\sim\mathcal{O}(5\times10^{-10})\,,
\end{align}
almost 10 orders of magnitude away from the aforementioned naive estimate (see Secs.~\ref{sec: BBN} and \ref{sec: CMB}).

To address this issues, one option is to assume a phase of the evolution of the universe between reheating and the decoupling of the particles $b$ where the particles-to-anti-particles ratio deviates from unity. If indeed one had $n_b/n_{\bar{b}}\neq1$ (and given that we live in a universe filled with baryons, $n_b/n_{\bar{b}}>1$) a relic number density would survive the annihilation catastrophe, which can occur only as long as both particles and anti-particles are available. The degree of the asymmetry between $n_b$ and $n_{\bar{b}}$ would then determine how many baryons survive the freeze out.\footnote{Essentially, the surviving baryon number density would be given by the initial excess of $b$ with no $\bar{b}$ left, given that the annihilation process conserves baryon number.}  In the context of the \lcdm model it is not important exactly which process leads to this baryon asymmetry, as long as it delivers $\eta\simeq5\times10^{-10}$ and it is completed early enough (typically at temperatures above the electro-weak phase transition occurring at $T\sim 100$~GeV and in any case before BBN). We will limit the discussion to mention the most prominent proposals which are leptogenesis and electroweak baryogenesis (see e.g., \cite{Bodeker:2020ghk} for a recent review).

Assuming this epoch has happened successfully, the correct initial relic abundances of all particles and anti-particles are set. The universe is then believed to go through a series of phase transitions and particle decouplings. First of all, without dwelling on the details, when the universe cools down roughly to the mass of the Higgs boson ($\sim \mathcal{O}(100$~GeV)) a process called spontaneous symmetry breaking (or electroweak phase transition) takes place, after which all particles obtain their mass either via the acquisition of a Goldstone boson (the $W^{\pm}$ and $Z$ bosons) or via a Yukawa coupling to the Higgs field expectation value (the fermions) \cite{peskin2018introduction}. Since the temperature of the universe is much larger than the mass of any SM particle at this epoch, however, all particles were still effectively massless at this stage, i.e., highly relativistic. Nevertheless, as soon as the universe cools down to the mass of a given particle the corresponding freeze-out happens and the particle becomes non-relativistic with a sizable production of photons that heat up the thermal bath. Given enough time these unstable particles with then further decay and disappear completely. After many of these phases, at temperatures of the order of $T\sim  \mathcal{O}(200$~MeV) the so-called quark-gluon plasma transitions to an hadronic gas since the quarks confine into hadrons at these energies. After this phase, the particle content of the universe reduces to protons, neutrons, electrons (and positrons), neutrinos, photons and the DM. 

At $T\sim\mathcal{O}(1$ MeV) the weak interaction-induced scattering process between electrons and neutrinos becomes inefficient and neutrinos decouple from the thermal bath composed of the other particles which maintain thermal equilibrium. From that moment on, neutrinos will free stream until today, without any further interaction with the other particles. While there is no direct cosmological evidence for the existence of the aforementioned electroweak and QCD phase transitions, neutrino decoupling is the first testable prediction of the \lcdm model.\footnote{Gravitational wave observations or the in-laboratory production of a quark-gluon plasma could possibly be able to shed light on the very early universe in the future, although it will still require significant theoretical and experimental development.}

In fact, since neutrinos decouple when electrons are still relativistic,  when electrons freeze out and heat up the medium the temperature increase will not be felt by the neutrinos. Also, while in the SM there are only three neutrino species, the non-instantaneous decoupling of the neutrinos leads to the definition of a parameter $N_{\rm eff}$ that represents the effective number of neutrino species, predicted to be $N_{\rm eff}=3.045$ within the SM \cite{deSalas:2016ztq}. The combination of these two effects determines the neutrino energy density after decoupling to be 
\begin{align}\label{eq: rho_nu}
	\frac{\rho_\nu}{\rho_\gamma}=\frac{7}{8}N_{\rm eff}\left(\frac{T_{\nu}}{T_{\gamma}}\right)^{4}\,,
\end{align}
where the temperature ratio takes into account the electron freeze out. The value of $\rho_\nu$, which contributes to the expansion rate during radiation domination, has a sizable impact on the BBN prediction, the anisotropies of the CMB and structure formation \cite{Steigman:2012ve,Aghanim2018PlanckVI, tanabashi2018particle} (see following sections). 
Should the aforementioned picture be correct, in analogy to the CMB one would also expect the existence of a cosmic neutrino background, which is however still eluding observations \cite{Betts:2013uya}. Should this change in the future, it would have deep consequences for our understanding of the neutrino sector.

\newpage
Finally, shortly after neutrino decoupling, processes such as $\beta$ decay and electron capture became inefficient, and the conversion of protons to neutrons and \textit{vice versa} stops. In this way, if it was not for the neutron decays the number of protons and neutrons would become constant (see left panel of Fig. \ref{fig: BBN_ev}). This means that the exact abundances of these particles, predicted to be of the order of $n_n/n_p\sim 1/5$ (due to their mass splitting), are strictly related to the time of neutrino decoupling \cite{Steigman:2012ve} and hence to the value of $N_{\rm eff}$.

In summary, after the decay of the inflaton field, a number of phase transitions, particle freeze outs and neutrino decoupling we are left at $T\sim\mathcal{O}(0.5$ MeV) with a baryon-photon fluid made of photons and non-relativistic protons, neutrons and electrons (with $n_p\sim n_e \sim n_n/5$), a decoupled neutrino fluid (with $\rho_\nu \sim 0.7\, \rho_\gamma$), and the DM (see below).

\subsection{Big Bang Nucleosynthesis}\label{sec: BBN}
In the previous sections we have discussed inflation, which has been constructed to explain the initial conditions of our universe, and the evolution of the early universe until neutrino decoupling, which is largely subject to speculation. In this section, instead, we focus on BBN, the earliest epoch we can really probe to date.

BBN describes the epoch of light element formation in the early universe (see e.g.,~\cite{Cyburt:2015mya,Hufnagel:2020nxa} for recent reviews). Fusion reactions leading to the creation of e.g., deuterium and helium have always been happening in the history of the early universe, but their products only begun to survive photo-disintegration for a sizable amount of time when the photon bath cooled down to\footnote{In principle, helium-4 could have formed already before that, at ${T\sim \mathcal{O}(0.3}$~MeV), but the negligible fraction of available deuterium at this time forbid a sizable production.} ${T\sim \mathcal{O}(0.1}$~MeV), \matteo{about one order of magnitude} below the deuterium binding energy.\footnote{The exact energy scale at which the production of an element can begin is given by the balance between the binding energy of the given nucleus and the large number of photons, $n_\gamma/n_b\sim10^{10}$, which can dissociate it. Once $T<B_i$ the number of photons with enough energy to dissociate nuclei is exponentially suppressed and a sizable production of the element can begin (relatively) shortly after.} After that, the stable deuterium initiated a chain of reactions leading to the formation of a non-negligible fraction of helium-3, helium-4, beryllium and lithium, among others. Finally, once the relevant nuclear reactions become inefficient with respect to the expansion rate, at $T\sim \mathcal{O}(10$~keV), the fusion stops and the light element abundances freeze out. By this time all remaining free neutrons have decayed. 

As such, BBN is very sensitive to the baryon-to-photon ratio $\eta$, i.e., to the amount of photo-disintegration which determines the beginning of deuterium production. As we discussed in the previous section, the determination of this quantity is crucial for the evolution of the very early universe. Furthermore, the process of neutrino decoupling sets the initial conditions for BBN by determining the neutron-to-baryon ratio and as such the latter is an important test of the neutrino sector. The value of $N_{\rm eff}$ also determines the expansion rate of the universe\footnote{In full generality, also the photon temperature $T_0$ plays a role in the expansion rate but it is considered to be exactly known in this context.} during RD, which affects the nucleosynthesis rates.

\begin{figure}[t]
	\centering
	\includegraphics[width=\textwidth]{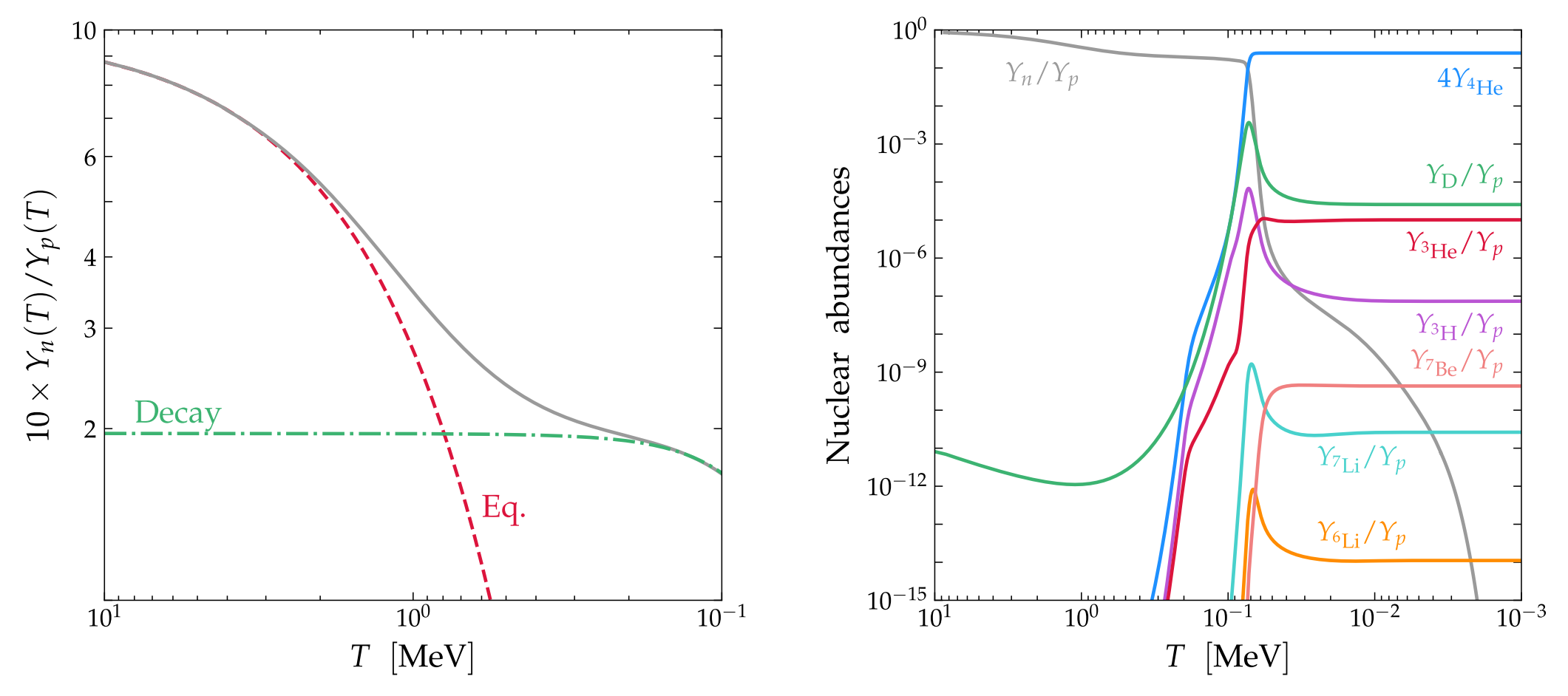}
	\caption{Time evolution of the neutron-to-proton ratio (left) and of the light element abundance yields (right). Figures taken from \cite{Hufnagel:2020nxa}.}
	\label{fig: BBN_ev}
\end{figure}

Given these dependencies, because of the fact that the nuclear physics happening during BBN takes place at relatively low temperatures, i.e., accessible on earth, its modeling is extremely predictive. A graphical representation of the evolution of the abundance of these elements is shown in Fig. \ref{fig: BBN_ev}, taken from \cite{Hufnagel:2020nxa}. Quantitatively, for typical values of the baryon-to-photon ratio ($\eta=6.138\times10^{-10}$ \cite{Aghanim2018PlanckVI}), of $N_{\rm eff}$ and of the neutron lifetime ($\tau_n=880.2$ s \cite{tanabashi2018particle}) one numerically obtains \cite{Hufnagel:2020nxa}
\begin{align}\label{eq: BBN_th}
	\frac{^4\text{He}}{\text{H}}=0.24745\pm0.00005\,,~~ \frac{\text{D}}{\text{H}}=(2.550\pm0.059)\times10^{-5} ~~ \text{and} ~~ \frac{^3\text{He}}{\matteo{\text{D}}}=0.40\pm0.02\,,
\end{align} 
where the errors are mainly due to the uncertainties of the nuclear rate measurements (see e.g., Sec. 3.2.2 of \cite{Hufnagel:2020nxa} for an overview).

An interesting question is then how to test the predictions of the standard BBN model. Obviously, we have no mean to directly observe the process of BBN at such early times and in our local environment star formation and galaxy dynamics have drastically modified the nuclear abundances. The right sweet-spot turns out to be young, low-metallicity galaxies where the late-time astrophysics has not onset yet and the primordial abundances (which froze out at the end of BBN) are still preserved.\footnote{To be exact, one attempts to observe the metal-poorest galaxies available and then performs a linear regression to extract the information at zero metallicity, with all the related extrapolation uncertainties.} From the observation of hydrogen and helium recombination and hyperfine transition lines as well as damped Lyman-$\alpha$ systems in such environments one infers \cite{Geiss2003Isotopic, tanabashi2018particle, zyla2020review}
\begin{align}\label{eq: BBN_obs}
	\frac{^4\text{He}}{\text{H}}=0.245\pm0.003\,,~~ \frac{\text{D}}{\text{H}}=(2.547\pm0.025)\times10^{-5} ~~ \text{and} ~~ \frac{^3\text{He}}{\matteo{\text{D}}}=0.83\pm0.15\,,
\end{align} 
the latter of which is commonly only used as upper bound due to the large astrophysical uncertainties underlying its determination \cite{Geiss2003Isotopic, Ellis2005Effects, Jedamzik2006Big}. 
\begin{figure}[t]
	\centering
	\includegraphics[width=0.45\textwidth]{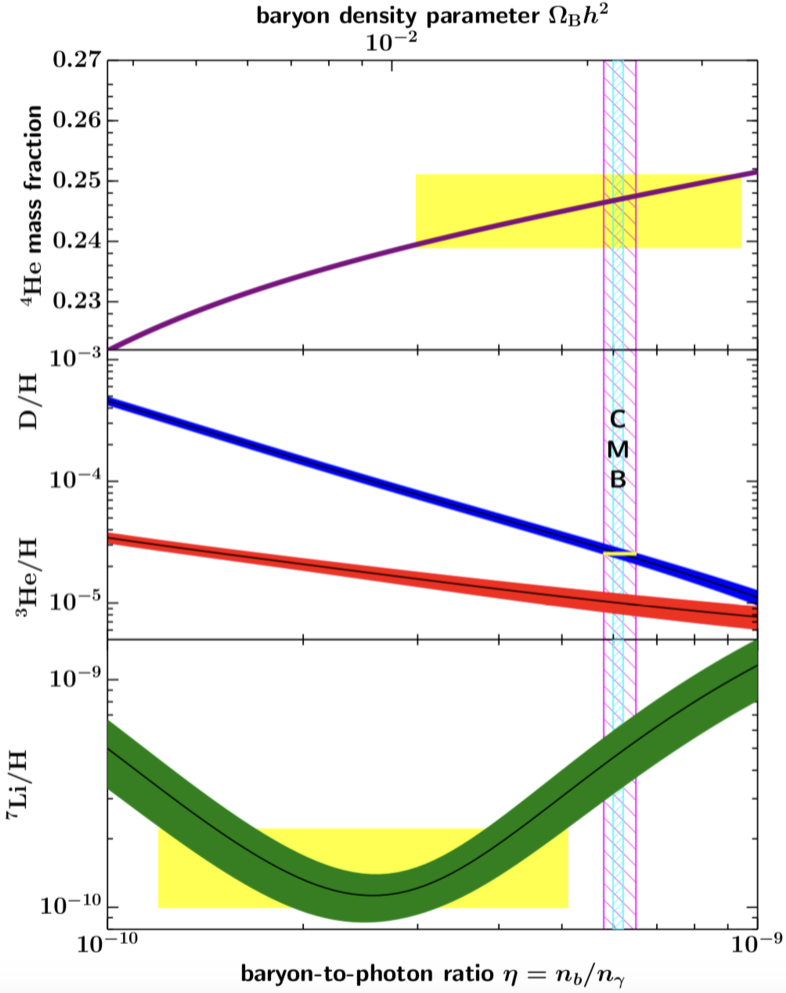}
	\caption{Dependence of the predicted abundances on the baryon-to-photon ratio $\eta$ (or, alternative, the baryon energy density, see Eq.~\eqref{eq: eta}) together with the respective experimental uncertainties (yellow bands). The vertical band corresponds to the value of $\eta$ inferred from CMB anisotropy data (see Sec. \ref{sec: CMB}). Figure taken from \cite{tanabashi2018particle}.}
	\label{fig: BBN_eta}
\end{figure}

Clearly, the observed abundances are in good agreement with the aforementioned theoretical values\footnote{One possible exception to this statement might regard the case of lithium \cite{Fields:2011zzb}, where numerical predictions overshoot the observed abundance by a factor of order three. Yet, this might be likely to a poor understanding of the late-time lithium nucleosynthesis (see e.g., Sec. IIIC of \cite{Cyburt:2015mya} for an overview).} and precisely this agreement allows to set constraints on the cosmological parameters that have an impact on the BBN predictions, i.e., $\eta$ and $N_{\rm eff}$. In fact, for instance, in Fig.~\ref{fig: BBN_eta}, taken from \cite{tanabashi2018particle}, one can directly compare the observed abundances listed in Eq.~\eqref{eq: BBN_obs} with the numerical predictions which are a function of the baryon-to-photon ratio $\eta$ (and hence of the baryon energy density via Eq. \eqref{eq: eta}) for a fixed value of $N_{\rm eff}$. A similar exercise can also be performed for the case of $N_{\rm eff}$ (see e.g., Eq. (5) of \cite{2013PhRvD..87h3008H}). As a result, ignoring information from the controversial lithium observations, one obtains \cite{Cyburt:2015mya}
\begin{align}\label{eq: BBN_res}
	\eta=(6.10\pm0.07)\times10^{-10} \quad \text{and} \quad N_{\rm eff}=2.85\pm0.28\,.
\end{align} 
The latter is in good agreement with the prediction of neutrino decoupling discussed in Sec. \ref{sec: neu} and both are perfectly compatible with the corresponding values inferred from CMB anisotropy data \cite{Aghanim2018PlanckVI}, displayed as vertical band in Fig. \ref{fig: BBN_eta} for the case of $\eta$ (see the following section for more details).

So far we have not touched upon the topic of DM. The \lcdm model assumes the DM to be a simple, cold, i.e., always non-relativistic, particle completely decoupled form the SM sector (aside from gravitational interactions). It has no ambition to explain where it came from and what its exact properties are, as long as its relic abundance results in $\Omega_{\rm cdm,0}=0.26$. With this value, it fits all the available data (cosmological and not) across a number of scales and epochs. 

These open questions can, however, be answered adopting a particle physics perspective, aiming for instance at determining how the DM could have been produced. To do so, one usually extends the cold, totally non-interacting DM of the \lcdm model, assuming for instance the DM to have specific interactions (and/or possibly be warm). Yet, these models are only allowed as long as their formation mechanism and characteristics do not significantly alter the standard \lcdm predictions which are in overall very good agreement with observations, starting with BBN. Therefore, when discussing the topic of BBN is also fundamental to be mindful of what DM model one is considering. For instance, if the DM had an interaction channel with electromagnetic (EM) particles, it could annihilate or decay into photons energetic enough to photo-disintegrate the BBN elements after BBN is complete and drastically modify its predicted abundances. At the same time, even if the DM was completely decoupled from the SM but relativistic (or at most with $m_\chi\lesssim1$~MeV, see e.g., \cite{Coy2021Domain}) at the time of BBN it would effectively act as a fourth neutrino and modify the BBN picture via its contribution to $N_{\rm eff}$. Many more such effects can be found in e.g., \cite{Jedamzik2009Big, Hufnagel:2020nxa}. For these reasons, BBN can be used to constrain the properties of DM. 

In summary, BBN covers the epoch of light element formation and its predictions are in very good agreement with the expected conditions resulting from neutrino decoupling as well as with late-time and CMB observations. This agreement makes of BBN a testament to our understanding of nuclear physics and of the early universe, as well as a valuable probe to constrain the non-standard properties of DM.

\subsection{Recombination and Cosmic Microwave Background}\label{sec: CMB}
\begin{figure}[t]
	\centering
	\includegraphics[width=0.65\textwidth]{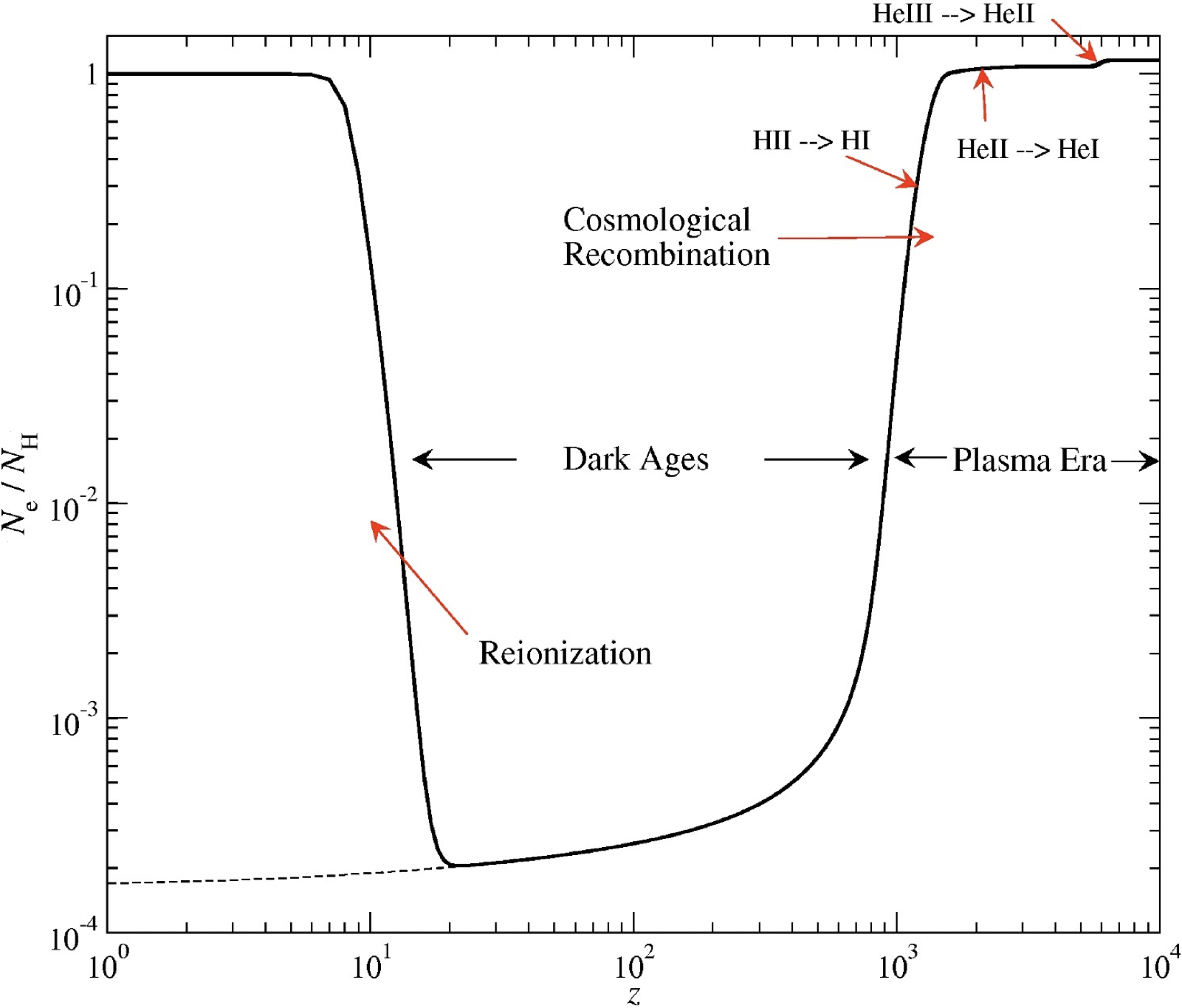}
	\caption{Evolution of the free electron fraction $x_e=n_e/n_H$. Figure adapted from \cite{Chluba_CosmoTools}.}
	\label{fig: CMB_xe}
\end{figure}

As discussed in the previous section, BBN finishes at around $T\sim\mathcal{O}(10$ keV), when the universe was more or less 3 \matteo{minutes} old. What happens in the following 40'000 years will be covered more in depth in Sec. \ref{sec: theory}, since this is the time where CMB SDs start to form.\footnote{In short, at a $T\sim\mathcal{O}(0.5$ keV), i.e., at $z\sim2\times10^6$, number changing processes such as bremsstrahlung and double Compton scattering become inefficient and the photon bath is not able anymore to perfectly restore the thermal (i.e., BB) distribution once distorted, leading to the creation of an effective chemical potential $\mu$ in the photon distribution. Furthermore, at redshifts of the order of $z\sim5\times 10^4$ also Compton scattering starts to become inefficient, so that at lower redshifts also the energy of the photons cannot be perfectly redistributed. The combination of these effects leads to the creation of CMB SDs.} Here we fast forward to redshifts of the order of $z\sim6000$, corresponding to $T\sim\mathcal{O}(1$ eV), where the photon bath has cooled down below the (second) ionization energy of helium. At these temperatures, the helium atoms can form bound states with the electrons without being immediately ionized by the surrounding photons. This process, called recombination, repeats at $z\sim2000$ with the second stage of helium recombination. Since helium is less abundant than hydrogen (see Sec. \ref{sec: BBN}), only a fraction of the electrons are captured. Yet, when the universe cools down to $T\sim\mathcal{O}(0.1$ eV) also hydrogen can recombine and then (almost) all the free electrons disappear. The universe has become (almost) exclusively made of neutral atoms. The evolution of the free electron fraction $x_e=n_e/n_H$, which is a good tracker of the recombination process, is graphically displayed in Fig. \ref{fig: CMB_xe}, adapted from \cite{Chluba_CosmoTools}. There one can see the sharp drop following the end of recombination, which lasts until $z\sim10$ when the first stars start to form and the universe ionizes again (see following section). 

A fundamental consequence of recombination is that, since all of the free electrons have disappeared, Compton Scattering (CS) suddenly cannot take place anymore and the photon decouple from the other particles. This leads to the creation of a so-called last-scattering surface after which photons are assumed to free stream until today. Since this process happens at the same time homogeneously over the sky at $T\sim0.1$ eV, it leads to a background of radiation with a temperature of the order of $T\sim0.1$ meV today. This lays in the microwave range and the resulting radiation is therefore referred to as CMB. The temperature of the CMB radiation has been assessed to an extremely high degree of precision with the FIRAS mission \cite{Fixsen1996Cosmic}, which measured the energy spectrum of the CMB photons and showed that this corresponds to a BB shape (up to a precision of order $10^{-6}$, see Sec.~\ref{sec: exp}) with a temperature $T_0=2.72548 \pm 0.00057$ K \cite{Fixsen2009Temperature}. This is in perfect agreement with the aforementioned picture.

\begin{figure}
	\centering
	\includegraphics[width=\textwidth]{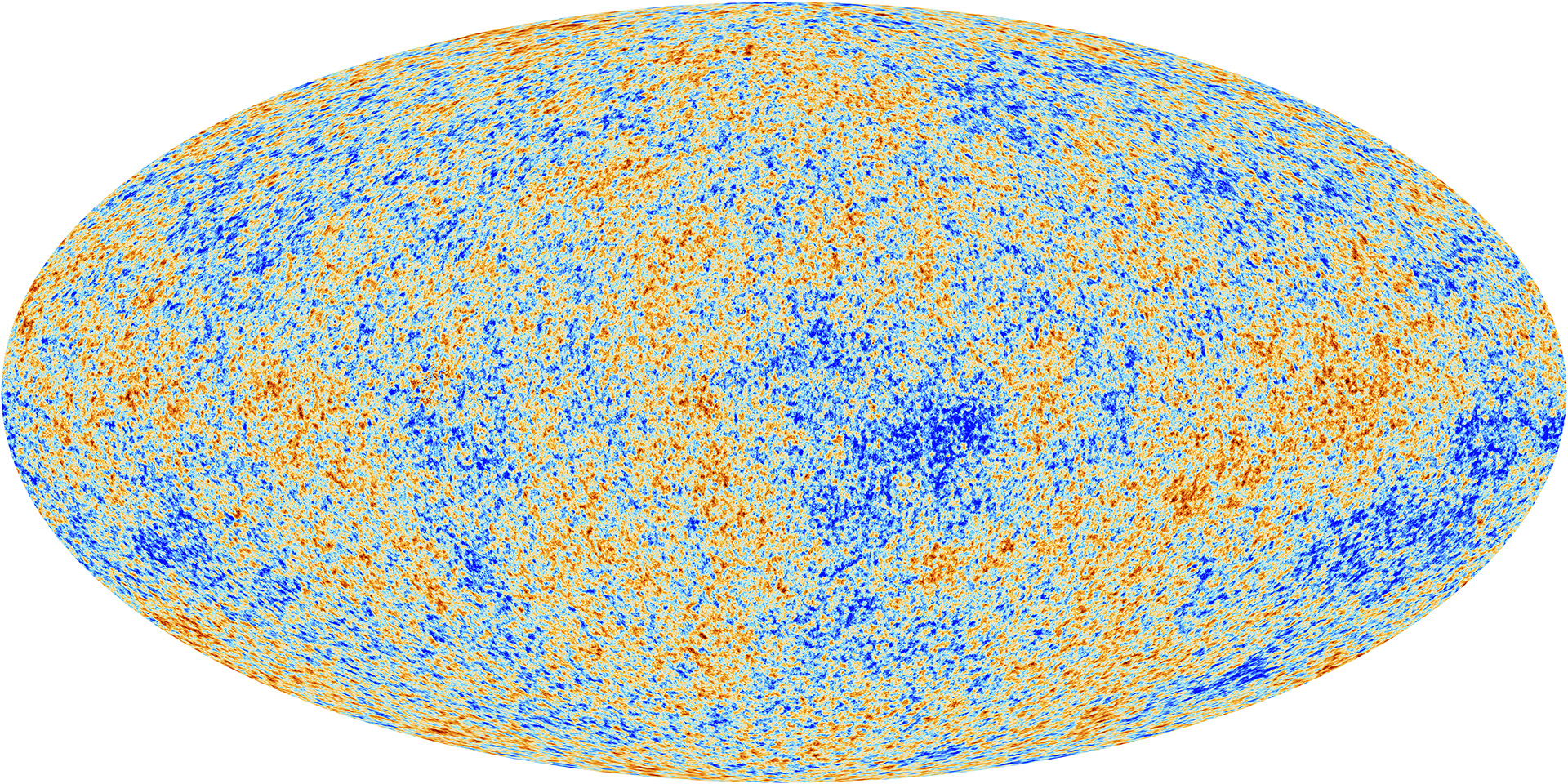}
	\caption{Sky map of the CMB temperature anisotropies as observed by the Planck mission (2018 release) \cite{Planck:2018nkj}. Figure taken from \cite{Planck_map}.}
	\label{fig: CMB_Planck_map}
\end{figure}

The CMB is, however, not perfectly homogeneous but is instead populated by temperature fluctuations of the order of $\Delta T/T\sim 10^{-5}$  \cite{Aghanim2018PlanckVI}, as shown in Fig. \ref{fig: CMB_Planck_map}, taken from \cite{Planck_map}. These anisotropies of the CMB are the richest source of information we currently have on the universe. 

\newpage
To understand why, we need to go back to the perturbations sourced at the time of inflation. As discussed in Sec.~\ref{sec: infl}, and in particular in reference to Fig.~\ref{fig: CMB_scales}, these perturbations are larger at the end of inflation the earlier they are created and eventually enter the super-Hubble regime during inflation (i.e., their wavelength becomes larger than the Hubble radius). During RD, however, the Hubble radius grows faster than the perturbations and at some point in time they ``re-enter'' the universe, the smaller perturbations first and the larger to follow. Perturbations larger than the size of the universe at the time of matter-radiation equality will enter the universe during MD, and similarly for $\Lambda$D. To get an intuition for the relation between scales at horizon re-entry and time, the QCD phase transition mentioned in Sec. \ref{sec: neu} corresponds to the crossing of scales of the order of $k\sim5\times 10^6$~Mpc$^{-1}$, while recombination to around $k\sim\matteo{1}\times 10^{-2}$~Mpc$^{-1}$.

Depending on when a given fluctuation re-enters the horizon it will experience a different expansion history. For instance, if this happens before recombination, when photons and baryons are still coupled, the baryons will be gravitationally attracted towards the overdense regions. Nevertheless, as this happens the temperature in these regions increases and the larger radiation pressure pushes the matter apart. Once the baryons have been sufficiently separated and cooled, the gravitational pull dominates again and the process re-starts. These oscillations effectively create sound waves in the primordial plasma that propagate at a (sound) speed
\begin{align}\label{eq: cs2}
	c_s^2=\frac{\delta p_\gamma + \delta p_b}{\delta \rho_\gamma + \delta \rho_b}\,,
\end{align}
where $\delta p$ and $\delta \rho$ represent the pressure and density perturbations of the photon and baryon fluid. The propagation length of these acoustic waves is given by the so-called sound horizon 
\begin{align}\label{eq: sound_hor}
	r_s=\int^\infty_{z_{\rm rec}} \frac{c_s\, \text{d}z}{H}\,.
\end{align}
Since the CMB is made of photons, let us focus here only on the impact of these oscillations on the spatial variation of the energy density (i.e., temperature) of the photons, parameterized with $\Theta_T(\tau, x_i, \hat{n})=\delta T/\bar{T}$ where $\bar{T}$ is the background temperature\footnote{The dependences are on the conformal time $\tau$, the coordinate $x_i$ and the direction of propagation $\hat{n}$.}. We will discuss the implications for the baryons in the following section.

Consider now the mode that enters the Hubble horizon just at the right time such that the baryon-photon fluid reaches the maximum compression, i.e., the maximum of the oscillation in energy density and hence of  $\Theta_T$, at the time of recombination, after which the fluids decouple and the oscillations freeze. This would correspond to the case in which $k\,r_s=\pi$. Here the universe is the hottest. Consider then the mode that leads to the maximum rarefaction, i.e., the minimum of the oscillation, obtained for $k\,r_s=2\pi$. Here the universe is the coldest. Repeating this for all multiples of $\pi$ would lead to an oscillatory pattern of $\Theta_T$ as a function of $k$, whose frequency is determined by $r_s$.\footnote{The true $\Theta_T$ evolution is more complex than this, but its exact description goes beyond the scope of the discussion.} Moreover, when the mean free path of the photons is smaller than the size of the over/underdensities the photons free stream and they will do so by following the pressure gradient, i.e., going from the overdense to the underdense regions, thereby homogenizing the medium and suppressing the oscillations of $\Theta_T$ at small scales.

As a consequence, $\Theta_T$ carries information \textit{i)} on the amount of the primordial fluctuations leading to the oscillations (i.e., to the PPS), \textit{ii)} on the balance between gravity and pressure in the early universe, \textit{iii)} on the expansion rate $H$ via Eq.~\eqref{eq: sound_hor} and specifically \textit{iv)} on the baryon energy density (which determines the photon mean free path) via the additional damping of the oscillations. In terms of cosmological parameters\footnote{Assuming $T_0$, and hence $\rho_\gamma$, is perfectly known and $N_{\rm eff}$ is fixed to its standard prediction.}, these effects allow to constrain \textit{i)} the PPS parameters $A_s$ and $n_s$,  \textit{ii\matteo{,iii})} the matter energy density \matteo{$\Omega_{m,0}h^2$} and  \textit{\matteo{iv})} the baryon energy density \matteo{$\Omega_{b,0}h^2$}.

But the shape of $\Theta_T$ can tell us even more than that. In fact, while the oscillations in $\Theta_T$ freeze at recombination, the photons still have to travel all the way towards us before we can observe them. This means that also the history of the universe after recombination can affect $\Theta_T$. For instance, the largest oscillations, i.e., the smallest $k$ modes, enter the horizon after recombination, where the photon mean free path is extremely large (always larger than the Hubble horizon). For this reason, at these large scales the oscillations of $\Theta_T$ are strongly damped. 

Furthermore, the size of the sound horizon (which determines how far apart the peaks of $\Theta_T$ are) can only be observationally determined as a projection on the sky, i.e., in the direction orthogonal to the propagation of the photons. In this way, the sound horizon is measured via its angular distance
\begin{align}\label{eq: theta_s}
	\theta_s=\frac{r_s}{D_a}\,,
\end{align}
where
\begin{align}\label{eq: Da}
	D_a = \int_0^{z_{\rm rec}} \frac{\text{d}z}{H}
\end{align}
is the comoving distance to the last-scattering surface. The observed distance between peaks is then given by $\theta_s$, which in turn depends on $D_a$ and hence on the late-time expansion history of the universe, allowing to determine parameters such as $\Omega_{\Lambda,0}$ and $h$. 

Additionally, as mentioned above, during the dark ages the universe is (almost) completely neutral, with very few free electrons, and it is therefore transparent to photons. At $z\sim10$, however, reionization begins (see following section) and the number of free electrons increases again (see Fig.~\ref{fig: CMB_xe}). The presence of these electrons acts as a sort of fog that hinders the diffusion of the photons. As such, reionization suppresses the overall amplitude of the oscillations of $\Theta_T$ since less photons make it through reionization. Mathematically, the ``thickness'' of the fog is given by the optical depth of reionization, defined as
\begin{align}\label{eq: tau_reio}
	\tau_{\rm reio}=\int \sigma_T a n_e x_e \text{d}\tau\,,
\end{align}
where $\sigma_T$ is the Thompson cross section, and the corresponding visibility function, i.e., the probability that a photon last scattered at a time $\tau$, is given by
\begin{align}
	g_{\rm reio}= -\dot{\tau}_{\rm reio}\,e^{-\tau_{\rm reio}}\,.
\end{align}
The visibility function determines the degree of the suppression, so that $\Theta_T$ becomes sensitive to $\tau_{\rm reio}$.

With $\Theta_T$ loaded with this much information, it is then fundamental to understand how to observe it. Imagine you observed a spot in the sky today (i.e., at $\tau=\tau_0$) with a given temperature difference $\delta T$ with respect to the sky average $T$, as it would be in Fig. \ref{fig: CMB_Planck_map}. On its own, that one spot does not tell you anything about $\Theta_T$ because there is no reference to compare it to. Imagine, however, you observed a second spot at a given distance $d$ from the first spot with the very same temperature difference $\delta T$. You could then wonder what is the probability that the two temperature differences you have observed are correlated, i.e., defined by some common history. Based on the discussion above, for instance, if $d$ is larger than the size of the universe at recombination all photons free stream and erase all oscillations. So you expect no correlation between the points. If, however, $d$ was of the same order of the sound horizon $r_s$, the probability that the two points are correlated increases, since both points could have been sourced by the compression of the baryon-photon fluid around the same overdensity. The probability becomes ``negative'' for $r_s/d \propto k\,r_s=2\pi$ since one would expect both spots to have a negative $\delta T$. If one moves to even smaller scales the probability becomes suppressed again due to photon free streaming and the damping of the oscillations. Therefore, the correlation between two values of $\delta T$ in directions $\hat{n}$ and $\hat{n}'$, which can be observed, contains the full information carried by $\Theta_T$ as a function of the Fourier modes $k$. The ensemble average over the sky of these correlations is referred to as two-point correlation function.

Mathematically, focusing on the dependence of $\Theta_T$ on space coordinates $x_i$ and direction of motion $\hat{n}=(1,\theta,\phi)$ in spherical coordinates, it is useful to realize that under the assumption of isotropy there is no preferred direction of motion and the dependence of one of the two angles can be suppressed. As explained before, considering the problem in Fourier space also facilitates its mathematical treatment. In this way, $\Theta_T(\tau_0,x_i,\hat{n})$ becomes $\Theta_T(\tau_0,k_i,\theta)$ or $\Theta_T(k_i,\theta)$ in short. The angle dependence can be further removed by expanding $\Theta_T(k_i,\theta)$ in Legendre polynomials $P_\ell(\cos\theta)$, i.e., 
\begin{align}
	\Theta_T(k_i,\theta) = \sum_{\ell} (-i)^\ell (2\ell+1) \Theta_{T,\ell}(k_i)P_\ell(\cos\theta)\,,
\end{align}
so that one only needs to focus on $\Theta_{T,\ell}(k_i)$. With this premises, the two-point correlation function intuitively becomes proportional to the initial size of the fluctuations (if the PPS was zero at a given mode $k$ there would be no oscillation there and hence no correlation) and to $\Theta_{T,\ell}(k)$, which would only depend on $k=|k_i|$ due to the average over the sky. Explicitly, this leads to
\begin{align}\label{eq: 2p_corr_th}
	\langle  \Theta_{T,\ell}(k_i)\Theta_{T,\ell}'(k_i') \rangle = \delta(k_i-k_i') \, \Theta_{T,\ell}^2(k) \, \mathcal{P}(k)\,,
\end{align}
where $\mathcal{P}(k)$ is the PPS defined in Sec. \ref{sec: infl}. In this way, $\Theta_{T,\ell}^2(k)$ assumes the role of a transfer function, determining how the primordial fluctuations have evolved until today. 

Observationally, however, one can only determine temperature differences as a function of $\hat{n}$, i.e., $\Theta_T(-\hat{n})$, where the $-\hat{n}$ comes from the fact that the direction of motion is opposite to the direction of observation, with no information on $k$. Expanded in spherical harmonics one has
\begin{align}
	\Theta_T(-\hat{n})=\sum_{\ell m} a_{\ell m} Y_{\ell m}(\hat{n})\,,
\end{align}
with
\begin{align}
	a_{\ell m}=(-1)^\ell \int \text{d}\hat{n} Y^*_{\ell m}(\hat{n})\Theta_T(\hat{n})\,.
\end{align}
The two-point correlation function of $\Theta_T(\hat{n})$ is then given by 
\begin{align}
	C_\ell\equiv\langle a_{\ell m} a^*_{\ell m} \rangle = \frac{1}{1+2\ell}\sum_{-\ell\leq m \leq \ell}|a_{\ell m}|^2\,,
\end{align}
which can be determined for each $\ell$ from observations of the CMB temperature anisotropies such as the one shown in Fig. \ref{fig: CMB_Planck_map}.\footnote{To go from $k$ to $\ell$ one can equate the wavelength corresponding to a given mode $k$, $\matteo{\lambda/2=a\pi/k}$, to the projection on the sky of a given length $\theta \matteo{a} D_a$, where $\theta=\pi/\ell$ is the corresponding angular distance (see previous discussion related to $r_s$). In this way, one obtains $\ell = k D_a$. Taking for instance the time of recombination, i.e., $a\simeq10^{-3}$ and $D_a=r_s/\theta_s\simeq 145\,\text{Mpc}/0.01\simeq 1.45 \times 10^4$ Mpc, gives $\ell = 1.45\times 10^{4}\,\text{Mpc} \,k$. Therefore, for instance $k=\matteo{0.01}$ Mpc$^{-1}$ (i.e., the scale corresponding to the size of the horizon at recombination, as given above) would correspond to $\ell \sim \matteo{150}$.} It is referred to as the CMB anisotropy temperature power spectrum. At the same time, using the properties of spherical harmonics and Legendre polynomials one can also show that
\begin{align}\label{eq: Cl}
	C_\ell\equiv\langle a_{\ell m} a^*_{\ell m} \rangle  \propto \int \frac{\text{d}k}{k} \Theta_{T,\ell}^2(k) \, \mathcal{P}(k)\,,
\end{align}
and hence proportional to the theoretical prediction of the two-point correlation function for $\Theta_{T,\ell}(k_i)$ introduced in Eq. \eqref{eq: 2p_corr_th}.

\begin{figure}[t]
	\centering
	\includegraphics[width=0.85\textwidth]{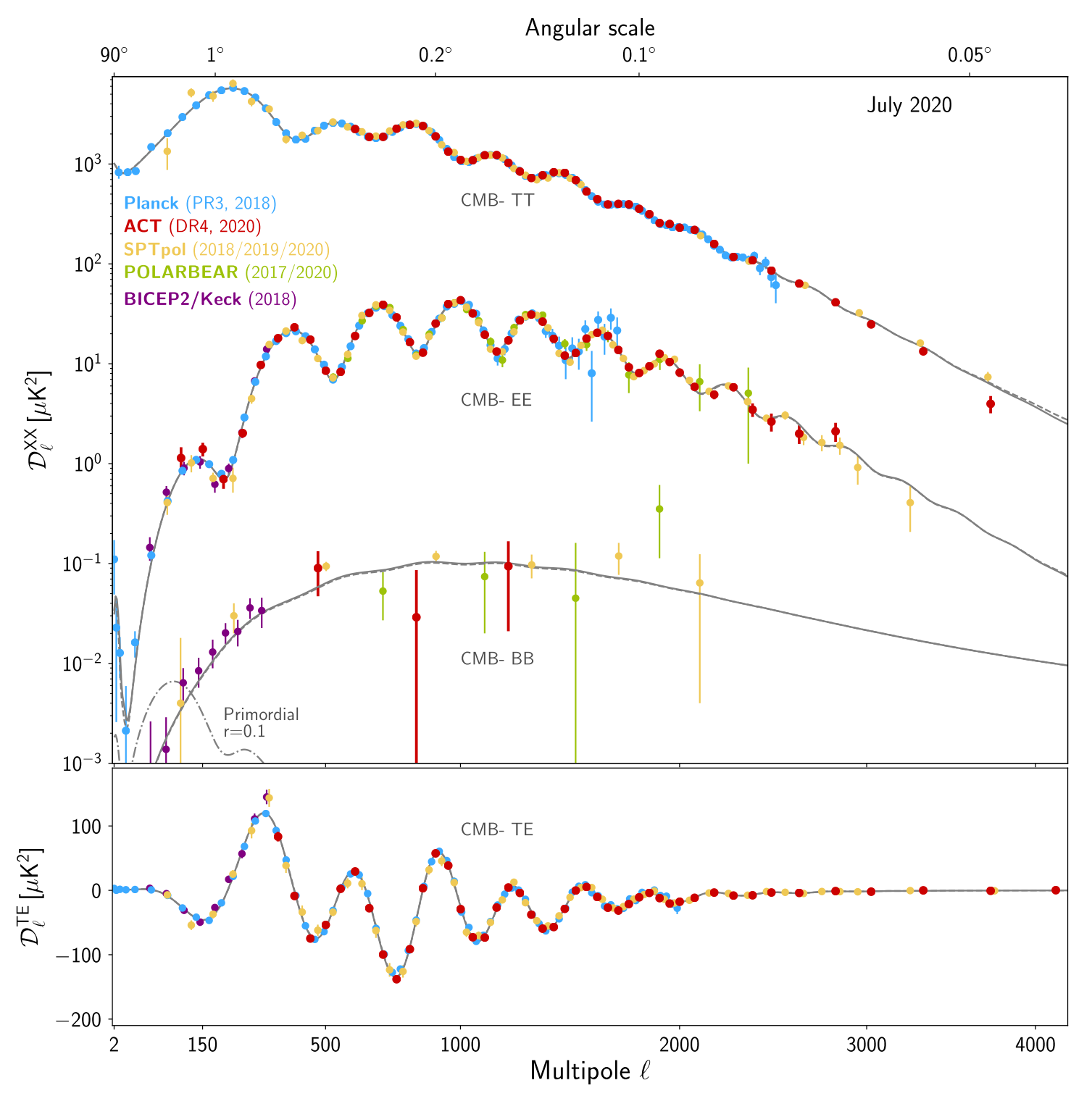}
	\caption{Anisotropy power spectra of the CMB (solid lines). The data points represent the latest measurements of a selection of modern experiments. Figure taken from \cite{Choi2020Atacama}.}
	\label{fig: CMB_spectra}
\end{figure}

The CMB temperature anisotropy power spectrum $C_\ell$ is thus the real observable linking the measurement of the temperature anisotropies of the CMB and to their theoretical prediction. In the context of the \lcdm model the two are compared in Fig. \ref{fig: CMB_spectra}, taken from \cite{Choi2020Atacama} (see also e.g., Fig. 5 of \cite{SPT-3G:2022hvq}), under the label TT (with a slightly different convention than the one we used above, but that does not change the overall conclusions). As one can see there, the exquisitely precise data gathered by the Planck mission \cite{Aghanim2018PlanckVI}, the Atacama Cosmology Telescope (ACT) \cite{Choi2020Atacama} and the South Pole Telescope (SPT) \cite{SPT-3G:2022hvq}, among others, perfectly overlap with the theoretical prediction (solid line), confirming the validity of the \lcdm model. 

Based of the aforementioned physical effects relating the shape of $\Theta_T$ and the six \lcdm parameters ($\Omega_{b,0}$, $\Omega_{\rm cdm,0}$, $h$, $\tau_{\rm reio}$, $n_s$ and $A_s$),\footnote{Using $\Omega_{\rm cdm,0}$ and $h$ or $\Omega_{\rm m,0}$ and $\Omega_{\Lambda,0}$ is equivalent when $\Omega_{b,0}$ is known.} one can also relate the latter to features in the shape of the power spectrum (see e.g., Sec. 2.6 of \cite{Lesgourgues:2013qba} for a full picture). For instance, we saw above that the position of peaks is related to the horizon angular scale $\theta_s$ (see Eq. \ref{eq: theta_s}). Therefore, a change in the parameters that affect $\theta_s$, i.e., $\Omega_{b,0}$, $\Omega_{\rm m,0}$ and $h$, will cause an horizontal shift of the peaks.  A similar discussion can be also carried out for the damping scale at which the anisotropies start to be suppressed, which allows to independently probe the same three parameters. Moreover, the balance between pressure and gravity that generates the acoustic oscillations in the RD era, fully determined by \matteo{$\Omega_{m,0}$}, affects the relative size of the even (maximum contraction) and odd (maximum rarefaction) peaks. From Eq. \eqref{eq: Cl} it is also clear that the PPS parameters directly influence the total amplitude and tilt of the power spectrum. A similar effect on the $C_\ell$ is given by the presence of reionization (see Eq. \eqref{eq: tau_reio}). Therefore, the observation of the temperature anisotropy power spectrum allows to constrain all of the \lcdm parameters (with only a mild degeneracy between $A_s$ and $\tau_{\rm reio}$).

Furthermore, so far we have only focused on the CMB temperature anisotropies. Yet, a similar discussion can also be carried out for their polarization and a pedagogical description of their origin is given in e.g., \cite{Kaplan:2003ye}. The CMB polarization power spectrum and its cross-correlation with the temperature power spectrum then become an additional source of information to constrain the aforementioned cosmological parameters. A graphical representation of these spectra and their agreement with the \lcdm prediction can be seen in Fig. \ref{fig: CMB_spectra}.

From the combination of all Planck spectra, one obtains within the \lcdm model that \cite{Aghanim2018PlanckVI}
\begin{align}\label{eq: planck_vals}
	\nonumber & 10^2\Omega_{b,0}h^2 = 2.236\pm0.015\,,~\Omega_{\rm cdm,0}h^2=0.1202\pm0.0014\,,~10^2\theta_s=1.04109\pm0.00030\,,\\  & 10^9A_s=2.101^{+0.031}_{-0.034}\,,~n_s=0.9649\pm0.0044~\text{and}~\tau_{\rm reio}=0.0544^{+0.0070}_{-0.0081}\,.
\end{align} 
Useful inferred quantities in terms of the expansion rate are e.g., 
\begin{align}
	\Omega_{\Lambda,0}=0.6834\pm0.0084~\text{and}~h=0.6727\pm0.0060\,,
\end{align}
while useful timescales are determined to be
\begin{align}
	z_{\rm reio}=7.68\pm0.79\,,~z_{\rm rec}=1089.95\pm0.27~\text{and}~z_{\rm MR-eq}=3407\pm31\,.
\end{align}
Compared to the discussion of the previous sections, it is important to underline that the CMB anisotropies are in \matteo{(almost)} perfect agreement with the idea of a scale invariant PPS (since $n_s\simeq 1$) at the scales probed by Planck\footnote{The lower bound is set by the size of the universe today, while the upper bound by the experiment resolution of $\theta\sim0.05^\circ$ -- corresponding to $\sim5$~Mpc at recombination and hence to $k\sim1$~Mpc$^{-1}$.}, i.e., $k\simeq10^{-4}-1$ Mpc$^{-1}$ \cite{akrami:planckinflation2018}\matteo{, thereby confirming the validity of the slow-roll paradigm at those scales (see Sec.~\ref{sec: infl}).} Also the CMB determination of $\Omega_{b,0}h^2$ and $N_{\rm eff}$ is compatible with the corresponding BBN value. 

The accurate match between observations of the CMB anisotropies and the \lcdm model prediction also implies that any deviation from it can be very well constrained. Furthermore, given the variety of dependencies on the expansion, thermal and perturbation history a number of different models can be explored. Here, we will mostly focus on those models that involve energy injections of EM particles, via e.g., the decay or annihilation of a dark component. The main way via which such process impact the CMB anisotropy power spectra is their inevitable modification to the evolution of the free electron fraction. In fact, if photons are injected in the recombining or neutral medium they could further ionize the hydrogen atoms and free more electrons than in the standard model. This would modify the visibility function introduced above and change the probability of the CMB photons to re-scatter along their way towards us, thereby changing the shape of the power spectra. We will use this numerous times in Sec.~\ref{sec: app} to constrain beyond-\lcdm models.

In summary, the recombination process of helium and hydrogen leads to the free streaming of the photons, producing the CMB. Its anisotropies carry an incredible amount of information on the early and late universe, and their measurement is in perfect agreement with the \lcdm model prediction.

\subsection{Structure formation}\label{sec: LSS}

While perturbations of the baryon-photon fluid were undergoing acoustic oscillations in the epoch prior to recombination, the DM perturbations were growing unimpeded under the force of gravity. This led to the formation of large-scale DM filaments and structures, well before the formation of galaxies and clusters thereof. In fact, only after the photons have decoupled from the baryonic matter, at $z\sim\mathcal{O}(200)$, the latter could fall into the potential of the DM, eventually heating up \matteo{and compressing} enough (due to adiabatic \matteo{pressure and heat release}) to initiate star formation and reionize the universe at $z\sim\mathcal{O}(10)$. 

\begin{figure}[t]
	\centering
	\includegraphics[width=0.41\textwidth]{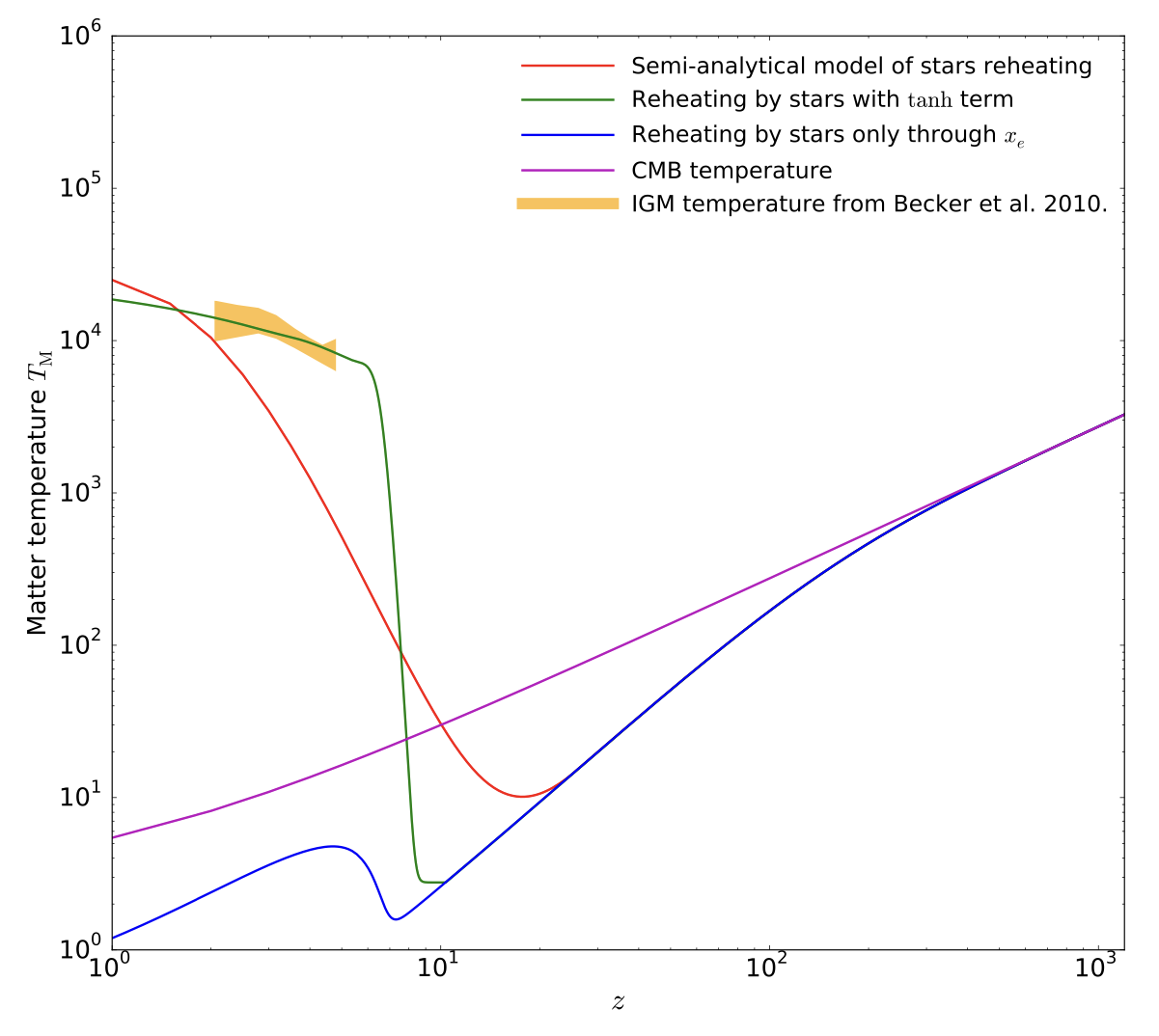}
	\includegraphics[width=0.55\textwidth]{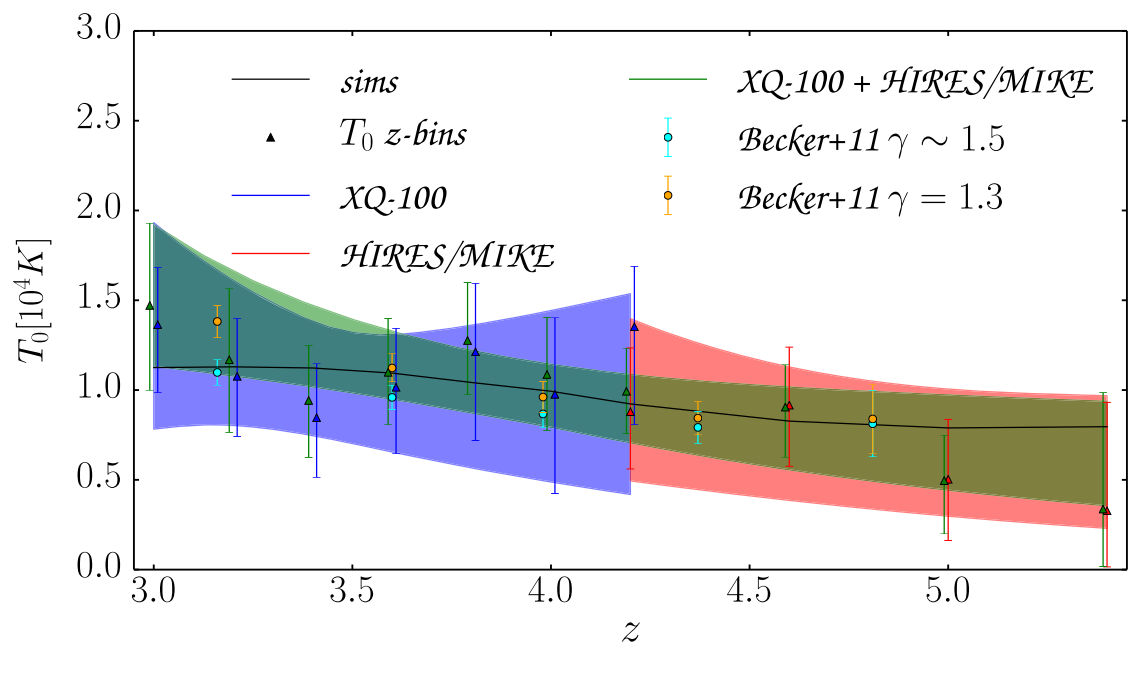}
	\caption{Evolution of the matter temperature for different late-time reheating models (left) compared to the Lyman-$\alpha$ observations of the same quantity (yellow shaded area in left panel, magnified in the right panel for graphical clearness). Figures taken from \cite{Poulin2015Dark, Irsic2017Constraints}.}
	\label{fig: LSS_Tm}
\end{figure}

The sequence of these events can be better understood on the basis of Fig. \ref{fig: LSS_Tm}, taken from \cite{Poulin2015Dark, Irsic2017Constraints}. In the left panel one can observe the evolution of the matter (or baryon) temperature $T_m$ (red, green and blue lines, depending on the underlying model) compared to the photon temperature $T_\gamma$ (purple line). In absence of any late-time astrophysics, the two quantities evolve according to 
\begin{align}\label{eq: T_b}
	\dot{T}_m = -2HT_m+\Gamma_{\rm int}(T_\gamma-T_m)\quad\text{and}\quad \dot{T}_\gamma = -HT_{\gamma}\,,
\end{align}
where $\Gamma_{\rm int}$ is the effective baryon-photon interaction rate (see e.g., Eq.~(3.4) of [\hyperlink{I}{I}] for a definition). The coefficients of the first terms (i.e., the expansion term) are due to the fact that, in absence of interactions, the temperature of non-relativistic matter scales as $T_{m} \propto (1+z)^2$, while that of the photons as $T_{\gamma} \propto (1+z)$. For $\Gamma_{\rm int}\gg H$ (i.e., at $z\gg 200$) the baryons and the photons are tightly coupled, so that $T_m\simeq T_\gamma$ and hence $\dot{T}_m = -HT_{\gamma}=-HT_{m}$. After the two fluids decouple at $z \sim 200$, the temperature scaling difference leads to a significant departure of $T_m$ from $T_\gamma$, with $T_m < T_\gamma$. The evolution of $T_m$ in this case is represented by the blue line in the left panel of Fig. \ref{fig: LSS_Tm}.

On top of this dynamics, once the two fluids decouple the baryons fall in the potential well of the DM, start to heat up, ignite star/galaxy formation and re-ionize the universe. In this way, the evolution of the matter temperature becomes tightly related to that of the free electron fraction $x_e$ in the late universe and much of the information learned on the one can be used to constrain the other. Nevertheless, the exact properties of these events are largely unknown (see e.g., \cite{Poulin2015Dark,Ahn:2020btj} for reviews), although the interplay of different probes allows to model their overall behavior \cite{greig2017global}.

In terms of the matter temperature, observations of the Lyman-$\alpha$ forest point to $T_m$ values of the order of $\sim10^4$ K at redshifts $z\simeq3-6$ \cite{2011MNRAS.410.1096B, Viel2013Warm, Irsic2017Constraints}, see right panel of Fig.~\ref{fig: LSS_Tm}, taken from~\cite{Irsic2017Constraints}. In short, along their way towards us, photons emitted from high-redshift luminous sources can get absorbed by eventual hydrogen clouds present along the LOS and be re-emitted in a (statistically) different direction. Depending on the size and number of these hydrogen clouds, this process creates a series of absorption features in the spectrum of the source, known as Lyman-$\alpha$ forest (see e.g., \cite{Viel2013Warm} for a review of the topic). Since the absorption probability depends on the amount of neutral hydrogen in the clouds, which in turn depends on the temperature of the environment, the Lyman-$\alpha$ forest is an ideal tool to measure the very late-time evolution of the hydrogen (and hence matter) temperature. These Lyman-$\alpha$ observations therefore imply that the baryons must have been significantly heated by e.g., star-forming galaxies before $z\simeq6$, which in turn suggests that the bulk of the reionization must have been completed by the same time \cite{Irsic2017Constraints}.

\begin{figure}[t]
	\centering
	\includegraphics[width=0.65\textwidth]{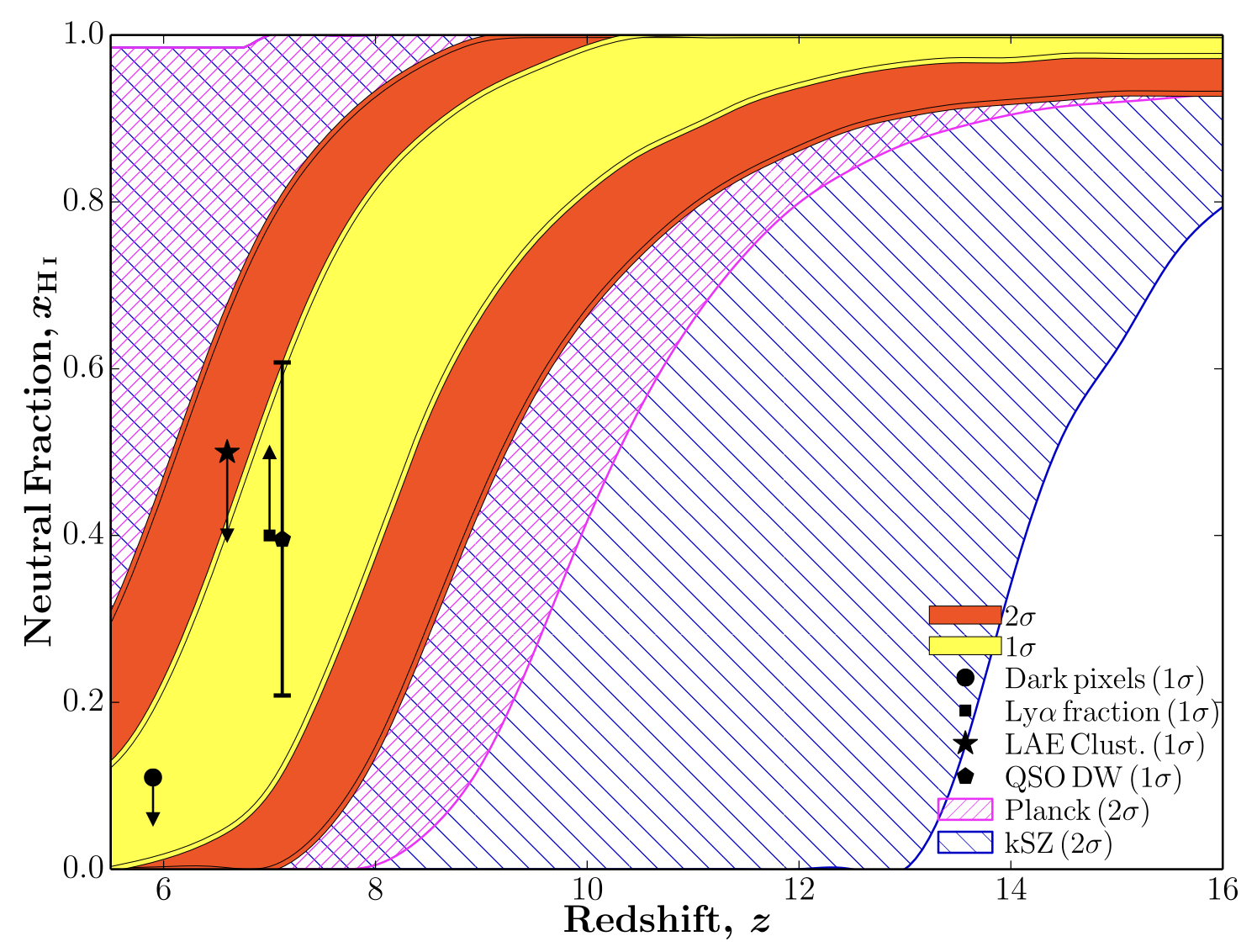}
	\caption{Range of reionization histories (parameterized with neutral hydrogen fraction) allowed by current data. Figures taken from \cite{greig2017global}.}
	\label{fig: LSS_reio}
\end{figure}

This estimate is consistent with the Planck measurement of the optical depth of reionization (see Sec. \ref{sec: CMB}), which can be expressed in term of the redshift of reionization\footnote{Formally defined as the mean of the late-time peak of the visibility function (see App. D of \cite{Poulin2015Dark} as well as Sec. \ref{sec: CMB}).} to find $z_{\rm reio}=7.82\pm0.71$. Another probe able to shed light on the reionization history is the so-called 21 cm cosmology (see e.g., \cite{Pritchard:2011xb}). This observational strategy relies on the emission of a photon with $\lambda\simeq21$ cm following the spin-flip transition of an electron in an hydrogen atom. Since the probability of this transition to occur depends on the thermal history of the universe (like e.g., on the temperature and abundance of the hydrogen atoms), the measurement of these HI lines can clarify the evolution of the neutral universe. A particularly interesting observation of the 21 cm signal has been delivered by the Experiment to Detect the Global Epoch of Reionization Signature (EDGES) \cite{Bowman:2018yin, Monsalve:2017mli, Monsalve:2019baw}, which favors reionization models lasting for more than $\Delta z \simeq 1$ (although currently matter of debate, see e.g., \cite{Ahn:2020btj}). Complementary probes (see e.g., \cite{greig2017global} for an overview) confirm this picture and their combination leads Fig. \ref{fig: LSS_reio}, taken from \cite{greig2017global}.

Based on the aforementioned picture, another quantity that is (chronologically) linked to the matter temperature and the reionization of the universe is the rate at which the baryons cluster around the pre-existing DM structures. In fact, on the basis of numerical simulations one can follow at the same time the amount and distribution of structures as well as the evolution of the associated thermal history (see e.g., \cite{Vogelsberger:2014dza}). By matching the former to observations of the related quantities today one can constrain the available parameter space for the latter.

\begin{figure}[t]
	\centering
	\includegraphics[width=0.8\textwidth]{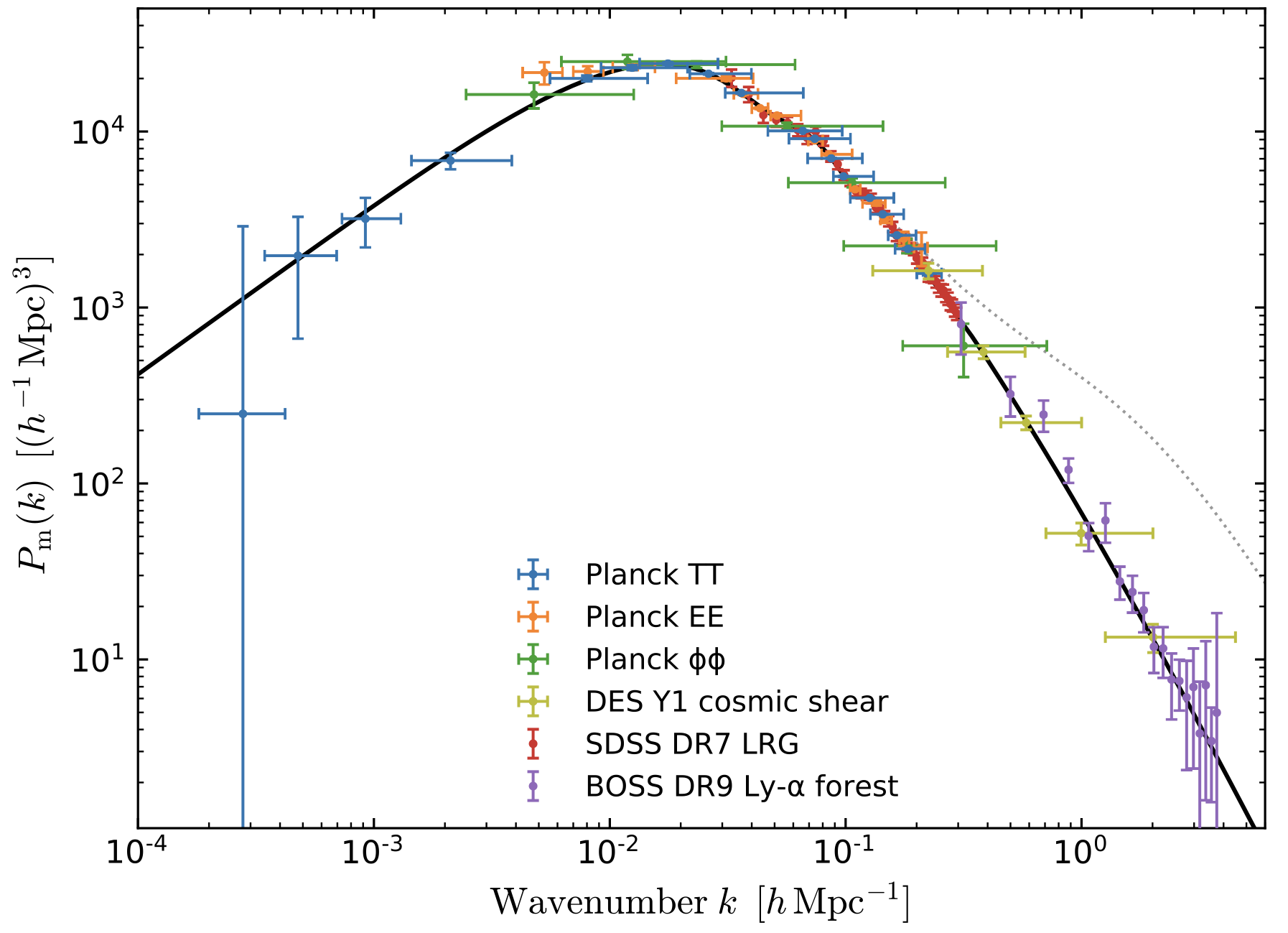}
	\caption{Same as in Fig. \ref{fig: CMB_spectra} but for the matter power spectrum. Figure taken from \cite{Planck:2018nkj}.}
	\label{fig: LSS_spectrum}
\end{figure}

In light of these considerations, the most important source of information on the process of structure formation is the so-called matter power spectrum $P_m(k)$. Similarly to the CMB anisotropy power spectra discussed in the previous section, the matter power spectrum contains information about the matter (DM+baryon) perturbations, i.e., on the abundance and correlation between massive structures. Its prediction using Planck best-fitting values for the cosmological parameters is shown in Fig. \ref{fig: LSS_spectrum}, taken from \cite{Planck:2018nkj}, and it can be tested in various ways. First of all, at very large scales, i.e., relatively small $k$ values, it can be inferred from observation of the CMB power spectra. Furthermore, weak gravitational lensing effects -- where the shape of a given luminous object is distorted by the presence of an intervening mass -- can be observed with surveys such as the Dark Energy Survey \cite{Troxel2017Dark} and the Kilo-Degree Survey \cite{Hildebrandt2018KiDS}. This allows to identify the distribution of the matter in the redshift range between $z\sim0.1-1.2$ \cite{Hildebrandt:2020rno, 10.1093/mnras/sty957} and scales of the order of $k\sim 0.1-1$~Mpc$^{-1}$. In a conceptually similar way, at much smaller scales observations of the Lyman-$\alpha$ forest allow to determine the hydrogen distribution, which can be recast in terms of the matter power spectrum at scales of the order of $k\sim 0.3-30$~Mpc$^{-1}$ with numerical simulations.\footnote{As mentioned above, one can make use of very computationally-expensive numerical simulations to determine at the same time the hydrogen distribution and the matter power spectrum for any combination of the underlying cosmological and astrophysical parameters. The combinations that reflect the observed Lyman-$\alpha$ data can be used to infer the corresponding matter power spectrum.} A further effect that can be used to test the matter power spectrum is the Sunyaev-Zeldovich (SZ) effect, which we will discuss in greater detail in Secs.~\ref{sec: SD_lcdm} and Sec. \ref{sec: res_lcdm} since it contributes to the creation of SDs. As can be seeen from Fig. \ref{fig: LSS_spectrum}, all of these data are in remarkably good agreement with the \lcdm prediction, although a mild inconsistency between the inferred cosmological parameters (at the level of $2-3\sigma$) exists, with the weak lensing surveys predicting a universe about 10\% more homogeneous than what would be expected from CMB data (see following section).

\begin{figure}[t]
	\centering
	\includegraphics[width=0.7\textwidth]{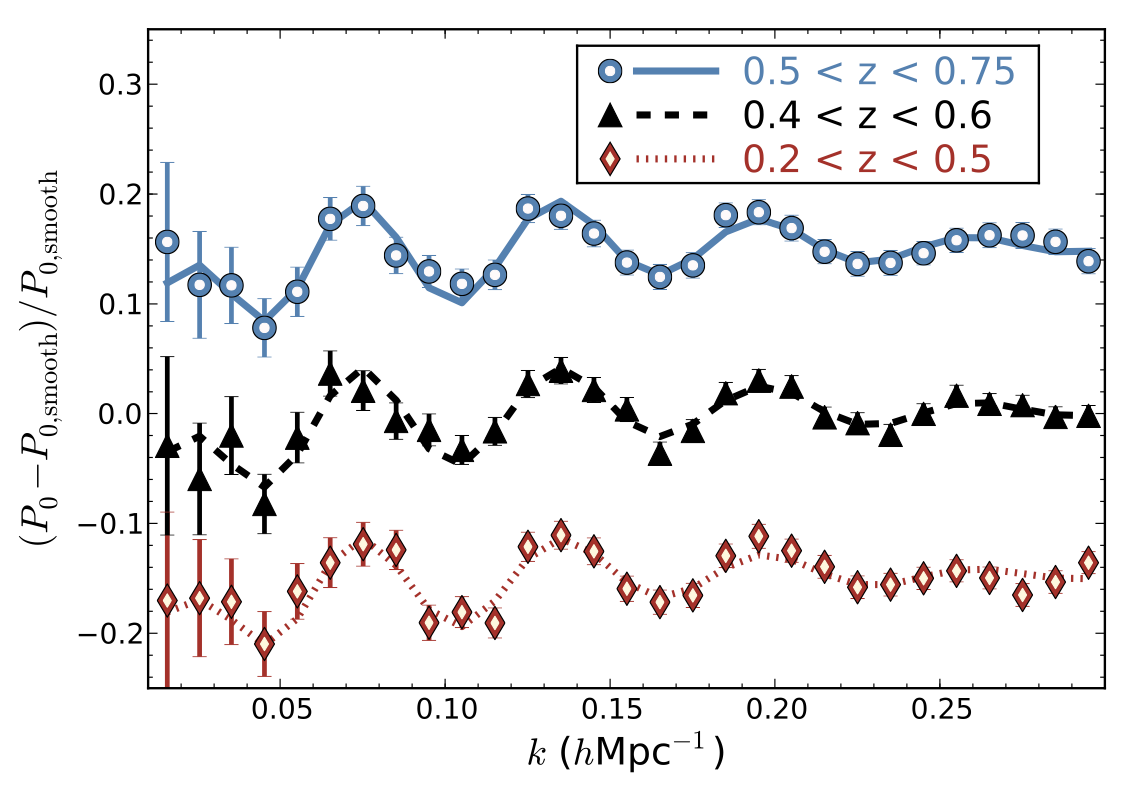}
	\caption{BAO (normalized with respect to a smooth matter power spectrum) as measured by SDSS. Figure taken from \cite{Alam2016Clustering}.}
	\label{fig: LSS_BAO}
\end{figure}

A particularly important feature in the matter power spectrum is given by the so-called Baryon Acoustic Oscillations (BAO). These represent the imprints on the matter (DM+baryon) perturbations of the pre-recombination oscillations between baryons and photons. Given the larger role of the DM, their overall contribution is suppressed but still observable. As a representative example, their measurement by the Sloan Digital Sky Survey (SDSS) III DR12 is shown in Fig. \ref{fig: LSS_BAO}, taken from \cite{Alam2016Clustering}. The figure represents the BAO normalized with respect to a (smooth) matter power spectrum with no features (the high and low redshift bins have been artificially shifted by $\pm0.15$ for graphical clearness). Similarly as for the photon perturbations and the CMB anisotropy power spectra discussed in Sec. \ref{sec: CMB}, also the BAO are a measure of the sound horizon $r_s$, although determined at much lower redshifts (typically of the order of $z\simeq0.1-1$, see e.g., \cite{Beutler2011Galaxy, Ross2014Clustering, Alam2016Clustering}). Therefore, the excellent agreement between various CMB anisotropy and BAO measurements represents a solid test of the \lcdm model.

As a final remark, the distribution of matter has only been determined at relatively low redshifts, mostly because relying on the presence of distant light sources which could only form in the very late universe. What happened between recombination and reionization is observationally largely unknown, although this will change in the future with the advent of the 21 cm cosmology (see e.g., \cite{Furlanetto201221CC}).

In summary, after photon decoupling the baryonic matter falls into the pre-existing DM overdensities to form large-scale structures, galaxies and stars. As a consequence, this re-ionizes the universe. Although much is still unknown about the period of reionization and structure formation, the evidence gathered by various complementary observational strategies delivers a consistent picture within the \lcdm model.

\subsection{Cosmological tensions}\label{sec: tensions}
Despite the many successes of the \lcdm model, some inconsistencies have emerged between the early-time inference and the late-time measurements of various cosmological parameters [\hyperlink{XV}{XV}$-$\hyperlink{XVIII}{XVIII},\,\hyperlink{XX}{XX}]. Simply put, the values today of the parameters derived from information coming from the early universe, as in the case of the CMB -- see Eq.~\eqref{eq: planck_vals}, are model-dependent projections (as it would be the case in e.g., \lcdm) of the values that these parameters had at the time when the physics happened. In other words, given a model, its evolution is anchored at the time of the observed process and projected forward until today. The same quantities can however also be observed locally in a model-independent way, which allows to test the prediction of the model used for the inference. When the early- and late-time values disagree at a statistical significant level one refers to the measurements involved as being in tension.

The so-called Hubble tension is the most (and possibly only) statistically significant of these inconsistencies, now above the $5\sigma$ level \cite{Riess:2021jrx}, and involves the early-time inference assuming \lcdm and late-time direct measurement of the $H_0$ parameter. The former has been consistently determined to be of the order of $\sim67-68$~km/(s Mpc) by a number of CMB experiences such as Planck, ACT and SPT as well as by CMB-independent combinations of data sets. The latter, however, suggests larger values, of the order of \matteo{$70-75$}~km/(s Mpc), that can be obtained e.g., via observations of the distance ladder (i.e., cepheids + Supernovae -- SN -- Ia) \matteo{\cite{Riess:2021jrx}} and of the tip of the red giant branch \matteo{\cite{Freedman2019Carnegie}}. A complete graphical summary can be seen in e.g., Fig. 2 of [\hyperlink{XX}{XX}], where the interested reader can also find more details and references. 

This tension points to the idea that the universe is expanding differently than what \lcdm predicts. A number of extension of the \lcdm model have been proposed to amend this issue, yet to date only a handful have been marginally successful (see e.g., \cite[\hyperlink{XX}{XX}]{DiValentino2021Realm, Schoneberg2021Olympics} for comprehensive reviews). Overall, one can differentiate two classes of models: those modifying the expansion history before and around recombination, and those increasing the Hubble rate at late times. Examples of the former class are Early Dark Energy (EDE)~\cite[\hyperlink{X}{X}]{Karwal2016Dark, Poulin2018Early, Smith2019Oscillating, Klypin2020Clustering, Hill:2021yec, Poulin:2021bjr, LaPosta:2021pgm, Poulin:2023lkg} (see however \cite{Hill2020Early, Ivanov2020Constraining, Goldstein:2023gnw}), self-interacting neutrinos \cite{Lancaster2017Tale, Kreisch2019Neutrino, Escudero2019CMB, Escudero2020Could, Archidiacono:2020yey, Corona:2021qxl} (see however \cite{Brinckmann:2020bcn, RoyChoudhury:2020dmd, Kreisch:2022zxp}) and Primordial Magnetic Fields (PMFs)~\cite{Jedamzik2020Relieving} (see however \cite{Galli:2021mxk, Thiele:2021okz}), while popular late-time models involve for instance DM-DE interactions \cite{DiValentino2017Interacting, DiValentino2019Minimal, Zhai:2023yny} (see however \cite[\hyperlink{II}{II}]{Nunes:2022bhn}). Of course, several other scenarios have been proposed that do not fall under this classification, such as models varying the electron mass \cite{Hart2019Updated}.

\begin{figure}[t]
	\centering
	\includegraphics[width=0.75\textwidth]{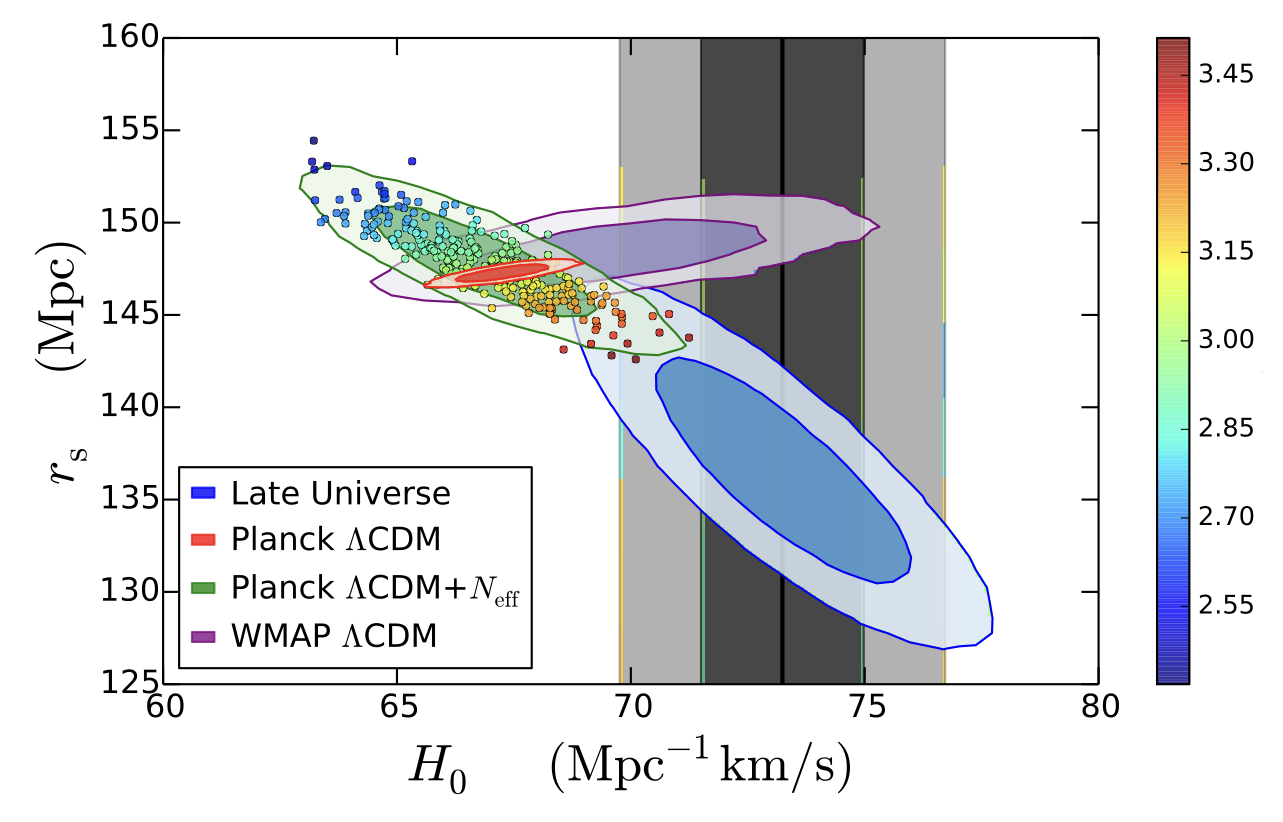}
	\caption{Representation of the Hubble tension in the $H_0-r_s$ plane (the color coding refers to the value of~$N_{\rm eff}$). Figure taken from \cite{Bernal2016Trouble}.}
	\label{fig: tensions_H0rs}
\end{figure}

However, it has been convincingly argued that models attempting to modify the expansion history of the universe at late times are incompatible with observations \cite{Bernal2016Trouble, Poulin2018Implications, Aylor2018Sounds, Knox2019Hubble}. On the one hand, the combination of BAO and SN Ia data agree with the \lcdm prediction and disfavor deviations from it (see e.g., Figs.~2 and 3 of~\cite{Poulin2018Implications} for a graphical depiction). On the other hand, the $H_0$ tension can be transposed in a physically more informative tension in the $H_0-r_s$ plane, where $r_s$ is the sound horizon introduced in Sec.~\ref{sec: CMB}. As it can be seen in Fig.~\ref{fig: tensions_H0rs}, taken from \cite{Bernal2016Trouble}, in this new plane the deviation from the \lcdm prediction needs to happen in both directions, i.e., both $H_0$ and $r_s$ have to be modified by a given model in order to successfully solve the tension \matteo{(see also~\cite{Jedamzik:2020zmd, ACT:2023kun} for related discussions)}. However, as $r_s$ is only affected by the history of the universe prior to recombination, late-time modifications of the $\Lambda$CDM model are intrinsically incapable of fully solving the $H_0-r_s$ tension. Therefore, scenarios predicting early-time deviations from \lcdm seem to be favored as possible solutions of the $H_0-r_s$ tension. As such, in Sec.~\ref{sec: res_tens} we will discuss the role that SDs could play in discerning and constraining these models.

Further, less significant tensions (at the level of $2-3\sigma$) involve the shape of the matter power spectrum and in particular its amplitude, parameterized via $\sigma_8$ (i.e., the root mean square of the linear matter perturbations within 8 Mpc/$h$ today) [\hyperlink{XX}{XX}], and its tilt at Lyman-$\alpha$ scales \cite[\hyperlink{VIII}{VIII}]{Palanque-Delabrouille2019Hints}. In this case, direct observations of the local environment, via e.g., weak lensing and Lyman-$\alpha$ forest measurements -- see Sec. \ref{sec: LSS}, prefer a matter power spectrum that is overall suppressed (both in terms of amplitude and tilt) with respect to the shape expected from CMB measurements (see e.g., [\hyperlink{XX}{XX}] for a review). Should this behavior be confirmed, it could point to the presence of non-standard DM interactions with e.g., DE \cite[\hyperlink{VI}{VI}, \hyperlink{VII}{VII}]{Poulin:2022sgp}, neutrinos [\hyperlink{VIII}{VIII}] and baryons \cite{Hooper:2022byl}.

In summary, the emergence of various cosmological tensions might hint to cracks in the otherwise extremely successful \lcdm model, although their implications are far from being fully understood.

\newpage
 \section{Up the mountain we go: the theory of CMB SDs}\label{sec: theory}
After having outlined the properties and evolution of the \lcdm model, we can turn our attention to CMB SDs. \matteo{While} the CMB temperature and polarization anisotropies introduced in Sec. \ref{sec: CMB} represent \matteo{first}-order deviations from the otherwise homogeneous and isotropic CMB photon field, CMB SDs represent second-order\footnote{As pointed out in \cite{Chluba2012CMB}, accurately deriving the SD signal -- in the reference singling out the contribution from acoustic wave dissipation, see below -- requires going to second-order in both the perturbations and the energy transfer by Compton scattering.} deviation from the pure BB shape of the CMB energy spectrum expected from a fluid in thermal equilibrium. As such, they carry invaluable information about the expansion and thermal history of the universe. 

In Sec. \ref{sec: ped} we present a graphical and pedagogical overview on their theory, aimed at providing an intuitive understanding of the topic without dwelling extensively on the underlying technical details. The exact calculation of the many quantities involved can be found in a series of previous works, see e.g., \cite[\hyperlink{I}{I}]{Hu1995Wandering, Chluba2005Spectral, DeZotti2015Another, Schoneberg:2021uak}. In Sec.~\ref{sec: sources} we then discuss all of the processes that can source CMB SDs within \lcdm, in Sec. \ref{sec: SD_lcdm}, and beyond it, in Sec. \ref{sec: SD_non_lcdm}, with particular emphasis on the concepts of energy injection and deposition. In Sec. \ref{sec: num} we present the numerical implementation employed to evaluate both the SD signal and the heating history. 

\subsection{Pedagogical introduction}\label{sec: ped}
To understand what CMB SDs are, in a first step we can momentarily set aside the whole universe and everything that there is in it, including the CMB itself. All we need is a box filled with photons and electrons.

\begin{figure}[t]
	\centering
	\includegraphics[width=\textwidth]{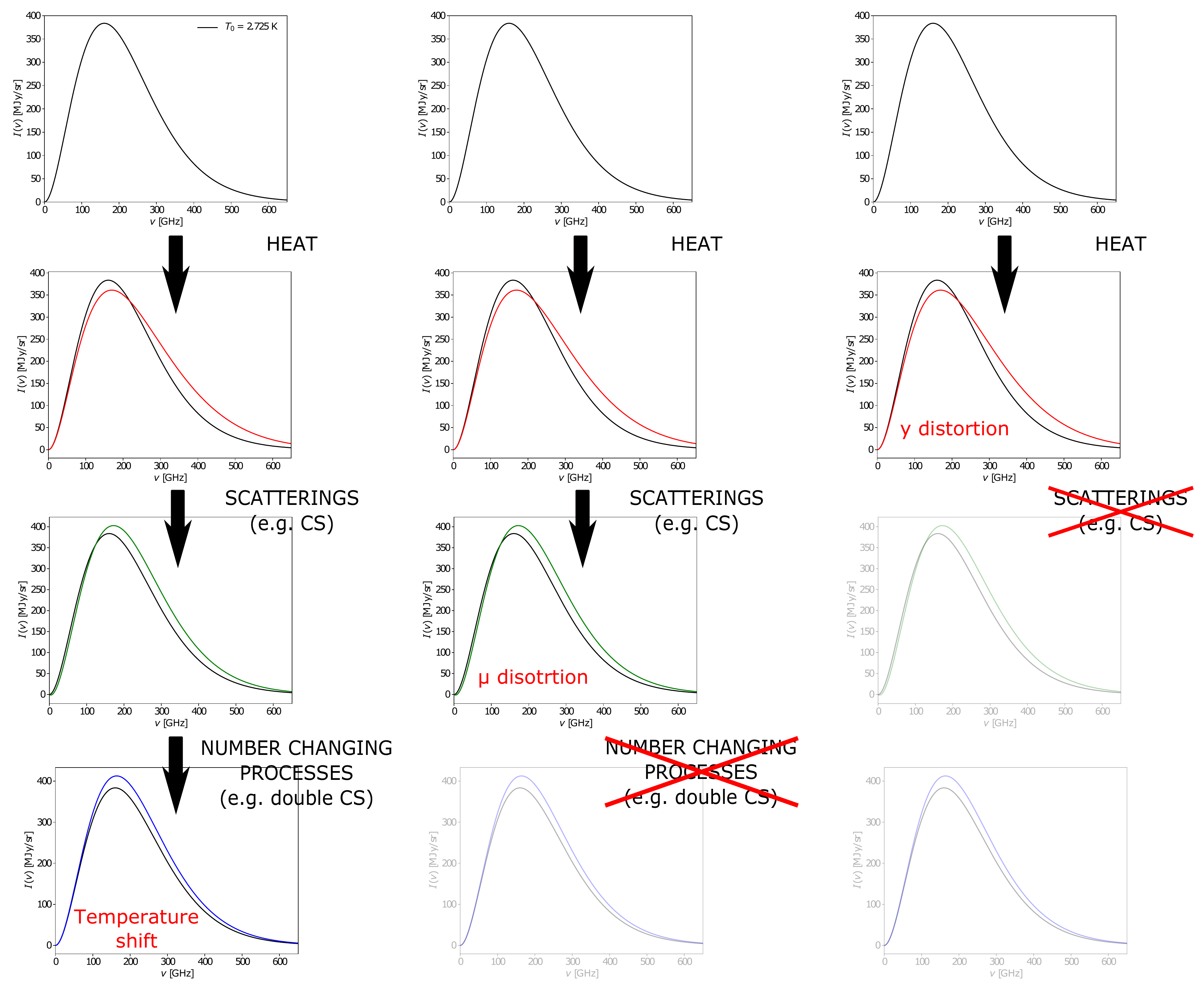}
	\caption{Graphical illustration of the evolution of the photon energy spectrum (and hence of the SD shapes) following a heating injection as a function of the relevant processes (from top to bottom). Figure adapted from [\protect\hyperlink{XXI}{XXI}].}
	\label{fig: SD_th_example_box}
\end{figure}

At first, let us consider the case where CS,
\begin{align}
	e^-+\gamma \leftrightarrow e^-+\gamma\,,
\end{align}
is very efficient and all particles are in kinetic equilibrium\footnote{In other words, electrons and photons follow Fermi-Dirac and Bose-Einstein momentum distributions.}, i.e., all particles share the same kinetic energy and hence temperature. Let us also assume that photons can be destroyed and created at any time via processes such as e.g., double CS, 
\begin{align}
	e^-+\gamma \leftrightarrow e^-+\gamma+\gamma\,.
\end{align}
This implies that the chemical potential of the photons $\mu_\gamma$ vanishes. In this setup, if we were to measure the energy of the photons from a hole in the box, we would observe a BB spectrum
\begin{align}
	B(x)=\frac{1}{e^x-1}
\end{align}
with $x=E/T_\gamma$ (black line in the top panel of the left column of Fig.~\ref{fig: SD_th_example_box}, adapted from~[\hyperlink{XXII}{XXII}]).

Let us now imagine to place a hand on the side of the box for a small amount of time and for sake of the narrative let us assume that the hand is hotter than the temperature of the box. In this way, we would be effectively heating up the photons that hit the wall while our hand is touching it. In terms of the energy spectrum of the photons, this implies that the photons from the low-energy tail get shifted towards the high-energy tail (red line in the second panel of the left column of Fig. \ref{fig: SD_th_example_box} -- to be compared to the same black line as before for reference). Since the majority of the photons reside at the peak of the distribution, the largest suppression will happen there. 

After enough time, the scattering processes will start to redistribute the injected energy equally among the photons and the result will be an overall horizontal shift of the spectrum towards higher energies due to the additional energy stored in the box (green line in the third panel of the left column of Fig.~\ref{fig: SD_th_example_box}). Effectively, this corresponds to the presence of a chemical potential $\mu_\gamma$. The excess of photons at high energies can nevertheless be corrected by the aforementioned number changing processes, which transform e.g., one high-energy photon into two lower-energy photons. Given enough time, this will drive the chemical potential to zero and restore a BB shape, although with an higher temperature than the original one (blue line in the fourth panel of the left column of Fig. \ref{fig: SD_th_example_box}). 

In full generality, the final shape of the spectrum can be parameterized as
\begin{align}\label{eq: f_decomp}
	f(x)=B(x)+\Delta f(x)\,.
\end{align} 
In the case of a temperature shift, we obtain
\begin{align}\label{eq: f_T_shift}
	f(x)=B\left(\frac{x}{1+\Delta T/T_\gamma}\right) \simeq B(x) - x \frac{\partial B(x)}{\partial x}\frac{\Delta T}{T_\gamma}\equiv B(x)+g\,G(x)\,,
\end{align} 
with $\Delta T \ll T_\gamma$. The deviation from the original BB is then given by
\begin{align}\label{eq: Df_T_shift}
	\Delta f(x) = g\, G(x)\,,
\end{align}
where we have defined the spectral shape of the temperature shift as
\begin{align}
	G(x)= - x \frac{\partial B(x)}{\partial x} = \frac{xe^x}{(e^x-1)^2}
\end{align}
and its amplitude as $g= \Delta T/T_\gamma$.

\begin{figure}[t]
	\centering
	\includegraphics[width=0.65\textwidth]{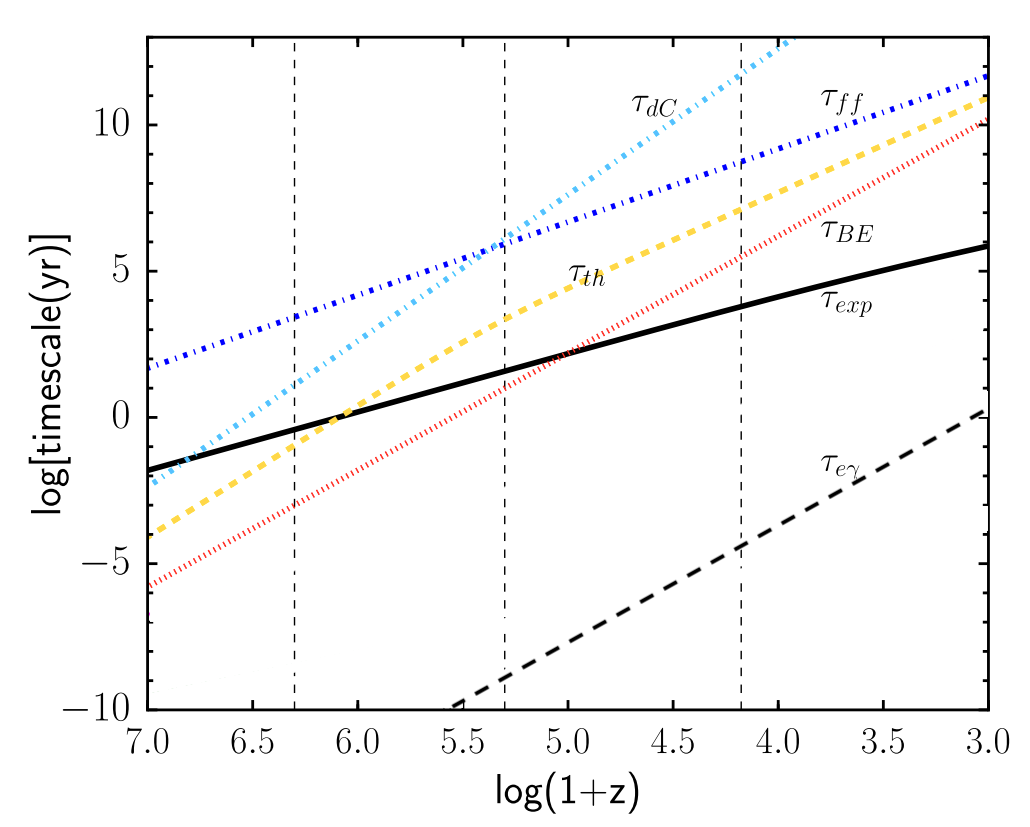}
	\caption{Timescales of the relevant processes involved in the thermalization ($\tau_{th}$) and Comptonization ($\tau_{BE}$) problems. Figure adapted from \cite{DeZotti2015Another}.}
	\label{fig: SD_th_rates}
\end{figure}

Let us now make the above example more realistic and assume that the box is expanding at a given rate $\Gamma_{\rm exp}=\tau_{\rm exp}^{-1}$. As it can be seen in Fig. \ref{fig: SD_th_rates}, adapted from \cite{DeZotti2015Another}, the timescales of the number changing processes (in the figure, ``$dC$'' for double CS and ``$ff$'' for bremsstrahlung) are always much larger than that of CS (``$e\gamma$'' in the figure). This implies that they are the first ones to become inefficient with respect to the expansion. When this happens the thermalization process (``$th$'' in the figure) cannot be completed and the BB shape cannot be recovered following an energy injection (see App. A2 of [\hyperlink{I}{I}] for more details). 

As shown in the middle column of Fig. \ref{fig: SD_th_example_box}, one is thus left with an effective chemical potential and the resulting distortion with respect to the original BB is therefore referred to as $\mu$ distortion. Because of the fact that CS is still very efficient in this case (and kinetic equilibrium is always maintained), in analogy to the temperature shifts one can define
\begin{align}\label{eq: f_mu}
	f(x)=B\left(x+\mu\right) \simeq B(x) + \mu M(x) \,,
\end{align}
where $\mu=\mu_\gamma/T$ is the amplitude of the distortion and 
\begin{align}\label{eq: G1}
	M(x)= -\frac{G(x)}{x}
\end{align}
is its spectral shape. As a technical note, this definition of $M(x)$ is actually a superposition of the ``pure'' $\mu$ distortion component (that is to say a distortion that can be traced back to a non-zero $\mu$) and a temperature shift, since the BB temperature $T$ itself depends on~$\mu$. For this reason, the spectral shape is instead commonly expressed as (see e.g., [\hyperlink{I}{I}] for more details)
\begin{align}\label{eq: G2}
	M(x)= G(x)\left(0.4561-\frac{1}{x}\right)\,,
\end{align}
where the numerical coefficient is obtained by imposing that the $\mu$ distortion conserves the photon number density.\footnote{When CS is efficient and the number changing processes are not, i.e., in the $\mu$ distortion regime, one expects the distribution function $f(x)$ to conserve the photon number density, i.e., $\int x^2 M(x)\text{d}x=0$. Using the form of $M(x)$ given in Eq.  \eqref{eq: G1} would not satisfy this condition, while the addition of a constant as done in Eq.  \eqref{eq: G1} does (see e.g., Eq. (2.16) of [\hyperlink{I}{I}]).}

With the box expanding more and more, eventually, also CS will become sufficiently inefficient such that the Bose-Einstein distribution of Eq. \eqref{eq: f_mu} cannot be perfectly attained (``$BE$'' in Fig. \ref{fig: SD_th_rates}).\footnote{Note that here the relevant timescale is the Comptonization timescale $\tau_{BE}$ and not the CS timescale~$\tau_{e\gamma}$. The difference between the two is that while the latter refers to the collision rate, the former represents the ability of the process to restore the quasi-equilibrium spectrum (see e.g., \cite{DeZotti2015Another} for a related discussion). As clear from Fig. \ref{fig: SD_th_rates}, $\tau_{BE}$ exceeds the expansion timescale much earlier than $\tau_{e\gamma}$, which is why $y$ distortions form already before recombination.} Once this moment is reached, the photon bath will not be able anymore to perfectly redistribute the injected energy, leading to a so-called $y$ distortion (see right column of Fig. \ref{fig: SD_th_example_box}). This stage will last ever after. In this case, to define the shape of the distortion one needs to follow the Boltzmann equation (see [\hyperlink{I}{I}] and in particular App. A1 therein for a derivation)
\begin{align}
	\frac{\der{f}}{\der{t}}=\frac{\partial f}{\partial t} = C(f)\,,
\end{align}
where
\begin{align}
	C(f)=\frac{T_e}{\tau_{e\gamma}m_e}\frac{1}{x^2}\left[x^4\left(\frac{\partial f}{\partial x}+\frac{T_\gamma}{T_e}f(1+f)\right)\right]
\end{align}
is the collision term of the CS process with $\tau_{e\gamma}=(n_e\sigma_T)^{-1}$ as the CS timescale. This particular Boltzmann equation is known as Kompaneets equation \cite{Kompaneets1957Establishment}. In the limit of small distortions, i.e., for $f(x)\simeq B(x)$, one can analytically solve the Kompaneets equation to obtain
\begin{align}
	\Delta f \simeq y\,Y(x)
\end{align}
where 
\begin{align}
	Y(x)=G(x)\left(x\frac{e^x+1}{e^x-1}-4\right)
\end{align}
is the spectral shape of the $y$ distortion and
\begin{align}\label{eq: y}
	y=\Delta \tau \frac{T_e-T_\gamma}{m_e}
\end{align}
its amplitude. Here, the optical depth due to CSs $\Delta \tau=n_e\sigma_T\Delta l= n_e\sigma_T\Delta t$ can be interpreted as the reference timescale. Intuitively, one can understand the $y$ parameter as being proportional to the amount of energy given to/up by the photons (depending on whether $T_\gamma$ is smaller or larger than $T_e$, respectively, and normalized to the electron mass)  in units of the number of collisions (proportional to $\Delta \tau$). Large energy differences and multiple scatterings lead to the same result.

Since the transition between $\mu$ and $y$ distortions is not instantaneous, at the cross-over hybrid distortions get created that do not fall under either prescription \cite{Chluba2011Evolution, Khatri2012Beyond, Chluba2013Green}. Furthermore, once CS has become fully inefficient, non-thermal energy injections can remain ``frozen'' at the injection frequency \cite{Chluba2015Green}, behavior that would not be captured by any of the aforementioned distortion types. These will be referred to as residual distortions~$R(x)$. Under this class of SDs will also fall the Cosmological Recombination Radiation (CRR), as we will see in Sec. \ref{sec: SD_lcdm}.

We thus end up with three types of SDs (although temperature shifts are not formally a distortion) which can each be broken down into an amplitude ($g$, $\mu$ and $y$) and a spectral shape ($G(x)$, $M(x)$ and $Y(x)$) as well as some residual distortion. Their respective contribution depends on whether the relevant processes have already become inefficient or not. The total spectrum can then be expressed as a superposition of these SD types\footnote{When doing so, the SD amplitudes are usually normalized such that an amplitude equal to one induces a relative variation of the photon energy density equal to one [\hyperlink{I}{I}]. For this reason, here we implicitly employ the transformations $g\to4g$, $\mu\to\mu/1.401$ and $y\to4y$.}, i.e., as
\begin{align}\label{eq: SD_signal_tot}
	f(x)=B(x)+g\,G(x)+\mu\,M(x)+y\,Y(x)+R(x)\,.
\end{align}
The corresponding intensity $I(x)$ is given by $I(x)= 2 T_0^3 x^3f(x)\equiv I_0 f(x)$ and henceforth we will use $\mathcal{B}=I_0 \, B$, $\mathcal{G}=I_0 \, G$, $\mathcal{M}=I_0 \, M$ and $\mathcal{Y}=I_0 \, Y$. The SD signal will be referred to as $\Delta I (x) = I(x)-\mathcal{B}(x)$.

This picture, although simplified, is not far from the reality of the universe. In fact, at $z\gtrsim2\times10^6$ \cite{Burigana1991Formation, Hu1993ThermalizationI} (when $\tau_{th}\lesssim\tau_{exp}$ in Fig. \ref{fig: SD_th_rates}) we expect the same conditions assumed initially for the box to be fulfilled, i.e., for CS, double CS and bremsstrahlung to be efficient, and for every energy injection to lead to a temperature shift. After that time, the number changing processes cannot thermalize the medium anymore and $\mu$ distortions start to form. Around a redshift of $z\sim5\times10^4$ (when $\tau_{BE}\simeq\tau_{exp}$ in Fig. \ref{fig: SD_th_rates}) CS also gradually starts to lose efficiency and the $y$ era begins, lasting until today. 

\begin{figure*}[t]
	\centering
	\includegraphics[width=0.48\textwidth]{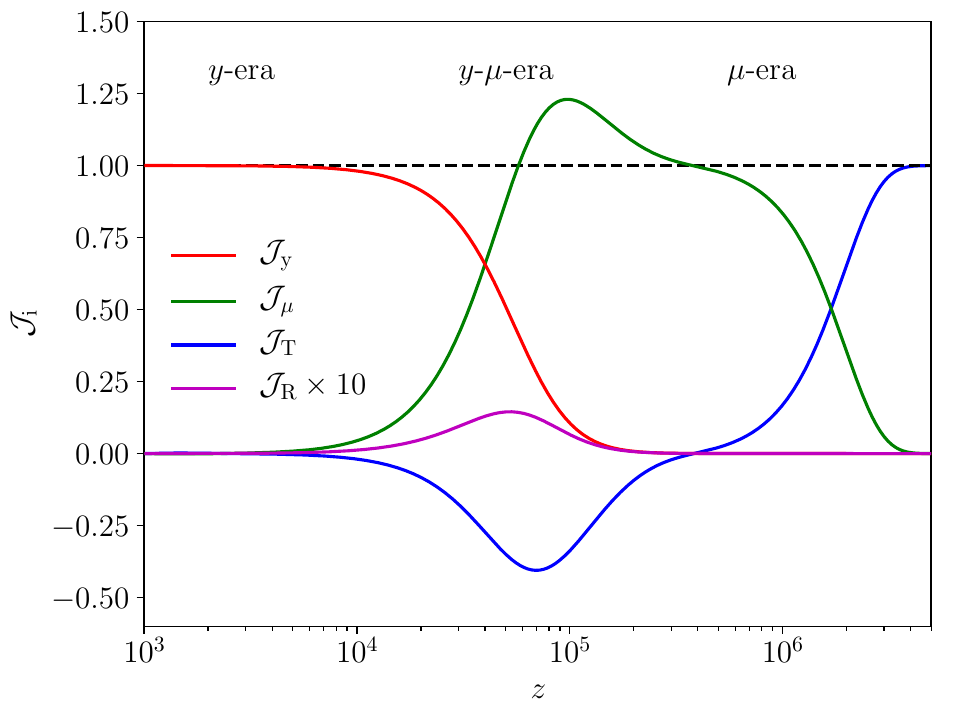}
	\includegraphics[width=0.48\textwidth]{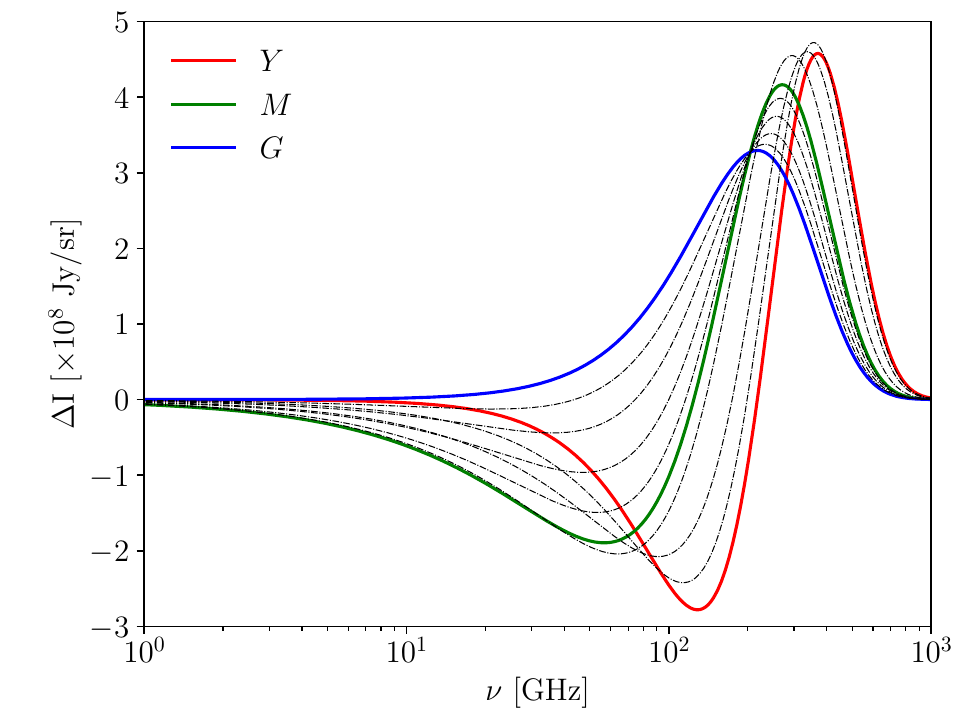}
	\caption{BRs (left) and spectral shapes (right) of the SD types. In the right panel the dashed black lines represent the shape of the SD signal in the transition phases between the three SD types. Figures taken from~[\protect\hyperlink{I}{I}].}
	\label{fig: SD_th_br_shapes}
\end{figure*}

One can thus define so-called Branching Ratios (BRs) $\mathcal{J}_i(z)$ for each distortion type determining when a given distortion forms. These are shown in the left panel of Fig.~\ref{fig: SD_th_br_shapes}, taken from [\hyperlink{I}{I}] (calculated according to the prescription discussed in Sec. \ref{sec: num}). Correspondingly, one can then intuitively see that the amplitude $a_i$ of a given SD type $i$ can be computed as
\begin{align}\label{eq: amplitude}
	a_i=\int \frac{\der{Q}/\der{z}}{\rho_\gamma}\mathcal{J}_i(z)\der{z}
\end{align}
for a given heating rate $\dot{Q}$.\footnote{Of course, $\dot{Q}$ also includes contributions from cooling processes, but for sake of succinctness we will just refer to it as heating rate implicitly assuming heating/cooling rate.} For instance, if the full energy injection happened during the $y$ era, the $y$ parameter would have a value proportional to the (integrated) amount of energy injected, while $\mu$ and $g$ would be zero. 

In turn, the BRs can be used to study the evolution of the SD shape. For instance, by assuming delta function-like energy injections at different redshifts one obtains the curves shown in the right panel of Fig.~\ref{fig: SD_th_br_shapes}. Starting from very high redshifts, these injections would lead solely to temperature shifts, whose shape is represented as a solid blue line in the figure (with the same color-coding as in the left panel). For redshifts lower than $z\simeq2\times10^6$ the spectrum starts to transition towards the $\mu$ era (dashed black lines between blue and green curves) until pure $\mu$ distortions are created (green solid line). At redshifts around $z\simeq5\times10^4$ the final transition to the $y$ era begins (dashed black lines  between green and red curves) leading to a pure $y$ shape (red solid line). From the figure, one can also note the important fact that the $y$ and $\mu$ shapes are very different, allowing to infer information about the time dependence of the injection depending, for instance, on where the peaks of the final spectrum lie.

Interestingly, the right panel of Fig.~\ref{fig: SD_th_br_shapes} can also be interpreted as tracking the evolution of the SD signal from the energy injection to the final temperature shift. In fact, coming back to the aforementioned analogy of the box, one can imagine the $y$ distortion to form immediately after the hand is placed at the side of the box. The $y$ shape then smoothly evolves towards the $\mu$ shape the more the scatterings redistribute the injected energy, as represented by the intermediate black dashed lines. Finally, if number changing processes are efficient the $\mu$ distortions becomes a temperature shift.

As a final note, the amplitude $g$ of the temperature shifts is determined by the difference between the true BB temperature $T_\gamma$ and the chosen reference temperature~$T_{\rm ref}$ (with $T_\gamma$ and $T_{\rm ref}$ coinciding at the earliest time). Since to date the only accurate measurement of $T_\gamma$ is from the observation by FIRAS of the BB spectrum today \cite{Fixsen1996Cosmic, Fixsen2009Temperature}, it is always possible to readjust the reference temperature to coincide with the observed one, so that the value of $g$ can always be fixed to zero. That is, however, only true as long as the only reference point is $z=0$. If, in fact, one had just as accurate measurements of $T_\gamma$ at the epochs of e.g., \matteo{BBN, recombination\cite{Ade2015PlanckXIII, Ivanov:2020mfr} and structure formation (see e.g., \cite{Hurier:2013ona} and references therein),} temperature shifts between different epochs could be measured. Nevertheless, since so far the experimental uncertainties are much larger than the predicted shifts we will just neglect the role of temperature shifts in the following discussion.

\vspace{0.3 cm}
\noindent \textbf{Follow-up idea 2}: To our knowledge no constraint on $T_{\gamma}$ from BBN data exists to date, although it should in principle be possible for a fixed value of $N_{\rm eff}$. Given the discussion above, having this information would be useful to determine the BB temperature at different epochs and constrain deviations from a universal power-law behavior of $T_\gamma$. Furthermore, also the analysis carried out in \cite{Ivanov:2020mfr} based on Planck 2018 data could be updated in combination with e.g., the latest ACT and SPT data.
\vspace{0.3 cm}

In summary, CMB SDs are deviations of the CMB energy spectrum from a pure BB shape and form when thermalizing processes become sufficiently inefficient. The total SD signal can be broken down into two informative components, $\mu$ and $y$ distortions, which form at redshifts $z\sim 10^4 -10^6$ and below, respectively. 

\subsection{Sources}\label{sec: sources}
In the previous section we have seen that all components of the SD signal can be decomposed in a spectral shape and an amplitude (aside from the residuals which we will come back to in Sec. \ref{sec: num}). The former are given by the respective deviations from a BB, and hence only depend on the particle physics processes at play. According to Eq. \eqref{eq: amplitude}, the latter can be further broken down into its dependence on the heating (via $\dot{Q}$) and cosmological (via the BRs) history. Since the spectral shapes and the BRs have already been introduced in the previous section, the only missing ingredient to determine the final SD spectrum is the heating history, which is the focus of this section.

Before diving into the different processes that might lead to heating, it is however important to discuss some generalities. First of all, one needs to distinguish two classes of sources of heating. Clearly, any direct injection of energy, via for instance the decay or annihilation of the DM into EM particles (see Sec. \ref{sec: SD_non_lcdm}), would heat up the medium and hence contribute to the heating rate. We will refer to this energy injection rate as $\dot{Q}_{\rm inj}$. However, it is also possible for photons to internally redistribute their energy or to exchange part of it with the baryons, via for instance the dissipation of acoustic waves or adiabatic cooling (see Sec. \ref{sec: SD_lcdm}). These effects do not involve any energy injection and will therefore be referred to as~$\dot{Q}_{\rm non-inj}$.

Focusing then more closely on energy injections, it is important to realize that the amount of injected energy does not necessarily coincide with the deposited energy, i.e., the energy injection effectively felt by the photon bath \cite{Slatyer2009CMB, Slatyer2013Energy, Slatyer2015IndirectII, Poulin2015Dark, Poulin2017Cosmological, Acharya2018Rich, Acharya2021CMB}. For instance, if part of the injected energy is in form of neutrinos that fraction of $\dot{Q}_{\rm inj}$ will not lead to the formation of SDs \cite[\hyperlink{IX}{IX}]{Slatyer2016General}. Furthermore, if for instance the energy injection happens after recombination, it might take some time for the injected photons to interact with the medium. During that time they will get redshifted and lose energy, modifying again $\dot{Q}_{\rm inj}$. These contributions are taken into account by the so-called deposition efficiency $f_{\rm eff}=f_{\rm em}(1-f_{\rm loss})$, where $f_{\rm em}$ is the fractional amount of $\dot{Q}_{\rm inj}$ in form of EM radiation and $f_{\rm loss}$ is the amount of energy lost between injection and deposition. Both $f_{\rm em}$ and $f_{\rm loss}$ are bound between 0 and 1. Additionally, if the energy is injected after recombination, not all of it will be spent to heat up the medium. In fact, it is possible for the EM radiation to ionize or excite the neutral hydrogen atoms or even for very low-energy photons that cannot initiate atomic reactions to be lost. One thus defines a so-called deposition fraction $\chi_c$ per channel $c$ with all contributions summing up to one at all times. Overall, one obtains the energy deposition rate per channel
\begin{align}\label{eq: Q_dep_1}
	\dot{Q}_{\text{dep},c}=\dot{Q}_{\rm inj}\,f_{\rm eff}\,\chi_c\,.
\end{align}

Before recombination, the density of the universe is high enough for scattering processes to be very frequent, so that $f_{\rm loss}$ can be neglected. This is known as the on-the-spot approximation, since the energy is deposited immediately. Furthermore, since the universe is completely ionized, the totality of the energy injection gets deposited in form of heat. One can thus simplify Eq. \eqref{eq: Q_dep_1} as
\begin{align}
	\dot{Q}_{\text{dep,tot}} = \dot{Q}_{\text{dep},h} = \dot{Q}_{\rm inj}\,f_{\rm em}
\end{align}
and obtain
\begin{align}
	\dot{Q}=\dot{Q}_{\rm inj}\,f_{\rm em}+\dot{Q}_{\rm non-inj}\,.
\end{align}
After recombination the picture is more complicated and one might need to resort to numerical tools such as \texttt{DarkAges} \cite{Stocker2018Exotic} and \texttt{DarkHistory} \cite{Liu2019Darkhistory} to calculate $f_{\rm loss}$ and~$\chi_c$. Nevertheless, in both cases various (semi-)analytical \cite{Chen2004Particle, Padmanabhan2005Detecting, AliHaimoud2017Cosmic} and tabulated \cite{Galli2013Systematic} approximations (with different degrees of precision) exist that can be employed without the need of the aforementioned advanced codes (see e.g., App. C1 of [\hyperlink{I}{I}] for a brief overview). When necessary, here we will employ the results of \cite{AliHaimoud2017Cosmic} for $f_{\rm loss}$ and Tab. V of \cite{Galli2013Systematic} for $\chi_c$. How these energy injections impact the thermal history of the universe beyond SDs has been qualitatively discussed in Sec. \ref{sec: lcdm} and has been detailed in e.g., Sec. 3.1 of [\hyperlink{I}{I}].

As a further disclaimer, unless stated otherwise in the reminder of the section we will only focus on the average SD signal when discussing the primordial SD signal, neglecting any anisotropies thereof (see e.g., most recently \cite{Chluba:2022xsd, Chluba:2022efq,Kite:2022eye} for a related discussion). This assumption is justified given that the latter are expected to be negligible within the \lcdm model \cite{Pajer:2012vz}. We will come back to the anisotropic signal in Sec. \ref{sec: res_PMFs}.

\subsubsection{Within \lcdm}\label{sec: SD_lcdm}
In this section, we will list all contributions to the heating history that are expected to exist within \lcdm. Assuming that the \lcdm model is correct in the relevant regimes, these effects are unavoidable and lead to a guaranteed SD signal. Related reviews can be found in e.g., \cite[\hyperlink{I}{I}]{Chluba2011Evolution, Chluba2013Distinguishing, Chluba2016Which}. 

\vspace{0.3 cm}
\noindent \textbf{Adiabatic cooling of baryons}

\noindent The evolution of the temperature of the baryonic fluid can be described by Eq.~\eqref{eq: T_b}. The interaction term implies that the colder baryons continuously extract energy from the photons at a rate
\begin{align}\label{eq: Q_b}
	\dot{Q}_b = \alpha_b \Gamma_{\rm int}(T_\gamma-T_b)\,,
\end{align}
where 
\begin{align}\label{eq:heat_capacity}
	\alpha_b  = \frac{3}{2} (n_\mathrm{H}+n_e+n_\mathrm{He}) =\frac{3}{2} n_\mathrm{H} (1+x_{e}+f_\mathrm{He})
\end{align}
is the heat capacity of the baryons with $f_{\rm He}=n_\mathrm{\rm He}/n_{\rm H}$.  In terms of the photons, this corresponds to a cooling rate $\dot{Q}_{\rm non-inj} = -\dot{Q}_b$. As explained in Sec. \ref{sec: LSS}, when the two fluids are coupled $T_b\simeq T_\gamma$ and hence $\dot{T}_b = -HT_{\gamma}$. By inserting this last result in Eq. \eqref{eq: T_b} one obtains
\begin{align}
	\Gamma_{\rm int}(T_\gamma-T_b) = HT_\gamma\,,
\end{align}
so that from Eq. \eqref{eq: Q_b} follows \cite{Seager2000Exactly, Chluba2005Spectral, Chluba2011Evolution}
\begin{align}\label{eq: heating_adiab_cool}
	\dot{Q}_{\rm non-inj}=-H\alpha_bT_{\gamma}\,.
\end{align}
After the two fluids decouple at $z \sim 200$ the cooling rate vanishes. 

\begin{figure}
	\centering
	\includegraphics[width=0.48\textwidth]{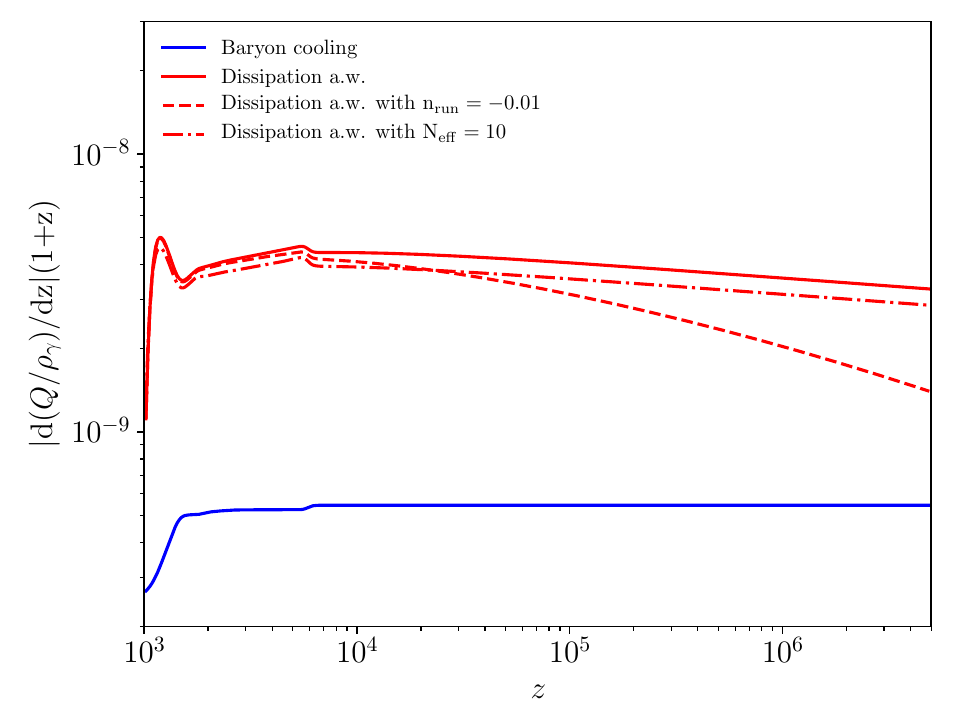}
	\includegraphics[width=0.48\textwidth]{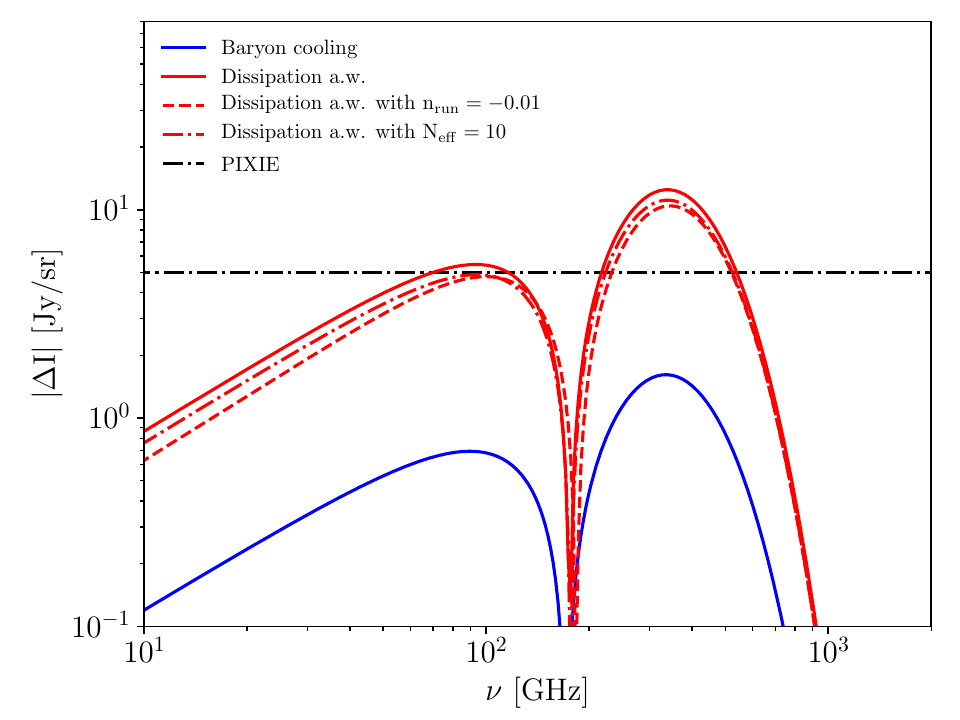}
	\caption{Heating rate (left) and SD signal (right) caused by the adiabatic cooling of baryons (blue) and by the dissipation of acoustic waves (red). The dashed and dash-dotted lines represent the impact of the running of the scalar spectral index ($n_{\rm run}$ in the figure) and of the number of relativistic degrees of freedom $N_{\rm eff}$. Figures taken from~[\protect\hyperlink{I}{I}].}
	\label{fig: SD_cool_diss}
\end{figure}

The evolution of the (absolute value of the) heating rate is shown in the left panel of Fig. \ref{fig: SD_cool_diss}, taken from [\hyperlink{I}{I}] (blue solid line). As expected from Eq. \eqref{eq: heating_adiab_cool} and clear from the figure, after recombination the heating rate (being proportional to $x_e$) significantly drops. The corresponding $\mu$ and $y$ parameters computed according to Eq.~\eqref{eq: amplitude} read $-3\times10^{-9}$ and $-5\times10^{-10}$, respectively, assuming the standard Planck values \cite{Aghanim2018PlanckVI} (see Sec. \ref{sec: CMB}). The absolute value of the cumulative SD signal calculated using Eq. \eqref{eq: SD_signal_tot} is displayed in the right panel of the same figure (the high-frequency peak is the negative one and the dip between the peaks corresponds to the zero-crossing). Since the main cosmological dependence of the heating rate is on $H\propto \Omega_m^{1/2}$ during the times at which the signal forms (assuming $T_0$ and hence $\Omega_\gamma$ to be known), the potential observation of the SD signal from the adiabatic cooling of baryons would allow to constrain $\Omega_m$. 

\vspace{0.3 cm}
\noindent\textbf{Dissipation of acoustic waves}

\noindent As discussed in Sec. \ref{sec: CMB}, the presence of density fluctuations in the pre-recombination era causes some regions of the universe to be hotter and denser than others. At scales below the photon mean free path, however, the photons can travel following the pressure gradient from the overdense to the underdense regions. This effect (known as diffusion damping or Silk damping~\cite{Silk1968Cosmic}) leads on the one hand to an isotropization of the photon field and on the other to a superposition of BBs with slightly different temperatures which creates SDs \cite{Sunyaev1970SmallI, Daly1991Spectral, Barrow199Primordial}.

Intuitively, one can imagine the resulting heating rate to be proportional to the (integrated) amount of damping, which in turn depends on the PPS $\mathcal{P}(k)$ and the damping scale \cite{Hu1995Wandering} (see Sec. \ref{sec: CMB})
\begin{align}\label{eq: kd}
	k_{d}=\frac{2\pi}{r_{d}}=2\pi\left[\int \mathrm{d}z \frac{c_{\rm s}^{2}}{2\dot{\tau}H} \left(\frac{R^2}{1+R}+\frac{16}{15}\right)\right]^{-1/2}~,
\end{align}
where $R=4\rho_\gamma/(3\rho_b)$ and $\dot{\tau}=n_e\sigma_T$. This is confirmed by the comprehensive calculation performed in~\cite{Khatri2012Mixing, Chluba2012CMB} (see also \cite{Chluba2015Spectral,Pitrou2019Radiative} for polarization corrections), which can be simplified to find \cite{Chluba2013CMB} (see also [\hyperlink{I}{I}])
\begin{align}\label{eq: heating_aw}
	\dot{Q}_{\rm non-inj}=4A^2\rho_{\gamma}\int \frac{\mathrm{d}k k^2}{2\pi^2} \mathcal{P}(k)k^2(\partial_t k_{d}^{-2})e^{-2(k/k_{d})^2}~,
\end{align}
where
\begin{align}
	A\simeq\left(1+\frac{4}{15}\frac{\rho_\nu}{\rho_r}\right)^{-1} 
\end{align}
(see \cite{Diacoumis2017Using} for a related discussion). The evolution of this heating rate is displayed in the left panel of Figure~\ref{fig: SD_cool_diss} (solid red line). Contrary to the case of the adiabatic cooling, the contribution of the dissipation of acoustic waves is positive as a result of the fact that in this case energy gets effectively dissipated into the system. For standard Planck values, the $y$ and $\mu$ parameters have values of approximately $4\times10^{-9}$ and $2\times10^{-8}$, respectively. The corresponding SD spectrum is shown in the right panel of Figure~\ref{fig: SD_cool_diss}. In the figure, the expected PIXIE sensitivity (see Sec. \ref{sec: future_exp}) is reported as horizontal dashed line to highlight the fact that a PIXIE-like mission \cite{Kogut2011Primordial} would in principle be sensitive enough to observe this contribution (as discussed in Sec. \ref{sec: exp}, however, the presence of (extra-)galactic foregrounds makes a direct measurement very challenging). 

As mentioned above, the SD signal from the dissipation of acoustic waves depends on the PPS, defined as in Eq. \eqref{eq: PPS_LCDM}, and the damping scales $k_{d}$, which in turn depends on the expansion rate $H$, the sound speed $c_s^2$, defined in Eq. \eqref{eq: cs2}, and the photon-to-baryon ratio $R$ according to Eq. \eqref{eq: kd}. In the context of the \lcdm model, the relation to the PPS allows to constrain its amplitude $A_s$ and tilt $n_s$. Intuitively, the more power is stored at small scales (i.e., the higher the values of the PPS parameters) the more energy will be dissipated, leading to a stronger SD signal.\footnote{\matteo{More technically, one can construct an \textit{effective} window function in $k$ space defining at which scales the dissipation of acoustic waves is sourcing SDs (see e.g., Eqs.~(24)-(25) of \cite{Chluba2013CMB} as well as Fig.~\ref{fig: GWs_window}). Since this in turn depends on the PPS, the shape of the latter directly determines the SD amplitudes.}} On the other hand, the expansion rate at the relevant times is entirely determined by $\Omega_m$ (assuming $T_0$ and hence $\Omega_\gamma$ to be known), while $c_s^2$ and $R$ allow to probe $\Omega_b$. The combined determination of $\Omega_m$ and $\Omega_b$ defines~$\Omega_{\rm cdm}$. As discussed in Sec. \ref{sec: CMB}, the higher the value of $\Omega_b$ (and hence of the number of electrons) the smaller the mean free path of the photons is going to be, which in turn reduces the amount of damping and dissipation. A similar dependence is also to be noted for the expansion rate and $\Omega_m$.

\begin{figure}[t]
	\centering
	\includegraphics[width=0.7\textwidth]{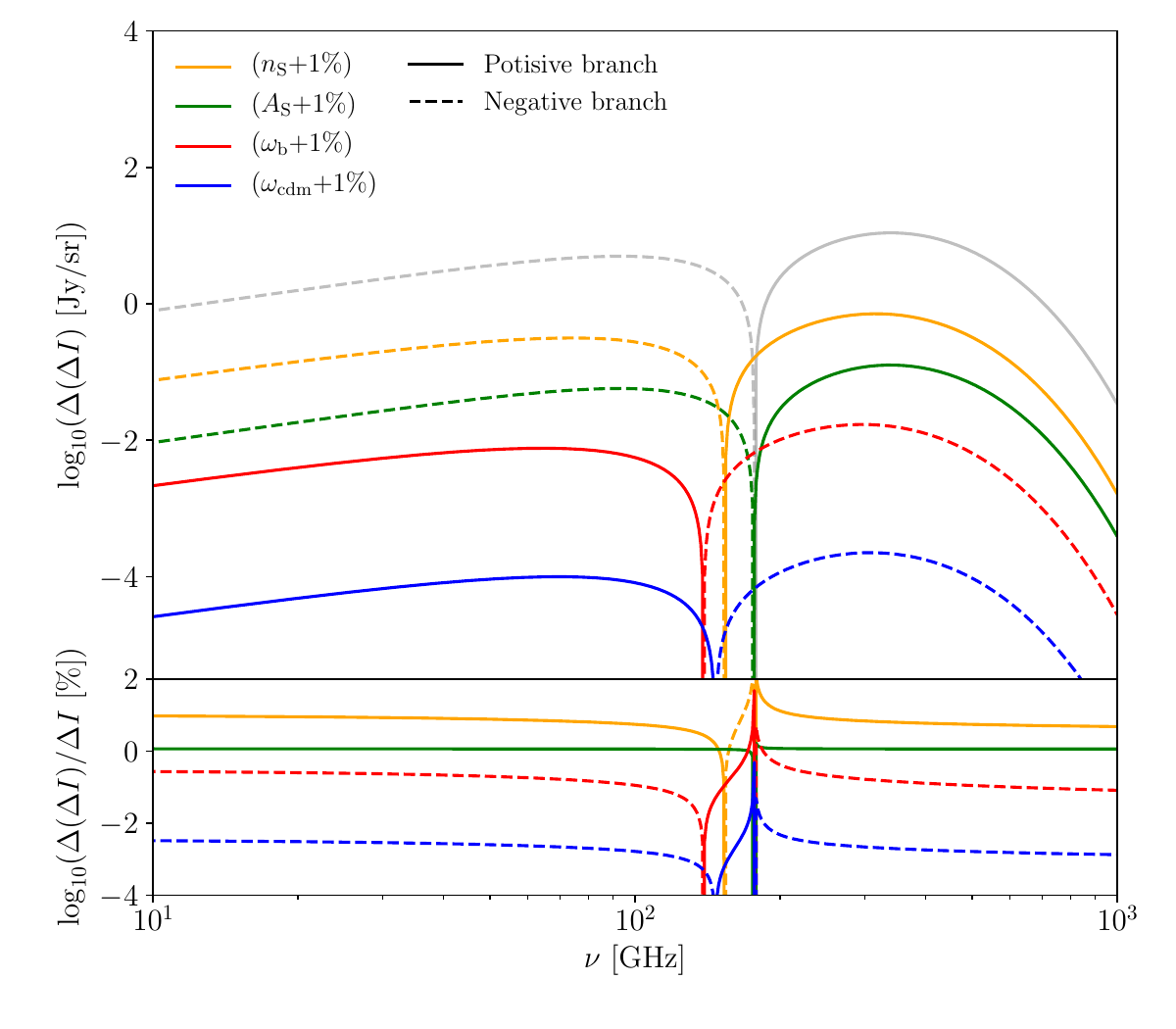}
	\caption{Absolute (top) and relative (bottom) impact of \lcdm parameter variations on the SD signal from baryon cooling and dissipation of acoustic waves. The gray line represents the fiducial spectrum, shown for reference. Figure taken from [\protect\hyperlink{III}{III}].}
	\label{fig: SD_cool_diss_dep}
\end{figure}

The impact of the variation of each of these parameters on the SD signal is shown in Fig. \ref{fig: SD_cool_diss_dep}, taken from [\hyperlink{III}{III}] (including also the sub-dominant contribution from the baryon cooling). As possible to see there, the largest effect is given by $n_s$, which induces variations of the order of $\Delta(\Delta I)/\Delta I \sim 10\Delta n_s/n_s$, followed by $A_s$ (directly determining the overall amplitude of the signal) and to a lower extent by the baryon and DM energy densities. The reason behind the big role of $n_s$ is that even a small change of this quantity (anchored at $k_*=0.05$~Mpc$^{-1}$, see Eq. \eqref{eq: PPS_LCDM}) strongly influences the amplitude of the PPS at small scales and hence the amount of damping at high redshifts.

More in general, the dependence on the PPS allows to constrain any inflationary model at scales of the order of $k\sim 1-10^4$ Mpc$^{-1}$ (see e.g., \cite[\hyperlink{III}{III},\,\hyperlink{V}{V}]{Chluba2012Inflaton, Pajer2012Hydrodynamical, Khatri2013Forecast, Chluba2013CMB, Chluba2014Teasing, Byrnes2018Steepest}), as we will extensively discuss in Sec. \ref{sec: res_infl}. This scales are perfectly complementary to those tested by the CMB anisotropies (see Sec. \ref{sec: CMB}). As an example, in Fig. \ref{fig: SD_cool_diss} we show the impact of adding a running of the spectral index $n_\mathrm{run}$ compatible with current limits (dashed red line). Furthermore, the presence of $\rho_\nu$ in the definition of the amplitude $A$ introduces a dependence of the signal on $N_\mathrm{eff}$ (via Eq. \eqref{eq: rho_nu}), which is however perfectly degenerate with the effect of $A_s$. This means that the signal from the dissipation of acoustic waves alone cannot constrain both parameters at the same time. Nevertheless, for reference we display in Fig. \ref{fig: SD_cool_diss} the impact of $N_\mathrm{eff}$ as dash-dotted red line. \matteo{Similarly, via the dependence of $\dot{Q}_{\rm non-inj}$ on $n_e$, the signal sourced by the dissipation of acoustic waves is also sensitive to variations of the helium abundance $Y_p$.}

As a final note, in Sec. \ref{sec: infl} we have briefly mentioned that inflation could source not only scalar but also tensor perturbations. Analogously to the discussion above, the latter can produce SDs as well \cite{Ota2014CMB, Chluba2015Spectral, Kite:2020uix}, although their contribution is expected to be subdominant due to the tight constraint of the tensor-to-scalar ratio $r_{0.002}\lesssim0.1$ \cite{Akrami2018PlanckX}. Even if only succinctly, the implications of this contribution will be discussed in more detail in Sec. \ref{sec: res_further}.

\vspace{0.3 cm}
\noindent \textbf{Cosmological Recombination Radiation}

\begin{figure}[t]
	\centering
	\includegraphics[width=0.75\textwidth]{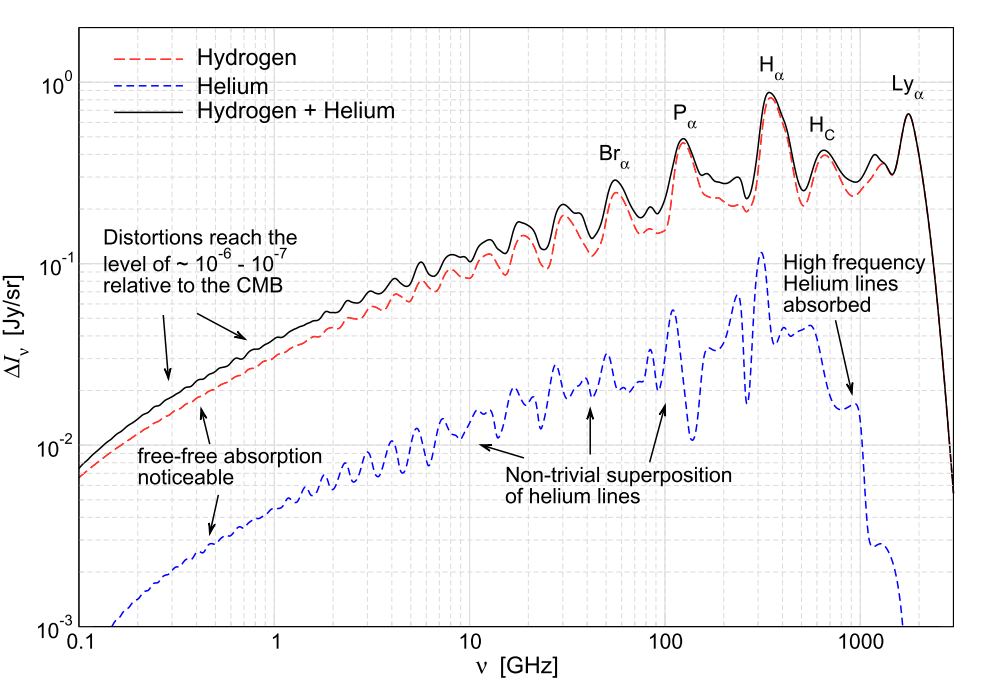}
	\caption{SD signal caused by the CRR. Figure taken from \cite{Hart2020Sensitivity}.}
	\label{fig: SD_CRR_spectrum}
\end{figure}

\noindent As helium and hydrogen recombinations evolve (see Sec.~\ref{sec: CMB}), photons get constantly emitted and absorbed, effect known as the cosmological recombination radiation (CRR) \cite[\hyperlink{XIV}{XIV}]{Zeldovich1969Interaction, Peebles1968Recombination, Dubrovich1975Hydrogen, Chluba2006Free, Rubino2008Lines, Chluba2008There, Chluba2009Pre, Sunyaev2009Signals, Glover2014Chapter}. This leads to a series of very peaked (in frequency) energy injections that cannot be thermalized because of the inefficient CS, resulting in residual SDs. The following SD spectrum is shown in Fig. \ref{fig: SD_CRR_spectrum}, taken from \cite{Hart2020Sensitivity}, whose intensity turns out to be about $6-7$ orders of magnitude lower than that of the CMB energy spectrum.

\newpage
Interestingly, from the figure it becomes clear that, although helium represents only a subdominant fraction of the total baryon abundance, its contribution can be sizable.\footnote{The main reason for this is that there have been two epochs of helium recombination, the first of which happening when the photon-baryon plasma was still in thermal equilibrium, so that the recombination process followed the Saha solution more closely. Furthermore, the number of photons related to helium atoms is enhanced by detailed radiative transfer effects and feedback processes \cite{Chluba2009Cosmological, Chluba2012HeRec}, following in more sharply peaked emission lines.} This means that, while for the aforementioned effects the time dependence of the heating rate could be inferred from the final signal via the different $\mu$ and $y$ spectral shapes, in the case of the CRR the different epochs of recombination could give access to the time dimension. This temporal information would bear important consequences for our understanding of the thermal history around the time of recombination \cite[\hyperlink{XIV}{XIV}]{Hart2020Sensitivity, Hart:2022agu} (see Sec. \ref{sec: app}).

\begin{figure}[t]
	\centering
	\includegraphics[width=0.75\textwidth]{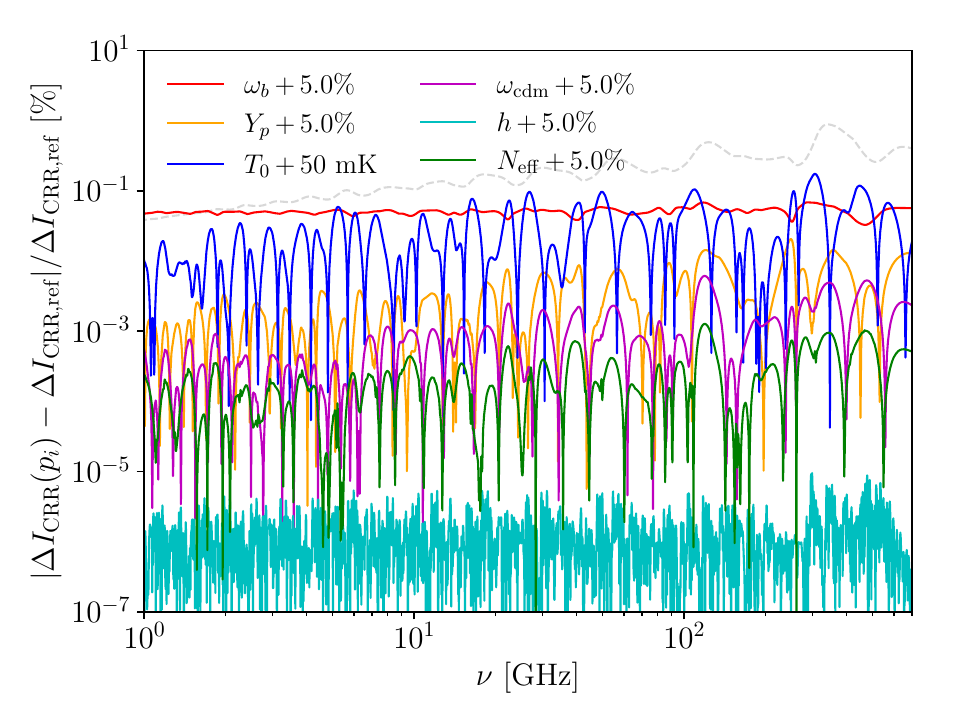}
	\caption{Same as in Fig. \ref{fig: SD_cool_diss_dep} but for the CRR. Figure taken from [\protect\hyperlink{XIV}{XIV}].}
	\label{fig: SD_CRR_dep}
\end{figure}

The CRR spectrum depends on the underlying cosmology in mainly three ways (see e.g., \cite[\protect\hyperlink{XIV}{XIV}]{Hart2020Sensitivity} for recent overviews). First of all, the amplitude of the spectrum is defined by the number of recombining atoms, and hence on the baryon and helium abundance, parameterized via $\omega_b\equiv\Omega_bh^2$ and $Y_p$. Secondly, the time of recombination, and hence the value of $T_0$, determines how much the photons have redshifted until today and hence the frequency (i.e., horizontal) dependence of the spectrum. Finally, the ratio of the atomic timescales to the expansion rate, determined by $N_{\rm eff}$ during RD and $\omega_m\equiv\Omega_m h^2$ during MD, influences the photon escape rate and thereby the relative amplitude of the peaks. As such, the CRR is sensitive to the \lcdm parameters\footnote{As a subtlety, in the parameterization with $\Omega_b$ and $\Omega_{\rm cdm}$ the spectrum would be proportional to $H_0$ as well, although this disappears once the physical energy densities $\Omega_bh^2$ and $\Omega_{\rm cdm}h^2$ are employed \cite[\hyperlink{XIV}{XIV}]{Hart2020Sensitivity}.} $\omega_b$, $T_0$ and $\omega_{\rm cdm}\equiv\Omega_{\rm cdm}h^2$ (employed here instead of $\omega_m$) as well as to $Y_p$ and $N_{\rm eff}$. The impact of these parameters on the CRR spectrum is graphically displayed in Fig. \ref{fig: SD_CRR_dep}, taken from [\hyperlink{XIV}{XIV}], from which it can be inferred that the strongest dependence is on $\omega_b$ and $T_0$. The aforementioned dependences could also manifest in beyond-\lcdm models, e.g., either by injections of energy \cite{Chluba2009Pre} or modifications to expansion history \cite{Hart:2022agu}.

Yet, this is assuming the recombination process happens homogeneously (or, equivalently, simultaneously) everywhere in the universe. Nevertheless, because of the density fluctuations introduced in Secs. \ref{sec: infl} and \ref{sec: CMB} this is not the case even within the \lcdm model and the sky-averaged CRR spectrum gets a second-order contributions proportional to the degree of the fluctuations.\footnote{In brief, the CRR spectrum can be expressed using a Taylor expansion up to second order in the difference $\Delta p$ between the value of a given parameter $p$ and its reference value $p_{\rm ref}$ (see Eq.~\eqref{eq: taylor} below). Assuming the presence of fluctuations in the parameter $p$, one can introduce a distribution function $P(p)$ normalized such that $\int P(p)\,\text{d}p=1$ and with mean value $\bar{p}=\int p\,P(p)\,\text{d}p$. To calculate the sky average of $\Delta I_{\rm CRR}$, one first of all needs to calculate the sky average of $\Delta p$, i.e., $\langle \Delta p \rangle=\int(p-p_{\rm ref})\,P(p)\,\text{d}p=\bar{p}-p_{\rm ref}$. This is equal to zero if the reference value is fixed to the mean value, i.e., $\langle \Delta p \rangle=0$ if $\bar{p}=p_{\rm ref}$, which can always be the case. So first-order contributions disappear in the sky average. For $\langle \Delta p^2 \rangle$ one has instead that $\langle \Delta p^2 \rangle=\langle p^2 \rangle-2\,\bar{p} \,p_{\rm ref}+p_{\rm ref}^2=\langle p^2 \rangle- \bar{p}^2=\langle p^2 - \bar{p}^2\rangle\neq0$. So second-order contributions survive, their amplitude depends on the free parameter $\langle \Delta p^2 \rangle$ and their spectral shape on the second-order derivative of the spectrum with respect to the given parameter $p$. In this way, starting from Eq.~\eqref{eq: taylor} one obtains~Eq.~\eqref{eq: DI_CRR}.} Concretely, in [\hyperlink{XIV}{XIV}] it was shown that the sky-averaged signal becomes
\begin{align}\label{eq: DI_CRR}
	\langle \Delta I_{\rm CRR} (p)\rangle 
	&\simeq \Delta I_{\rm CRR}(\bar{p}) 
	+ \frac{1}{2}\sum_{i,j}\frac{\partial^2 \Delta I_{\rm CRR}}{\partial \bar{p}_i\partial \bar{p}_i}
	\,\langle \Delta p_i \Delta p_j\rangle\,,
\end{align}
where 
\begin{align}\label{eq: params}
	p\equiv \{\omega_b,\, Y_p,\, T_0,\, \omega_{\rm cdm},\, N_{\rm eff}\}
\end{align}
is the set of parameters that can affect the CRR and $\Delta p_i$ is the difference of a given parameter (in e.g., the overdensities) from the respective reference value (i.e., the sky average) $\bar{p}$. As explained in the reference, the inhomogeneous contribution is nothing else but the second-order term of the Taylor expansion of the CRR spectrum around the sky-averaged spectrum. In the \lcdm model, these contributions are at the sub-percent level (and can be neglected for all missions considered in Sec.~\ref{sec: exp})~[\hyperlink{XIV}{XIV}], but beyond-\lcdm scenarios might predict more sizable overdensities, as we will see in Sec. \ref{sec: res_PMFs}.

\vspace{0.3 cm}
\noindent\textbf{Sunyaev-Zeldovich effect}

\begin{figure*}
	\centering
	\includegraphics[width=0.47\textwidth]{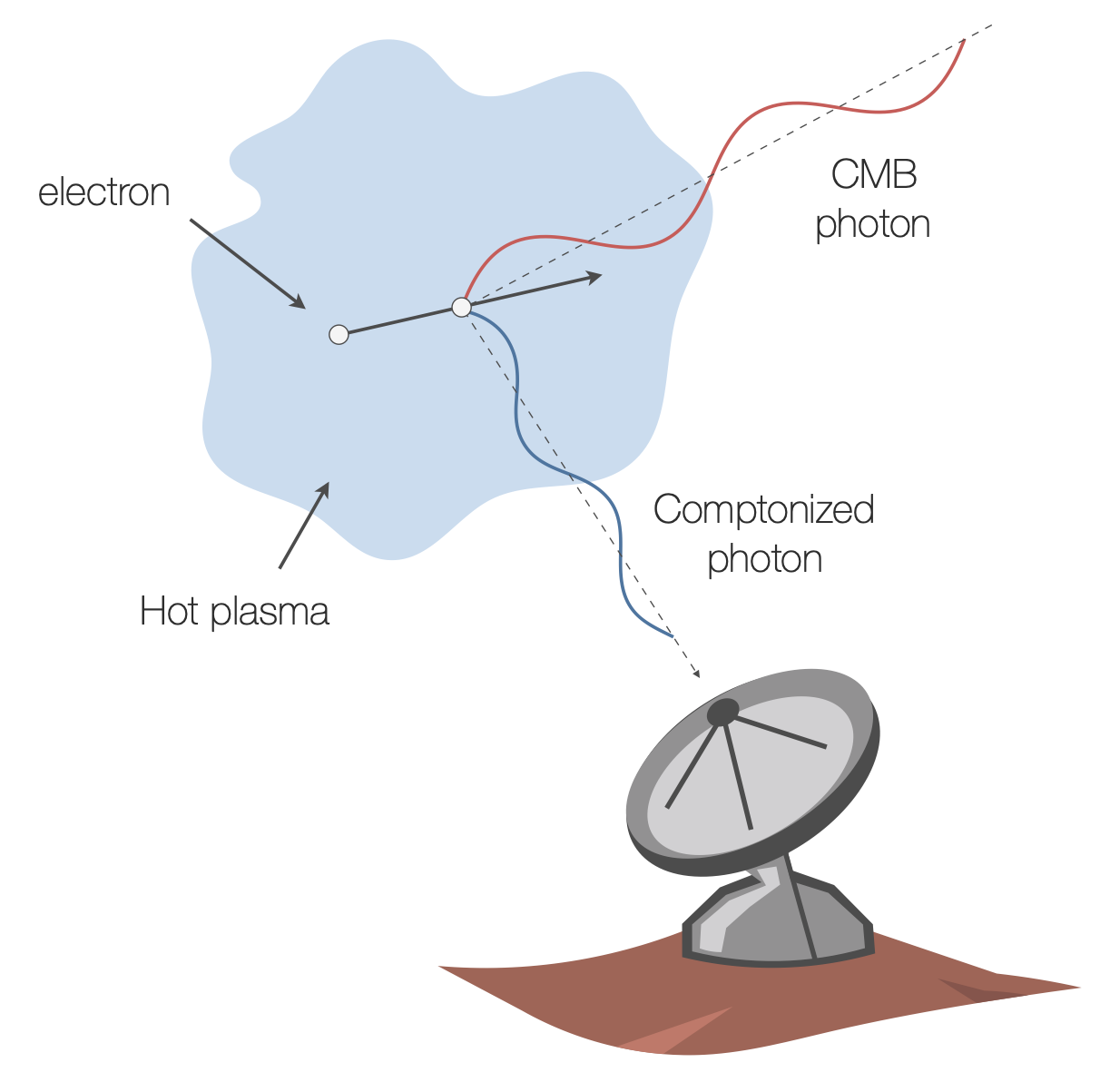}
	\includegraphics[width=0.49\textwidth]{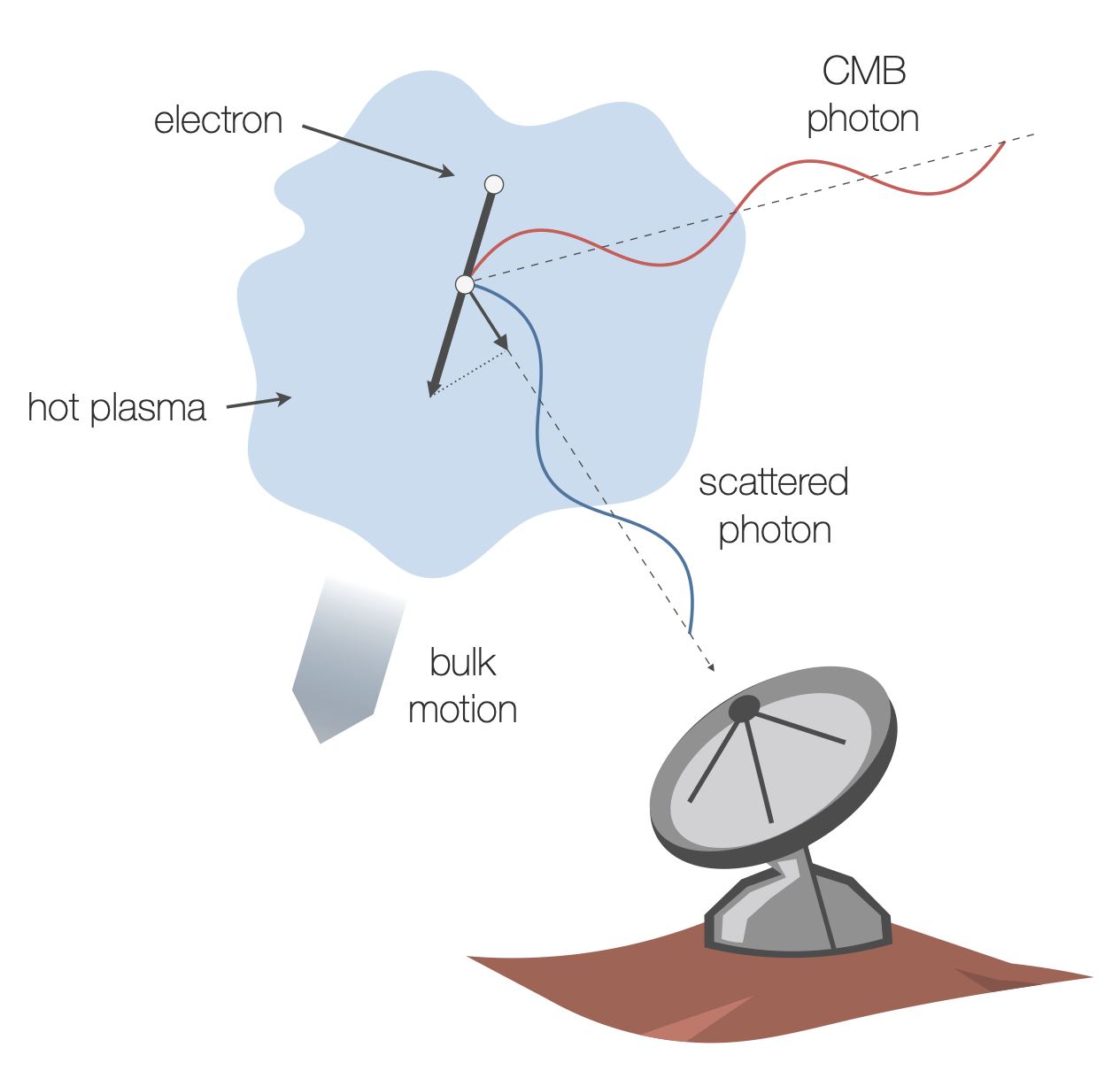}
	\caption{Illustration of the tSZ (left) and kSZ (right) effect. Figures adapted from \cite{Mroczkowski2019Astrophysics}.}
	\label{fig: SD_SZ}
\end{figure*}

\noindent So far, we have solely focused on sources of SDs in the pre-recombination era, i.e., the so-called primordial SDs. The dominant contribution to the final SD signal comes, however, from the late universe, via the Sunyaev-Zeldovich (SZ) effect \cite{Zeldovich1969Interaction} (see also e.g., \cite{Mroczkowski2019Astrophysics} for a recent review). The SZ effect refers to the inverse CS of CMB photons by the comparatively hot free electrons present in the intracluster and intergalactic medium as well as during the epoch of cosmic reionization (see Sec. \ref{sec: LSS}). As such, there are two main contributions to the SZ signal: the thermal and the kinematic SZ (tSZ and kSZ, respectively) effects. The former effect arises due to the thermal (i.e. following a Maxwell-Boltzmann distribution) electrons, while the latter due to the proper motion of the hosting environment along the LOS. Both effects are graphically represented in the left and right panels of Fig. \ref{fig: SD_SZ}, adapted from \cite{Mroczkowski2019Astrophysics}, respectively. The total SZ signal can therefore be decomposed in
\begin{align}\label{eq: Delta_I_reio}
	\Delta I_{\rm SZ}=\Delta I_{\rm tSZ}+\Delta I_{\rm kSZ}\,.
\end{align}

\begin{figure}[t]
	\centering
	\includegraphics[width=0.7\textwidth]{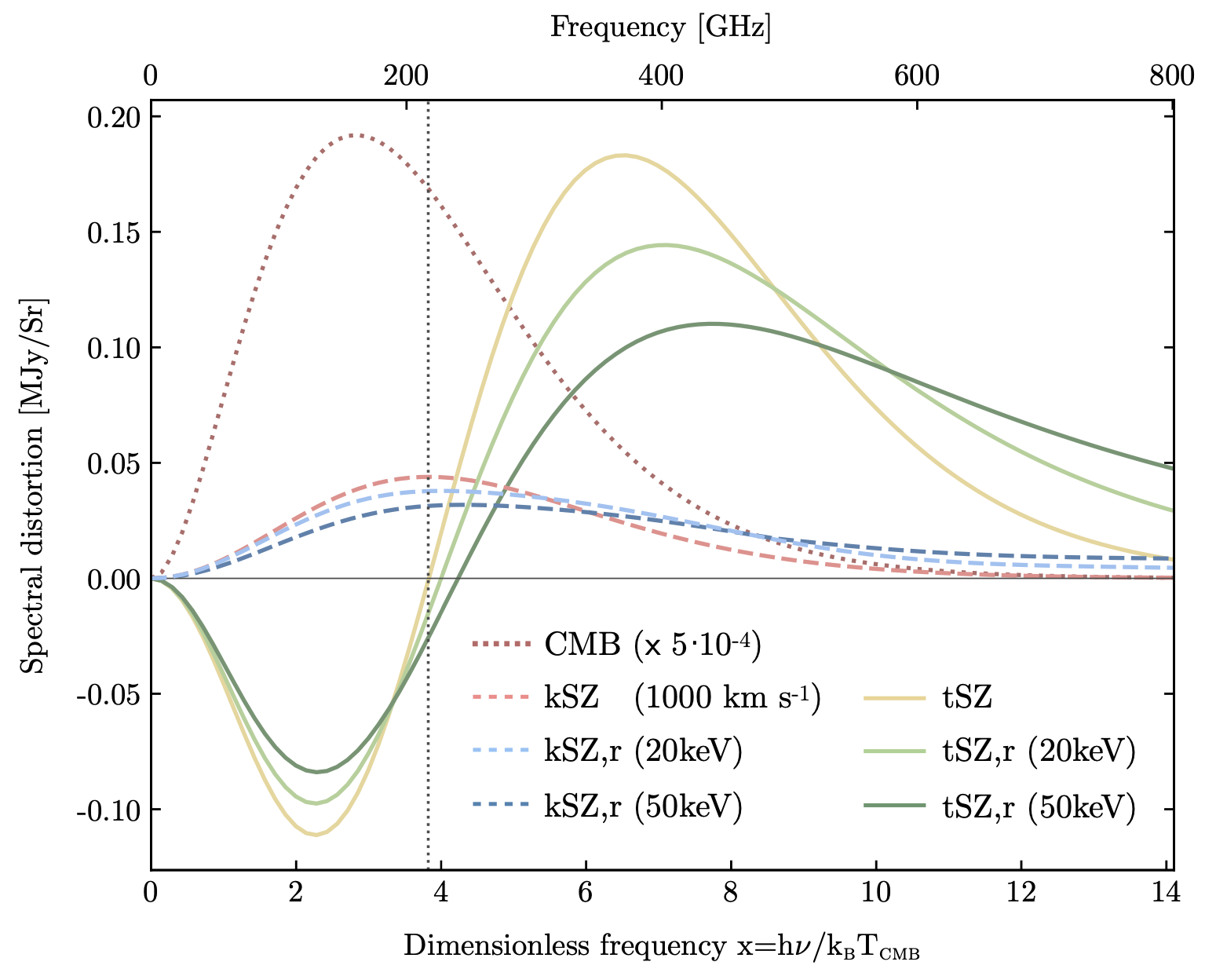}
	\caption{SD signal caused by the tSZ (solid) and kSZ (dashed) effects including relativistic corrections. Figure taken from \cite{Mroczkowski2019Astrophysics}.}
	\label{fig: SD_SZ_spectrum}
\end{figure}

Focusing first of all on the tSZ term, the thermal nature of the scatterings (in an epoch when CS is inefficient) leads to the formation of $y$-type distortions.\footnote{Non-thermal distributions can also exist, but will not be discussed here. The interested reader can find more information in e.g., Sec. 2.4 of \cite{Mroczkowski2019Astrophysics} and references therein.} Intuitively one can imagine the magnitude of the effect to be directly proportional to the temperature of the electrons $T_e$ in the given medium (see Eq.~\eqref{eq: y} with $T_e\gg T_\gamma$), for convenience often expressed in dimensionless units as $\theta_e=T_e/m_e$. Since $T_e$ has been determined to be of the order of $\sim1-10$~keV \cite{Hill2015Taking, Erler2017Planck, AtacamaCosmologyTelescope:2020wtv}, and hence $\theta_e\sim 10^{-2}-10^{-3}$, $\Delta I_{\rm tSZ}$ can be expanded in multipoles of $\theta_e$ \cite{Chluba2012Fast, Chluba2012Sunyaev} finding
that
\begin{align}
	\Delta I_{\rm tSZ}=y \, \mathcal{Y}+Y_{\rm rel}\,,
\end{align}
where $\mathcal{Y}$ is defined in Sec. \ref{sec: ped} and $y\simeq \Delta \tau \, \theta_e$ (from Eq. \eqref{eq: y}). Here $\Delta\tau$ is the scattering optical depth along the LOS in the cluster frame (a proxy for the number of collisions), which is typically of the order of $\Delta \tau\sim10^{-3}-10^{-4}$ \cite{Hill2015Taking, Vavagiakis:2021ilq, SPT-3G:2022zrq}. In terms of the $y$ parameter, one obtains $y\sim 10^{-6}-10^{-7}$ \cite{Hill2015Taking, Vavagiakis:2021ilq}, several orders of magnitude above the aforementioned contributions to the SD signal. Furthermore, for $\theta_e\sim10^{-2}$ the electrons have typical speeds of the order of $\beta\simeq\sqrt{3\theta_e}\simeq 0.2$, which makes the inclusion of relativistic corrections, captured in $Y_{\rm rel}$, necessary. Following the aforementioned asymptotic expansion one obtains \cite{Chluba2012Fast, Chluba2012Sunyaev}
\begin{align}\label{eq: Y_rel}
	Y_{\rm rel}= \sum_{k=1}^{k_{\rm max}}\Delta \tau\,\theta_{e}^{k+1}\,Y_k\,, \quad \text{with} \quad Y_k=\sum_{n=1}^{2k+2}a_{n}^{(k)}x^n\partial_{x}^{n}\mathcal{B}(x)\,.
\end{align}
The numerical coefficients $a_{n}^{(k)}$ are found in Table B1 of \cite{Chluba2012Fast}. The corresponding tSZ spectrum is shown in Fig. \ref{fig: SD_SZ_spectrum}, taken from \cite{Mroczkowski2019Astrophysics} (solid lines). The different displayed electron temperatures highlight the role of the relativistic corrections. The overall effect is to broaden and shift the signal towards higher frequencies.

On the other hand, since due to the bulk motion of the galaxy cluster, the kSZ effect imprints an overall temperature increase to the passing CMB photons, which is proportional to the peculiar velocity $\beta$ of the cluster and its direction with respect to the LOS, parameterized via the angle $\theta$. As such, instead of a $y$-type distortion it will lead to a localized (to the size of the galaxy cluster) temperature shift\footnote{Without complementary astrophysical information, the resulting effect is indistinguishable from the temperature anisotropies of the CMB.}. At leading order in $\beta$, $\Delta I_{\rm kSZ}$ can then simply be determined to be
\begin{align}
	\Delta I_{\rm kSZ} \simeq g\, \mathcal{G}+Y_{\rm rel}\,,
\end{align}
where $g=\Delta T/T\simeq\Delta \tau \,\beta \cos\theta$ is the amplitude of the signal and 
\begin{align}
	Y_{\rm rel}= \sum_{k=0}^{k_{\rm max}} \Delta \tau \, \beta \, \theta_e^{k+1} \, D_k^{\rm low}\,,
\end{align}
where $D_k^{\rm low}$ is given in App. A of \cite{Chluba2012Sunyaev}. Conceptually similarly to the $y$ parameter above (which depends on temperature instead of velocity), the value of $g$ can be interpreted as being proportional to the velocity of the galaxy cluster (and the alignment with respect to the LOS) or the number of scatterings.\footnote{The value of $g$ is positive or negative for LOS velocities pointing towards or away from the observer.} Since typical values of $\beta$ are of the order of $\sim10^{-3}$, $g$ is about one order of magnitude smaller than $y$ and higher-order corrections in $\beta$ to $\Delta I_{\rm kSZ}$ can be neglected \cite{Chluba2012Fast, Chluba2012Sunyaev}. The kSZ spectrum is shown in Fig.~\ref{fig: SD_SZ_spectrum} (dashed lines). The effect of the relativistic contribution is qualitatively similar to the tSZ case.

As a final note, we point out that so far we have only discussed the spectral information carried by the SZ effect and its link to the thermal history of the intergalactic medium as well as to the velocity distribution of galaxy clusters. Nevertheless, once the sources of the distortions have been localized, they can be used to perform number count analyses or (including also un-localized sources) to build the respective SZ power spectrum (see Sec.~\ref{sec: res_lcdm} and e.g., \cite{Bolliet:2018yaf, Bolliet:2020kwa} for concise pedagogical overviews). This allows to study the spatial distribution of the emitting objects on cosmological scales and determine the parameters describing matter power spectrum such as $\sigma_8$ and $\Omega_m$. We will touch upon the ramifications of these analyses in Sec.~\ref{sec: res_lcdm}.

\vspace{0.3 cm}
\noindent \textbf{CMB dipole}

\noindent The last effect that needs to be taken into account in \lcdm is the CMB dipole, which arises due to the proper motion of the earth. In fact, while the CMB is isotropic, through the relativistic Doppler effect one observes CMB photons to be blueshifted along earth's direction of motion or redshifted in the opposite direction. The superposition of these two BBs with different temperatures -- measured by FIRAS to be $3.381\pm0.007$ \matteo{mK} with respect to the sky average -- is not itself a BB and leads to the formation of $y$ distortions \cite{Zeldovich1972Effect, Chluba2004Superposition, Stebbins2007CMB}.\footnote{Intuitively, the superposition of two BBs with temperatures, say, $T_{1}$ and $T_2$ can be seen as a temperature shift with respect to a reference BB with $T_{\rm ref}$ (which could be either of the two BBs), so that at first order one has Eq. \eqref{eq: Df_T_shift} with $g=(T_i-T_{\rm ref})/T_{\rm ref}$. As summarized in e.g., App. B1 of [\hyperlink{I}{I}], by expanding Eq.~\eqref{eq: f_T_shift} at second order one obtains $\Delta f(x)=G(x)g(1+g)+Y(x)g^2/2$, which leads to the additional  formation of $y$ distortions.} The corresponding $y$ parameter has been determined to be \cite{Chluba2004Superposition} $y={(2.525\pm0.012)\times10^{-7}}$. On top of the dipole also the contribution from higher multipoles can be determined, but since it turns out to be of sub-percent level \cite{Chluba2004Superposition, Ade2015PlanckXIII} we will neglect it here. 

Being due a well-understood aberration of the signal, the CMB dipole contribution can be removed up to its uncertainty, of the order of $10^{-9}$. The final spectrum will however need to be marginalized over this residual unknown, masking part of the information that can be extracted from the aforementioned physical sources of SDs.  

\vspace{0.3 cm}
\noindent \textbf{Summary}

\begin{figure}[t]
	\centering
	\includegraphics[width=0.8\textwidth]{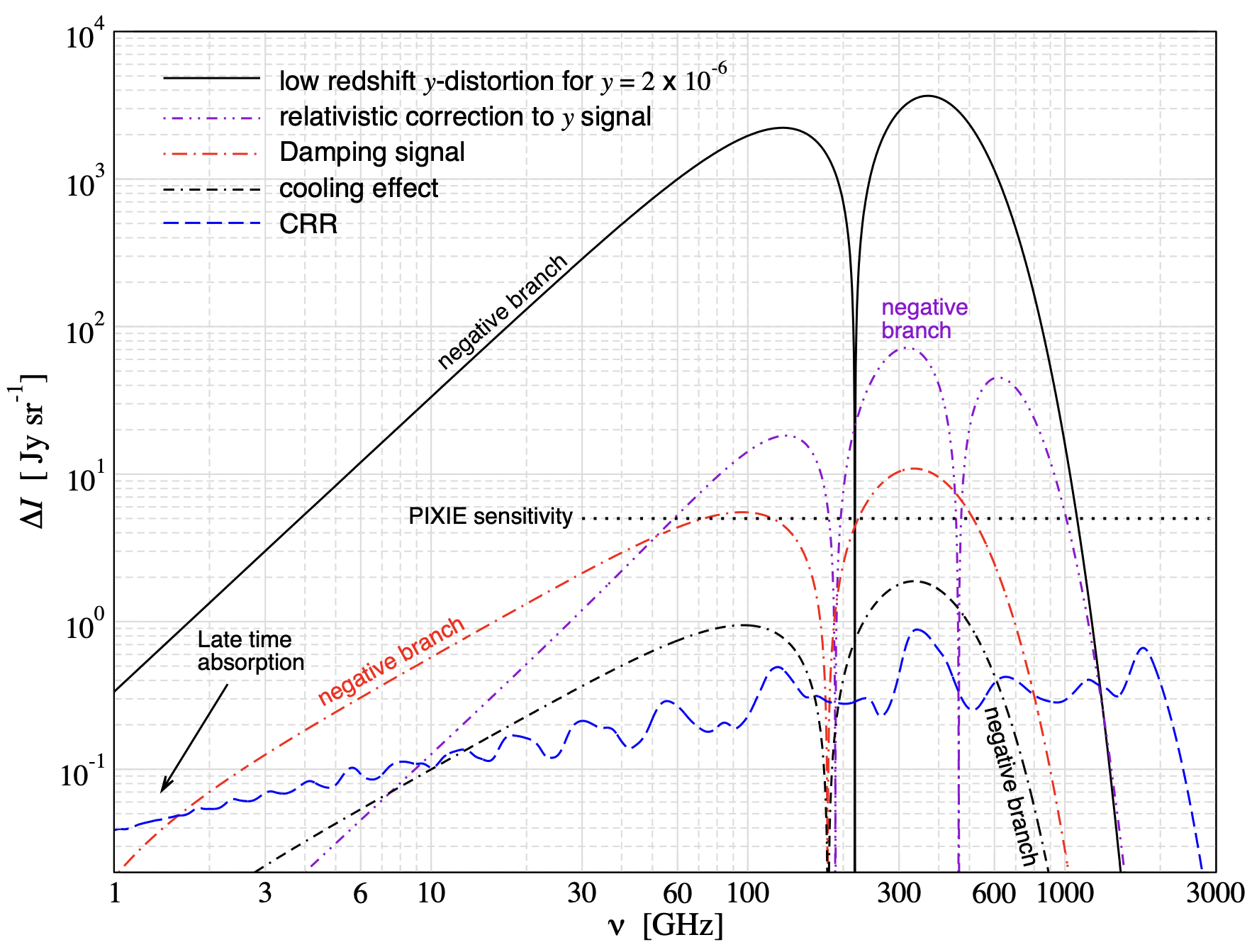}
	\caption{Complete collection of the \lcdm sources of SDs. Figure taken from \cite{Chluba2016Which}.}
	\label{fig: SD_LCDM_summary}
\end{figure}

\noindent In summary, even within the \lcdm model a number of effects are expected to lead to the formation of SDs, with very different amplitudes and spectral shapes. All combined, they allow to constrain the inflationary, expansion and thermal history of the universe at otherwise inaccessible scales. A summary plot comparing all of them is reported in Fig.~\ref{fig: SD_LCDM_summary}, taken from \cite{Chluba2016Which}.

\subsubsection{Beyond \lcdm}\label{sec: SD_non_lcdm}
In the previous section we have discussed several processes expected to source SDs within the \lcdm model. Via their general dependence on the inflationary, thermal and expansion history they can of course also be used to constrain several beyond-\lcdm scenarios, as we will see in Sec. \ref{sec: app}. There are, however, also other non-standard processes that might lead to the production of SDs whose impact cannot be captured by the aforementioned effects. These will be overview below in a general fashion, to be discussed on a model-by-model basis in Sec. \ref{sec: app}.

\vspace{0.3 cm}
\noindent \textbf{Non-standard energy injections}

\noindent The first, large class of examples is given by models predicting energy injections of photons (or EM radiation more in general), such as it could be the case for the decay \cite[\hyperlink{I}{I}]{Dimastrogiovanni:2015wvk, Poulin2017Cosmological, Acharya2019New, Chluba2020Thermalization, Bolliet:2020ofj}, annihilation \cite{Chluba2010Could, Chluba2013Distinguishing} and cooling (in the macroscopic limit) \cite{Kumar:2018rlf} of massive relics as well as the evaporation \cite[\hyperlink{I}{I}]{Carr2010New, Chluba2020Thermalization} and accretion of matter \cite{AliHaimoud2017Cosmic} onto Primordial Black Holes (PBHs). Intuitively, the related heating rate can be parameterized as
\begin{align}\label{eq: E_inj 1}
	\dot{Q}_{\rm inj} = n_{\rm \chi} L_{\chi}\,,
\end{align}
where $n_\chi=\rho_\chi/m_\chi$ represents the number density of sources $\chi$ of energy injection ($\rho_\chi$ and $m_\chi$ are the respective energy density and mass) and $L_{\chi}$ is the luminosity of the process (to be understood as the released energy per source). Expressed in its fractional form with respect to the DM energy density, i.e., $f_\chi=\rho_\chi/\rho_{\rm cdm}$, the rate above can be re-written as
\begin{align}\label{eq: E_inj}
	\dot{Q}_{\rm inj} = \rho_{\rm cdm}\, f_\chi\, \frac{L_\chi}{m_\chi}\,.
\end{align}
Depending on the considered model, the luminosity $L_\chi$ can assume various time-dependences (see Secs. \ref{sec: res_DM} and \ref{sec: res_PBH}).

\vspace{0.3 cm}
\noindent \textbf{Non-standard scattering processes}

\noindent Another way via which the presence of new physics might create SDs is if the particles involved were directly or indirectly coupled with the photons via scattering processes \cite{AliHaimoud2015Constraints, Ali-Haimoud:2021lka}. In fact, assuming for instance the presence of a non-relativistic new particle $\chi$ that interacts with the photons at a rate $\Gamma_{\chi\gamma}$, in analogy to the case of baryons (see Eq.~\eqref{eq: T_b}) its temperature would evolve as
\begin{align}\label{eq: T_chi}
	\dot{T}_\chi = -2HT_\chi+\Gamma_{\chi\gamma}(T_\gamma-T_\chi)\,.
\end{align}
The related heating rate due to the scatterings would be given by (see Eq.~\eqref{eq: heating_adiab_cool})
\begin{align}\label{eq: Q_chi}
	\dot{Q}_{\rm non-inj} = -H\alpha_\chi T_\gamma
\end{align}
in the tight-coupling regime and by
\begin{align}
	\dot{Q}_{\rm non-inj} = -\alpha_\chi \Gamma_{\chi\gamma}(T_\gamma-T_\chi)
\end{align}
afterwards, where $\alpha_\chi=(3/2)n_\chi$ is the heat capacity of the $\chi$ fluid. As shown in \cite{AliHaimoud2015Constraints, Ali-Haimoud:2021lka}, a similar discussion also applies to e.g., DM-electron and DM-nucleon interactions, since the baryons are themselves tightly coupled to the photons and can transfer the cooling.

\vspace{0.3 cm}
\noindent \textbf{Non-standard energy dissipation mechanisms}
\enlargethispage{\baselineskip}

\noindent Just as for the acoustic waves, also Primordial Magnetic Fields (PMFs) \cite{Subramanian:2015lua}, cosmic strings \cite{Vachaspati:2015cma} and primordial bubbles \cite{Deng:2017uwc}, if existing, would get dissipated in the early universe and generate SDs (see e.g., \cite{Subramanian:1997gi, Jedamzik:1999bm, Kunze:2013uja, Jedamzik:2018itu}, \cite{ostriker1987distortion, Tashiro:2012pp, Ramberg:2022irf} and \cite{Deng:2020pxo}, respectively). In fact, for instance, in close analogy with the case of the dissipation of acoustic waves, the heating rate due to the dissipation of PMFs can be determined to be \cite{Kunze:2013uja}
\begin{align}\label{eq: Q_PMFs}
	\dot{Q}_{\rm non-inj} = \frac{n_B+3}{2} \int \frac{\mathrm{d}k k^2}{2\pi^2} P_{B}(k)k_d^2(\partial_t k_{d}^{-2})e^{-2(k/k_{d})^2}\,,
\end{align} 
where $P_B=A_B \, k^{n_B}$ is the PMF power spectrum.

\vspace{0.3 cm}
\noindent \textbf{Non-standard photon mixings}

\noindent The last example we propose involves scenarios where photons can mix with Axion-like (ALPs) or light scalar particles \cite{Tashiro:2013yea, Ejlli:2013uda, Mukherjee2018Polarized} or with gravitons \cite{Dolgov:2013pwa} in the presence of a magnetic field. In the case of an ALP $a$, for instance, the conversion probability is given~by
\begin{align}\label{eq: P_ALP}
	P(\gamma\to a) = \sin^2(2\theta)\sin^2(\Delta_{\rm osc}s/2)\,,
\end{align}
where $\theta$ is the mixing angle, $\Delta_{\rm osc}$ is the (effective) mass difference between photon and axion and $s$ is the distance traveled by the photon. This is perfectly analogous to the case of neutrino oscillations. The corresponding SD of the CMB spectrum would then read
\begin{align}
	\Delta I = - \mathcal{B}\, P(\gamma\to a)\,,
\end{align}
where $\mathcal{B}$ has been defined in Sec.~\ref{sec: ped} (below Eq.~\eqref{eq: SD_signal_tot}), imprinting therefore a very characteristic oscillatory behavior (see e.g., Fig. 1 of \cite{Mukherjee2018Polarized} for a graphical representation). Furthermore, SDs produced by photon-ALP conversions would be polarized (due to the dependence on the alignment with respect to the magnetic field) which would also be a unique signature of this mixing.

\subsection{Numerical implementation}\label{sec: num}
Considering all aforementioned sources of CMB SDs, the cumulative spectrum can be broken down into
\begin{align}\label{eq: DI}
	\Delta I = \Delta I_{g,\mu,y} + \Delta I_{\rm CRR} + \Delta I_{\rm SZ}\,,
\end{align} 
where $\Delta I_{g,\mu,y}$ accounts for those contributions whose shape can be decomposed into $g$, $\mu$ and $y$ distortions (such as dissipation and cooling effects), $\Delta I_{\rm CRR}$ and $\Delta I_{\rm SZ}$ are given in Eqs. \eqref{eq: DI_CRR} and \eqref{eq: Delta_I_reio}, respectively. Below we will discuss how each of these components can be numerically evaluated. 

We will focus in particular on the implementation in the publicly-available, cosmological Boltzmann-solver \texttt{CLASS}, which has been carried out in [\hyperlink{I}{I},\,\hyperlink{XIV}{XIV}]. As we will see, the overarching theme behind the \texttt{CLASS} implementation is to employ accurate and fast approximations of exact solutions obtained with dedicated (slower) codes, so as to be able to efficiently perform data-based statistical analyses (see Sec. \ref{sec: num_2}). As such, \texttt{CLASS} and the original codes are very complementary, both in terms of development and scope. Furthermore, the \texttt{CLASS} implementation of SDs unlocks the synergy between this observable and all the others that can be computed by the code, such as the CMB anisotropy power spectra [\hyperlink{I}{I},\,\hyperlink{III}{III}].

\vspace{0.3 cm}
\noindent\textbf{Generalities of \texttt{CLASS}}

\noindent The Cosmic Linear Anisotropy Solving System (\texttt{CLASS}) \cite{Lesgourgues2011CosmicI, Blas2011Cosmic, Lesgourgues2011CosmicIV} is a \texttt{C} code designed for the computation of cosmological observables, such as the CMB anisotropy power spectra and the matter power spectrum, as well as other physical quantities, such as the free electron fraction and the PPS. It is particularly optimized for the solution of the Einstein-Boltzmann equations, i.e., the cosmological perturbations equations, and can model the evolution of a number of cosmological scenarios and particles.

\begin{figure}[t]
	\centering
	\includegraphics[width=0.6\textwidth]{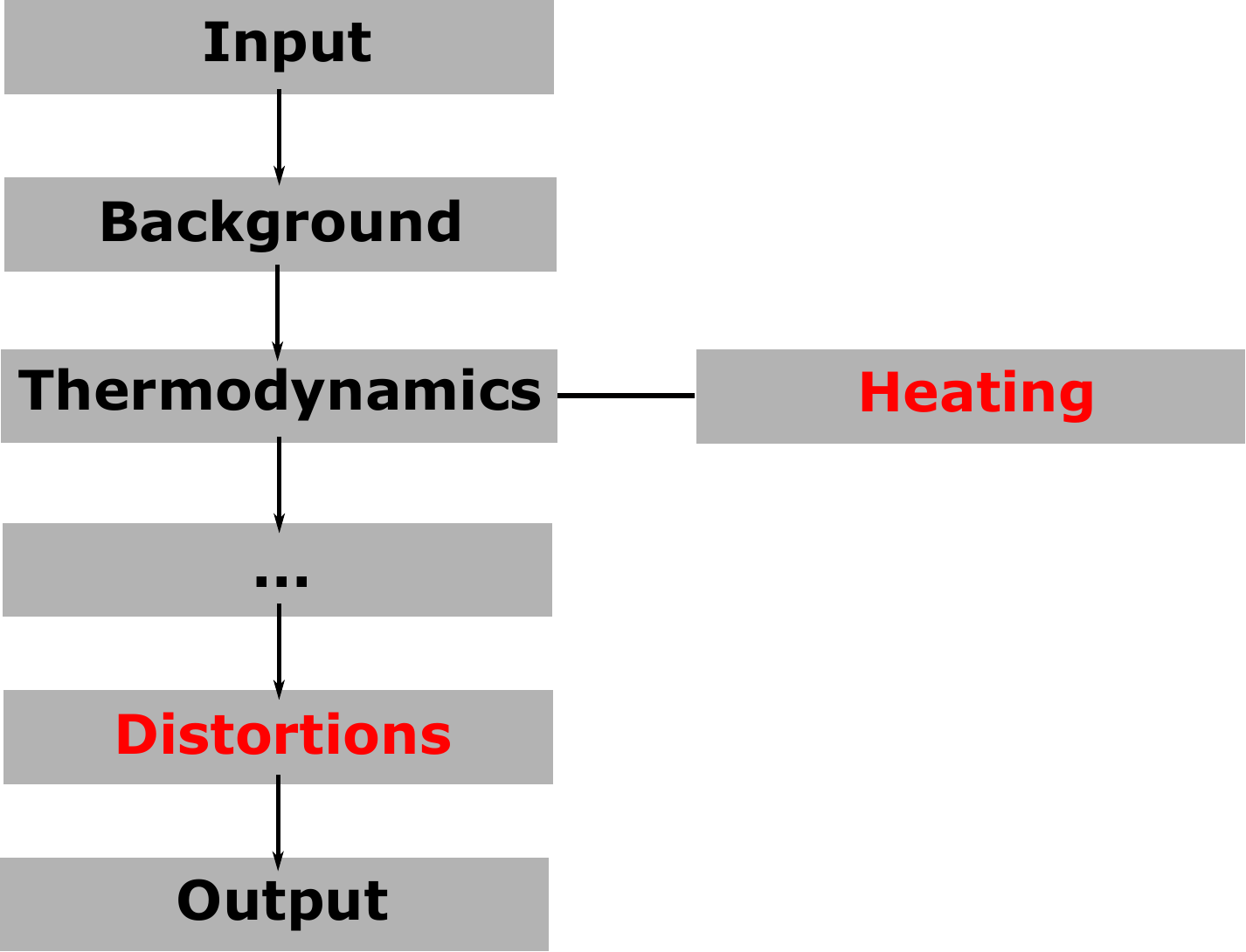}
	\vspace{0.2 cm}
	\caption{Graphical illustration of the (simplified) structure of \texttt{CLASS}. The modules in red have been created and developed in the course of the thesis. Figure taken from~[\protect\hyperlink{XXII}{XXII}].}
	\label{fig: CLASS_Overview}
\end{figure}

The code has a modular structure, which is graphically illustrated in Fig. \ref{fig: CLASS_Overview}, taken from [\hyperlink{XXIII}{XXIII}]. After having read and initialized the user-defined input parameters, it computes all relevant background quantities once (the fact that there is no redundant calculation is a peculiarity of the code) and stores them in the related structure for later use. It then computes in a similar way the thermodynamics of the universe, in particular the free electron fraction and the matter temperature evolution, which only rely on the background evolution. In the following modules, among others, the code calculates the PPS (either using the slow-roll approximation or for a given shape of the inflationary potential), evolves the perturbation equations and calculates the requested power spectra. If required, non-linear scales can also be modeled with different approaches. All outputs are saved in dedicated files. More detailed overviews can be found in the dedicated lecture notes at \cite{CLASS_website}.

The \texttt{CLASS} implementation of CMB SDs has been conducted in [\hyperlink{I}{I},\,\hyperlink{XIV}{XIV}], amounting to several modifications of existing modules and to the development of one entirely new module dedicated to CMB SDs. Among other minor restructurings of the input and background modules, the main update with respect to the previously available modules regards the thermodynamics, which now includes an appended structure dedicated to the calculation of energy injections (\texttt{injection.c}). This accounts for several approximations of the deposition efficiency and deposition fraction, as discussed in Sec.~\ref{sec: sources}, as well as many of the energy injection processes outlined in Sec. \ref{sec: SD_non_lcdm}. It has been designed to facilitate the inclusion of further similar effects~(see e.g., [\hyperlink{XI}{XI},\,\hyperlink{XII}{XII},\,\hyperlink{XV}{XV}] for practical examples). 

\vspace{0.3 cm}
\noindent \textbf{Follow-up idea 3}: Implement the possibility to use dedicated codes such as \texttt{DarkAges} and \texttt{DarkHistory} (see Sec. \ref{sec: sources}) for a more refined calculation of the deposition function.  
\vspace{0.3 cm}

The new \texttt{distortions.c} module has been added at the very end of the chain of modules and its implementation follows the prescription outlined below. Of note, the sources of ``non-injected'' heating have been grouped in a separate appendix, \texttt{noninjection.c}, in analogy to the \texttt{injection.c} module. So far, the appendix includes all \lcdm sources of SDs that fall under the classification $\Delta I_{g,\mu,y}$ and its extension with beyond-\lcdm sources (see Sec. \ref{sec: SD_non_lcdm}) would be straightforward. The other contributions are calculated directly in the main module.

\vspace{0.3 cm}
\noindent\textbf{$g$, $\mu$ and $y$ contribution}

\noindent According to Eq. \eqref{eq: SD_signal_tot}, $\Delta I_{g,\mu,y}$ can be decomposed in
\begin{equation}\label{eq: DI 1}
	\Delta I_{g,\mu,y} = g\, \mathcal{G} + \mu\, \mathcal{M} + y\,\mathcal{Y} + R
\end{equation}
and there are in principle two ways to calculate it for a given heating history. The first, straightforward approach is to implement all the necessary Boltzmann equations and solve the thermalization problem directly, from $z\simeq10^6$ to today. Although time-consuming, this is possible with the \texttt{CosmoTherm} code \cite{Chluba2011Evolution}, which is however currently not publicly available, and more recently with the publicly-available \texttt{DarkHistory} code \cite{Liu:2023nct}, which however only focuses on $z\lesssim3000$. The second approach, relies on the idea that the thermalization problem can be linearized \cite{Chluba2013Green, Chluba2015Green} (see in particular Sec. 4.1.3 of \cite{Schoneberg:2021uak} for more details)\footnote{In brief, the photon distribution function can be decomposed as in Eq. \eqref{eq: f_decomp} to isolate the role of SDs via $\Delta f(x)$. In this way, $f(x)$ becomes linear in $\Delta f(x)$, which is only true as long as ${\Delta f(x)\ll1}$, as it is the case here. When written as differential equation to follow its time-evolution (see e.g., Eq.~(4.24) of \cite{Schoneberg:2021uak}), it becomes a linear differential equation. As concisely explained in App.~D1 of [\hyperlink{I}{I}], any linear differential equation of the type $\mathcal{D}y(x)=x(x)$, where $\mathcal{D}$ is a given differential operator, can be solved to find $y(x)=\int G(x,x')x(x')dx'$, where $G(x,x')$ is the Green's function of the problem. Neglecting for simplicity the frequency dependence of the Green's function (see \cite{Chluba2015Green, Schoneberg:2021uak} for generalizations), $\Delta f(x)$ can then be written as $\Delta f(x,z)=\int G(x,z,z')S(z')dz'$, where $G(x,z,z')$ can be found by solving Eq.~(4.40) of \cite{Schoneberg:2021uak} and $S(z')$ is a given source term. Since here we are only interested in the SD spectrum today, we can fix $z=0$ and write $G(x,z')=G(x,0,z')$.} so that $\Delta I_{g,\mu,y}$ can be expressed as 
\begin{equation}\label{eq: DI 2}
	\Delta I_{g,\mu,y} = \int^{\infty}_{0} \id z \, G_{\rm th}(z)\, \frac{\id Q(z)/\id z}{\rho_{\gamma}(z)}\,,
\end{equation} 
where $G_{\rm th}$ is the Green's function of the (linear) thermalization problem and evolves the distortion caused by an heating at a redshift $z$ to today. This formulation of $\Delta I_{g,\mu,y}$ is very useful as it allows to completely decouple the cosmological model-independent thermalization problem from the cosmological model-dependent heating history. This means that, while $\dot{Q}$ can vary depending on which cosmological model is assumed, $G_{\rm th}$ is largely independent of it and can be in principle pre-computed once and for all. This second avenue to calculate $\Delta I_{g,\mu,y}$ is therefore significantly faster than the direct one and it is the one followed in \texttt{CLASS}. 

To find $\Delta I_{g,\mu,y}$ we thus need to determine $G_{\rm th}$ and $\dot{Q}$. In terms of $\dot{Q}$, both the baryon cooling and the dissipation of acoustic waves (see Sec. \ref{sec: SD_lcdm}) have been implemented in \texttt{CLASS} [\hyperlink{I}{I}]. Several beyond-\lcdm models have also been included as representative examples [\hyperlink{I}{I}] (largely based on the previous implementation of the same processes in the \texttt{ExoCLASS} version of \texttt{CLASS} \cite{Stocker2018Exotic}), namely the annihilation and decay of DM and the evaporation and accretion of matter onto PBHs (see Sec. \ref{sec: SD_non_lcdm}). The addition of all other processes discussed in Sec. \ref{sec: SD_non_lcdm} is relatively straightforward and left for future work (see follow-up ideas in Sec.~\ref{sec: app}).


Moreover, similarly as before, there are two possibilities to find $G_{\rm th}$. The first, which has been performed in \cite{Chluba2013Green}, is to calculate $G_{\rm th}$ by computing the response of the plasma to $\delta$-like heating terms approximated as narrow Gaussian peaks. The resulting tabulated $G_{\rm th}$ function has been made publicly available at \cite{Chluba_Greens} and its sole knowledge allows to calculate $\Delta I_{g,\mu,y}$ for a given heating history. One can, however, employ a more physically intuitive form of $G_{\rm th}$ using the BRs $\mathcal{J}_i$ introduced in Sec. \ref{sec: ped} and this is the approach adopted in \texttt{CLASS}. In fact, by inserting Eq. \eqref{eq: amplitude} in Eq. \eqref{eq: DI 1}, one obtains Eq. \eqref{eq: DI 2} with 
\begin{equation}\label{eq: Greens function 1}
	G_{\rm th}(z) =  \mathcal{J}_g(z)\,\mathcal{G} + \mathcal{J}_\mu(z) \,  \mathcal{M} + \mathcal{J}_y(z) \, \mathcal{Y} + R(z)\,.
\end{equation}
The BRs can then be determined in multiple ways, with increasing levels of accuracy (see App.~B3 of [\hyperlink{I}{I}] for a complete list). For instance, based on the discussion had in Sec.~\ref{sec: ped} one could naively assume $\mathcal{J}_T(z) = 1$ for $z \geq z_{\rm th}$, $\mathcal{J}_\mu(z) = 1$ for $z_{\mu y} \leq z \leq z_{\rm th}$, $\mathcal{J}_y(z) = 1$ for $z \leq z_{\mu y}$ and otherwise all functions to be zero, where $z_{\rm th}=2\times10^6$ and $ z_{\mu y}=5\times10^4$. 

An exact form of the BRs, however, exists and derives the BRs directly from the exact known form of $G_{\rm th}$ using a Gram-Schmidt orthogonalization procedure \cite{Chluba2014Teasing}. In simpler terms, one performs a least-square fit of $G_{\rm th}$ (discretized in frequency space) based on the $\mathcal{G}$, $\mathcal{M}$ and $\mathcal{Y}$ shapes, and the time-dependent coefficients are the BRs (the residuals define $R$). The resulting BRs are shown in the left panel of Fig.~\ref{fig: SD_th_br_shapes} (for a PIXIE-like mission, see below). In this case, the final value of $\Delta I_{g,\mu,y}$ computed with the tabulated form of $G_{\rm th}$ or with the BR decomposition is identical. The latter approach has however the advantage that it allows to calculate the $y$ and $\mu$ parameters using Eq. \eqref{eq: amplitude}, which are often useful for the characterization of the SD signal. 

\begin{figure}
	\centering
	\includegraphics[width=0.48\textwidth]{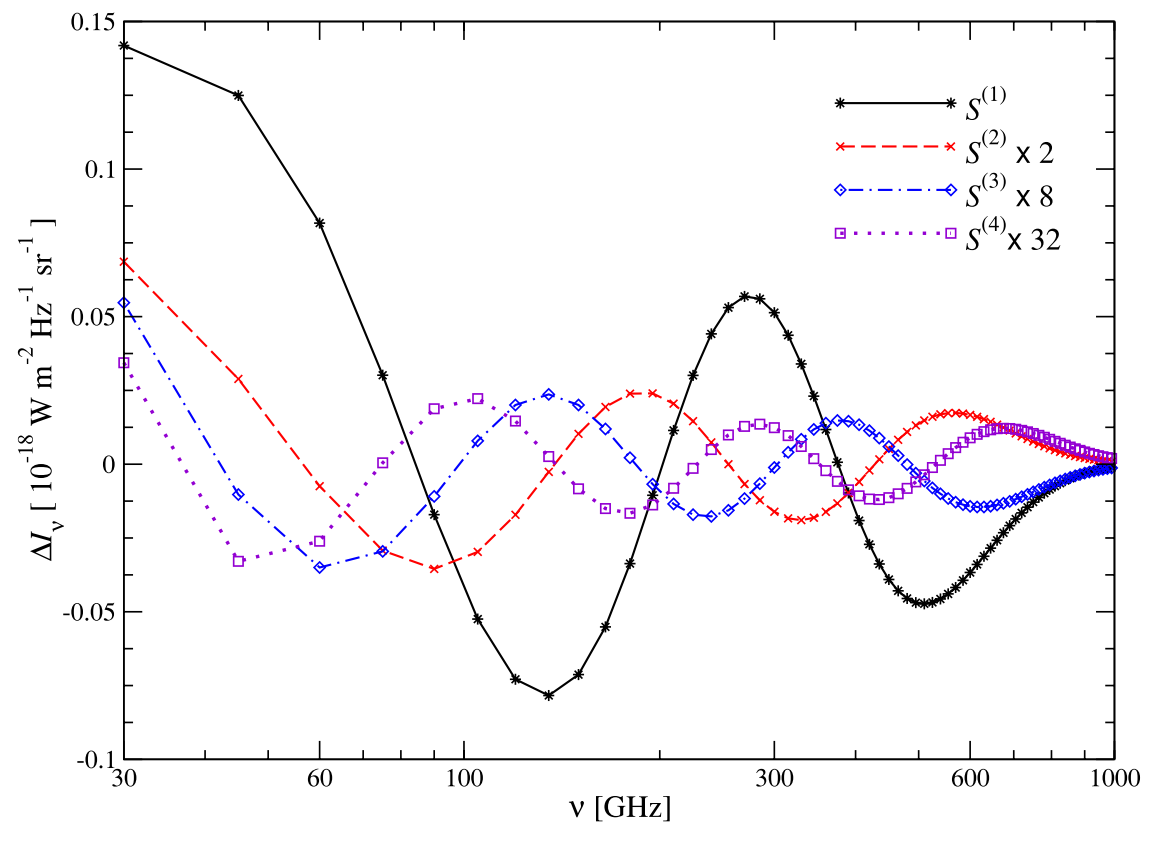}	\includegraphics[width=0.48\textwidth]{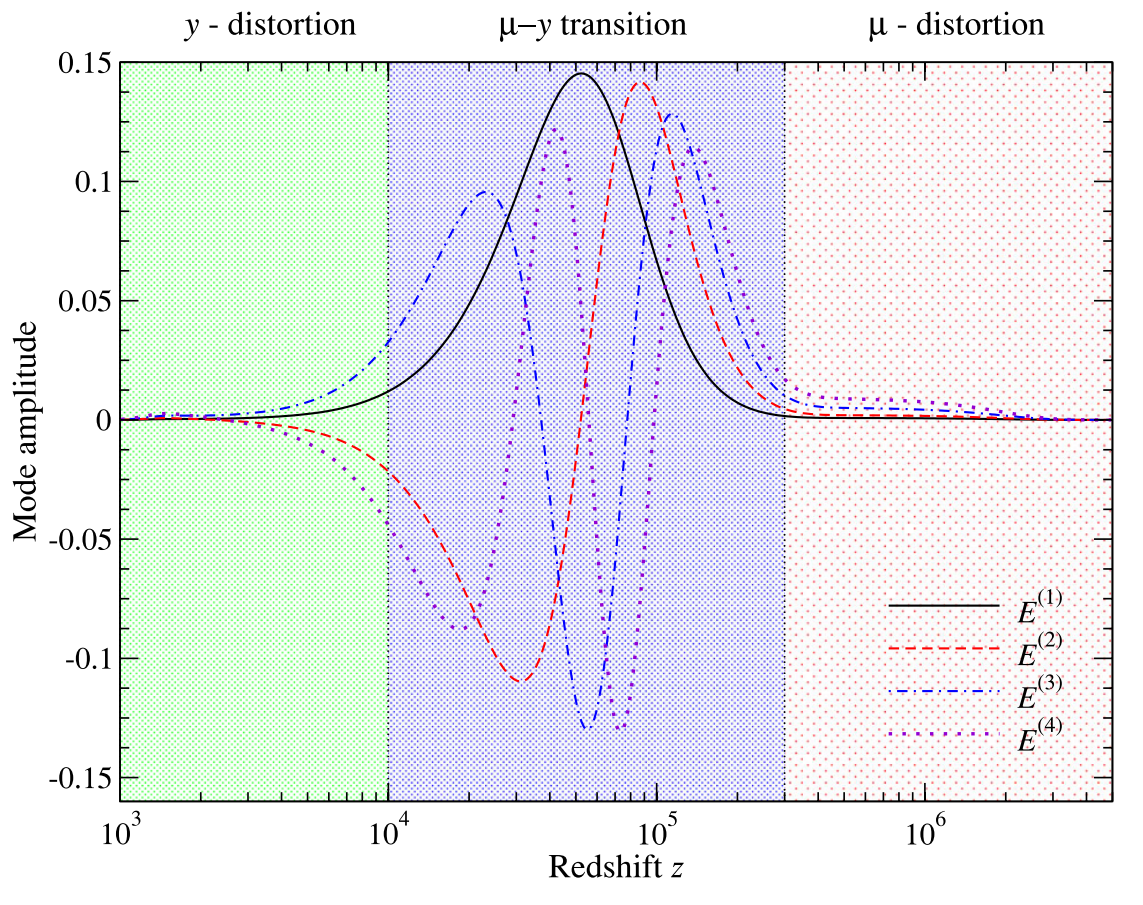}
	\caption{PCA spectral shapes of the residual distortion signal (left) and eigenvectors of the Fisher matrix (right). Figures taken from \cite{Chluba2014Teasing}.}
	\label{fig: SD_PCA}
\end{figure}

Furthermore, the definition of the residuals in the second approach (the BR decomposition) can be a very valuable information. In fact, as shown in \cite{Chluba2014Teasing}, $\Delta I_{\rm R}$ can be decomposed using a Principal Component Analysis (PCA) as
\begin{equation}\label{eq: DI residuals 2}
	\Delta I_{\rm R}(x_i)\simeq\sum_k \mu^{(k)} S^{(k)}(x_i)\,.
\end{equation}
Here $k$ represents the eigenmode of the decomposition, while $\mu^{(k)}$ and $S^{(k)}(x_i)$ are the respective amplitude (in the sense of Eq. \eqref{eq: amplitude}) and spectral shape discretized in frequency space (see Sec. 3.2.2 of [\hyperlink{I}{I}] for more details on notation and derivation). The behavior of the latter is shown in the left panel of Fig. \ref{fig: SD_PCA}, taken from \cite{Chluba2014Teasing} (a similar representation can also be found in Fig.~6 of~[\hyperlink{I}{I}]). As expected the first eigenmodes carry the most information and lead to the largest amplitudes. By construction, both $\mu^{(k)}$ and $S^{(k)}$ are proportional to the orthonormal eigenvectors $E^{(k)}_\alpha=E^{(k)}(z_\alpha)$ of the Fisher information matrix discretized in redshift space (see e.g., Eq. (9) of \cite{Chluba2014Teasing}), which parameterizes the time-dependent information content carried by the signal. The shapes corresponding to the first four eigenmodes are shown in the right panel of Fig. \ref{fig: SD_PCA}, taken from \cite{Chluba2014Teasing}. Not surprisingly, the eigenvectors $E^{(k)}_\alpha$ have non-zero values in the transition era between $y$ and $\mu$ distortions (blue region in the right panel of Fig.  \ref{fig: SD_PCA}), consistently with the discussion surrounding Fig. \ref{fig: SD_th_br_shapes}.
	
To be more explicit, and limiting the discussion to $\mu^{(k)}$, one has that
\begin{align}\label{eq: mu_k}
	\mu^{(k)}=\sum_\alpha E^{(k)}_\alpha dQ_\alpha\,,
\end{align}
where $dQ_\alpha=dQ(z_\alpha)$ is the heating rate discretized in redshift space. In a crude way, this definition of $\mu^{(k)}$ can be compared to the definition of the $y$ and $\mu$ amplitudes given in Eq.~\eqref{eq: amplitude}, with the difference that in the case of Eq.  \eqref{eq: mu_k} it is $E^{(k)}_\alpha$ that plays the role of the BRs for each eigenmode $k$. In this way, the larger the heating around the transition era the larger the values of $\mu^{(k)}$ are going to be. A similar analogy also applies to $S^{(k)}$ and the $Y$ and $M$ spectral shapes. In qualitative terms, this decomposition can be very useful since it breaks down the temporal and spectral information into related components, so that the observation of a given feature in the residual distortion spectrum can be very telling of the time at which that distortion has been sourced. The various components are conceptually similar to the $y$ and $\mu$ distortions, only localized in the intermediate era between the two SD types and with a more complex spectral dependence.

As a technical but important note, at the moment of observation (and hence of information extraction) the spectral shapes are fixed vectors in frequency space, with frequency range and binning defined by the observing experimental setup, and so is $G_{\rm th}$, which is projected onto these vectors. However, since these vectors are not orthogonal, the projection is not unambiguous and hence the BRs depend on the frequency array of the experiment \cite[\hyperlink{I}{I},\,\hyperlink{III}{III}]{Chluba2014Teasing} (see in particular Sec. 4.1 of the latter for an in-depth treatment of the problem). As a consequence, for the same heating history \matteo{(which would lead to unique values of the spectral amplitudes)} one can have different SD amplitudes depending on the observational setup (which would correspond to the true\matteo{, physical} amplitudes only in the limit of an infinitely precise frequency coverage). This is consistent with the idea that a realistic experiment will never be able to perfectly disentangle the different contributions to the total signal, which are usually fitted to the total signal itself. In this way, the derived $y$ and $\mu$ parameters really correspond to those that would be observed by a given experiment and their numerical evaluation is consistent with the eventual experimental determination. This does not mean that also $\Delta I_{g,\mu,y}$ -- the true observable -- is dependent on the experimental characteristics, since it is the weighted sum of all shapes and it is therefore independent of the details of the projection.

Concretely, in the \texttt{CLASS} implementation there are several operating options to calculate the BRs and hence $G_{\rm th}$. Either one chooses one of the many available approximated forms of the BRs or one specifies the frequency settings of the observing experimental setup. In the latter case, \texttt{CLASS} will automatically calculate the corresponding BRs and PCA components (which are stored for later use). The representative cases of FIRAS and PIXIE (see Sec. \ref{sec: exp}) are provided by default. The former is an example for missions with variable frequency binning, for which a file needs to be provided with the $\{\nu,\,\delta I_{\rm noise}\}$ array, while the latter is an example of mission with constant frequency binning, for which it is sufficient to specify $\{\nu_{\rm min},\,\nu_{\max},\,\Delta\nu,\,\delta I_{\rm noise}\}$ in the \texttt{CLASS} input file.

\vspace{0.3 cm}
\noindent\textbf{CRR}

\noindent Just as for $\Delta I_{g,\mu,y}$, the numerical computation of the CRR is extremely challenging. Not only the recombination process needs to be modeled very accurately, but also the atomic physics of hydrogen and helium (e.g., absorption/emission rates) has to be meticulous. Of course, on top of this the thermalization process needs to be solved as well. 

Of the three recombination codes currently available, \texttt{Recfast} \cite{Wong2007How}, \texttt{HyRec} \cite{AliHaimoud2010HyRec, Lee2020HYREC} and \texttt{CosmoRec} \cite{Chluba2011Towards} (see e.g., \cite{Chluba_CosmoTools} for a comparison between the codes), only the  last two employ the required effective multi-level approach to eventually calculate the CRR. In particular, the latter was taken as the basis for the development of \texttt{CosmoSpec} \cite{Chluba2016Cosmospec}, which allows for the precise computation of the CRR spectrum with the unique feature of also solving the two stages of helium recombination accurately on top of the other properties of \texttt{CosmoRec}. With respect to previous attempts to calculate the CRR (see e.g, \cite{Chluba2006Free, Rubino2008Lines, Chluba2009Cosmological, 2013PhRvD..87b3526A}), \texttt{CosmoSpec} delivers similar results in terms of accuracy although being significantly faster (a run with all relevant processes activated and precision settings for a \lcdm cosmology with typical Planck values takes $\sim 50$~s). 

\begin{figure}[t]
	\centering
	\includegraphics[width=\textwidth]{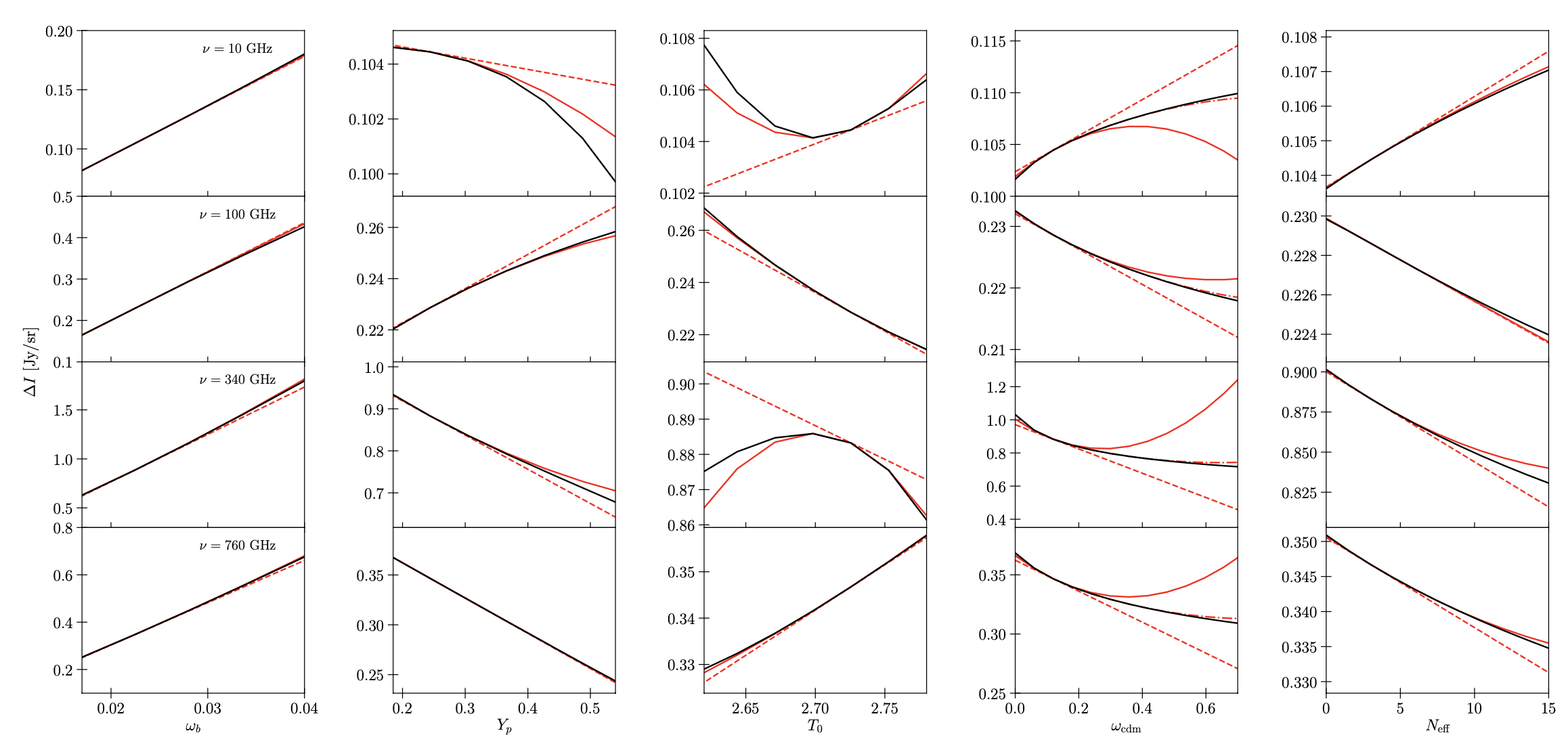}
	\caption{Fractional change of the CRR spectrum as a function of the relevant cosmological parameters for four arbitrary frequencies. The different lines refer to the prediction of \texttt{CosmoSpec} (black) and the Taylor expansion approximation (red) at first (dashed), second order (solid) and with the multi-pivotal approach (dash-dotted). Figure taken from [\protect\hyperlink{XIV}{XIV}].}
	\label{fig: SD_CRR_p}
\end{figure}

To further reduce the computational speed (by about a factor 250), [\hyperlink{XIV}{XIV}] developed a method based on the Taylor expansion of the CRR spectrum $\Delta I_{\rm CRR}$ around a reference\footnote{In [\hyperlink{XIV}{XIV}] assumed to be the spectrum calculated within \lcdm for parameters set to the Planck+BAO mean values found by \cite{Aghanim2018PlanckVI}.} $\Delta I_{\rm CRR,ref}$ for all relevant cosmological parameters $p$ listed in Eq. \eqref{eq: params} (see Sec. \ref{sec: SD_lcdm}). In fact, using the exact calculations of \texttt{CosmoSpec} one can vary the CRR spectrum with respect to each of the parameters $p$ singularly, finding the almost linear ($\omega_b$ and $N_{\rm eff}$) and parabolic ($Y_p$, $T_0$ and $\omega_{\rm dm}$) behaviors shown in Fig.~\ref{fig: SD_CRR_p}, taken from [\hyperlink{XIV}{XIV}], as black lines for four representative frequencies. For comparison, the full frequency dependence can be seen in Fig. \ref{fig: SD_CRR_dep}. The simplicity of these scalings suggests that the CRR spectrum can be approximated as a second-order Taylor expansion around the reference spectrum, i.e., as
\begin{align}\label{eq: taylor}
	\Delta I_{\rm CRR} (\nu,p) = \Delta I_{\rm CRR, ref}(\nu) + \sum_i \frac{\partial(\Delta I_{\rm CRR})}{\partial p_i}\Delta p_i+ + \sum_{i,j}\frac{\partial^2(\Delta I_{\rm CRR})}{\partial p_i\partial p_i}\frac{\Delta p_i \Delta p_j}{2}.
\end{align}
The derivatives can then be tabulated for a fine grid of frequencies and stored for later use. The predicted spectrum at first and second order is shown in Fig. \ref{fig: SD_CRR_p} as dashed and solid red line, respectively. 

\vspace{0.3 cm}
\noindent \textbf{Follow-up idea 4}: As discussed in Sec. \ref{sec: SD_lcdm}, the cosmology dependence of the CRR spectrum is not limited to the five parameters of Eq. \eqref{eq: params}, but extends also to many other beyond-\lcdm models. The method introduced here can be very simply applied to all of them as long as their impact on the CRR spectrum can be parameterized with a second-order expansion. One would just need to implement the model in \texttt{CosmoSpec} and compute the frequency-dependent Taylor coefficients (i.e., the derivatives). Given the advanced state of  \texttt{CosmoSpec}, the addition of e.g., DM decay/annihilation, PMFs, variation of fundamental constants and EDE would be straightforward. Since many of these models are already implemented in \texttt{CLASS}, the study of the interplay between CRR and e.g., CMB anisotropies would be a guarantee.
\vspace{0.3 cm}

\noindent \textbf{Follow-up idea 5}: As pointed out in Sec. \ref{sec: SD_lcdm}, the different contributions to the total CRR spectrum from the three recombination eras can be in principle disentangled and used to provide a time dimension to the observable \cite[\hyperlink{XIV}{XIV}]{Hart:2022agu}. Extending the Taylor expansion approximation (which at the moment only parameterizes the total spectrum) to include separately these three contributions would allow to open this line of research.
\vspace{0.3 cm}

To quantitatively judge the validity range of the approximation, one can then keep increasing the value of a given parameter until the difference between the CRR spectrum predicted by the Taylor expansion and by \texttt{CosmoSpec} exceeds 1\%. Furthermore, importantly, one also needs to ensure that the allowed range of variations is at least as wide as the error bars predicted by \cite{Hart2020Sensitivity} for Voyage 2050 (the least sensitive mission with enough accuracy to observe the CRR, see Sec. \ref{sec: future_exp}), so as not to bias the forecasts for this mission (and those with better sensitivities) with numerical artefacts. By doing so, as also expected from Fig. \ref{fig: SD_CRR_p}, one obtains that the second order corrections deliver a sufficiently accurate representation of the true dependence of the CRR over the range of variations that can be probed by future missions for all parameters but $\omega_{\rm cdm}$ [\hyperlink{XIV}{XIV}]. To amend for this limitation~[\hyperlink{XIV}{XIV}] suggested a multi-pivotal approach with three different reference values of $\omega_{\rm cdm}$, the closest of which to be chosen automatically depending on the selected value of $\omega_{\rm cdm}$. The corrected prediction of the Taylor expansion method is shown as dash-dotted line in Fig.~\ref{fig: SD_CRR_p} and significantly improves upon the simple second-order expansion.

\vspace{0.3 cm}
\noindent \textbf{Follow-up idea 6}: Perform a similar validity test also for the Green's function approximation discussed in the previous subsection.
\vspace{0.3 cm}

The fast calculation of the CRR spectrum with the Taylor expansion method can then be safely included in the \texttt{CLASS} pipeline. The tables of the Taylor coefficients, released together with the public version of the code, are read, interpolated and (for a given choice of the input parameters) the resulting CRR spectrum $\Delta I_{\rm CRR}$ is added on top of the other contributions according to Eq. \eqref{eq: DI}. 

As a final remark, the calculation of the inhomogeneous contributions given in Eq.~\eqref{eq: DI_CRR} can be performed using the very same second-order derivatives calculated for the Taylor expansion of Eq. \eqref{eq: taylor}. The aforementioned numerical setup therefore intrinsically allows to also account for density inhomogeneities in the recombination process.

\vspace{0.3 cm}
\noindent\textbf{SZ effect}

\noindent The SD signal caused by the SZ effect can be determined numerically very accurately (far below experimental sensitivity) with the \texttt{SZpack} code \cite{Chluba2012Fast, Chluba2012Sunyaev} either by explicit numerical integration or by using the asymptotic expansion described in Sec. \ref{sec: SD_lcdm}. The latter (faster) approach has been followed in \texttt{CLASS} as well for both the tSZ effect (implemented up to fourth order in $\theta_e$) and for the kSZ effect (up to fourth order in the linear combination of $\theta_e$ and $\beta$). As pointed out in Sec. \ref{sec: SD_lcdm}, this implementation assumes the observer to be in the cluster frame. For sake of completeness, the \texttt{CLASS} implementation also includes the calculation of the distortion in the CMB frame following the parameterization of \cite{Nozawa2005Improved}.

\vspace{0.3 cm}
\noindent\textbf{Summary}

\noindent In summary, the \texttt{CLASS} implementation of SDs -- which is the one employed throughout the thesis -- unifies a number of (computationally) efficient approximations of dedicated codes and is publicly available (which is not the case for some of the codes it relies upon). It includes all sources of SDs predicted to exist within \lcdm and several examples beyond it, and it can calculate the SD signal consistently with the experimental configurations designed to observe it.

\newpage
\section{Clouds in sight: the experimental status}\label{sec: exp}
In Sec. \ref{sec: theory} we have seen how to characterize the SD signal and what can source it. In this section we focus instead on how to measure it. 

The greatest obstacles to the observation of SDs are atmospheric absorption (which can however be avoided with satellites) as well as galactic and extra-galactic foregrounds, which we review in Sec. \ref{sec: fore}. These need to be understood and removed to an extremely challenging degree, one part per $10^{5}-10^{6}$, and in the following we also briefly discuss how this can be achieved. The second technical difficulty is to attain the required experimental sensitivity to measure deviations from the BB spectrum of the order of $10^{-9}-10^{-10}$ (for the primordial signal). Although it has been argued that such precision would be realistic with today's technology, in practice it has never been reached before and no CMB SD has ever been measured. To provide a more complete picture about \textit{Where do we come from? What are we? Where are we going?}\footnote{P. Gauguin, 1898.} in the experimental development of SDs we discuss in Sec.~\ref{sec: firas} the state of the art of the CMB energy spectrum observations, in Sec.~\ref{sec: current_exp} the SD missions that are going to collect data in the up-coming future and in Sec.~\ref{sec: future_exp} more futuristic proposals aimed at extracting the most information from the expected SD signal. We conclude in Sec. \ref{sec: num_2} with a presentation of the numerical setup employed to perform the statistical analysis of these (real or mock) data sets.

\subsection{Atmospheric windows and (extra-)galactic foregrounds}\label{sec: fore}
As we will see in the following sections, there are mainly three strategies for the observation of CMB SDs (and incoming radiation more in general): ground-based antennas, balloons and satellites. Each of them has specific limitations and advantages, involving most notably frequency and sky coverage as well as costs and accessibility (see e.g., \cite{Rocha:2019nof} for an overview). 

\begin{figure}[t]
	\centering
	\includegraphics[width=\textwidth]{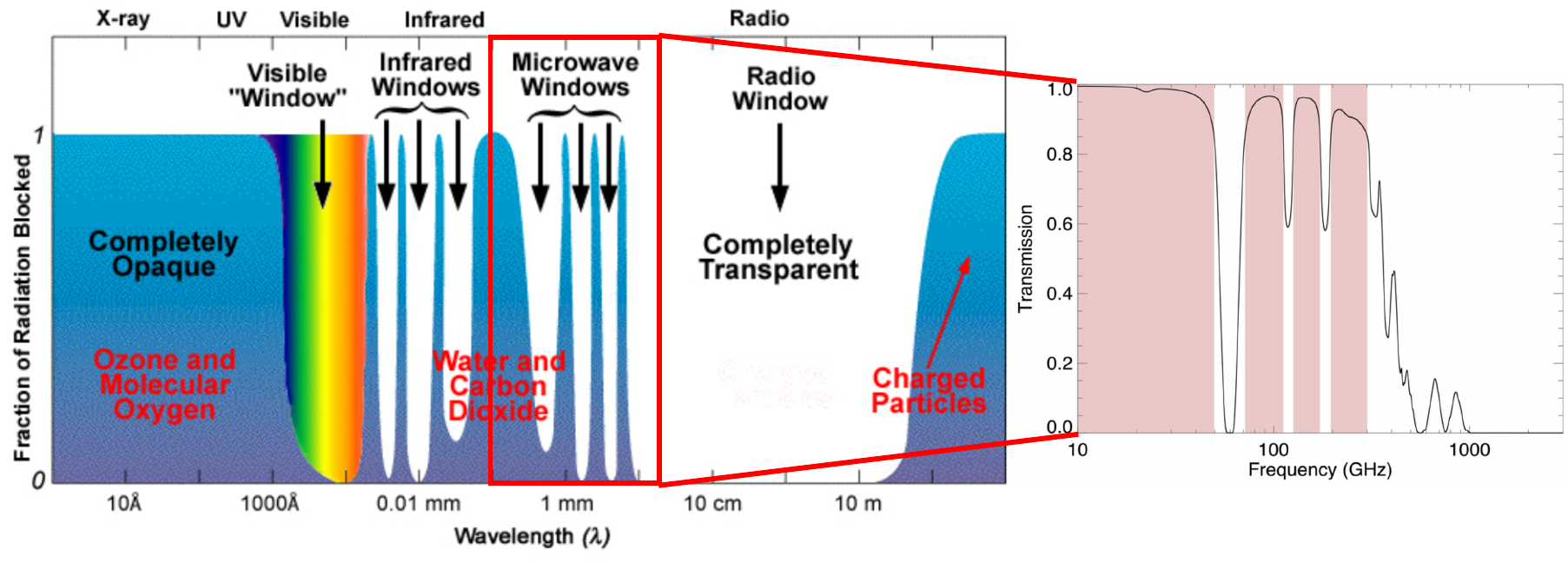}
	\vspace{-0.3 cm}
	\caption{Graphical illustration of the atmospheric \textit{absorption}, with zoom in on the \textit{transmission} in the microwave region. Figures adapted from~\cite{Atmospheric_window, Rocha:2019nof}.}
	\label{fig: SD_fore_atm}
\end{figure}

In the case of ground-based experiments (and partially also for balloons), for instance, one of the greatest restrictions comes from the atmosphere. In fact, because of the diverse composition of the atmosphere the incoming radiation is reflected or absorbed in different regions of the EM spectrum, which forbids observations at those frequencies. As graphically summarized in Fig. \ref{fig: SD_fore_atm}, adapted from~\cite{Atmospheric_window, Rocha:2019nof}, this leaves only a limited number of so-called atmospheric windows where measurements can be carried out. Since a BB spectrum with $T=2.725$~K peaks at around $\sim 160$ GHz, for CMB observations the 150 GHz and 220 GHz windows (see zoomed-in panel of Fig.~\ref{fig: SD_fore_atm}) are of particular importance. The only way to minimize or remove these constraints is to employ stratospheric balloons or satellites, respectively. For instance, forecasts for the proposed SD mission COSMO \cite{2021arXiv211012254M} (see Sec. \ref{sec: current_exp}) show that the balloon implementation would lower the atmospheric emission by $4-5$ orders of magnitude with respect to the ground-based counterpart (see Fig. 1 of the reference). 

\newpage
On top of this, temporal and spatial atmospheric fluctuations have to be taken into account as well. These can be mainly minimized by placing the ground-based experiment in dry, high-altitude locations such as the South Pole and the Atacama desert in Chile (see \cite{2011RMxAC..41...87R} for a review). Furthermore, one can also modulate the fluctuations with various techniques (see e.g., \cite{POLARBEAR:2015mbo, Morris:2021jed} for examples in the two sites).

\begin{figure}[t]
	\centering
	\includegraphics[width=0.8\textwidth]{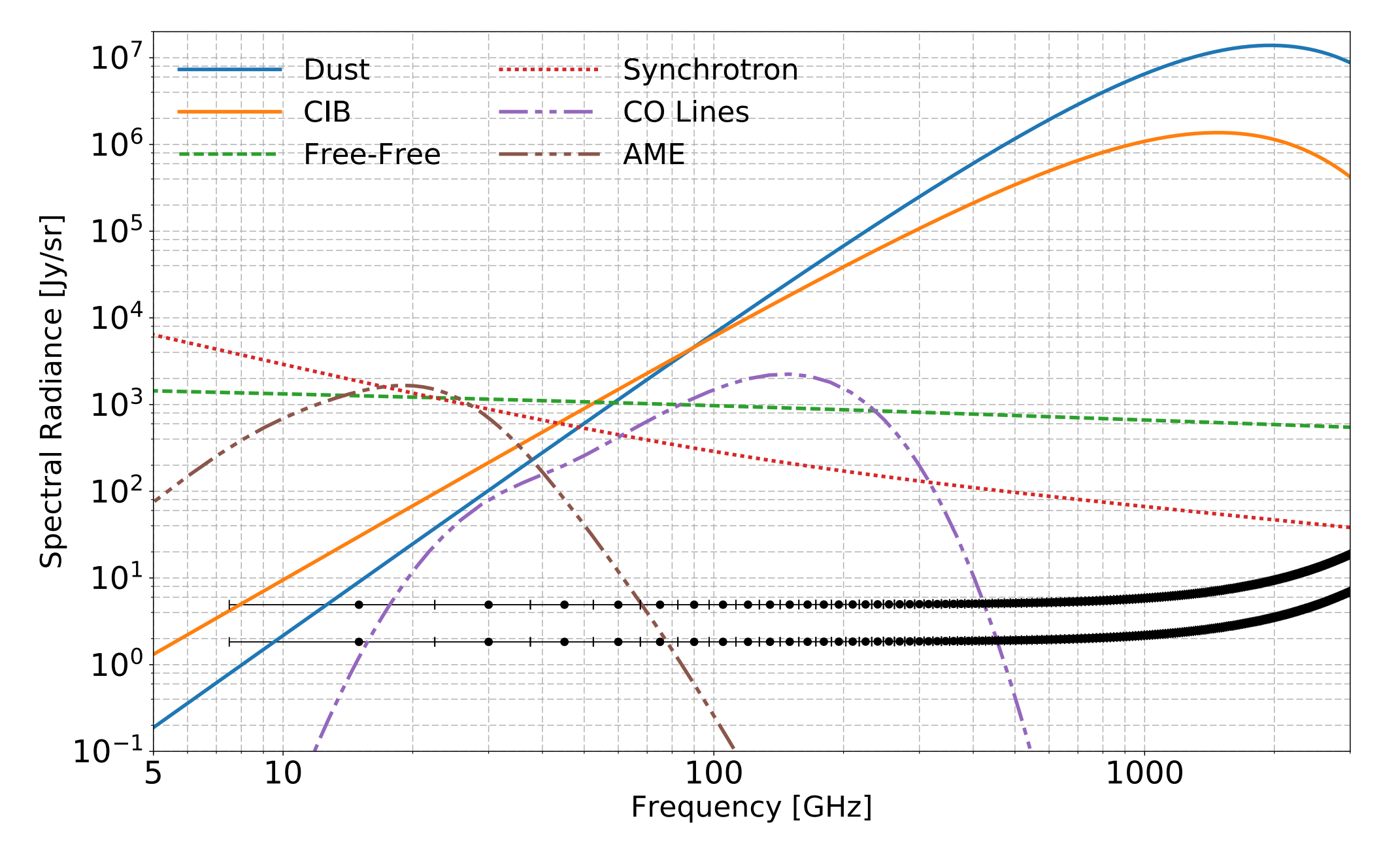}
	\caption{Collection of galactic and extra-galactic foregrounds relevant for the CMB SDs. The horizontal black lines represent the projected PIXIE sensitivity curves. Figure taken from~\cite{Abitbol2017Prospects}.}
	\label{fig: SD_fore_gal}
\end{figure}

Regardless of the experimental environment, however, galactic and extra-galactic foregrounds are unavoidable. In the frequency range relevant for the observation of CMB SDs (i.e., 1 GHz $-$ 1 THz, see Fig. \ref{fig: SD_LCDM_summary}), the relevant foregrounds are \textit{i)} galactic and extra-galactic thermal dust emission (the latter often referred to as Cosmic Infrared Background -- CIB), \textit{ii)} synchrotron emission from cosmic rays traveling through the galactic magnetic field, \textit{iii)} bremsstrahlung (or free-free) emission from electron-ion collisions in the galactic environment, \textit{iv)} the cumulative extra-galactic Carbon monoxide (CO) emission and \textit{v)} the Anomalous Microwave Emission (AME) due to the presence of spinning dust grains with an electric dipole moment (see e.g., \cite{Adam2015PlanckX, Akrami2018PlanckIV, Abitbol2017Prospects} for recent reviews). Further secondary contributions include the inverse CS of the CIB \cite{Sabyr:2022lhj} (in analogy to the SZ effect for CMB photons), intergalactic dust emission \cite{imara2016distortion} and additional spectral lines \cite{Yue:2015sua,serra2016dissecting}. Moreover, since the thermal history of the late universe is not homogeneous, the signal of the single components can present spatial variations \cite{Chluba:2017rtj}. All of these effects lead to different spectral shapes and intensities, as summarized in Fig.~\ref{fig: SD_fore_gal}, taken from \cite{Abitbol2017Prospects} (with more details given in Sec. \ref{sec: num_2}). Fig.~4 of \cite{Adam2015PlanckX} showcases the dependence of each contribution on the underlying parameters.

The question is then to which extent these galactic and extra-galactic foregrounds need to and can be removed from the observed signal. To answer the former point, the required degree of subtraction depends on the target signal. For instance, the (spatially-averaged) parametric modeling shown in Fig. \ref{fig: SD_fore_gal} suffices for the purposes of observational strategies involving differential measurements, as it is the case for all CMB anisotropy missions such as Planck \cite{Adam2015PlanckX, Akrami2018PlanckIV}, ACT and SPT. A similar discussion also applies to missions targeting the CMB energy spectrum such as FIRAS. However, for the observation of CMB SDs this simpler approach would likely not be enough (being the target signal up to $5-6$ orders of magnitude below the foregrounds) and a more realistic treatment might become necessary. 

This therefore calls for the second point raised above, i.e., how much of the foreground signal can actually be subtracted. The obvious answer is that it can be only as much as our understanding of their (combined) properties. In a broad way, \cite{Rocha:2019nof, Kogut2019CMB} argue that there are mainly three ways to improve upon the current status and push the limits to an acceptable range. Firstly, a better characterization (and hence removal) of the foregrounds can only be achieved with a significant refinement of their theoretical modeling beyond the aforementioned parametric models (see e.g., \cite{draine2010physics, Mashian:2016bry, rao2016gmoss} for examples of works in this direction). Secondly, the development of more and more sophisticated numerical tools to simulate and parameterize the complexities of the foregrounds will become necessary (see e.g., \cite{delabrouille2013pre,Chluba:2017rtj,SathyanarayanaRao:2016ope, Novikov:2023kvb}). Finally, the synergy between complementary data sets will offer the possibility to constrain the foreground properties over a wide range of frequencies and thereby contribute to the definition of their overall spatial and spectral dependence (see e.g., \cite{Kusiak:2023hrz}).

We note, however, that while these steps are necessary to be able to extract the maximum amount of information from the eventual observation of CMB SDs, the foreground model developed in \cite{Adam2015PlanckX, Akrami2018PlanckIV, Abitbol2017Prospects} still serves several important purposes. For instance, it allows to perform rapid forecasts that might in turn point towards particularly sensitive aspects of the foreground-subtraction problem while discarding others as non-issues. Furthermore, it is also very useful to estimate the realistic reach of proposed future experiments, which is fundamental for their design.

As a final remark, we point out that when targeting the primordial SD signal also the late-time contribution from the SZ effect needs to be removed. Although complementary information on the localized sources might allow to identify them and partly subtract their emission (contributing to the intracluster and intergalactic component), the residual diffuse contribution from cosmic reionization can hardly be avoided. Since at leading order the SZ spectrum can be parameterized by a $y$ distortion, as explained in Sec. \ref{sec: SD_lcdm}, once marginalized over this foreground completely masks the primordial $y$ signal, thereby significantly reducing the amount of information that can be extracted from it. For this reason, when considering SD missions targeting the primordial SDs the discussion will be mostly phrased in terms of the $\mu$ parameter. The CRR signal also becomes an ideal target as it presents a very characteristic shape, not degenerate with any of the aforementioned foregrounds.

In summary, the atmosphere presents a formidable challenge for the observation of CMB SDs from the ground and with balloons, but its impact can be mitigated either by placing the experiment in suitable sites or by employing satellites. On the other hand, galactic and extra-galactic foregrounds are unavoidable and their cumulative contribution lays several orders of magnitude above the target SD signal. How to best subtract them is still an open question, although significant effort has been and is being dedicated to the task.

\subsection{State of the art}\label{sec: firas}
Before discussing the observational status of CMB SDs, let us first overview our current understanding of the CMB energy spectrum. Focusing first of all on ground-based measurements, after pioneering attempts in the '70s \cite{weiss1980measurements} only in the '80s and '90s the effort of a number of groups led to sufficiently precise measurements of the low-frequency tail of the CMB energy spectrum. In particular, an Italian-American (mainly based the Lawrence Berkeley Laboratory -- LBL) collaboration conducted several observations from the South Pole and the White Mountains (California) in the approximate frequency rage between $1-10$ GHz as well as at 33 and 90 GHz (limited by the aforementioned atmospheric windows)  \cite{smoot1983low, sironi1986temperature, mandolesi1986measurements, levin1988measurement, 1989ApJ...339..632B, de1989temperature, sironi1991brightness, amici1991temperature, 1993ApJ...409....1B, bersanelli1994absolute, 1995ApL&C..32....7B}. A subset of the members of the collaboration later followed up on these results with the TRIS experiment designed to improve the accuracy of the measurements below 1 GHz \cite{zannoni2008tris, Gervasi:2008eb}. Over the same years it was also realized that the amount of excited Cyanogen (CN) in the interstellar medium is proportional to the CMB temperature. Therefore, by measuring CN spectra in different locations one can infer the corresponding CMB temperature at the excitation frequencies (113 and 226~GHz) \cite{meyer1985precise, crane1986cosmic, kaiser1990precise, palazzi1990precise, 1993ApJ...413L..67R, crane1989cosmic}. 

Furthermore, in the '80s and '90s also balloon flights from the South Pole have been conducted by independent groups (mainly based in Princeton) and set limits on the spectrum at frequencies between $1-25$ GHz \cite{1987ApJ...313L...1J, staggs1995measurement, staggs1996absolute}. In the early 2000s the Absolute Radiometer for Cosmology, Astrophysics and Diffuse Emission (ARCADE) and ARCADE 2 balloon-borne experiments collected high-precision data in the $3-90$~GHz frequency range \cite{fixsen2004temperature, 2011ApJ...734....5F}.

Nevertheless, as explained in the previous section, because of the limiting atmospheric conditions accurate measurements of the peak of the spectrum between $60-600$ GHz can almost exclusively be conducted from space. This lead in the '90s to the launch of a series of spectrometers in the outer space. The first to successfully measure the CMB energy spectrum has been the Far-InfraRed Absolute Spectrophotometer (FIRAS) on the Cosmic Background Explorer (COBE) satellite \cite{1990ApJ...354L..37M}, which confirmed the CMB spectrum to perfectly follow a BB shape with $T=2.735\pm 0.06$ K (a following re-analysis of the data delivered $T=2.728\pm 0.004$ K \cite{Fixsen1996Cosmic}). This Nobel-prize winning measurement is the most accurate determination of the CMB energy spectrum available to date \cite{Fixsen2009Temperature}. The Differential Microwave Radiometer (DMR) on COBE also delivered complementary data at lower frequencies \cite{1996ApJ...464L...1B}. Usually without enough credit in the literature, as close second came the COBRA instrument launched with a sounding rocket only a week after the announcement of the FIRAS observation in January 1990 \cite{gush1990rocket} (see \cite{Grant2018Experiment} for the extremely fascinating story of Herbert Gush and COBRA). In 1990 not only the COBRA measurement confirmed the FIRAS determination of the CMB temperature with ${T=2.736\pm0.017}$~K, but at the time it also became its most accurate determination.

\vspace{0.3 cm}
\noindent\textbf{Follow-up idea 7:} To our knowledge, the technology employed by the COBRA experiment has not seen any follow-up, but is certainly deserving of more attention. This observational strategy would in fact present many of the advantages of satellites (like the lower atmospheric contamination) and of ground-based/balloon experiments (like lower costs and easier accessibility) at the same time. Of course, it would also share the respective limitations to the same degree. A dedicated work with upgraded experimental design and resulting sensitivity curves would be of major interest.
\vspace{0.3 cm}

\begin{figure}[t]
	\centering
	\includegraphics[width=0.87\textwidth]{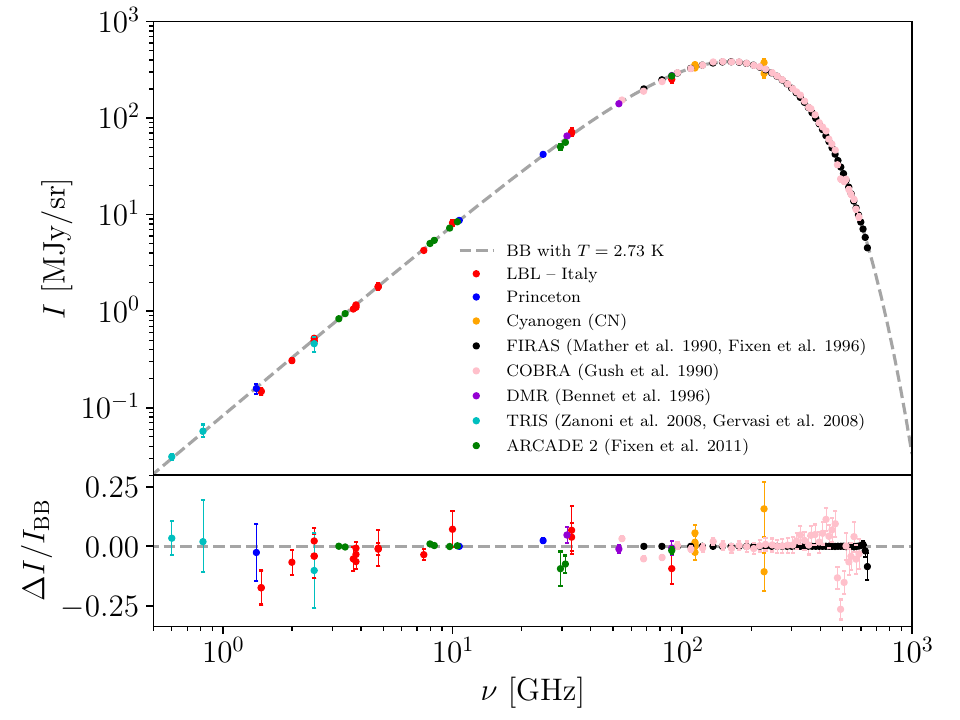}
	\caption{Energy spectrum of the CMB (dashed gray line). The data points represent a comprehensive sample of measurements (see text). The lower panel shows the relative difference with respect to the reference BB spectrum. Original figure.}
	\label{fig: SD_exp_spectrum}
\end{figure}

The comprehensive collection of all these ($\sim130$) data points is shown in Fig. \ref{fig: SD_exp_spectrum}, an original figure made for the thesis.\footnote{A few disclaimers for sake of preciseness. All error bars in the figure are reported at $1\sigma$. All data points with uncertainties larger than 10\%, i.e., $\sim0.25$~K, have been neglected in the figure except for the case of TRIS. For the case of COBRA, in the original paper (as well as in any follow-up of our knowledge) no error on the single data points is given, with only a conservative estimate of the uncertainty on the mean temperature of 17~mK. Here we adopt this number for every data point to illustrate the overall accuracy of the experiment, keeping in mind that these are however not the actual error bars. For the case of the COBE/DMR, only the average value of $T$ is given in the original paper \cite{1996ApJ...464L...1B} (to our knowledge). The data points have been taken from Fig. 1 of \cite{singal2006cosmic} instead. For the case of ARCADE 2, the data points are only given in terms of the total thermodynamic temperature, which also gets a contribution from the radio foregrounds (dominating at $\nu<10$ GHz, see e.g., Fig. 1 of \cite{Abitbol2017Prospects}) on top of the CMB. Here the former contribution has been removed using Eq.~(6) of \cite{2011ApJ...734....5F} with the best fitting values listed just below the equation. No error propagation has been performed and therefore the error bars reported here are slightly optimistic (the increase should be $\sim10-20\%$ for the low-frequency data points and negligible otherwise) but should suffice for the graphical comparison.} As clear from the figure, analogously to Figs. \ref{fig: CMB_spectra} and~\ref{fig: LSS_spectrum}, there is excellent agreement between the BB shape predicted by \lcdm and the observational data over more than three orders of magnitude in frequency. The BB temperature obtained by averaging over several of the shown data points has been determined to be $T=2.72548 \pm 0.00057$ K \cite{Fixsen2009Temperature}, a precision of 0.02\%.

\begin{figure}[t]
	\centering
	\includegraphics[width=0.8\textwidth]{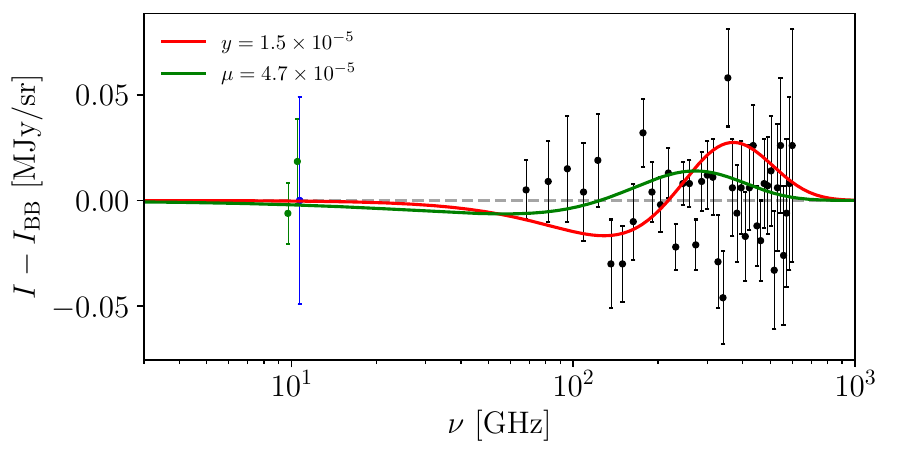}
	\caption{Comparison between a selection of the data points shown in Fig.~\ref{fig: SD_exp_spectrum} and the maximum $y$ and $\mu$ distortions allowed by FIRAS (at $95\%$ CL). Original figure.}
	\label{fig: SD_exp_y_mu}
\end{figure}

\newpage
These very accurate measurements have therefore established the shape of the CMB energy spectrum to a sub-percent confidence, but this has not been sufficient to detect CMB SDs. In fact, based on FIRAS data \cite{Fixsen1996Cosmic} derived upper bounds on the $y$ and $\mu$ parameters that are one order of magnitude above the expected late-time signal and $3-4$ orders of magnitude above the expected primordial signal, namely $y<1.5\times 10^{-5}$ and $\mu<9\times10^{-5}$ (at 95\% Confidence Level -- CL). A recent re-analysis of the FIRAS data reported $\mu<4.7\times10^{-5}$ (at 95\% CL) \cite{Bianchini:2022dqh}. The improvement comes from a more accurate cleaning of the foregrounds, achieved using complementary information from Planck. In Fig. \ref{fig: SD_exp_y_mu}, an original figure made for the thesis, the maximum allowed $y$ and $\mu$ distortions are graphically compared to the same data points shown in Fig. \ref{fig: SD_exp_spectrum} in the frequency range relevant for CMB SDs ($\sim 3-3000$ GHz).

In summary, three decades of measurements have confirmed with great confidence that the CMB energy spectrum follows a BB shape over a wide range of frequencies. Yet, the achieved sensitivities have not been enough to detect any SD, for which only upper bounds ($y<1.5\times 10^{-5}$ and $\mu<4.7\times10^{-5}$ at 95\% CL) have been derived.

\subsection{Up-coming missions}\label{sec: current_exp}

As discussed in the previous section, (with the marginal exception of the ARCADE experiment) no observational progress has been made towards the detection of CMB SDs since the early '90s. Nevertheless, in the up-coming $1-3$ years several ground-based and balloon experiments are expected to be deployed.

\begin{figure}[t]
	\centering
	\includegraphics[width=0.75\textwidth]{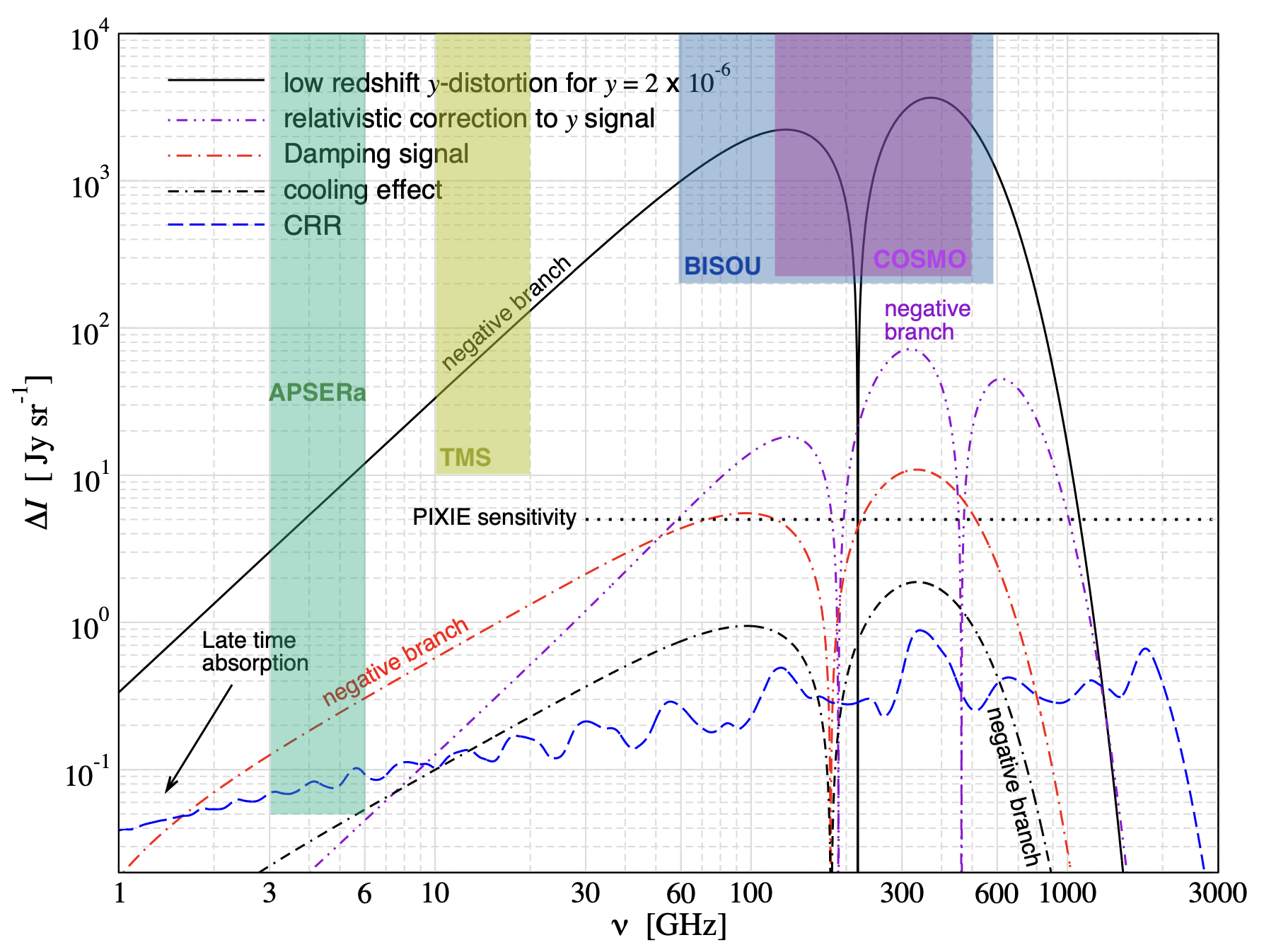}
	\caption{Same as in Fig. \ref{fig: SD_LCDM_summary} but with up-coming experiment sensitivities over-imposed. Figure adapted from \cite{Chluba2016Which}.}
	\label{fig: SD_exp_upcoming}
\end{figure}

Starting from the low frequency end, the ground-based Array of Precision Spectrometers for the Epoch of Recombination (APSERa) \cite{APSERa, 9726484} will target the $3-6$~GHz frequency range and aims at reaching sensitivities of the order of nK, forecasted to be enough to detect the CRR signal with high significance \cite{SathyanarayanaRao:2015vgh,SathyanarayanaRao:2016ope}. In the complementary frequency window between $10-20$ GHz, the Tenerife Microwave Spectrometer (TMS) \cite{2020SPIE11453E..0TR} will test the CMB energy spectrum to the $10-20$ Jr/sr level, well below the expected SZ signal. The commissioning of the mission is planned to happen early 2024. At higher frequencies, the Balloon Interferometer for Spectral Observations of the Universe (BISOU) \cite{Maffei:2021xur,trappe2022optical} will test the FIRAS frequency range with a precision up to two orders of magnitude higher. The first test flights are expected to begin in 2026. A similar frequency range will also be the focus of the COSmic Monopole Observer (COSMO) \cite{2021arXiv211012254M}. This experiment is planned to have a first observational phase from the ground (limited to the $110-170$~GHz and $200-300$ GHz frequency bands) starting in 2025 and a following second phase on a stratospheric balloon (widening the frequency coverage up to 500 GHz). The projected sensitivity on the $y$ distortion signal is of the order of $2.8\times10^{-7}$ from the ground (guaranteeing a Signal-to-Noise Ratio -- SNR -- of about 5 with respect to the expected late-time SD signal) and a factor of about 2 lower with the balloon implementation. Since both BISOU and COSMO will be operated from the South Pole with similar experimental setups, the two experiments will evolve synergetically. The forecasted sensitivities and frequency windows of these experiments are graphically summarized in Fig. \ref{fig: SD_exp_upcoming}.

\newpage
This generation of SD missions will bring us two fundamental steps forward in the quest to observe the full SD signal. First of all, they will explore the galactic and extra-galactic foregrounds with sensitivities and at frequencies far beyond those probed by FIRAS. Secondly, they will provide a first spectroscopic detection of the SZ effect with high significance. As such, they will unveil much of the complexity of the late-time contributions, setting up the stage for the next generation of SD missions targeting the primordial~signal.

\vspace{0.3 cm}
\noindent\textbf{Follow-up idea 8:} With the newly developed \texttt{CLASS}+\texttt{MontePython} pipeline (Secs. \ref{sec: num} and~\ref{sec: num_2}) it would be straightforward to perform a combined analysis of all the aforementioned missions including foregrounds as well as the late-time and primordial signal (all missions specifics are publicly available in the respective papers). The resulting forecasted constraints on foregrounds and SZ parameters could in turn be used as priors to more realistically predict the sensitivity of future missions to the primordial SD signal.
\vspace{0.3 cm}

On top of these mission targeting as primary goal the isotropic SD signal, in the coming years much more will be understood about the relevant foregrounds (see Sec.~\ref{sec: fore}) as the byproduct of experiments such as BLAST \cite{marsden2009blast, lowe2020balloon}, FYST \cite{CCAT-Prime:2021lly}, LSPE \cite{LSPE:2020uos}, PILOT \cite{mangilli2018pilot, mangilli2019geometry} and TIME \cite{crites2014time}, whose primary goal is the observation of e.g., CMB polarization or galaxy evolution. This will further help in reducing the impact of these contributions. Moreover, also line-intensity mapping\footnote{In brief, the idea in this case is to probe the cumulative, integrated emission of spectral lines from unresolved galaxies and the intergalactic medium.} \cite{Bernal:2022jap} experiments such as CONCERTO \cite{CONCERTO:2020ahk, Monfardini:2021mxj, Catalano:2021umn}, EXCLAIM \cite{cataldo2020overview} and DESHIMA/TIFUUN \cite{endo2019first, 2022JLTP..209..278T} or other experiments such as OLIMPO \cite{masi2019kinetic, presta2020first} will collect spectroscopic information on numerous localized SZ sources. This will both widen the current catalogs of SZ sources (see Sec.~\ref{sec: res_lcdm}) and allow to better characterize the related signal.

In summary, although little concrete progress has been made in the last 35 years towards the observation of CMB SDs, the next $1-3$ years will see the deployment of a number of ground-based and balloon experiments with enough sensitivity to detect the late-time SD signal with high significance. In this way, they will also become the ideal pathfinders for more advanced satellites targeting the primordial signal.

\subsection{High-sensitivity missions}\label{sec: future_exp}

Reaching the required sensitivity to observe the primordial SD signal is expected to be possible only from space. The disadvantage of satellites are however the costs, which make the funding of these missions more complicated than for the cases discussed in the previous section and this is the reason why no satellite targeting CMB SDs has ever been launched since FIRAS.

\begin{figure}
	\centering
	\includegraphics[width=0.8\textwidth]{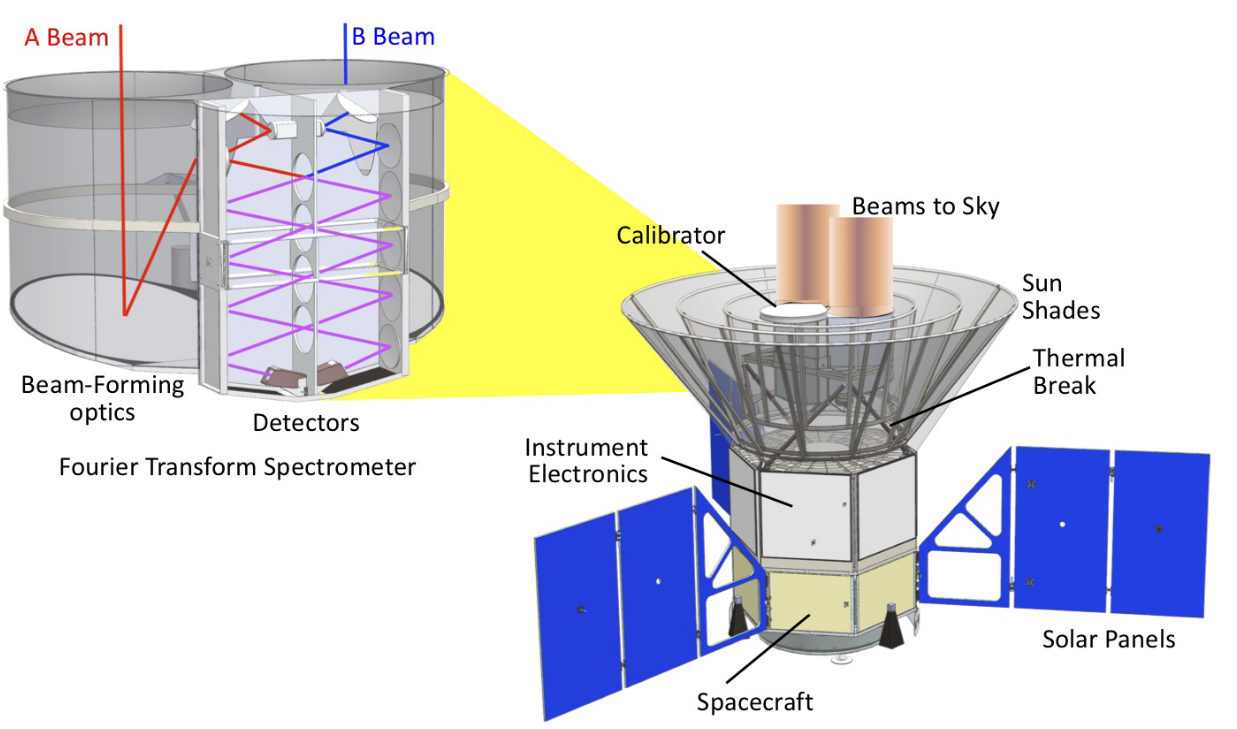}
	\caption{Illustration of the PIXIE mission concept. Figure taken from \cite{Kogut2019CMB}.}
	\label{fig: SD_PIXIE}
\end{figure}

The prime example of this is the Primordial Inflation Explorer (PIXIE) \cite{Kogut2011Primordial}, a broad-band Fourier transform spectrometer \cite{davis2001fourier}. Its design is graphically summarized in Fig.~\ref{fig: SD_PIXIE}, taken from \cite{Kogut2019CMB} (see also Fig. 2 of \cite{Kogut2011Primordial}). The PIXIE frequency array would span the range between 30 GHz and 3 THz with a bin size of 15 GHz ($\sim 400$ spectral channels in total). Although the SD signal is expected to vanish above 1 THz, the sensitivity to the highest frequencies is fundamental for an accurate removal of the foregrounds (the spectrum of dust and CIB emissions peaks \matteo{roughly} at $\sim2$ THz). The overall sensitivity of PIXIE is $\delta I = 5$ MJy/sr (see e.g., App. E1 of [\hyperlink{I}{I}] for a summarized derivation), corresponding to a SNR for the detection of the $\mu$ distortion signal of 0.5 (explicitly, $\delta \mu = 4\times 10^{-8}$) and of the amplitude of the CRR of 0.3  \cite{Abitbol2017Prospects, Chluba2019Voyage}.\footnote{As mentioned in Sec.~\ref{sec: fore}, when targeting the primordial SD signal the late-time SZ effect masks much of the primordial $y$ information. For this reason, here we will not focus on $y$ distortions, noting however that such advanced missions will observe the $y$ signal coming from the SZ effect at hundreds or even thousands of $\sigma$ \cite{Chluba2019Voyage}.}

Although these bounds would improve by orders of magnitude the current and up-coming ones, they would not lead to a statistically significant detection of the primordial SD signal. To amend for this limitation, an advanced version of the mission, Super-PIXIE \cite{Kogut2019CMB}, has been designed with an additional high-sensitivity, low-frequency module on top of a 10 fold improved sensitivity at PIXIE frequencies. Super-PIXIE is forecasted to lead to $2-3\sigma$ detection of both the $\mu$ and CRR signal. The role of the low-frequency module has been further investigated in [\hyperlink{XIV}{XIV}], finding however that its presence might not be essential (in terms of constraining-power gain), at least assuming its original design.

\vspace{0.3 cm}
\noindent\textbf{Follow-up idea 9:} Given the freedom in the \texttt{CLASS}+\texttt{MontePython} implementation discussed in Secs. \ref{sec: num} and \ref{sec: num_2} to design any experimental setup, the study of such additional modules can now be performed systematically with an high degree of realism in the resulting forecasts. Among others, this would allow to perform cost-reward analyses that might help support the case for future missions. Depending on the target signal, one can also choose to focus solely on the CRR or the $\mu$ distortions, if not both.
\vspace{0.3 cm}

\begin{figure}[t]
	\centering
	\includegraphics[width=0.57\textwidth]{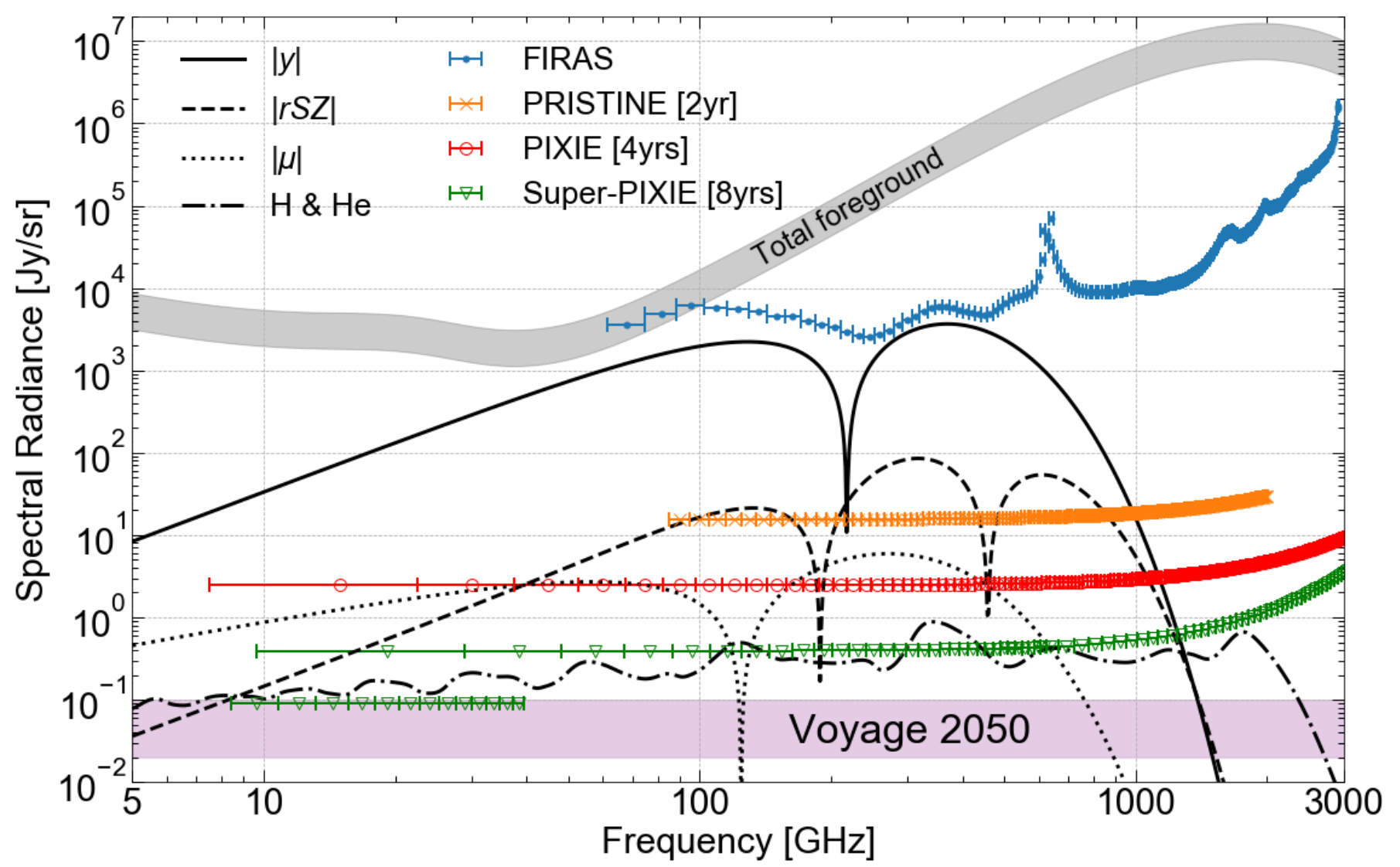}
	\includegraphics[width=0.36\textwidth]{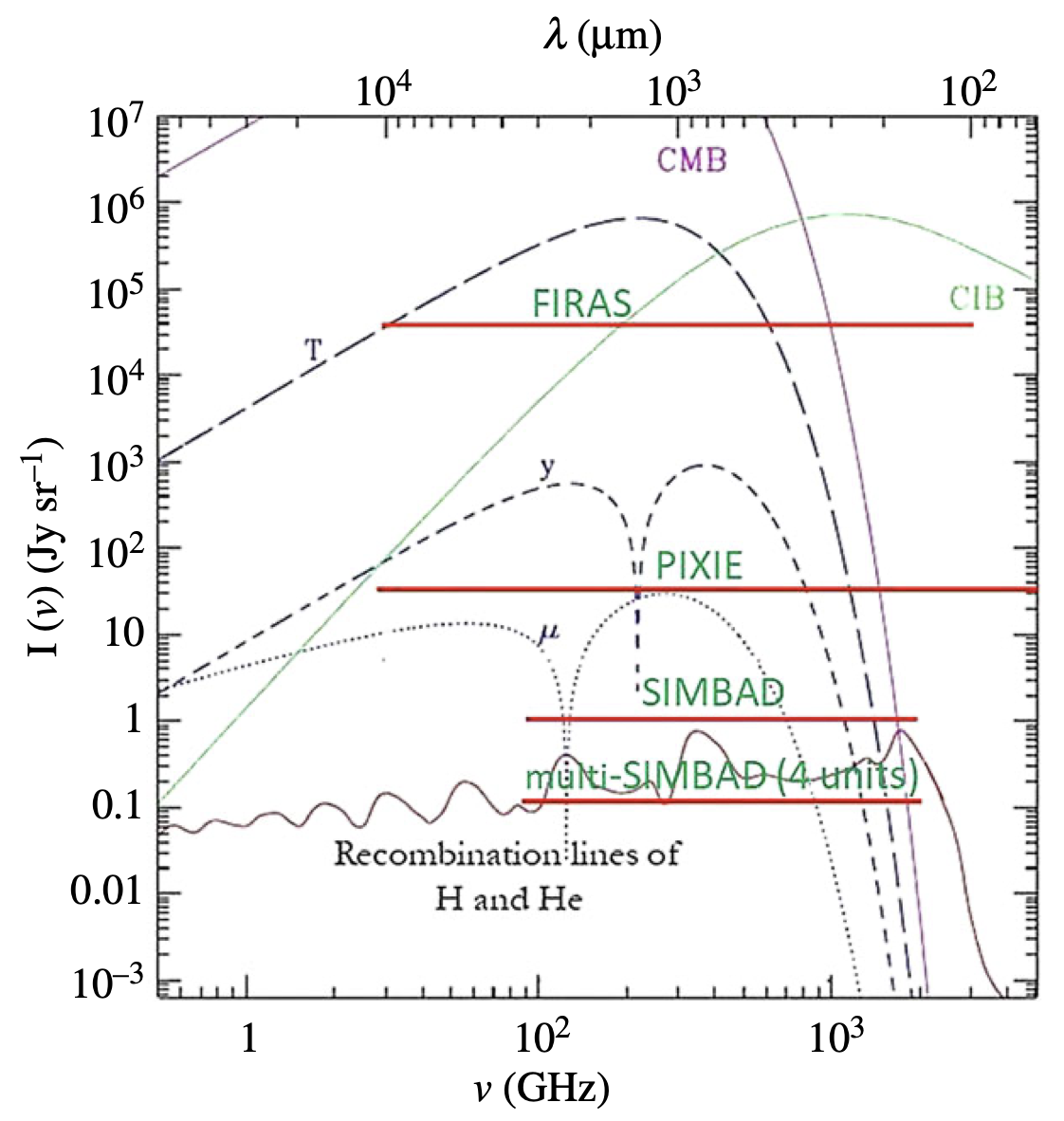}
	\caption{Estimated sensitivity of proposed high-sensitivity SD missions. Figures taken from \cite{Chluba2019Voyage, maillard2021moon}.}
	\label{fig: SD_exp_future}
\end{figure}

Nevertheless, the (Super-)PIXIE proposal has been submitted and rejected three times (in 2011, 2016 and 2019) by NASA's Explorer program. A similar fate met also other PIXIE-like proposals such as the PRISTINE, PRISM \cite{Andre2014Prism}, SIMBAD and FOSSIL\footnote{The author of the thesis was a member of the proposal, which is however not publicly-available.} missions, the former with a sensitivity roughly 10 times lower than PIXIE and the latter three with one about $5-10$ times higher. The sensitivity curves of these missions are graphically compared in Fig. \ref{fig: SD_exp_future}, taken from \cite{Chluba2019Voyage, maillard2021moon}.

The tides, however, seem to be changing with the new report of the Senior Committee appointed by ESA to ``develop a plan, Voyage 2050, to follow Cosmic Vision and establish ESA’s Space Science Programme up to 2050'' \cite{ESA_report}. In the report, in fact, the committee expresses the recommendation ``for a Large mission deploying gravitational wave detectors or precision microwave spectrometers to explore the early Universe at large redshifts'', following the case presented in \cite{Chluba2019Voyage}. The long term future of ESA's scientific program is therefore a \textit{two-horse race} between Gravitational Waves (GWs) and CMB SDs. The final decision to be made depending on either's theoretical and experimental readiness as well as wider community interest. 

In this sense, the decade-long development of PIXIE, although so far unsuccessful, would not be wasted and its experimental setup remains a useful benchmark. The Voyage 2050 proposal for SDs could in fact be a Super-PIXIE-like mission with a factor $\sim5$ improved sensitivity \cite{Chluba2019Voyage} (see Fig. \ref{fig: SD_exp_future}) and all technical aspects already carefully worked out \cite{nagler2015systematic, nagler2016multimode, pan2019compact, kogut2019systematic, kogut2020calibration,Kogut:2023lru}. Furthermore, the up-coming generation of SD missions would provide a solid understanding of the late-time contributions, thereby allowing the Voyage 2050 mission to reach its expected sensitivity to the primordial signal with a larger degree of confidence. The solid theoretical foundation has been extensively discussed in Sec. \ref{sec: theory} and, as we will see in Sec. \ref{sec: app}, the perspectives are rich and broad. CMB SDs
thus make an excellent candidate for the Voyage 2050 initiative, although future work will be needed in order to strengthen their case. 

As a final remark, we point out that, while all of the aforementioned experimental setups rely on spectrometers, also imaging telescopes equipped with an inter-frequency calibrator could be used to measure CMB SDs \cite{Mukherjee2019How}. As argued in the reference, on Voyage 2050 timescales even this alternative strategy could be pushed to achieve sensitivities of the order of 0.1 Jy/sr. Furthermore, for how futuristic it might sound, the installation of lunar astronomical facilities might also be a realistic possibility to explore in the coming future \cite{silk2021limits}. In fact, for instance, an adapted version of SIMBAD has been forecasted to reach just as high sensitivities if installed on the Moon \cite{maillard2021moon} (see left panel of Fig. \ref{fig: SD_exp_future}). This mission would also have the advantage of being relatively cost-friendly and simple to install and operate, making of it an ideal first step for this novel observational site.

In summary, despite the advanced technical development of satellites for the observation of CMB SDs, all proposals have been pushed back by the funding agencies in the past decade. The situation has changed, however, with the new report of ESA's Voyage 2050 initiative, which put SDs among the two top priorities of the European agency for its Large class mission program. 

\subsection{Numerical implementation}\label{sec: num_2}
In the previous sections we have outlined many experimental setups for the observation of CMB SDs. In the following we present some of the most advanced numerical tools that can be used to evaluate their sensitivity to the various SD signals discussed in Sec.~\ref{sec: theory}. Given the \texttt{CLASS} implementation of SDs introduced in Sec. \ref{sec: num}, here we mostly focus on the publicly-available parameter extraction code \texttt{MontePython}, since the two codes are tightly interfaced.

\vspace{0.3 cm}
\noindent\textbf{Generalities of \texttt{MontePython}}

\begin{figure}
	\centering
	\includegraphics[width=0.65\textwidth]{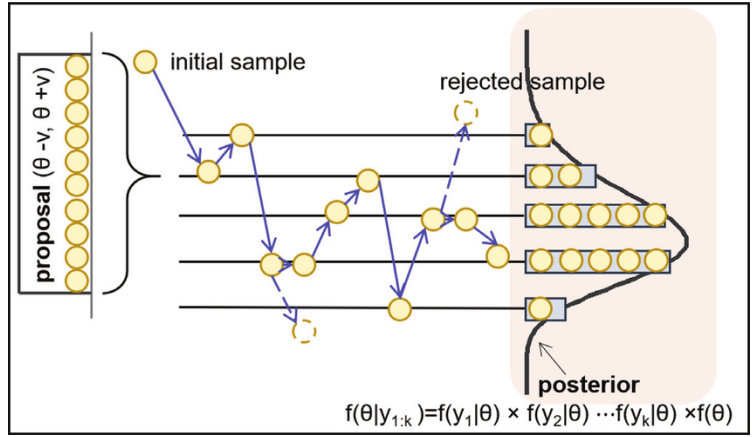}
	\vspace{0.2 cm}
	\caption{Graphical illustration of an MCMC sequence. Figure taken from \cite{dong2019prognostics}.}
	\label{fig: MP_illustration}
\end{figure}

\noindent \texttt{MontePython} \cite{Audren2013Conservative, Brinckmann2018MontePython} is a Markov Chain Monte Carlo (MCMC) sampler written in~\texttt{Python} and designed for the cosmological parameter inference. In a very crude way, the operating process is that, for some given data and a set of parameters $\theta$, the code randomly (hence Monte Carlo) attributes values to the parameters within a pre-defined (prior) range, creating a set $\theta_i$, and calculates the respective goodness of fit to the data, i.e., the likelihood that the parameters $\theta_i$ are the true ones compared to a set of best-fitting values. After each point the code updates the cumulative likelihood (or posterior probability) function of the whole chain for each parameter in $\theta$. Then, based on that information (hence Markov chain) the code chooses in which direction to ``explore'' the parameter space and decides whether to accept the next point or not following a given acceptance criterium (here defined by a Metropolis-Hastings algorithm). If accepted, the point will be included in the chain and the process continues. After a sufficient number of explorations, the multi-dimensional likelihood function of the chain converges to the its ``true'' form, from which statistical properties such as mean values and CLs can be extracted. The convergence of the chain can be determined via e.g., the Gelman-Rubin (also known as $|R-1|$) criterium \cite{Gelman1992Inference}, usually required to be below $\sim10^{-2}$. A graphical representation of this process is given in Fig. \ref{fig: MP_illustration}, taken from \cite{dong2019prognostics}. This approach is standard in cosmology (as well as in many other fields) as it allows to efficiently determine \textit{a priori} unknown probability distributions in high-dimensional parameter spaces (since for MCMCs the convergence time scales approximately linearly with the dimension of $\theta$).

In practice, it is enough to specify the array of parameters to scan over (including nuisance parameters related to e.g., foreground marginalization) and the data sets to be used. At each point of the chain \texttt{MontePython} calls \texttt{CLASS} to go from the given set $\theta_i$ to the required observable, which is then compared to the data to infer the likelihood (see the related lectures at \cite{CLASS_website}). For instance, in the case of SDs, the likelihood $\mathcal{L}$ is calculated as [\hyperlink{I}{I}]
\begin{align}
	-2\log\mathcal{L}=\chi^2=\sum_{\nu_i}\left(\frac{\Delta I(\nu_i)-\matteo{\Delta I_{\rm data}}(\nu_i)}{\delta I_{\rm noise}(\nu_i)}\right)^2\,,
\end{align}
where $\nu_i$ are the components of the frequency array, $\Delta I$ is the spectrum calculated at each point of the chain, \matteo{$\Delta I_{\rm data}$ represents the data points} and $\delta I_{\rm noise}$ is the noise level of the given experimental setup.

The chosen data sets can either be real or mock. The latter rely on the arbitrary definition of a so-called fiducial observable, which qualitatively substitutes the best fitting curve in the case of a real data set. In the following we will always assume the fiducial to be the given observable defined within \lcdm assuming Planck+BAO mean values (see right column of Tab. 2 of \cite{Aghanim2018PlanckVI}).\footnote{The parameters ruling the presence of beyond-\lcdm physics will be set so that the model recovers \lcdm (i.e., they will most often be set to zero). In this way, the sensitivities will be interpreted as constraints, although in a more general sense they are to be taken as the discovery space potentially probed by the given mission.} Depending on the experimental sensitivity, error bars are then set around the fiducial and used to derive the posterior distribution. As such, mock likelihoods are particularly suited for forecasts and to compare the raw sensitivity of different missions (completed or not) to a given observable or parameter. However, since the choice of the fiducial is arbitrary, the only meaningful quantities to be extracted from the posterior distributions are the CLs (the mean values should recover the fiducial). More details on mock likelihoods can be found in e.g., \cite[\hyperlink{I}{I}]{Perotto2006Probing, Brinckmann2018MontePython}.

\vspace{0.3 cm}
\noindent\textbf{Experimental setups}

\noindent Thanks to the versatility of the \texttt{CLASS} implementation of SDs in terms of experimental frequency settings, any SD mission can be easily implemented in \texttt{MontePython} and evaluated consistently [\hyperlink{I}{I}]. It suffices to specify same characteristics of the given experiment as explained in Sec. \ref{sec: num}, i.e., either via the external file with $\{\nu,\,\delta I_{\rm noise}\}$ for missions with varying frequency arrays or via $\{\nu_{\rm min},\,\nu_{\max},\,\Delta\nu,\,\delta I_{\rm noise}\}$ if the binning is constant, and the code automatically generates the likelihood. Since none of the aforementioned SD missions has yet taken place\footnote{Although this is not the case for FIRAS, the reported error bars are so large that assuming a mock instead of the actual residuals reported in Tab. 4 of \cite{Fixsen1996Cosmic} only marginally affects the results. One can therefore safely extend the use of mock likelihoods to FIRAS as well. Private conversation with Sebastian Hoof, Felix Kahlh\"ofer and Nils Sch\"oneberg, who we thank for bringing this point to our attention.}, mock likelihoods are automatically generated for any new mission. The first run of the code creates the fiducial. As for \texttt{CLASS}, the examples of FIRAS and PIXIE are provided by default.

\begin{figure}
	\centering
	\includegraphics[width=0.48\textwidth]{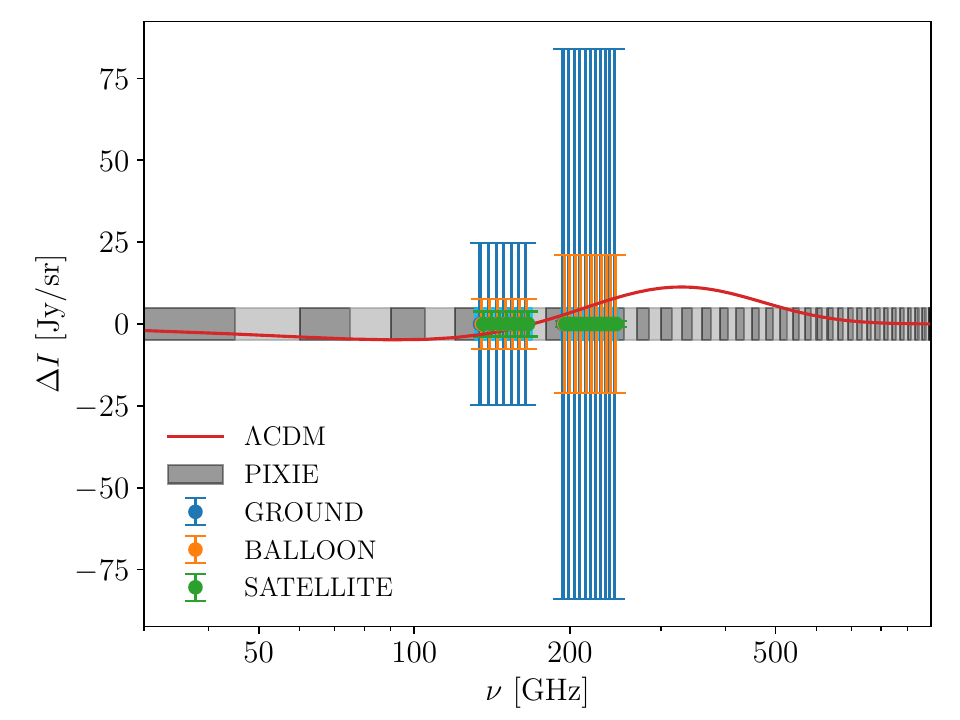}
	\includegraphics[width=0.48\textwidth]{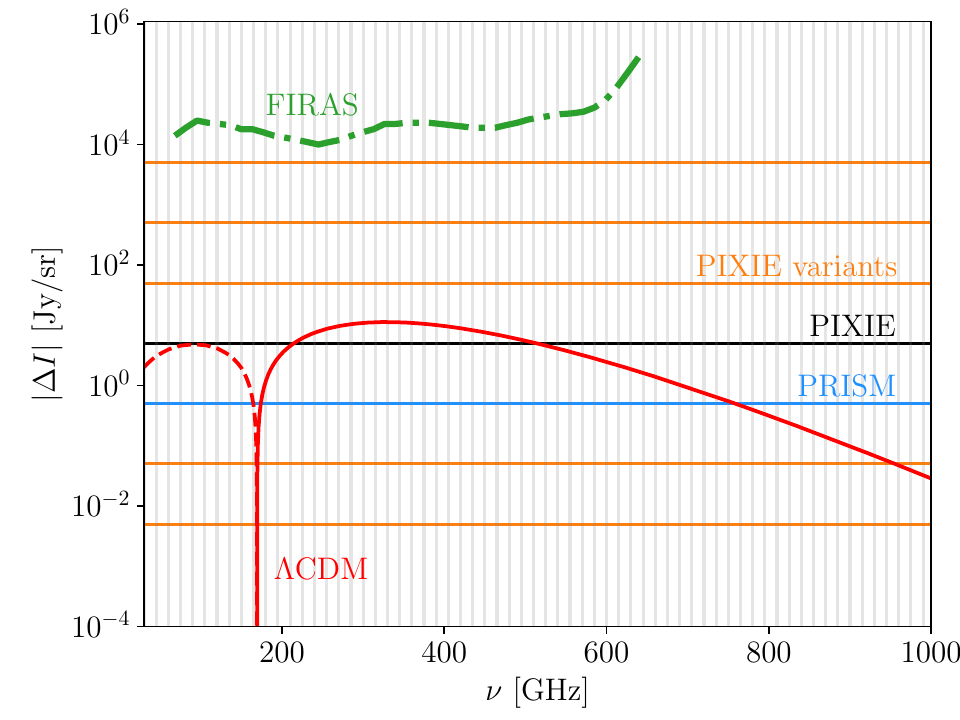}
	\caption{Representative examples of experimental configurations that can be explored with the \texttt{CLASS}+\texttt{MontePython} pipeline. Figures taken from [\protect\hyperlink{III}{III}].}
	\label{fig: SD_exp_num}
\end{figure} 

To showcase the possible applications of this setup we report the experimental configurations employed in [\hyperlink{III}{III}] and graphically represented in Fig. \ref{fig: SD_exp_num}, taken from the same reference. The left panel shows the sensitivity of a realistic ground-based (COSMO-like) setup, which is then compared to the sensitivities achievable with stratospheric balloons and satellites assuming the same frequency coverage. In this way, one can study what gains are to be obtained in a specific frequency window in the different configurations (although more realistically one would expect the frequency coverage to increase the higher the experiment -- case that the code can also easily accommodate). The right panel represents a similar situation, but focused on the comparison between satellite sensitivities.

\vspace{0.3 cm}
\noindent\textbf{Foregrounds and other nuisance parameters}

\noindent For the forecasts to deliver realistic results, however, also the galactic and extra-galactic foregrounds discussed in Sec. \ref{sec: fore} need to be taken into account. The foregrounds implementation of \texttt{MontePython} has been conducted in [\hyperlink{V}{V}] and is largely based on the work carried out in \cite{Adam2015PlanckX, Akrami2018PlanckIV, Abitbol2017Prospects}. As mentioned in Sec. \ref{sec: fore}, this modeling of the foregrounds could be improved in several directions, but still serves a useful purpose in terms of forecasting realism.

As in \cite{Adam2015PlanckX}, the contribution from galactic dust emission and the CIB is parameterized by a power law with spectral index $\beta$ at low frequencies and a BB at high frequencies, i.e., 
\begin{align}
	\Delta I_i =  A_i \left(\frac{x_i}{x_{i,\mathrm{ref}}}\right)^{\beta_i+3}
	\frac{\exp\left(x_{i,\mathrm{ref}}\right)-1}{\exp\left(x_i\right)-1}~,
\end{align}
where $i\in \{\mathrm{D, CIB}\}$,  $A_i$ is the amplitude of the respective emission, $x_i=\nu/T_i$ and $\nu_\mathrm{ref} = 545$~GHz. The various free parameters have been determined to be $\beta_D=1.53\pm0.05$, $T_D=21\pm2$~K \cite{Adam2015PlanckX} (see Tab. 5 therein as well as \cite{Erler2017Planck}), $\beta_{\rm CIB}=0.86\pm0.12$, $T_{\rm CIB}=18.8\pm1.2$~K \cite{Fixsen1998Spectrum, Abitbol2017Prospects} (see also App. F of \cite{Akrami2018PlanckIV}) and are marginalized over\footnote{By ``marginalizing'' here we mean accounting for the possibility that the foreground parameters can degrade the sensitivity of a given mission to the SD signal. This can happen, for instance, if a given parameter affects the shape of a foreground signal in the same way as a cosmological parameter would affect the shape of the SD signal.} the respective uncertainties. In full generality, both amplitudes should be left free to vary around their central values 1.49 and 0.41 MJy/sr, respectively \cite{Abitbol2017Prospects}. In practice, however, at frequencies below 1~THz the two contributions follow almost perfectly the power law behavior (see Fig.~\ref{fig: SD_fore_gal}) and their amplitudes are therefore largely degenerate (see e.g., Fig.~A1 of~\cite{Abitbol2017Prospects}). Without much loss of generality, one can therefore fix one of the two contributions to its best-fitting values and this is the approach taken as default in the \texttt{MontePython} implementation.

Following \cite{Abitbol2017Prospects}, we describe the synchrotron emission as a power law with a logarithmic dependence on the spectral curvature $\omega_S$ (which accounts for the average over multiple spectral energy distributions with slightly different power laws \cite{Chluba:2017rtj}), i.e., 
\begin{align}
	\Delta I_{\rm S} = A_{\rm S} \left(\frac{x_{\rm S}}{x_{{\rm S},\mathrm{ref}}}\right)^{\alpha_{\rm S}}\left[1+\frac{1}{2}\omega_{\rm S}\ln^2\left(\frac{x_{\rm S}}{x_{{\rm S},\mathrm{ref}}}\right)\right]\,,
\end{align}
where in this case $\nu_\mathrm{ref} = 100$~GHz. As in \cite{Abitbol2017Prospects}, we adopt $A_{\rm S}=288\pm29$ Jy/sr and $\alpha_{\rm S}=-0.82\pm0.08$. While in the reference $\omega_{\rm S}$ is fixed to 0.2, here we instead marginalize over it.

The spectrum of the bremsstrahlung (free-free) emission has been determined in \cite{draine2010physics} and can be parameterized in various ways. Here we follow \cite{Adam2015PlanckX} and use
\begin{align}
	\Delta I_{\rm ff} = A_{\rm ff} T_e (1-e^{-\tau_{\rm ff}})\,,
\end{align}
where $A_{\rm ff}=2 \nu^2$, $T_e$ is the electron temperature and $\tau_{\rm ff}$ the optical depth of the process which depends on the integrated squared electron density along the LOS (the so-called emission measure -- $EM$ \cite{Draine1997Diffuse, Draine2011Physics}). A complete definition of this quantities can be found in e.g., \cite[\hyperlink{V}{V}]{Adam2015PlanckX}. For $T_e$ one has that $T_e=7000\pm500$ K, while $EM=15\pm35$~pc/cm$^6$~\cite{Adam2015PlanckX}.

The spectra of the CO emission and AME are accounted for using spectral templates, which are adapted from \cite{Adam2015PlanckX} (the former in turn based on \cite{Mashian:2016bry}). We neglect the secondary contributions mentioned in Sec. \ref{sec: fore} as well as possible spatial variations. 

Finally, one also needs to marginalize over the uncertainty on the determination of the CMB monopole temperature (which determines the BB with respect to which $\Delta I$ is defined) as well as over the SZ effect, if interested in observing the primordial signal. The former can be achieved by including temperature shift distortions up to second order in $\Delta_T=\Delta T/T$, i.e., 
\begin{align}
	\Delta I_T = \Delta_T (1+\Delta_T)\mathcal{G}+\frac{\Delta_T^2}{2}\mathcal{Y}\,,
\end{align}
and marginalizing over $\Delta_T=0\pm0.00022$ \cite{Fixsen2009Temperature}. The SZ contribution can in principle be marginalized over using the \texttt{CLASS} implementation presented in Sec. \ref{sec: num}. However, to reduce the number of nuisance parameters\footnote{Nuisance parameters are the subset of parameters that need to be marginalized over (see previous footnote), in opposition to the cosmological parameters of interest.} by default we approximate the SZ signal to
\begin{align}
	\Delta I_{\rm SZ}=y_{\rm SZ}\mathcal{Y}\,,
\end{align} 
i.e., neglecting the kSZ effect and the relativistic contribution to the tSZ effect. Doing so only affects the forecasted sensitivity to $\mu$ by $\sim15\%$ \matteo{(for a PIXIE-like mission)}, with however a great reduction of the computation time [\hyperlink{V}{V}]. The amplitude $y_{\rm SZ}$ is varied in the range $y_{\rm SZ}=(1.77\pm1.57)\times 10^{-6}$ \cite{Hill2015Taking, Dolag2015SZ}. This automatically also accounts for the residual contribution from the CMB dipole, which leads to the same type of distortion (see Sec. \ref{sec: SD_lcdm}). 

\matteo{If wished, for more realistic forecasts the inclusion of higher-order contributions to the SZ signal would be rather straightforward. On one hand, as mentioned above, one could employ the multipole expansion of the SZ effect discussed in Sec.~\ref{sec: SD_lcdm} to account for both the relativistic tSZ and the kSZ effects (as done in e.g., [\hyperlink{III}{III}]). This would simply require the inclusion of the underlying quantities ($T_e$, $\Delta\tau$, $\beta$ and $\cos\theta$) as cosmological parameters at the level of \texttt{MontePython} to effectively treat them as nuisance parameters. Alternatively, one could make use of the fact that the kSZ effect averages to almost zero over the sky (because of the random distribution of the galaxy cluster velocities) and that for relatively low electron temperatures the first order correction to the tSZ effect captures rather accurately the overall relativistic contribution (see e.g., Fig.~1 of~\cite{Chluba2012Sunyaev}). In this way, it would be enough to include a single extra parameter, i.e., the amplitude of the first-order correction (or equivalently the $k=1$ term in Eq.~\eqref{eq: Y_rel}), to reproduce with good accuracy the complexity of the SZ effect.\footnote{\matteo{We thank Colin Hill for pointing out this second possibility.}} A quantitative estimation of the impact on the forecasts of these different approaches is left for future work.}

\vspace{0.3 cm}
\noindent\textbf{Summary}

\noindent In summary, the \texttt{MontePython} implementation of SDs (interfaced with the \texttt{CLASS} code presented in Sec.~\ref{sec: num}) has been designed to be able to deal with any experimental configuration and accounts for galactic and extra-galactic foregrounds. As such, it represents an ideal tool to perform realistic sensitivity forecasts for the missions to come.

\newpage
\section{The view from the top: perspectives for the future}\label{sec: app}

Very broadly, there are two ways via which CMB SDs can bring forward our understanding of cosmology: test and discover. Both are fundamentally important and both need to be considered in synergy with the other available complementary probes of the universe. The former applies to e.g., models or effects whose scale- or time-dependence we believe we have fully explored via observations of e.g., BBN  and the CMB anisotropies. An example thereof is \lcdm, as seen in Sec.~\ref{sec: lcdm}. In fact, should \lcdm be the ``true'' model, SDs would only further test its parameterization and the question becomes which degree of precision one needs to reach in order to improve upon current bounds. On the other hand, there are many other models with large portions of unexplored parameter space because of observational or physical limitations of the aforementioned probes. For instance, CMB anisotropies are blind to the shape of the PPS at scales below 1 Mpc$^{-1}$ or to scenarios where most of the heating happens before recombination (like it could be the case in many DM models, including PBHs), which could both be probed with SDs. This creates an enormous discovery space for CMB SDs as we will see below.

In the following sections we review many of these scenarios. We cover the case of \lcdm in Sec. \ref{sec: res_lcdm}. In Sec. \ref{sec: res_infl} we discuss what CMB SDs are and will be able to tell us about inflation at scales inaccessible by any other probe. On a similar note, in Sec. \ref{sec: res_further} we emphasize the connection between CMB SDs and primordial GWs, which could be sourced during or right after the inflationary epoch. In Secs. \ref{sec: res_DM} and \ref{sec: res_PBH} we then present two representative cases of how CMB SDs could constrain the properties of DM, be it under the form of a particle or PBHs, respectively. While in these last two sections the energy injection is assumed to be instantaneous and homogeneous, Sec.~\ref{sec: res_PMFs} focuses on the fact that many of these processes might instead lead to anisotropic energy injections, inducing potentially observable spatial fluctuations of the SD signal. In relation to inhomogeneities in the early universe, in the same section we also treat various sources of inhomogeneous recombination, as it would be the case in the presence of PMFs. Finally, to highlight the fact that SDs do not only constrain the thermal but also the expansion history of the universe, in Sec.~\ref{sec: res_tens} we furthermore place CMB SDs in the context of the Hubble tension using EDE as a proof of principle. 

For sake of completeness, the results obtained during the period of the thesis will be placed in the more general context of CMB SDs, emphasizing also the relevance of other works. Adequately introducing these related studies is fundamental in order to set the stage for either the analyses developed for the thesis or the possible follow-ups that might stem from them (for instance in combination with the referenced works). As a useful byproduct, the following sections might then serve as a helpful review of the status and perspectives of CMB SDs. We remark, however, that although rather comprehensive the following discussion in not all-inclusive. Complementary information can be found in previous reviews such as e.g.,~\cite{Chluba2019Spectral, Chluba2019Voyage}. Furthermore, many of the models discussed below implicitly serve as representative examples for other scenarios that might induce similar effects as discussed in Secs.~\ref{sec: SD_lcdm} and~\ref{sec: SD_non_lcdm}.

\subsection{\lcdm}\label{sec: res_lcdm}

There is mainly a twofold contribution that CMB SDs can deliver in the context of the \lcdm model. On the one hand, as mentioned in Sec.~\ref{sec: SD_lcdm}, the observation of the late-time SZ effect provides useful information on the thermodynamics and spatial distribution of structures. On the other hand, the observation of the primordial signal (i.e., that generated before recombination) would allow to study the thermal and expansion history of the universe prior and around recombination. In the following discussion we cover both of these aspects, noting, however, that while the latter possibility has been abundantly studied in the context of the thesis, the former is mostly presented for sake of completeness (and will therefore be treated more succinctly, with more details to be found in the given references).

\vspace{0.3 cm}
\noindent \textbf{Late-time contribution}

\noindent Starting from the former aspect, although no spectral observation of the SZ effect has ever been made to date, catalogs of thousands of SZ sources have been produced by several CMB anisotropy missions (such as Planck \cite{Planck:2015koh}, ACT \cite{ACT:2017dgj, ACT:2020lcv} and SPT \cite{SPT:2014wbo, SPT:2019xsj}) on the basis of a matched (spatial-spectral) filter approach.\footnote{In a crude way, in this approach the correlation of the noise (such as the CMB at large scales and white noise at very small scales) and of the SZ signal between the observed maps at different frequencies is used to suppress the former and enhance the latter (see e.g., Fig. 3 of \cite{ACT:2020lcv} as well as \cite{2011ApJ...738..139W}).} For many of these sources, the redshift has also been determined via complementary X-ray and optical observations. With this information available, two types of analyses can be carried out. 

\begin{figure}
	\centering
	\includegraphics[width=0.45\textwidth]{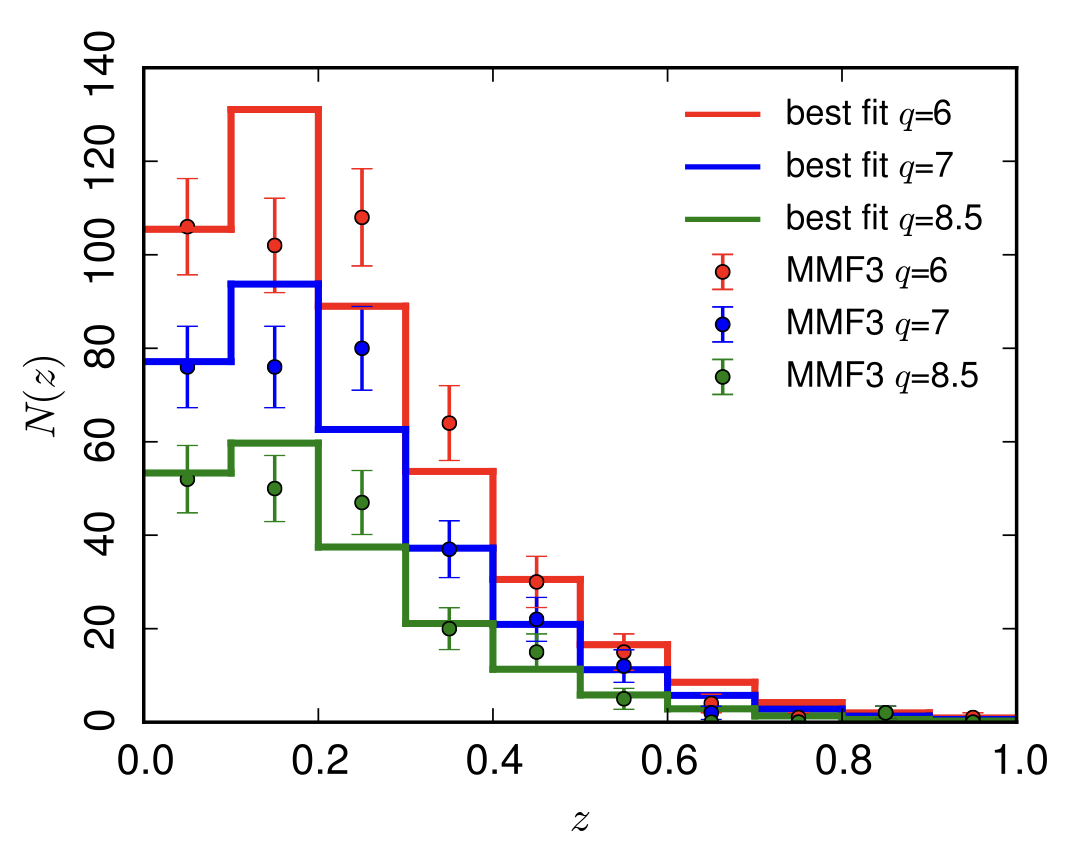}	\includegraphics[width=0.52\textwidth]{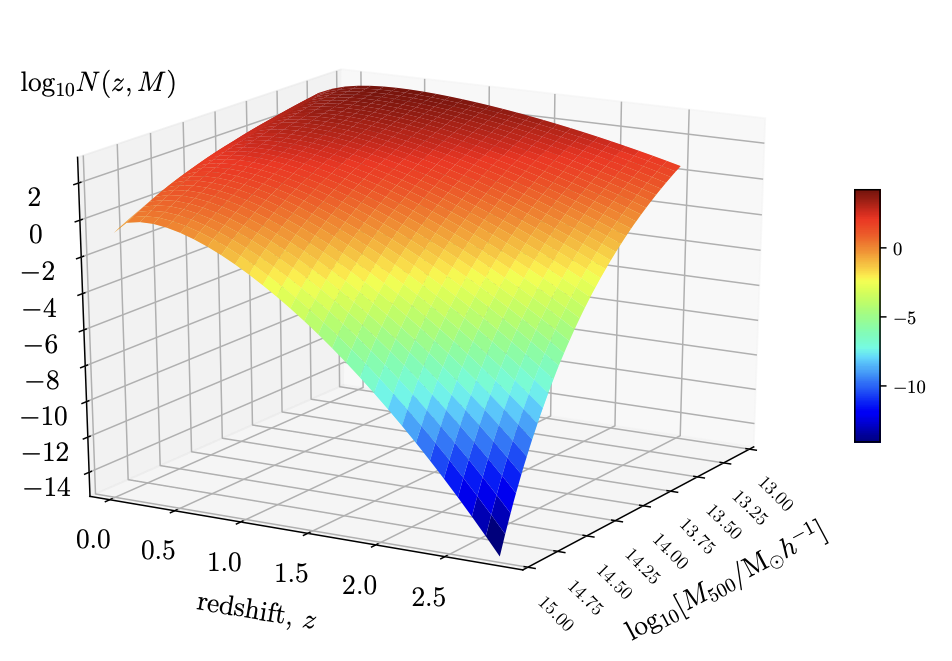}
	\caption{Representative illustration of the distribution functions $N(z,q)$ (left) and $N(z,M)$ (right) resulting from CNC analyses. Figures taken from \cite{Planck:2015lwi, Bolliet:2019zuz}.}
	\label{fig: SD_SZ_CNC}
\end{figure}

On the one hand, the observed sources can be binned (or ``stacked'') as a function of redshift $z$ and SNR $q$ (a proxy for the intensity of the source), defining the distribution function $N(z,q)$. The observed counts for the Planck 2015 sample are reported in the left panel of Fig. \ref{fig: SD_SZ_CNC}, taken from \cite{Planck:2015lwi}, as illustration. Intuitively, the higher the intensity of the signal (i.e., the SNR) the larger the SZ effect (and the related $y$ parameter) must have been, and hence the larger the mass $M$ of the emitting cluster (which is in turn related to the temperature of the environment, see e.g., \cite{reichert2011observational}). A similar correlation also exists with the (angular) size of the emitter $\theta$. One can therefore re-write $N(z,q)$ as $N(z,M)$, which is graphically displayed in the right panel of Fig. \ref{fig: SD_SZ_CNC}, taken from \cite{Bolliet:2019zuz}. In turn, the abundance of clusters of a given mass and at a given redshift can be determined on the basis of the so-called halo-mass function, which can be theoretically predicted given (among others) $\Omega_m$ and $\sigma_8$ \cite{cooray2002halo} (see Sec.~\ref{sec: LSS}). Inverting then the logic, one can use cosmological information to define the halo-mass function, determine $N(z,M)$ and infer $N(z,q)$, which can finally be compared to the data (solid lines in the left panel of Fig.~\ref{fig: SD_SZ_CNC}). This approach is referred to as Cluster Number Count (CNC).

\begin{figure}
	\centering
	\includegraphics[width=0.67\textwidth]{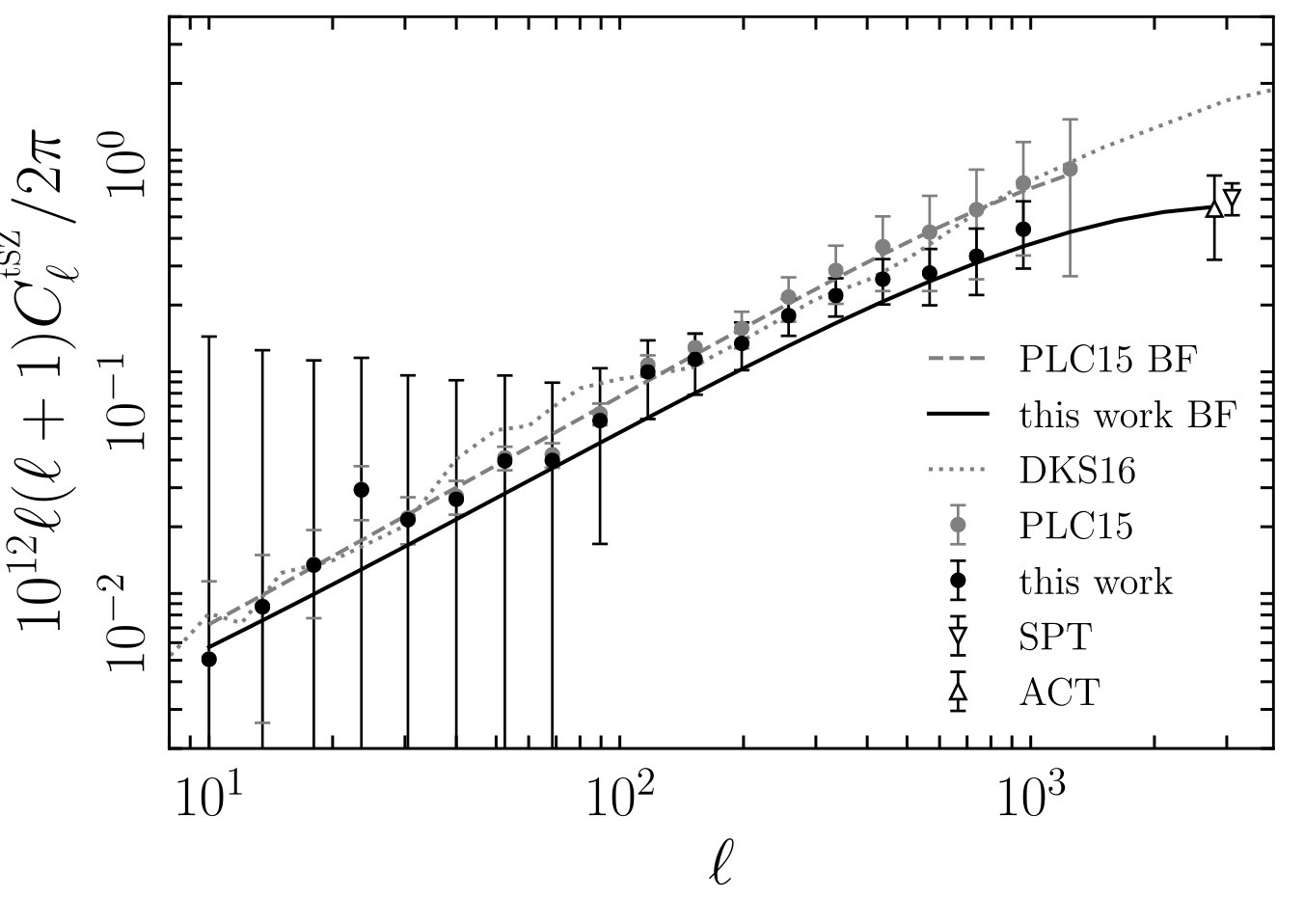}
	\caption{Same as Fig. \ref{fig: CMB_spectra} but for the SZ power spectrum. Figure taken from \cite{Bolliet:2017lha}.}
	\label{fig: SD_SZ_power_spectrum}
\end{figure}

On the other hand, in full analogy with the CMB power spectra one can construct the SZ power spectrum (i.e., the two-point correlation function of the SZ sources) directly from the observed maps. In this way all clusters (resolved and unresolved) are contributing to the spectrum. A graphical representation of its determination from various experiments is shown in Fig. \ref{fig: SD_SZ_power_spectrum}, taken from \cite{Bolliet:2017lha}. Similarly to CNC analyses, also in this case the theoretical prediction builds upon the halo-mass function as well as on the electron pressure profile. In terms of the relevant cosmological parameters, these dependences boil down to \cite{Bolliet:2017lha}
\begin{align}
	C_\ell\propto \sigma_8^{8.1} \left(\frac{\Omega_m}{B}\right)^{3.2} h^{-1.7} \quad \text{for} \quad \ell<10^3\,,
\end{align}
where $B$ is the mass bias, and are graphically summarized in Fig. 2 of the same reference.

\begin{figure}[t]
	\centering
	\includegraphics[width=0.49\textwidth]{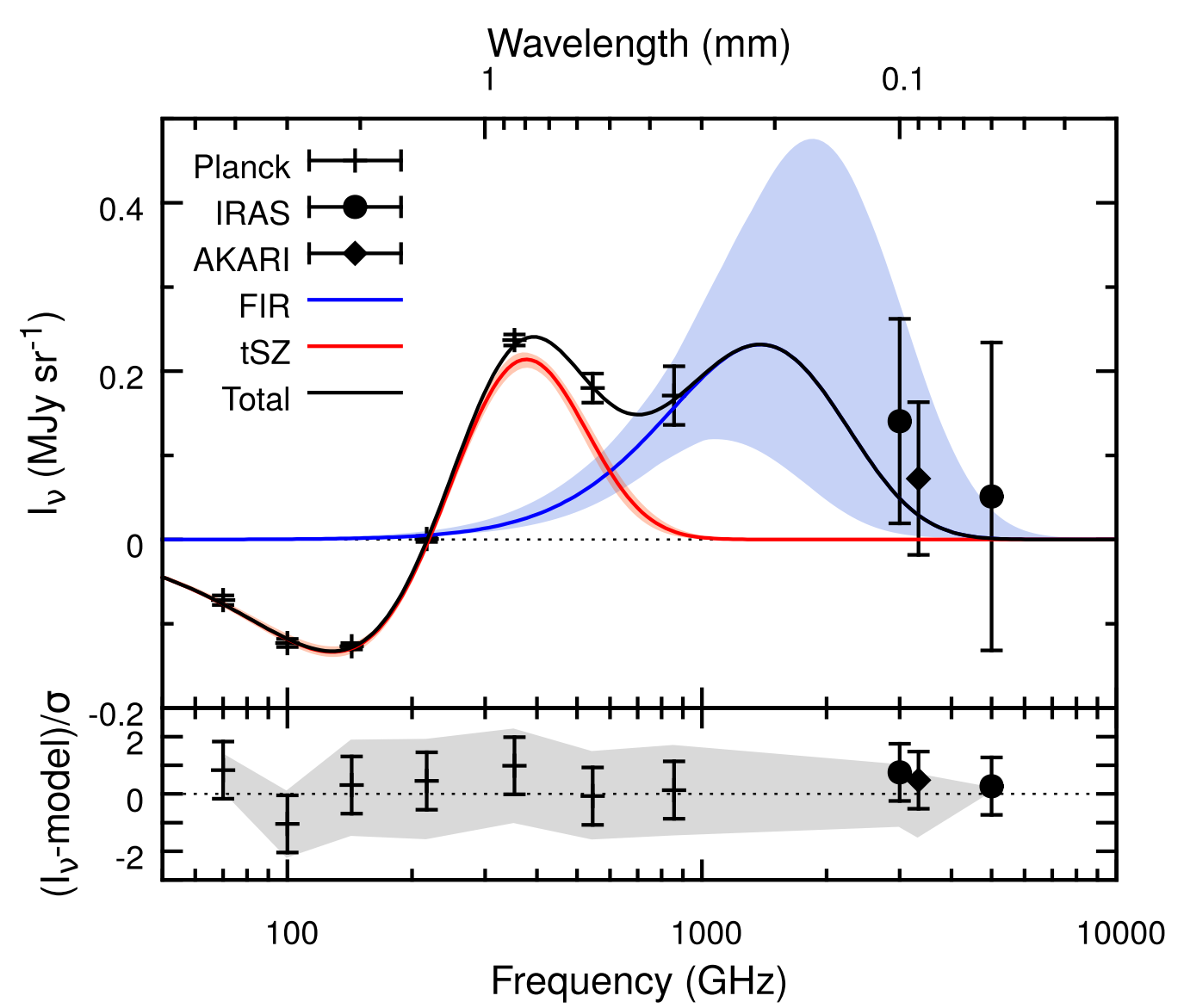}
	\includegraphics[width=0.47\textwidth]{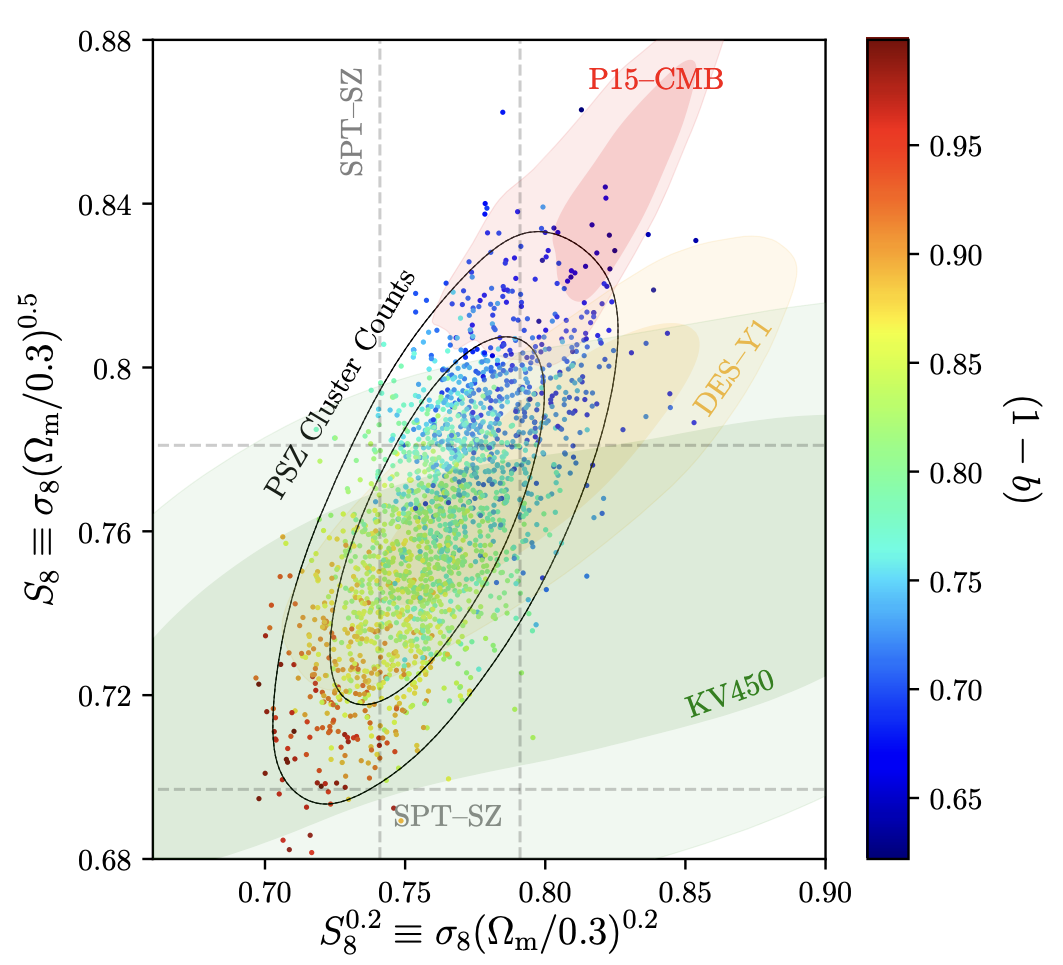}
	\caption{Reconstructed SZ signal (left) and constrains on the $\sigma_8-\Omega_m$ plane (right) resulting from CNC analyses. Figures taken from \cite{Erler2017Planck, Bolliet:2019zuz}.}
	\label{fig: SD_SZ_signal}
\end{figure}

Via the complementary of these two approaches, it is then possible to constrain several cosmological parameters and this exercise has been conducted in a number of references (see e.g., \cite{Planck:2013lkt, Planck:2015lwi, Erler2017Planck, SPT:2018njh, Vavagiakis:2021ilq, SPT:2021efh} and \cite{Hill:2013baa, Planck:2015vgm, Bolliet:2017lha, Bolliet:2019zuz, Rotti:2020rdl} for examples of CNC and power spectrum analyses, respectively). Particularly interesting is the fact that, although not directly observing the SD spectrum, the dependence of these data on the characteristics of the SZ signal (via e.g., the transformation $q\to M$ in CNC studies) allows to reconstruct the shape of the underlying SZ signal, as shown for instance in the left panel of Fig. \ref{fig: SD_SZ_signal}, adapted from \cite{Erler2017Planck}. Since the signal depends of the electron temperature (see Sec. \ref{sec: SD_lcdm}), this bears important consequences for our understanding of the thermal history of the late universe. In terms of $\sigma_8-\Omega_m$, the results are overall compatible with weak lensing surveys with comparable precision, consistently underpredicting the CMB estimates. The right panel of Fig. \ref{fig: SD_SZ_signal}, taken from \cite{Bolliet:2019zuz}, reports the cases of Planck 2015 (solid lines) and SPT (dashed lines).

\begin{figure}
	\centering
	\includegraphics[width=0.52\textwidth]{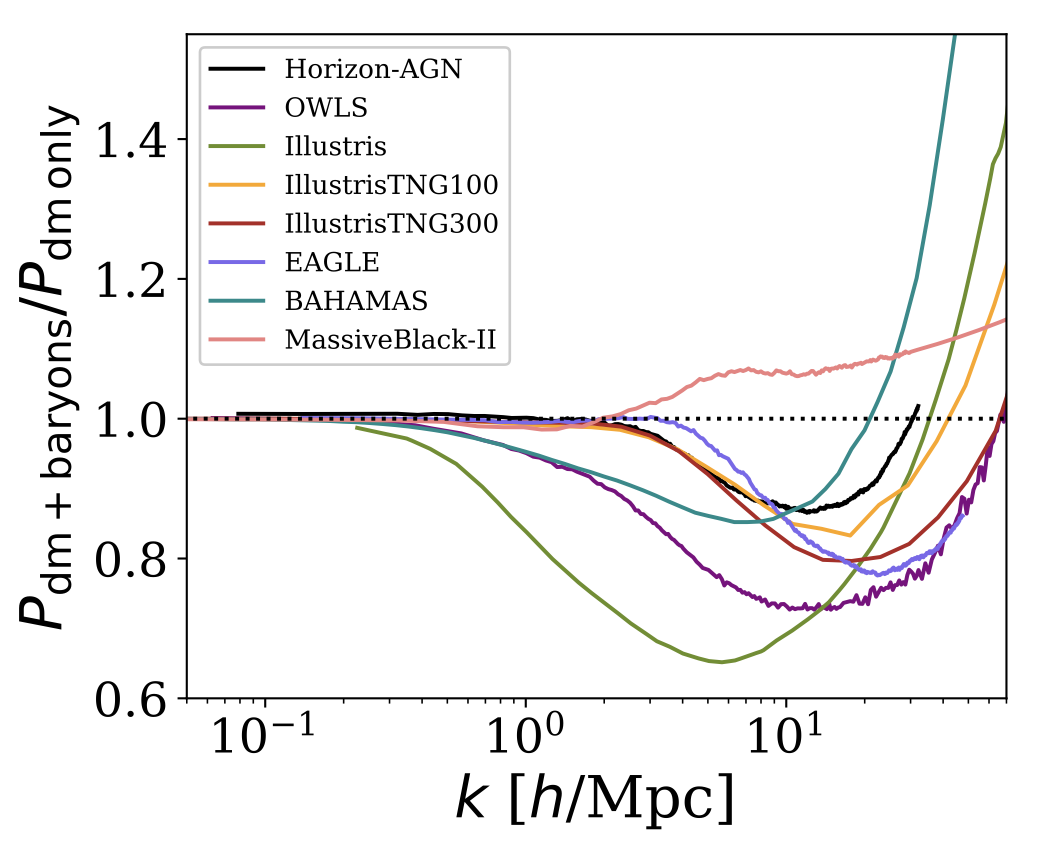}
	\includegraphics[width=0.46\textwidth]{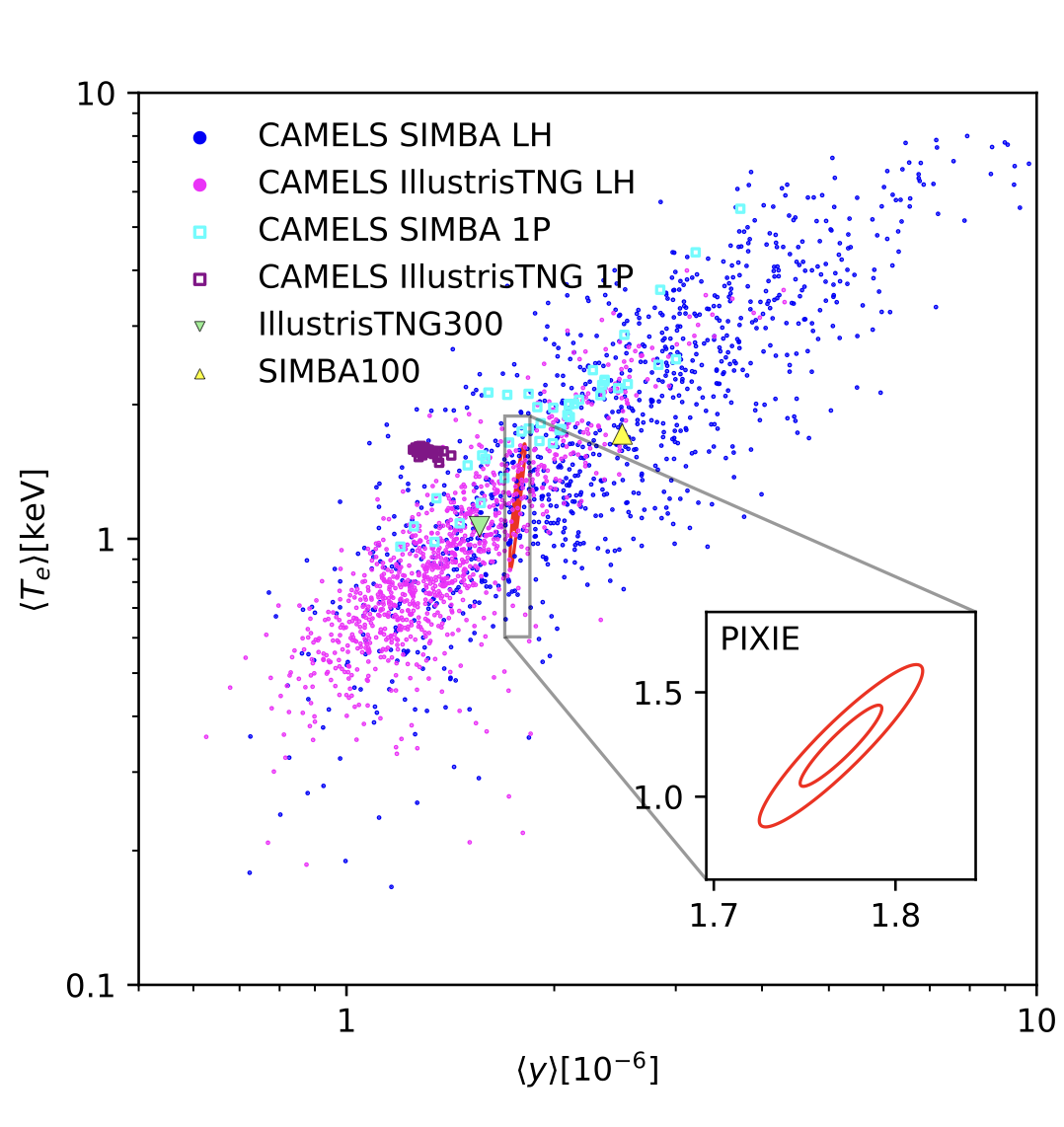}
	\caption{Uncertainty on the matter power spectrum due to baryonic feedback effects (left) and forecasted constraining power of a PIXIE-like mission (right). Figures taken from \cite{Chisari:2019tus, Thiele:2022syq}.}
	\label{fig: SD_SZ_b_feed}
\end{figure}

Up-coming spectroscopic observations of the SD signal (see Sec.~\ref{sec: current_exp}) will be able to target both localized sources and the background signal with significantly improved precision. This will not only widen the current catalogs, but it will also strengthen the constraints on the inferred parameters. A particularly promising prospect is, for instance, the one considered in \cite{Thiele:2022syq}, where the authors systematically investigate the dependence of the SZ parameters ($y$ and $T_e$) on (baryonic) feedback effects due to e.g., star formation and active galactic nuclei (see Fig.~4 of the reference). Although in the (numerical) modeling of structure formation it can be assumed that the dynamics is fully determined by gravity at relatively large scales, at smaller scales the back xreaction from galactic astrophysics needs to be taken into account (see e.g., \cite{Angulo:2021kes} -- and in particular Sec. 9.9.5 therein -- for a review). Nevertheless, the impact of the latter is notoriously difficult to predict at scales below 0.1 $h$/Mpc, as clearly shown in the left panel of Fig. \ref{fig: SD_SZ_b_feed}, taken from \cite{Chisari:2019tus}, and this uncertainty propagates to all analyses that rely upon the simulations, as it is the case for weak lensing surveys (see e.g., \cite{Chisari:2019tus}). The work of \cite{Thiele:2022syq} demonstrated that the future observation of the SD signal by a PIXIE-like mission would significantly constrain the available parameter space for the simulations, as shown in the right panel of Fig.~\ref{fig: SD_SZ_b_feed}, taken from \cite{Thiele:2022syq}, and thereby indirectly reduce the numerical uncertainty on the role of the baryonic feedback. This would have important consequences for both our understanding of galactic dynamics and the accuracy of future cosmological surveys. 

\vspace{0.3 cm}
\noindent \textbf{Primordial contribution}

\noindent As discussed in Sec. \ref{sec: SD_lcdm}, the combination of the primordial sources of SDs depends on four \lcdm parameters: the amplitude and tilt of the PPS, $A_s$ and $n_s$, and the baryon and DM energy densities, $\omega_b$ and $\omega_{\rm cdm}$. The former two can only be constrained via their effect on the dissipation of acoustic waves, while the latter via the baryon cooling and the CRR as well (see in particular Figs.~\ref{fig: SD_cool_diss_dep} and \ref{fig: SD_CRR_dep}). As calculated in \cite{Chluba2019Voyage} (see also Sec.~\ref{sec: future_exp}), only SD missions as sensitive as Super-PIXIE or better would be sufficient to directly measure the full primordial signal.

\begin{figure}[h!]
	\centering
	\includegraphics[width=0.7\textwidth]{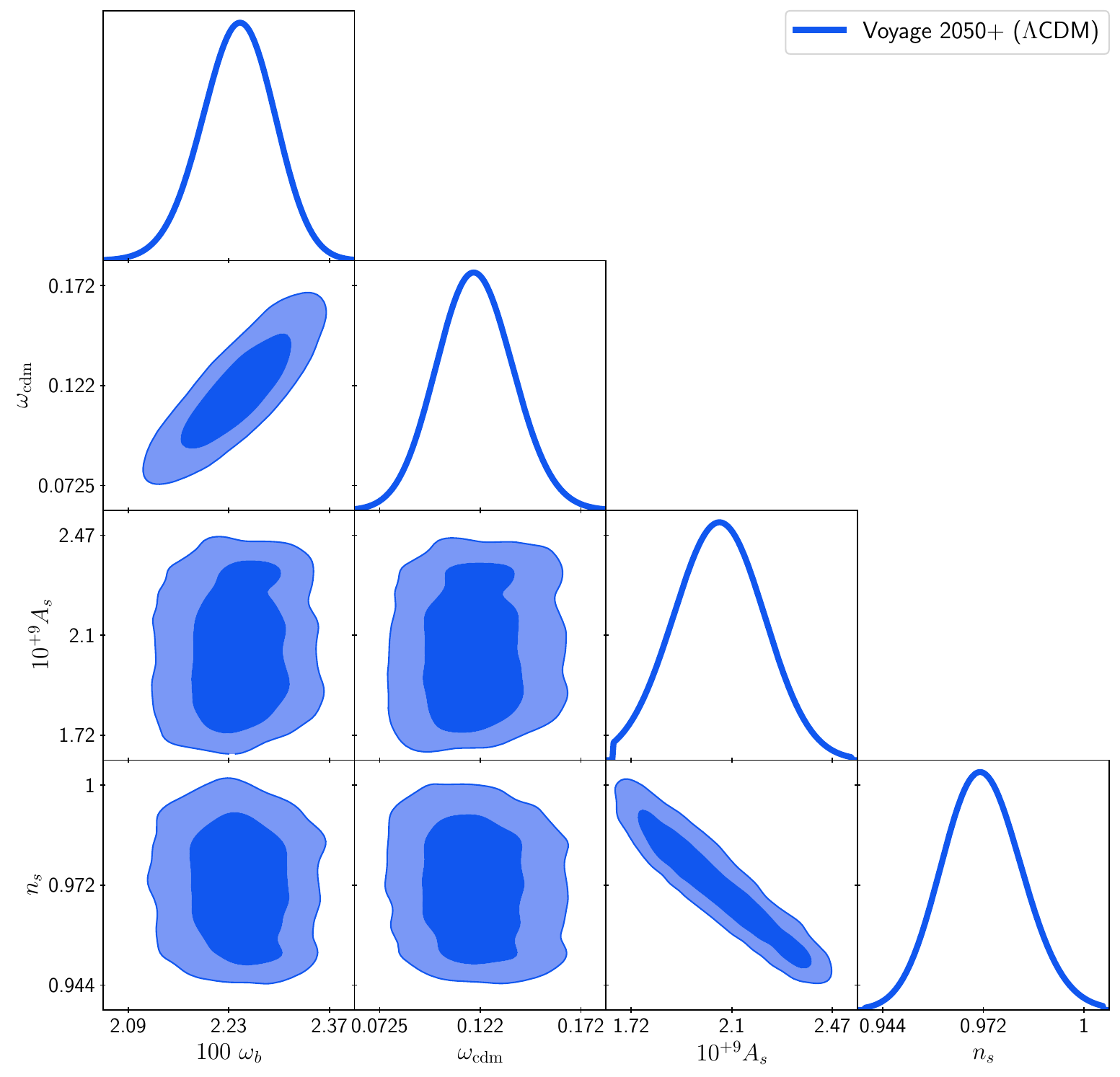}
	\caption{1D posteriors and 2D contours (at 68\% and 95\% CL) of the cosmological parameters that can be constrained with CMB SDs within the \lcdm model. Figure taken from [\protect\hyperlink{XIV}{XIV}].}
	\label{fig: SD_MCMC_res_1}	
\end{figure}

The first sensitivity forecast to the combination of all of these cosmological parameters including all \lcdm sources of SDs (see Sec.~\ref{sec: num}) and with a realistic treatment of the foregrounds (see Sec.~\ref{sec: num_2}) has been conducted in [\hyperlink{XIV}{XIV}], although analyses dedicated to subsets of these quantities have been presented in various works before (see e.g, \cite[\hyperlink{III}{III}]{Chluba2016Which, Hart2020Sensitivity}).\footnote{Most works, such as the one carried out in \cite{Chluba2019Voyage}, focus solely on the SD parameters, such as $\mu$ and the signal amplitudes, without drawing the connection to the underlying \lcdm parameters.}~The analysis of [\hyperlink{XIV}{XIV}] has been developed in the context of the Voyage 2050 mission and its advanced version Voyage 2050+ with a 10-fold sensitivity improvement. The resulting 1$\sigma$ uncertainties and corresponding SNRs are reported in Tab. \ref{tab: SD_SNRs}, adapted from~[\hyperlink{XIV}{XIV}], while the constraints for the case of Voyage 2050+ are graphically displayed in Fig.~\ref{fig: SD_MCMC_res_1}, taken from [\hyperlink{XIV}{XIV}].  

\begin{table}[t]
	\centering
	\def\arraystretch{1.2}
	\scalebox{1.0}{
		\begin{tabular}{l|cccc} 
			& $\omega_b$ & $\omega_{\rm cdm}$ & $10^9A_s$ & $n_s$ \\
			\hline
			& \multicolumn{4}{c}{$1\sigma$} \\
			\hline 
			Voyage 2050 & $4.9\times10^{-3}$ & $0.19$ & $2.1$ & $0.14$ \\
			\hline
			Voyage 2050+ & $4.9\times10^{-4}$ & $0.019$ & $0.21$ & $0.014$ \\
			\hline
			\hline
			& \multicolumn{4}{c}{SNR} \\
			\hline
			Voyage 2050 & $4.9$ & $0.68$ & $1.0$ & $6.9$ \\
			\hline
			Voyage 2050+ & $49$ & $6.8$ & $10$ & $69$ \\
			\hline
	\end{tabular} }
	\caption{Forecasted $1\sigma$ sensitivities (top) and SNRs (bottom) of the relevant \lcdm parameters for the Voyage 2050 and Voyage 2050+ missions. Table adapted from [\protect\hyperlink{XIV}{XIV}].\vspace{0.5 cm}}
	\label{tab: SD_SNRs}
\end{table}

As argued in the reference on the basis of these results (and those obtained in \cite{Hart2020Sensitivity}), the constraining power of a Voyage 2050-like mission on the two energy densities would come entirely from the CRR, with sensitivities of the order of $0.005$ and $0.2$ for $\omega_b$ and $\omega_{\rm cdm}$, respectively. On the other hand, the sensitivity to the PPS parameters $A_s$ and $n_s$ is of about $2 \times 10^{-9}$ and 0.15, respectively. In both instances, the error bars are inflated by the strong degeneracies that exist between $\omega_b-\omega_{\rm cdm}$ and $A_s-n_s$, as evident from Fig.~\ref{fig: SD_MCMC_res_1}. The reason behind the presence of the former is that, when $\omega_b$ is increased, the whole spectrum increases proportionally to $\Delta I_{\rm CRR}$, so that the peaks are enhanced more than the dips. When increasing $\omega_{\rm cdm}$ the peaks are suppressed and the dips become less pronounced (see \cite{Hart2020Sensitivity} and in particular Fig. 2 therein). As a result, the effects of these two quantities compensate when both are increased, leading to the degeneracy shown in the respective subplot of Fig. \ref{fig: SD_MCMC_res_1}. In turn, as discussed in [\hyperlink{III}{III}], the second degeneracy can be explained by the fact that at very small scales (far away from the chosen pivot value -- see Sec. \ref{sec: infl}) the considered SD mission cannot distinguish an increase of power due to a larger amplitude or a larger tilt, causing to the strong anti-correlation visible in the respective subplot of Fig. \ref{fig: SD_MCMC_res_1}.

Although these constraints would not be competitive with those imposed on the same quantities from e.g., CMB anisotropy measurements (see Sec. \ref{sec: CMB}), there are several reasons why the observation of primordial CMB SDs would be important even within the \lcdm model. First of all, it is \textit{per se} remarkable that CMB SDs could in principle place bounds on four of the six \lcdm parameters independently of any other probe. No other cosmological observable beside the CMB anisotropies \matteo{(and possibly 21 cm cosmology in the future \cite{Liu:2019awk})} can do so. As such, CMB SDs would be able to deliver a totally independent and complete picture of the expansion and thermal history of the early universe.

Furthermore, an improved understanding and modeling of the foregrounds, which is not unrealistic given the discussion had in Sec. \ref{sec: fore}, could improve the sensitivity to the $\mu$ signal by a factor 10 \cite{Abitbol2017Prospects}, which would correspondingly lower the errors on $A_s$ and $n_s$. A qualitatively similar discussion would also apply to the CRR and the inferred constraints on the energy densities. 

\begin{figure}
	\centering
	\includegraphics[width=0.7\textwidth]{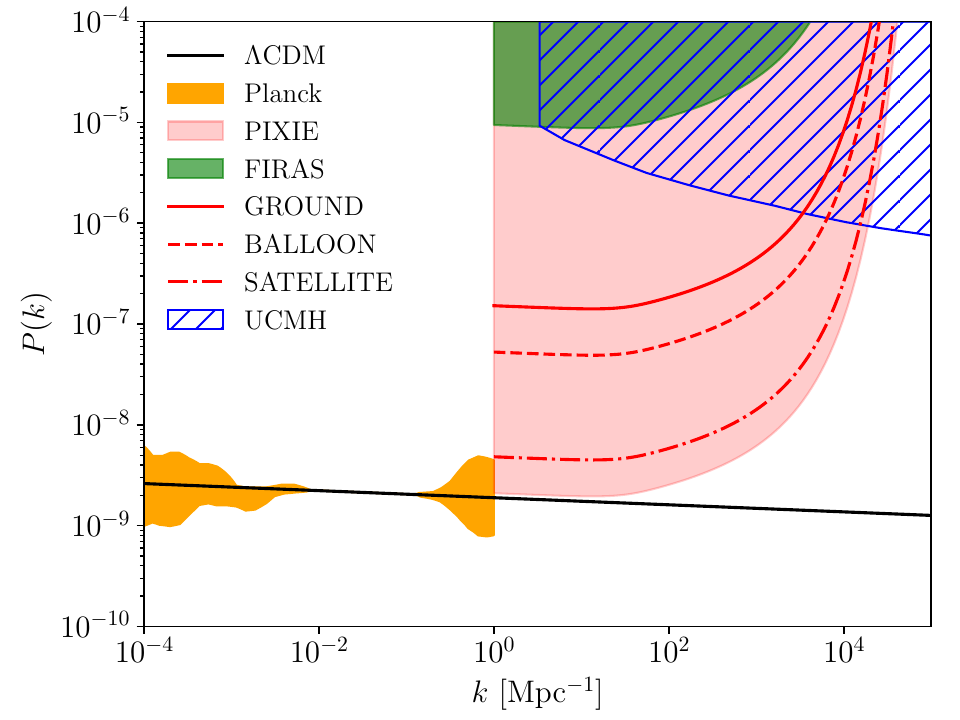}
	\caption{Current and forecasted constraints on the PPS. Figure taken from [\protect\hyperlink{III}{III}].}
	\label{fig: LCDM_Pk_constraints}
\end{figure}

Even more importantly, CMB SDs would test scales that are unexplored by CMB anisotropy measurements (as discussed in Sec. \ref{sec: SD_lcdm} and as we will see in the following sections in more details) and therefore the SD constraints on the PPS parameters would not only test the CMB anisotropy results, but significantly extend their validity range. In this sense, even the non-observation of the primordial SD signal by lower-sensitivity mission would deliver relevant constraints, as shown in Fig. \ref{fig: LCDM_Pk_constraints}, taken from [\hyperlink{III}{III}]. As it can be seen in the figure, SD constraints would almost double the number of decades of $k$ modes that can be tested and improve by orders of magnitude upon the current bounds even with ground-based setups. As a further remark in this regard, it has to be pointed out that the relative improvement bought by the addition of CMB SDs on top of CMB anisotropies is mainly due to the choice of pivot scale used as default to define the PPS parameters (see Eq. \eqref{eq: PPS_LCDM}). If instead of $k_*=0.05$ Mpc$^{-1}$ one used, say, $k_*=50$ Mpc$^{-1}$ the relative constraining power of the two CMB probes would tilt towards SDs.

A similar point can also be made about the observation of parameters such as $N_{\rm eff}$, $Y_p$ and $T_0$ which are commonly fixed in \lcdm analyses to their theoretical values (as in the case of the former) or to their experimental determination (as in the case of the latter two). Nevertheless, the related determinations are either inferred or anchored today. The direct observation of the CRR would provide an invaluable and independent test of these values at $z\sim\mathcal{O}(10^3)$. In the case of $Y_p$, this would also be the first direct measurement of this quantity. 

Moreover, of course, CMB anisotropy and SD data do not necessarily have to be competing, but can also be combined, although it might seem that \textit{a priori} the latter would not have much to add to the former. This exercise has been performed most recently in \cite[\hyperlink{III}{III}]{Hart2020Sensitivity} in the context of \lcdm, showing that a Voyage 2050-like mission would most notably be able to improve the Planck determination of $n_s$ by a factor 3 [\hyperlink{III}{III}]\footnote{In [\hyperlink{III}{III}], a Voyage 2050-like mission with foregrounds -- which were not included in the analysis -- would be realistically (i.e., conservatively) comparable to the PIXIE100 case of the reference.} and of $\omega_b$ and $\omega_{\rm cdm}$ by about $20-30\%$\cite{Hart2020Sensitivity}. The strong improvement in $n_s$ is due to the larger level arm in $k$ space that the inclusion of SD at small scales could offer \cite[\hyperlink{III}{III},\,\hyperlink{V}{V}]{Powell:2012xz, Chluba2014Teasing, Clesse:2014pna, Cabass:2016ldu}. By including a CMB anisotropy-independent constraining on $A_s$, also the well-known degeneracy between $A_s$ and $\tau_{\rm reio}$ would be broken, thereby indirectly also strengthening the CMB anisotropy constraints on $\tau_{\rm reio}$ by about $25\%$. Although these improvements are expected to be less significant with respect to future CMB anisotropy experiments such as CMB-S4 \cite{Abazajian2016CMB, Abitbol2017CMB, Abazajian2019CMB} and LiteBird \cite{Matsumura2013Mission, Suzuki2018LiteBIRD}, the constraints on $n_s$ would still be a factor 2 more stringent after the inclusion of CMB SD data [\hyperlink{III}{III}].

Finally, although the primordial SD signal does not depend on $H_0$, a high-sensitivity mission designed to observe the former might also be able to determine the latter by observing the cooling of the CMB over a long period of time \cite{Abitbol2019Measuring}. In fact, from $\dot{T}_\gamma = - HT_\gamma$ one has that
\begin{align}
	\dot{T}_{\gamma,0} = - H_0T_{\gamma,0}\simeq -0.2~\text{nK~yr}^{-1}\,,
\end{align}
(assuming $h=0.7$) which would lead to a temperature shift distortion of the order of 1~Jy/sr after 10 years (see Fig. 1 of the reference). By measuring sufficiently accurately $T_{\gamma,0}$ and $\dot{T}_{\gamma,0}$ in real-time one could then in principle infer the value of $H_0$. The longer the integration time or the better the sensitivity of the observing mission, the lower the uncertainty on $H_0$ is going to be. Reaching a precision of the percent level might be achieved with (effective) sensitivities of the order of 0.2 nK$\sqrt{\text{yr}}$ or below and an integration time of 10 years or longer. A Voyage 2050-like mission can be designed to operate for $8-10$ years and have an effective sensitivity of the order of 10 nK$\sqrt{\text{yr}}$. Although not sufficing to accurately determine $H_0$, given the potential improvements in terms of foregrounds and mission design, this perspective is still worth bearing in mind when discussing future CMB SD missions.

\vspace{0.3 cm}
\noindent \textbf{Summary}

\noindent In summary, the future observation of CMB SDs would allow to test the \lcdm model at both early and late times. In the former case, a Voyage 2050-like SD mission would be able to constrain four (or, optimistically, five) of the six \lcdm parameters, with particular emphasis on $\omega_b$ and $n_s$. The SD information would also be relevant in combination with present and future CMB anisotropy experiments. In the context of the matter power spectrum, SDs would build on the current SZ measurements and deliver strong constraints on $\sigma_8$ and $\Omega_m$. They would also deepen our understanding of galactic thermodynamics and astrophysics, with the latter playing an important role in the numerical determination of the matter power spectrum at small-scales.

In the context of this section, the work carried out in the course of the thesis has been particularly significant for the characterization of the constraining power coming from the primordial signal, thanks to the very general numerical setup developed in [\hyperlink{I}{I},\hyperlink{XIV}{XIV}] (see Secs.~\ref{sec: num} and \ref{sec: num_2}). First of all, performing all-inclusive sensitivity forecasts such as the one shown in Fig.~\ref{fig: SD_MCMC_res_1} is now possible for any SD mission, with any combination of sourcing effects and with the inclusion of (extra-)galactic foregrounds. Although similar analyses had already been conducted for a limited subset of effects and cosmological parameters, forecasts as comprehensive as the one by~[\protect\hyperlink{XIV}{XIV}] are the first of their kind and represent now the state of the art. Furthermore, this general setup also opens the possibility to perform combined analyses with e.g., CMB anisotropy data, as systematically done in~[\protect\hyperlink{III}{III}]. These aspects have been considered in the context of the \lcdm model in this section, but more generally apply to all of the following discussion.

\subsection{Inflation}\label{sec: res_infl}
As mentioned in the preamble of the section, as soon as one allows for deviations from the standard \lcdm model large sections of the parameter space become unexplored by the very precise CMB anisotropy measurements. When this is the case, the complementary between different probes becomes fundamental and a particularly striking example is given by the scale dependence of the PPS, where the CMB anisotropies can constrain scales in the approximate range $10^{-4}-0.5$ Mpc$^{-1}$ (see Sec. \ref{sec: CMB}), while CMB SDs are sensitive to the physics happening at $1-10^{4}$ Mpc$^{-1}$ (see Sec. \ref{sec: SD_lcdm} in the context of the dissipation of acoustic waves). 

One can thus consider extensions of the \lcdm model where the PPS $\mathcal{P}(k)$ is allowed to vary more freely than in Eq. \eqref{eq: PPS_LCDM} and study what implications the future observation of SDs would have for the resulting constraints. It is possible, for instance, to introduce a running (or multiple runnings) of the scalar spectral index, or features such as kinks and steps in the PPS. Via the dependence of $\mathcal{P}(k)$ on the variance of the inflation field one can also use this approach to place bounds on the shape of the inflationary potential beyond the slow-roll approximation.\footnote{We remind here that the shape of the inflationary potential affects the PPS via Eq. \eqref{eq: PPS_phi} (see related discussion). In turn, the PPS directly determines the SD signal coming from the dissipation of acoustic waves, as given in Eq.~\eqref{eq: heating_aw}.} These models are rather general and have important implications for inflationary scenarios with features \cite{Starobinsky:1992ts, Adams:2001vc, Hazra2014Inflation, Hazra2014Wiggly, Hazra2016Primordial} or inflection points \cite{Polnarev:2006aa, Kohri:2007qn, Ben-Dayan:2009fyj, Choudhury:2013woa, Clesse:2015wea, Germani:2017bcs} in the potential (the latter particularly relevant for PBH formation, see also Sec.~\ref{sec: res_PBH}), particle production during inflation \cite{Barnaby:2009dd,Cook:2011hg, Dimastrogiovanni:2016fuu, Domcke:2018eki}, waterfall phase transitions \cite{Linde:1993cn, Lyth:1996kt, Abolhasani:2010kn, Bugaev2011Curvature} and axion inflation \cite{Barnaby:2010vf, Barnaby:2011qe, Meerburg:2012id} (see e.g., \cite{Ade2013PlanckXXII, Chluba2015Features} for reviews). In terms of CMB anisotropy constraints, the state of the art has been extensively discussed in \cite{Akrami2018PlanckX, Forconi:2021que}, while that in combination with SDs has been most recently presented in [\hyperlink{V}{V}] (see also \cite{Chluba2012Inflaton}). 

\begin{figure}[t!]
	\centering
	\includegraphics[width=\textwidth]{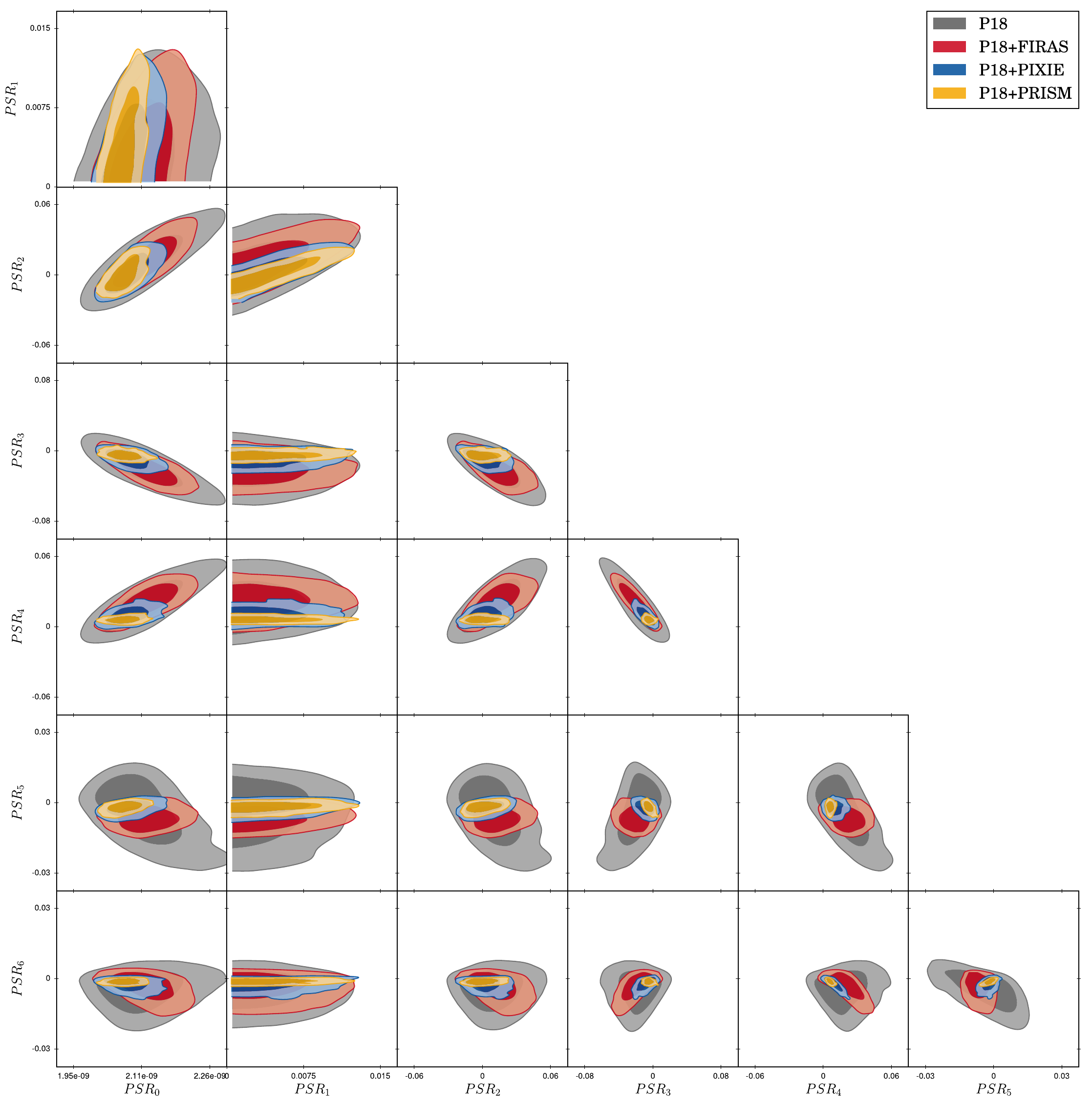}
	\caption{2D contours (at 68\% and 95\% CL) of the PSR parameters up to 6$th$ order in the Taylor expansion of the inflationary potential. Figure taken from [\protect\hyperlink{V}{V}].}
	\label{fig: infl_P18_V6}
\end{figure}

A first, very general\footnote{This expansion is only valid as long as the inflationary potential is smooth enough to be described in terms of a Taylor expansion. More complex shapes of the potential (such as those predicted by waterfall phase transitions) would not be captured.} parameterization of the PPS can by obtained at the level of the inflationary potential by Taylor-expanding the latter around $\phi=0$, i.e., 
\begin{align} \label{eq: V_expansion}
	V(\phi) = V_0 + V_1 \phi + \frac{V_2}{2!} \phi^2 + \frac{V_3}{3!} \phi^3 + ... + \frac{V_N}{N!} \phi^N = \sum\limits_{n=0}^N \frac{V_n}{n!} \phi^n\,,
\end{align}
not imposing any slow-roll approximation. For sake of both computational efficiency (i.e., to avoid parameter degeneracies) and physical intuition (with respect to the limiting slow-roll case), the $V_i$ coefficients are often rephrased in terms of Potential Slow-Roll (PSR) parameters
\begin{align}\label{eq:SR_params}
	PSR_{1} = \epsilon_{V} =\frac{1}{2} \left(\frac{V_1}{V_0}\right)^{2}\,, \quad PSR_{2} = \eta_{V} = \frac{V_2}{V_0}\,, \quad  PSR_{n} = \frac{V_1^{n-1}V_{n+1}}{V_0^n}\,,
\end{align}
with $PSR_0=A_s$. When assuming slow roll one has $\epsilon_V \ll 1$ and $|\eta_V| \ll 1$. The resulting constraints on the PSR parameters up to 6$th$ order are shown in Fig. \ref{fig: infl_P18_V6}, taken from~[\hyperlink{V}{V}], comparing Planck 2018 constraints with and without the additional SD information.\footnote{As mentioned in Sec. \ref{sec: num_2}, when considering beyond-\lcdm (and in this case beyond-slow roll) models, the fiducial SD signal and anisotropy power spectra are fixed to the \lcdm prediction.} The corresponding $1\sigma$ uncertainties are given in Tab. 4 of the reference. As clear from the figure, the inclusion of SDs significantly tightens the constraints the higher the order of the expansion. On the one hand, this is due to the fact that, since SDs probe smaller scales than the CMB anisotropies, inflation has to last longer when the two probes are combined with respect to the anisotropy-only case (see Sec. \ref{sec: infl}). This intrinsically limits the higher-order PSR parameters regardless of the order of the expansion and of the SD mission sensitivity. On the other hand, as discussed in the previous section in reference to Fig.~\ref{fig: SD_MCMC_res_1}, the inclusion of CMB SDs extends the lever arm in $k$ space placing an additional anchor at small scales. In this case the tightening of the constraints is proportional to the sensitivity of the given SD mission. A similar type of improvement is still to be expected even in combination with future CMB anisotropy experiments [\hyperlink{V}{V}] or other current CMB anisotropy experiments~\cite{Forconi:2021que}.

\begin{figure}[t!]
	\centering
	\includegraphics[width=\textwidth]{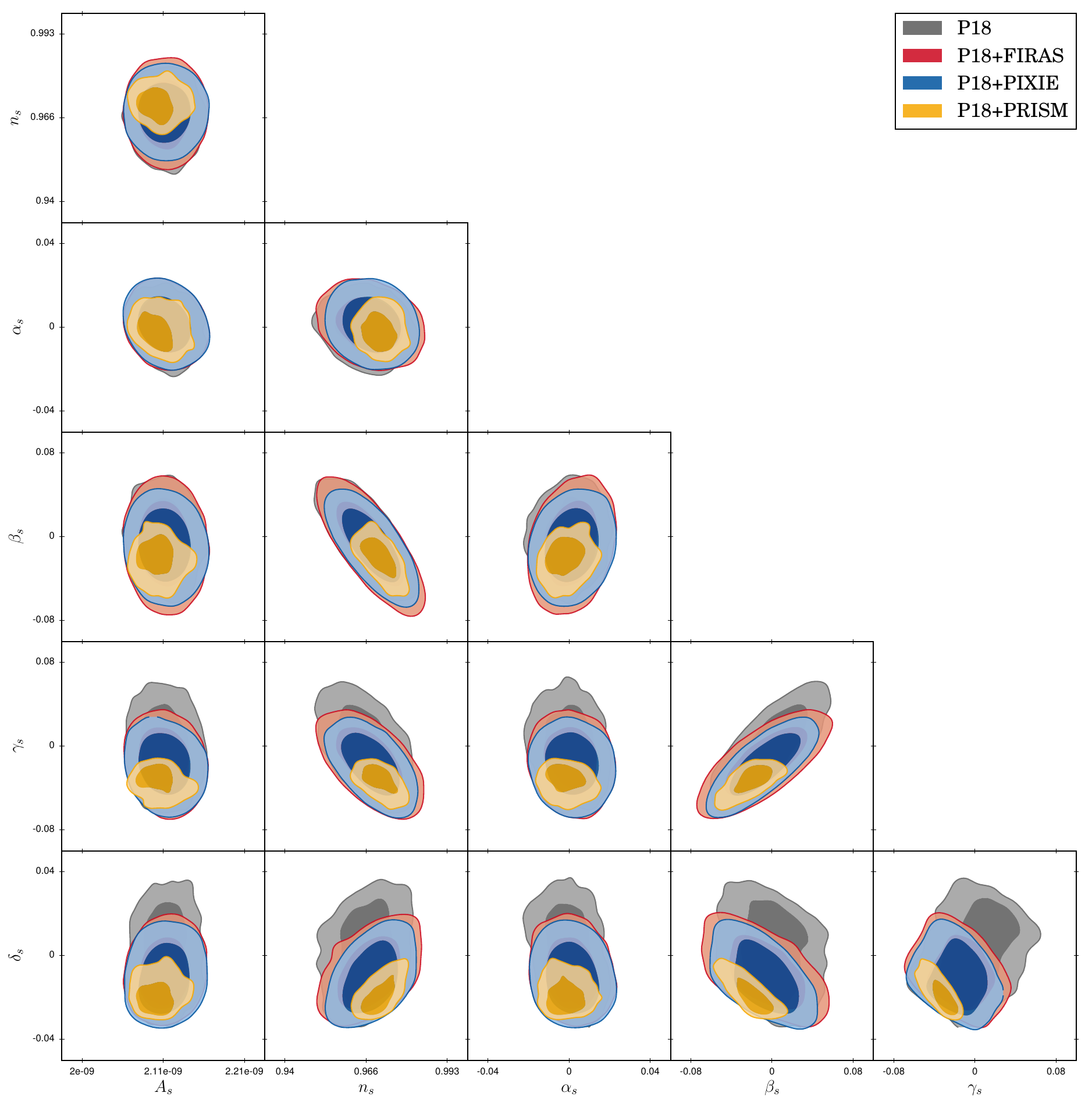}
	\caption{2D contours (at 68\% and 95\% CL) of the slow-roll parameters up to 4$th$ order. Figure taken from [\protect\hyperlink{V}{V}].}
	\label{fig: infl_P18_SR}
\end{figure}

\newpage
The slow-roll conditions can be however explicitly imposed and this is commonly done directly at the level of the PPS, which becomes
\begin{align}\label{eq: PPS_SR_expansion}
	\mathcal{P}(k) = A_s \left(\frac{k}{k_*}\right)^n \quad \text{with} \quad n=  (n_s-1) + \alpha_s \frac{1}{2!} \ln \left(\frac{k}{k_*}\right)  +  \beta_s \frac{1}{3!} \ln^2 \left(\frac{k}{k_*}\right) + \dots
\end{align}
Here $k_*$ is a given pivot scale which determines the corresponding pivot field and scale factor values $\phi_*$ and $a_*$, respectively. The parameters $\alpha_s$, $\beta_s$ and the higher order coefficients are referred to as running and running of the running of the scalar spectral index. They are connected to the PSR parameters via the relations given in e.g., \cite{Ade2013PlanckXXII}. The impact that, for instance, $\alpha_s$ ($n_{\rm run}$ in the figure) has on the heating rate from acoustic waves dissipation and on the respective SD signal is graphically represented in Fig. \ref{fig: SD_cool_diss}. 

The assumption of slow roll implies that successive terms should be smaller than the preceding terms. This intrinsically limits the shape of the PPS, reducing the freedom of the previous Taylor expansion, and enhances the constraining power of observations around~$k_*$. For this reason, as evident from Fig. \ref{fig: infl_P18_SR}, taken from [\hyperlink{V}{V}], in this case the addition of SDs on top of Planck 2018 data yields a minor improvement with respect to the previous looser scenario, although still a considerable one even for the running of the scalar spectral index. In fact, \cite[\hyperlink{III}{III}]{Chluba2019Voyage} considered how the constraints on $\alpha_s$ scale as a function of the SD mission sensitivity showing that a Voyage 2050-like mission would improve the constraints on this quantity by a factor $3-5$ even with respect to future CMB measurements.

\begin{figure}[t!]
	\centering
	\includegraphics[width=0.48\textwidth]{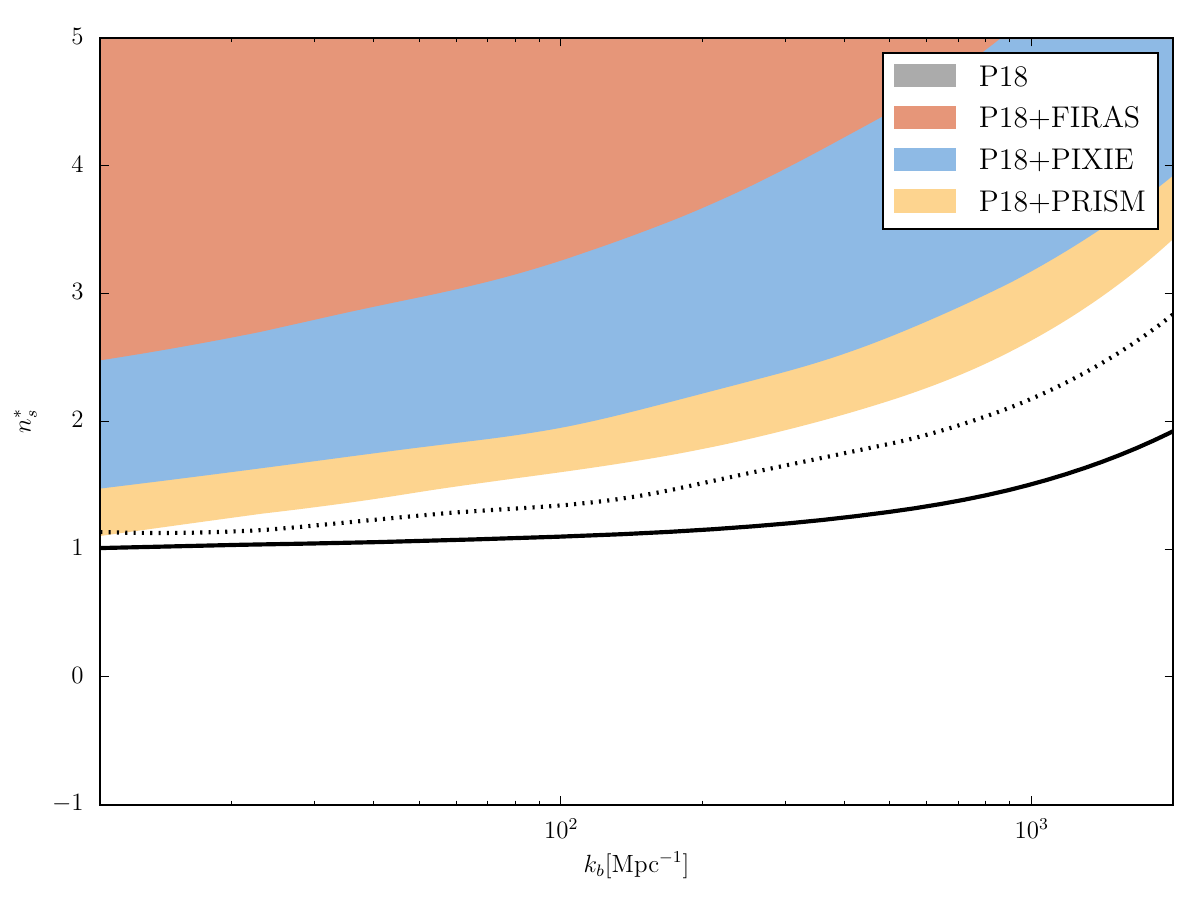}
	\includegraphics[width=0.48\textwidth]{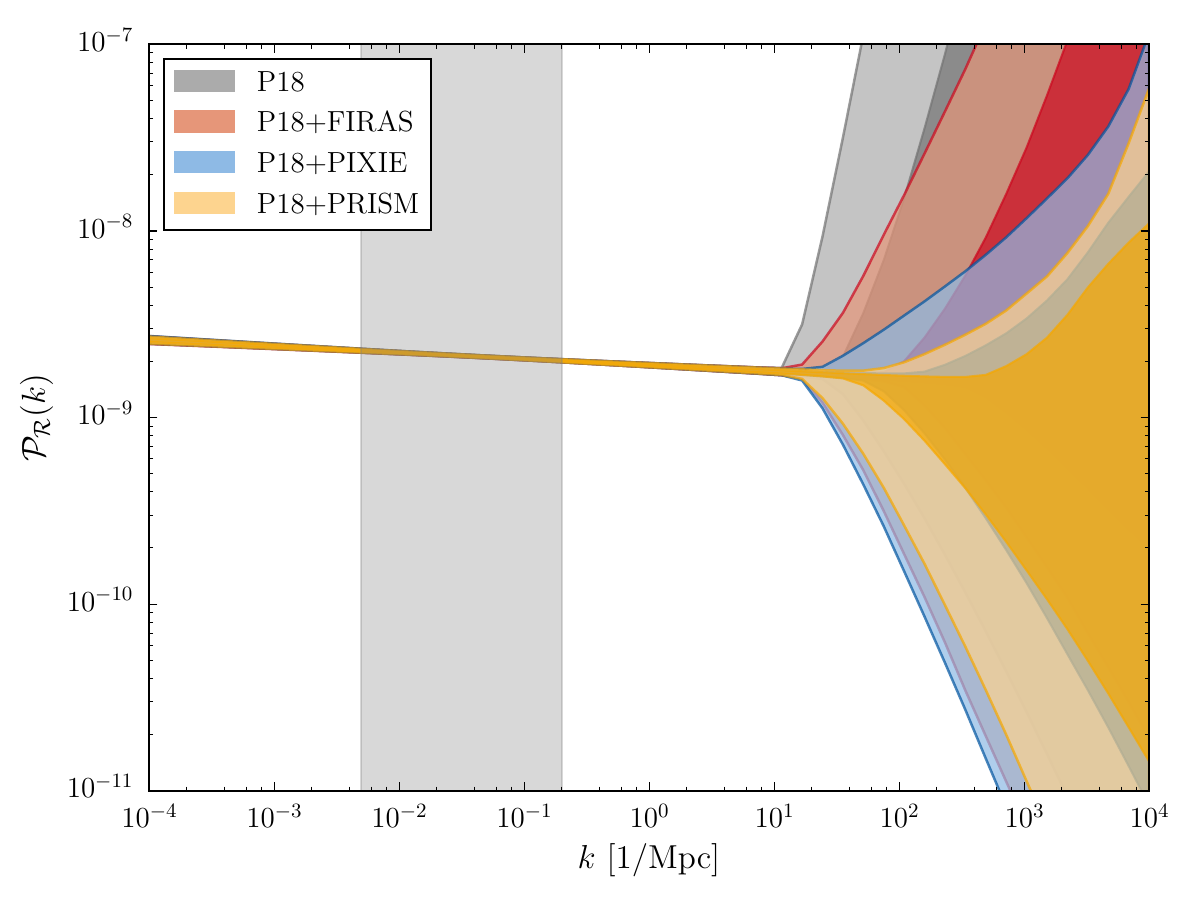}
	\\
	\includegraphics[width=0.48\textwidth]{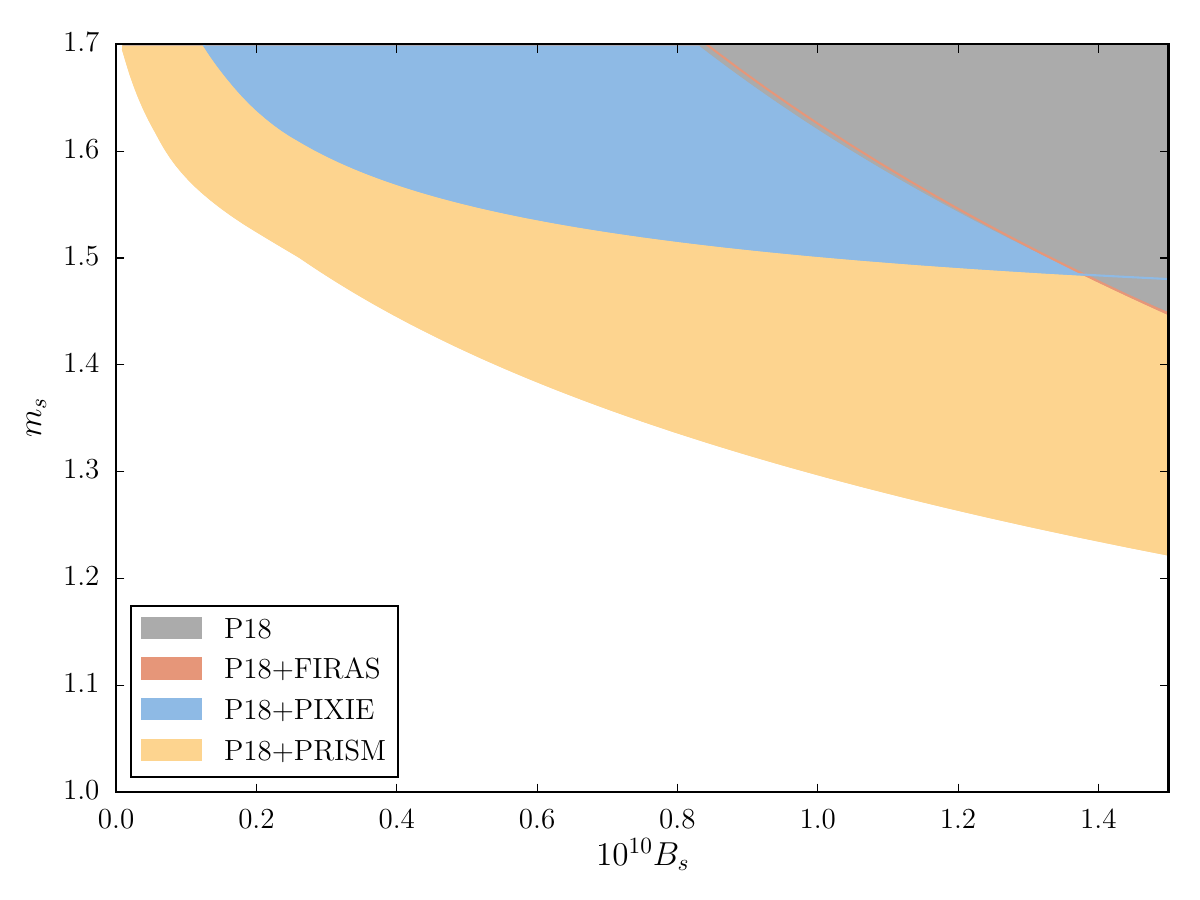}
	\includegraphics[width=0.48\textwidth]{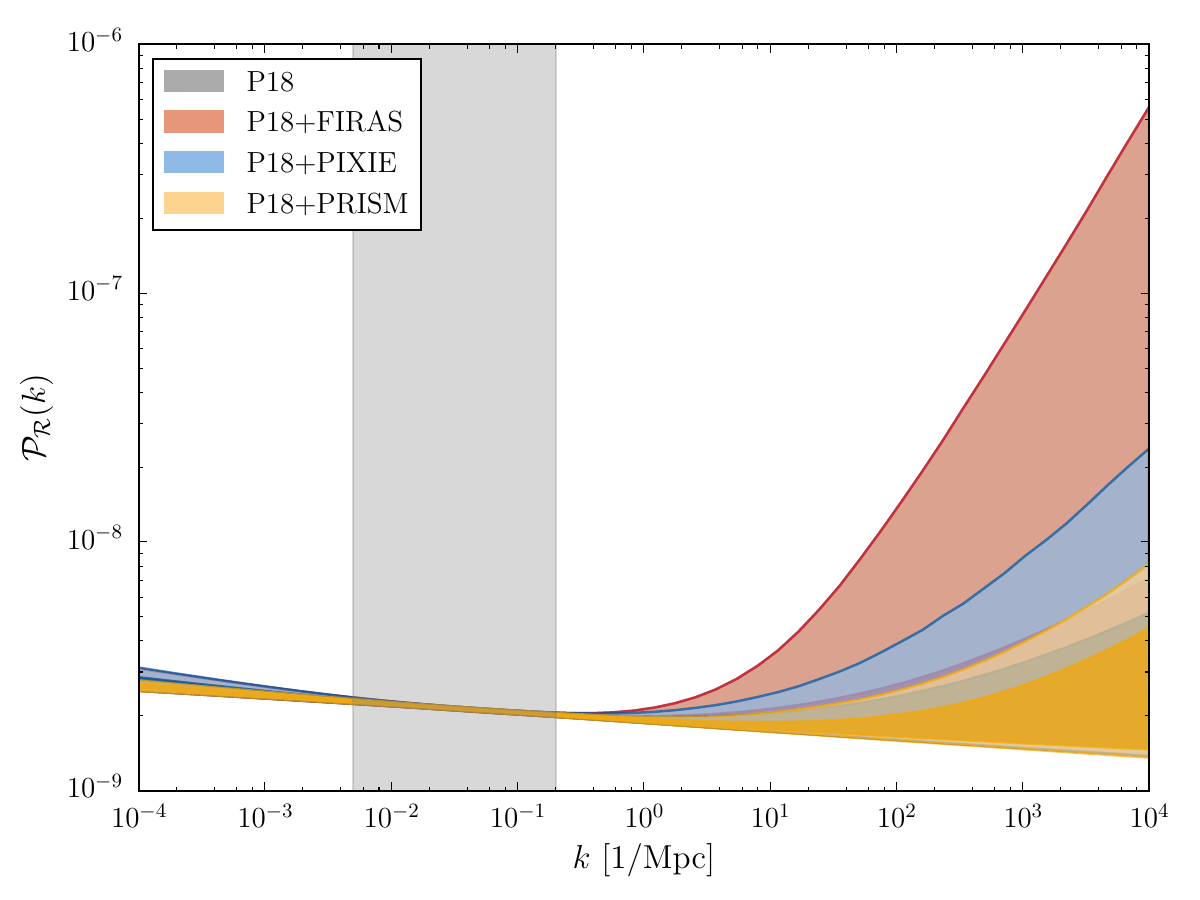}
	\\
	\includegraphics[width=0.48\textwidth]{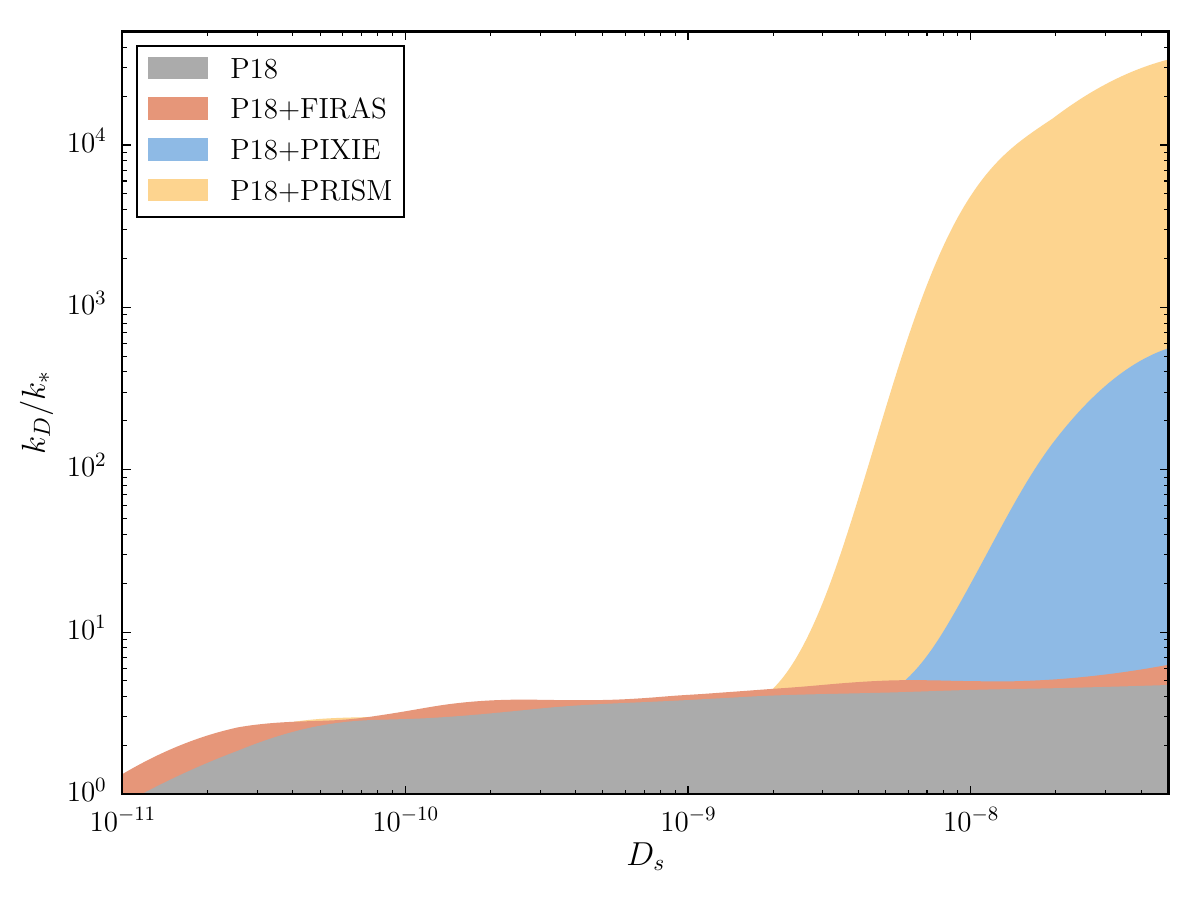}
	\includegraphics[width=0.48\textwidth]{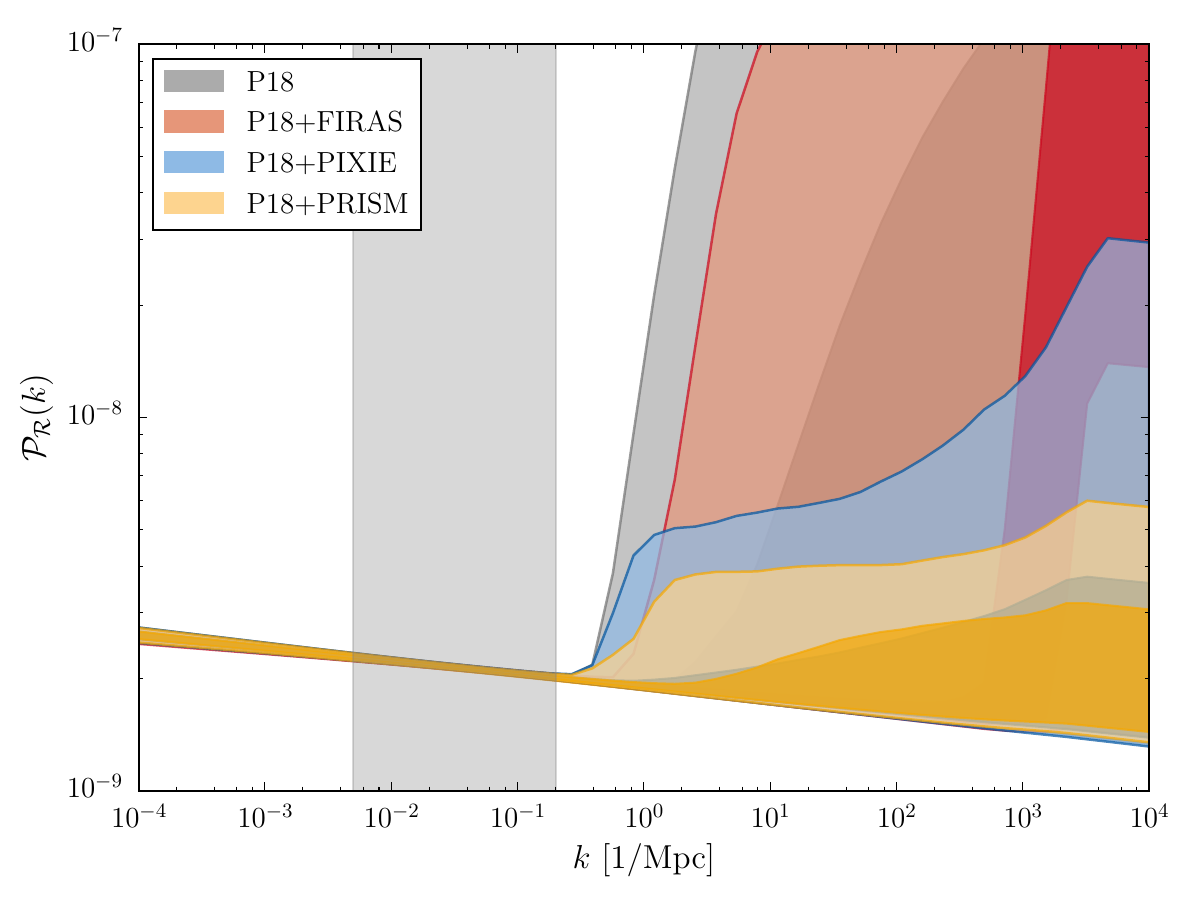}
	\caption{Constraints (at 95\% CL) on the models parameters (left) and the reconstructed PPS (at 68\% and 95\% CL) allowing for kinks (top two rows) and steps (bottom row) in the PPS. Figures taken from [\protect\hyperlink{V}{V}].}
	\label{fig: infl_P18_features}
\end{figure}

Finally, one can go beyond these smooth parameterizations of the PPS and introduce features that mimic the effect of specific inflationary scenarios. One popular option (see e.g., \cite{Chluba2012Inflaton}) is to introduce a kink at a given scale $k_b$, so that
\begin{align}
	\mathcal{P}_{\mathcal{R}}(k)=\left\{\begin{array}{ll}
		\mathcal{P}_{\mathcal{R}}(k) & \text { for } k<k_{\mathrm{b}} \\
		\mathcal{P}_{\mathcal{R}}\left(k_{\mathrm{b}}\right)\left(\frac{k}{ k_{\mathrm{b}}}\right)^{n_{\mathrm{s}}^{*}-1} & \text { for } k \geq k_{\mathrm{b}}
	\end{array}\right.\,,
\end{align} 
where $n_s^*$ defines the tilt of the spectrum at scales $k>k_b$. The resulting constraints (at 95\% CL) on $n_s^*$ and $k_b$ are given in the top left panel of Fig. \ref{fig: infl_P18_features}, taken from [\hyperlink{V}{V}]. The right top panel of the same figure shows the reconstructed PPS (at 68\% and 95\% CL). Alternatively, continuous kinks at arbitrary scales are also present in models that feature two independent, scale-invariant components of the PPS. Such models can be described by two \lcdm-like PPSs as in Eq. \eqref{eq: PPS_LCDM} with amplitudes $A_s$ and $B_s$ and spectral indices $n_s$ and $m_s$ (assuming $B_s>0$ and $n_s>0$ to avoid cancellations). The exclusion plot in the $B_s-m_s$ plane is shown in the middle left panel of Fig. \ref{fig: infl_P18_features}, with the reconstructed PPS on the right. A last example inspired by the results of \cite{Byrnes2018Steepest, Cole:2022xqc} (see in particular Figs.~1-2 of \cite{Cole:2022xqc} for a graphical connection between the modified inflationary potential and the PPS as well as Sec. \ref{sec: res_PBH} for a related discussion) is given by steps in the PPS of the form
\begin{equation}\label{eq: PPS_step}
	\mathcal{P}_\mathcal{R}(k) = 
	\begin{cases}
		A_s (k/k_*)^{n_s-1} & \text{ for $k<k_D$} \\ 
		A_s (k_D/k_*)^{n_s-1} (k/k_D)^4 & \text{ for $k_D<k<\widetilde{k}_D$} \\ 
		(A_s+D_s) (k/k_*)^{n_s-1} & \text{ for $\widetilde{k}_D < k$}
	\end{cases}
\end{equation}
where the growth of the PPS is assumed to be at most proportional to $k^4$. The results are displayed in the bottom panels of Fig. \ref{fig: infl_P18_features} analogously to the cases above. In all instances, the impact of SDs is evident.

On top of being able to tightly constraining the PPS, CMB SDs also offer the unique possibility to test the presence primordial non-Gaussianities in the correlation between cosmological perturbation (see e.g., \cite{Pajer:2012vz, Ganc:2012ae, Biagetti:2013sr, Ota:2014iva, Khatri:2015tla, Emami:2015xqa, Dimastrogiovanni:2016aul, Chluba:2016aln, Ota:2016mqd, Ravenni:2017lgw, Cabass:2018jgj, Remazeilles:2018kqd, Remazeilles:2021adt, Rotti:2022lvy}). In Sec. \ref{sec: infl} we have discussed the formation of Gaussian perturbations in a single-field inflation scenario. Imagine, however, the existence of also other fields (such as in the curvaton scenario \cite{Bartolo:2003jx}) that function as ``spectators'' to the inflaton field during inflation and that decay after the end of inflation, i.e., when the inflationary perturbations are super-Hubble and are gradually re-entering the horizon. In this case, the decay of e.g., the curvaton would generate additional small-scale perturbations that evolve on top of the much larger inflationary perturbations, thereby modifying the Gaussian nature of the latter. Many other non-standard scenarios (including several of the ones listed at the beginning of the section) would lead to a qualitatively similar result (see e.g., \cite{Bartolo:2004if} for a comprehensive review). In the context of CMB SDs, this would lead to inhomogeneities in the amplitude of the small-scale power and hence of the SD signal. These anisotropies can then be correlated with traces of the large perturbations, such as the CMB anisotropy power spectra.

The amount of such non-Gaussianities is often expressed in terms of the (local\footnote{One can typically define so-called equilateral and orthogonal $f_{\rm NL}$ parameters in the case of single-field inflation and a local $f_{\rm NL}$ parameter in multi-field inflationary scenarios (see e.g., \cite{Planck:2019kim} and references therein). Here we only focus on the latter \cite{Byrnes:2010em} since it is the one that can be probed via CMB SDs.}) non-linearity parameter $f_{\rm NL}$ (see e.g., \cite{Bartolo:2004if}), which has been constrained by Planck 2018 to be $f_{\rm NL}\lesssim5$ at CMB scales \cite{Planck:2019kim}. In terms of this parameter, the temperature-SD cross-correlation\footnote{The amplitude of the SD-SD power spectra is much lower and will be therefore neglected henceforth.} power spectrum $C^{XT}_\ell$, where $X$ stands for any of the SD types, can then be shown (see e.g., Eq.~(17) of \cite{Chluba:2016aln}) to be proportional to
\begin{align}
	C^{XT}_\ell \propto C^{TT}_\ell\, \langle X\rangle \, f_{\rm NL}\,,
\end{align}	
where $C^{TT}_\ell$ is the CMB anisotropy temperature power spectrum (in the Sachs-Wolf limit) and $\langle X\rangle$ the corresponding mean SD amplitude (i.e., $\mu$ or $y$ depending on the chosen cross-spectrum). A similar result can be derived also for the case of the polarization-SD power spectra $C^{XE}_\ell$. The non-observation of $C^{XT}_\ell$ and $C^{XE}_\ell$ from imaging experiments such as Planck\footnote{The way imaging experiments can measure spatial variations in the primordial $\mu/y$ parameters is analogous to how they can detect the late-time SZ effect sources, as explained in the first part of Sec.~\ref{sec: res_lcdm}.} imposes bounds on $f_{\rm NL}$ at $k\simeq 740$~Mpc$^{-1}$ of the order of $f_{\rm NL}\lesssim 6800$ (fixing the value of $\mu$ to its \lcdm value, see Sec. \ref{sec: SD_lcdm}) \cite{Rotti:2022lvy}. In this way, as for the discussion related to the PPS, CMB SDs are very valuable to test the running of $f_{\rm NL}$ \cite{Emami:2015xqa, Biagetti:2013sr, Remazeilles:2018kqd}. Future CMB anisotropy missions such as LiteBIRD \cite{Matsumura2013Mission, Suzuki2018LiteBIRD} would further strengthen the constrain down to $f_{\rm NL}\lesssim 800$ \cite{Remazeilles:2021adt}, while cosmic variance-limited experiments could reach the level of $f_{\rm NL}\lesssim 10^{-4}$.

In summary, CMB SDs are fundamental to constrain the inflationary potential and the shape of the PPS as well as the Gaussianity of primordial perturbations at scales of the order of $1-10^4$ Mpc$^{-1}$, with wide-ranging consequences for a number of inflationary scenarios. The combination with current and future CMB anisotropy data will be able to place tight bounds on these models over eight decades in $k$ space.

In the context of this section, the contribution of the thesis has broadly been to extend, update and review the landscape of possible inflationary scenarios and PPS parameterizations that can be constrained with SDs. This has been done in [\protect\hyperlink{V}{V}], with the full numerical setup discussed in Secs.~\ref{sec: num} and \ref{sec: num_2}. Specifically, the constraints on the Taylor expansion of the inflationary potential (up to 6$th$ order in the PSR parameters, see Eq. \eqref{eq: V_expansion}) shown in Fig. \ref{fig: infl_P18_V6} have been presented in the reference for the first time. Similarly, also those on the slow-roll parameters (up to 4$th$ order in the running of the scalar spectral index, see Eq. \eqref{eq: PPS_SR_expansion}) shown in Fig. \ref{fig: infl_P18_SR} have been derived for the first time in [\protect\hyperlink{V}{V}]. Finally, although steps and kinks in the PPS had already been considered before, the derived constraints where not including the latest Planck data and a marginalization over (extra-)galactic foregrounds. The corresponding limits have been recomputed in [\protect\hyperlink{V}{V}], and reported in Fig.~\ref{fig: infl_P18_features}, improving upon the previous works in these directions.

\subsection{Primordial gravitational waves}\label{sec: res_further}

Should primordial tensor perturbations, i.e., primordial GWs, be created in the very early universe (see e.g., \cite{Caldwell:2022qsj} for a review of scenarios that might lead to the formation of primordial GWs), they would dissipate their energy by inducing an anisotropic stress of the baryon-photon fluid. Similarly to the acoustic wave dissipation in the case of scalar perturbations (see Sec. \ref{sec: SD_lcdm}), also the mixing of BBs following the damping of tensor perturbations would lead to the formation of SDs \cite{Ota2014CMB, Chluba2015Spectral, Kite:2020uix}.

\begin{figure}
	\centering
	\includegraphics[width=0.48\textwidth]{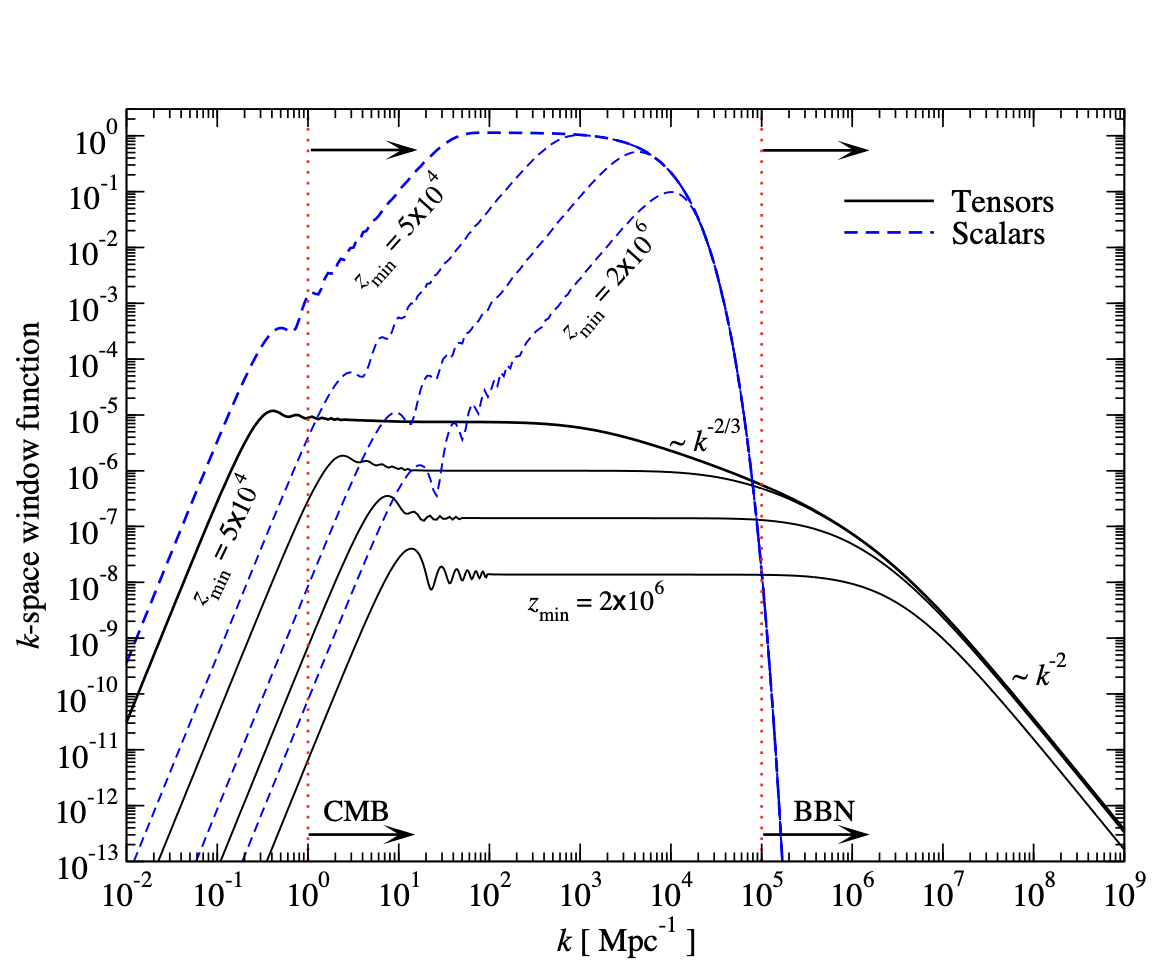}
	\includegraphics[width=0.48\textwidth]{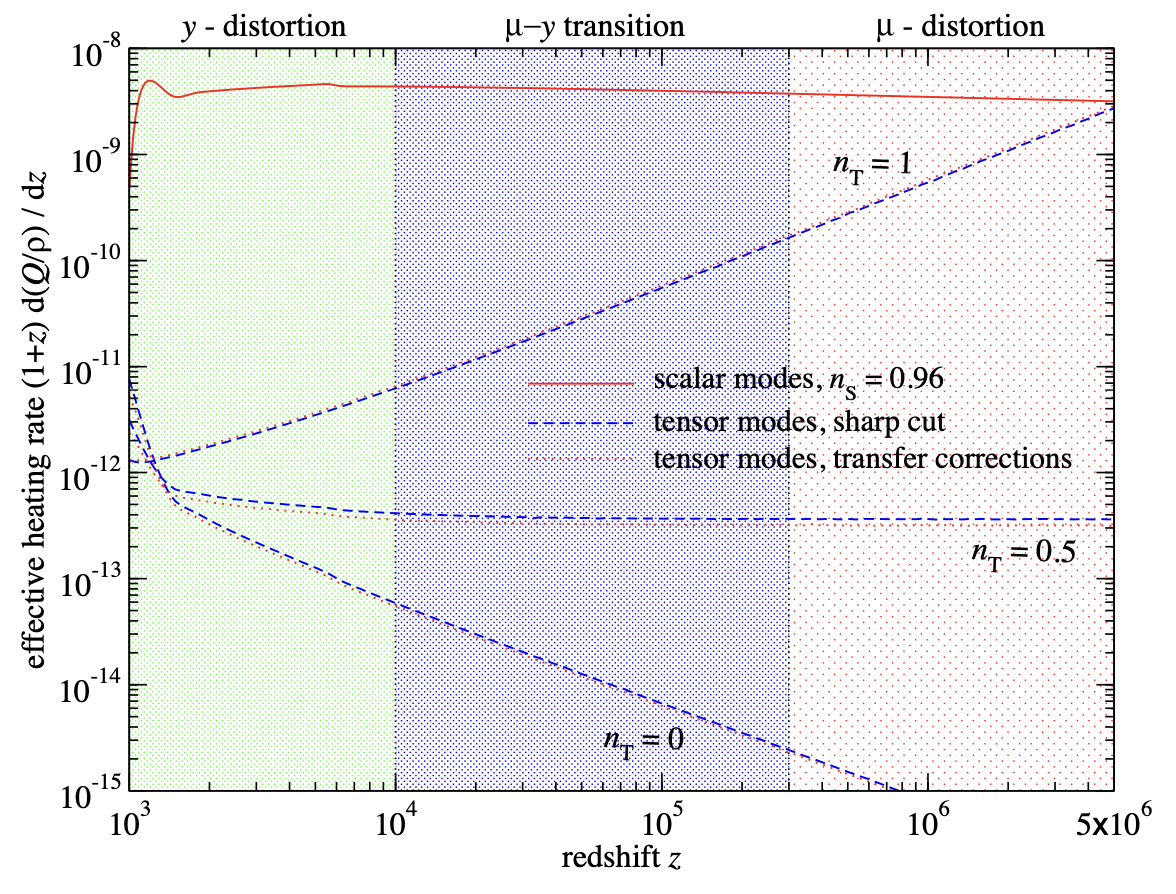}
	\caption{Comparison between the window function (left) and heating rate (right) of the scalar (i.e., adiabatic) and tensor mode dissipation. Figures taken from \cite{Chluba2015Spectral}.}
	\label{fig: GWs_window}
\end{figure}

The main difference with respect to the scalar perturbations is two-fold and it is in both instances related to the weaker coupling of tensor perturbation with the photon fluid than the scalar perturbations. On the one hand, this implies that is takes longer for the perturbations to dissipate, meaning that the range of scales that can source the corresponding SD is larger. In other words, while e.g., a scalar perturbation with wave mode $k=10^5$ Mpc$^{-1}$ would have dissipated completely before the $\mu$ era (creating no SD, see Sec. \ref{sec: SD_lcdm}), a tensor perturbation that entered the horizon at the same time would still have energy to dissipate during the $\mu$ era (creating SDs). This is reflected in the so-called window function $W(k)$ that determines which scales contribute to the creation of $\mu$ distortions according to
\begin{align}
	\mu = \int \frac{\text{d}k k^2}{2\pi^2}\mathcal{P}_i(k)W_i(k)\,,
\end{align}
where $i$ can stand either for scalar or tensor perturbations, and $\mathcal{P}(k)$ is the respective PPS. As displayed in the left panel of Fig. \ref{fig: GWs_window}, taken from \cite{Chluba2015Spectral}, tensor perturbations entering the horizon at scales as small as $k=10^9$ Mpc$^{-1}$ can still be observed via SDs.

The other side of the coin is, however, that the heating rate from the dissipation of tensor perturbations is much lower than in the scalar case, as represented in the right panel of Fig. \ref{fig: GWs_window}, taken from \cite{Chluba2015Spectral}. Although this is mostly due to the aforementioned weaker coupling of the tensor perturbations to the photon fluid, the subdominant contribution is also partly due to the lower PPS of the tensor sector. In fact, in analogy to the dissipation of acoustic waves (see Eq.~\eqref{eq: heating_aw}), one can intuitively imagine the heating from the dissipation of primordial GWs to be proportional to the tensor PPS, which typically (within slow-roll inflation) takes the form
\begin{align}
	\mathcal{P}_T \propto A_T \left(\frac{k}{k_*}\right)^{n_T}\,,
\end{align}
where $A_T$ and $n_T$ are its amplitude and tilt. Of note, the amplitude $A_T$ is related to the amplitude of the scalar PPS by the so called tensor-to-scalar ratio $r=A_T/A_s$, which has been determined to be $r<0.06$ (at 95\% CL) at $k = 0.002$ Mpc$^{-1}$ \cite{BICEP2:2018kqh}. Therefore, the amplitude of the tensor dissipation is much smaller than that of the scalar sector. As a result, one obtains $\mu$ distortions of the order of $\mu\simeq 2\times 10^{-13}\,r$ (for a scale-invariant tensor PPS, i.e., $n_T=0$), meaning that the contribution from this effect can be neglected in the standard slow-roll inflation cosmology.

\begin{figure}
	\centering
	\includegraphics[width=\textwidth]{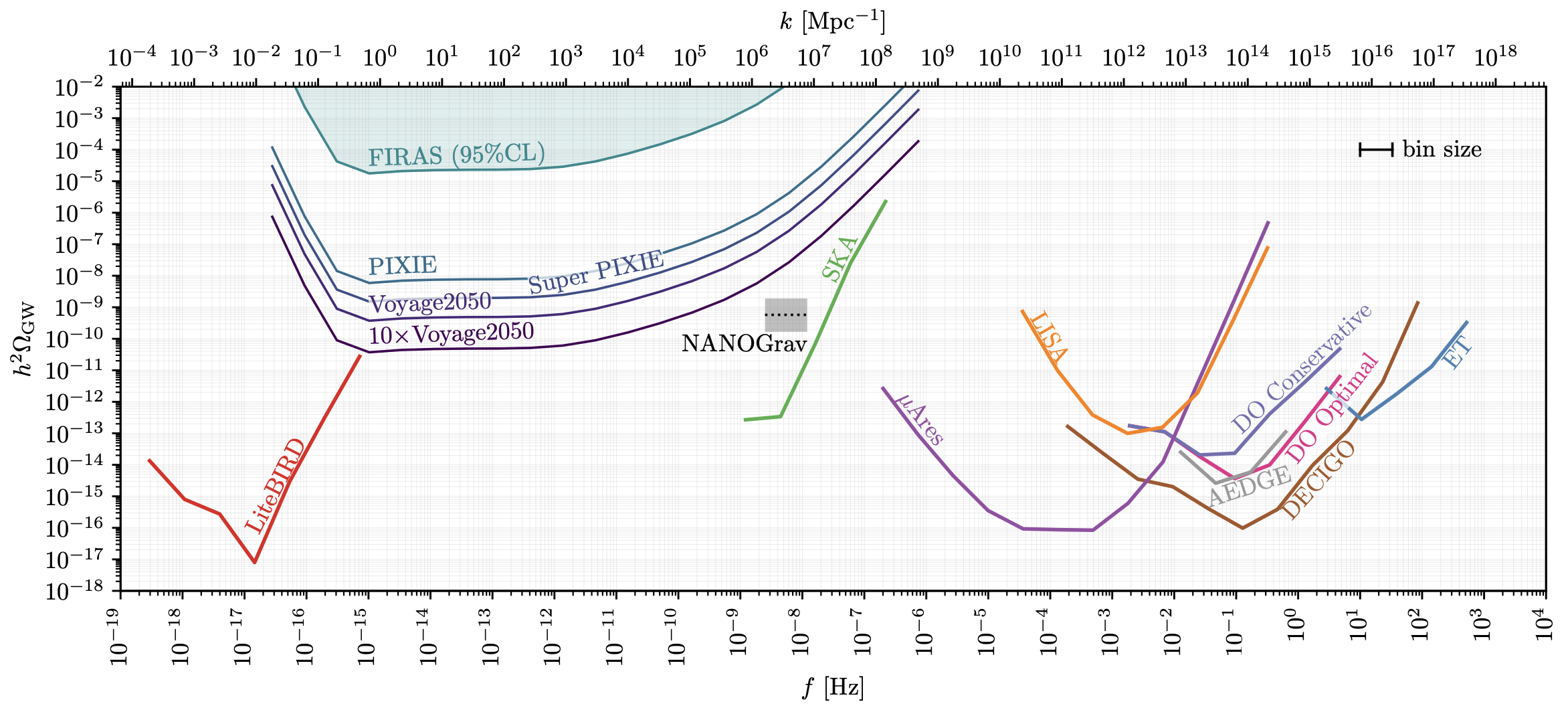}
	\caption{Summary plot of (current and future) constraints on the energy density of~GWs. Figure taken from \cite{Kite:2020uix}.}
	\label{fig: GWs_constr}
\end{figure}

Nevertheless, scale invariance is by no means the only possibility for the tensor PPS and a number of other scenarios might predict an enhanced GW spectrum at the relevant scales for SDs (see e.g., Sec. 5 of \cite{Kite:2020uix} for a list of examples), making of them a unique opportunity to test them. As overviewed in Fig. \ref{fig: GWs_constr}, taken from \cite{Kite:2020uix}, SDs turn out to be sensitive to the frequency range between $10^{-16}-10^{-7}$ Hz (with $k/\text{Mpc}^{-1} = 6.5 \times 10^{14}\,f/$Hz) and as such they perfectly complement other cosmological and astrophysical observational strategies. The improved sensitivity reachable by future missions such as Voyage 2050 would also largely compensate for the weakness of the signal and deliver constraints more competitive with the other probes.

\vspace{0.3 cm}
\noindent \textbf{Follow-up idea 10:} \texttt{CLASS} can by default account for a very rich tensor sector. Including the calculation for the heating rate from the dissipation of GWs (which can be done straightforwardly as explained in Sec. \ref{sec: num}) would allow to immediately take advantage of this versatility and to be applied to a large class of models.
\vspace{0.3 cm}

In summary, should they be sourced in the very early universe, the dissipation of primordial GWs would lead to the formation of CMB SDs over a wide range of scales. As such, CMB SDs are perfectly complementary to the other known observational strategies and could play an important role in constraining models predicting GW spectra peaking at frequencies in the $10^{-15}-10^{-9}$ Hz range. Although not directly developed during the course of the thesis, the physics related to primordial GWs could be easily accounted for in the numerical setup introduced in Sec. \ref{sec: num}.

\subsection{Particle dark matter}\label{sec: res_DM}
As outlined in Sec. \ref{sec: SD_non_lcdm}, CMB SDs are sensitive to a number of beyond-\lcdm effects, many of which are predicted to exist in extensions of the cold DM assumption. It is possible, for instance, for the DM particles to mix and scatter with, annihilate and decay into electromagnetically interacting SM particles such as photons and electrons. Below we review the consequences that current and future SD measurements imply for these models. Specifically, while the cases of DM decay and annihilation have been studied and developed in the course of the thesis, the other two examples are proposed for sake of completeness.

As a note, in the following discussion (as well as the one carried out in Sec. \ref{sec: res_PBH} in the context of PBHs) the constraints are only relying on $\mu$ and $y$ distortions, although the considered energy injections would also affect the CRR \cite{Chluba2010Could}. With the eventual implementation of these effects in the numerical pipeline to compute the CRR (see Sec.~\ref{sec: num} and the ``follow-up idea 4'' there) it would be straightforward to study the interplay between the different sources of SDs.

\vspace{0.3 cm}
\noindent\textbf{DM decay}

\noindent Should the DM (or any other massive relic -- as we will imply throughout) decay in the early universe, it would inject energy in the system at a rate (see e.g., \cite{Poulin2016Fresh, Slatyer2016General})
\begin{align}
	\dot{Q}_{\rm inj}=  \rho_{\rm cdm} f_\chi \Gamma_{\rm dec} e^{-\Gamma_{\rm dec}t}\,,
\end{align}
where $\Gamma_{\rm dec}=1/\tau_{\rm dec}$ is the decay rate and $\Gamma_{\rm dec}e^{-\Gamma_{\rm dec}t}$ the decay probability (combining in the luminosity $L_\chi$ together with the released energy per decay $E=m_\chi$, to be inserted in Eq. \eqref{eq: E_inj}). In general, the two independent free parameters are $\tau_{\rm dec}$ and $f_\chi$, which can be constrained by a number of cosmological probes. The decay probability becomes nevertheless proportional to $\Gamma_{\rm dec}$ when the lifetime $\tau_{\rm dec}$ is larger than the age of the universe, becoming fully degenerate with $f_\chi$ in this case. In this regime one then usually imposes $f_\chi=1$ (since all the DM is still present) and only places constraints on $\Gamma_{\rm dec}$. The energy at which the DM decays (i.e., its mass) does not play a significant role for cosmological probes \matteo{in the approximate keV$-$TeV mass range \cite{Acharya2019CMB} (regime that we assume henceforth), although it might become important for very soft injections \cite{Bolliet:2020ofj}. The DM mass becomes furthermore} a key quantity for late-time measurements since it determines which observational strategies can detect the decay products \cite{Esmaili2012Probing, ElAisati2015New, Slatyer2016General, Garcia-Cely2017Neutrino, Aartsen2018Search, Bhattacharya2019Update, Coy2020Neutrino, Coy2021Seesaw}.

\begin{figure}
	\centering
	\includegraphics[width=0.48\textwidth]{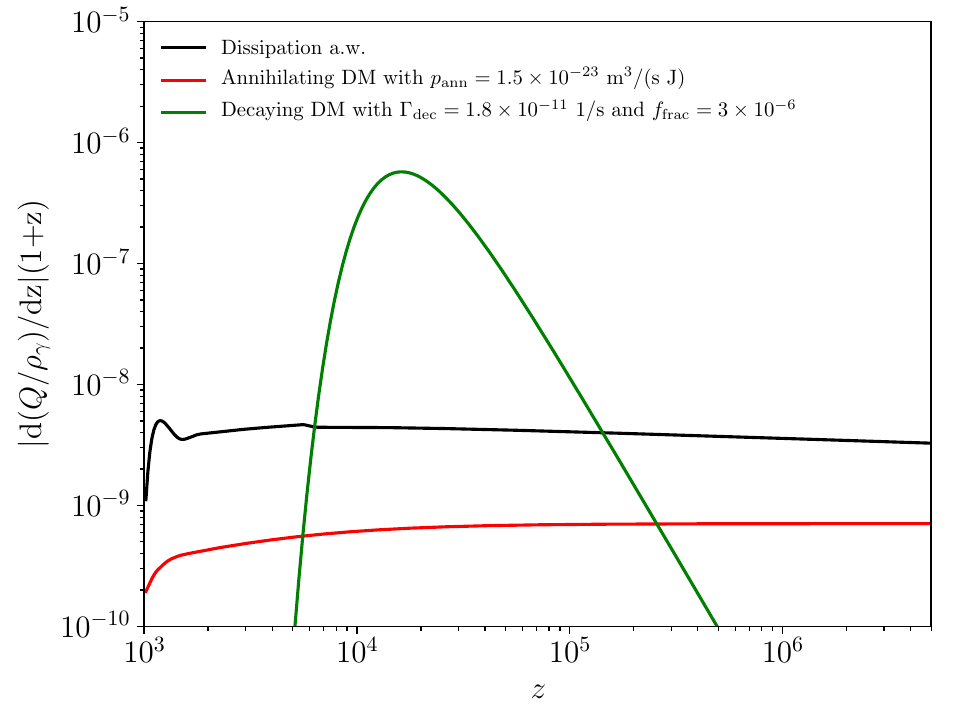}
	\includegraphics[width=0.48\textwidth]{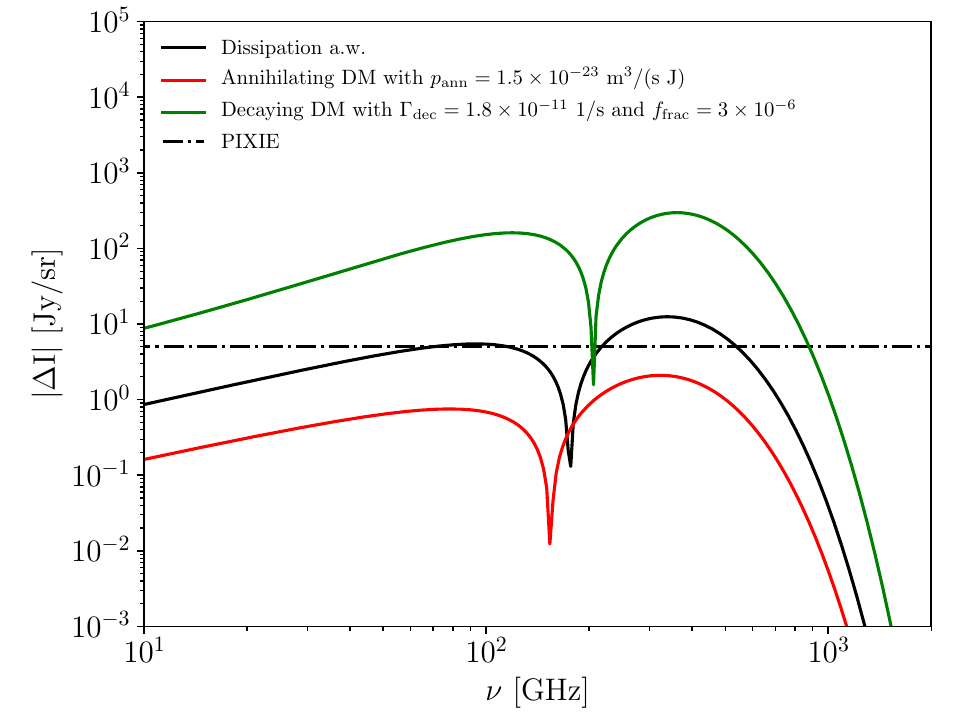}
	\caption{Same as in Fig.~\ref{fig: SD_cool_diss} but for two representative examples of DM annihilation and decay. Figures taken from [\protect\hyperlink{I}{I}].}
	\label{fig: DM_heating_ann_dec}
\end{figure}

A representative graphical example of the corresponding heating rate is shown in the left panel of Fig.~\ref{fig: DM_heating_ann_dec}, taken from [\hyperlink{I}{I}] (green line, to be compared with the black line representing the solid red line in Fig. \ref{fig: SD_cool_diss} for reference). The figure, where one assumes $\Gamma_{\rm dec}=1.8\times 10^{-11}$ 1/s (in agreement with observations, see below), confirms that the heating rate follows a linear behavior at first and then is exponentially suppressed as soon as $t>\tau_{\rm dec}$. The corresponding SD signal is shown in the right panel of the same figure.

\begin{figure}[t]
	\centering
	\includegraphics[width=0.75\textwidth]{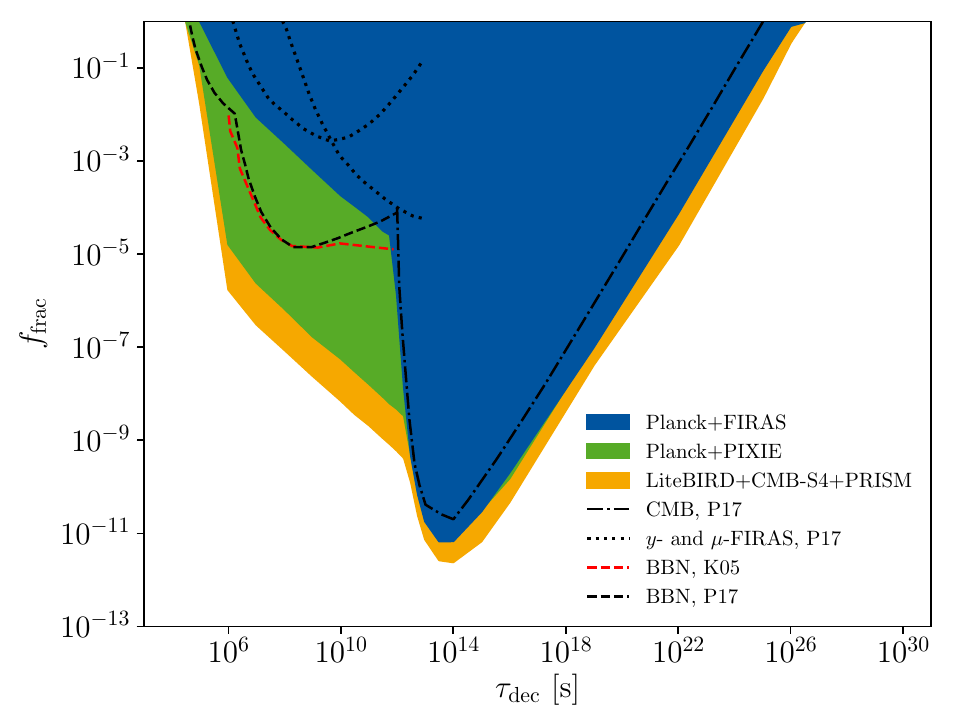}
	\caption{95\% CL exclusion contours on the fractional abundance of decaying DM as a function of the DM lifetime. Figure taken from [\protect\hyperlink{I}{I}].}
	\label{fig: DM_decay_exclusion_region}
\end{figure}

Assuming for instance $f_{\rm eff}=1$ in a first approximation (i.e., assuming the whole energy to be released in form of EM radiation and the on-the-spot approximation at all times), one obtains the exclusion contours (at 95\% CL) displayed in Fig. \ref{fig: DM_decay_exclusion_region}, taken from [\hyperlink{I}{I}], which are a combination of CMB anisotropy and SD data.\footnote{With the \texttt{CLASS}+\texttt{MontePython} pipeline, this type of runs can be performed in principle in one run, as it has been the case in the analysis performed in \cite{Balazs:2022tjl} discussed below. In Fig. \ref{fig: DM_decay_exclusion_region} the parameter space has been instead sliced along the $\tau_{\rm dec}$ axis to reduce the number of free parameters to be scanned over~[\hyperlink{I}{I}].} BBN constraints are also shown for reference and probe the same region of parameter space as CMB SDs. From the figure it can be seen that the combination of CMB data can constrain almost 20 decades in the decaying DM lifetime. BBN constraints are currently more stringent than those imposed by SDs, but future SD mission will improve the bounds by orders of magnitude (while BBN observations are currently saturating the amount of precision that can be reached). A PRISM-like mission would already be able to set limits competitive with current and future CMB anisotropy missions.

More accurate treatments of the deposition history lead to the similar plots derived in e.g., \cite{Poulin2017Cosmological, Slatyer2016General, Acharya2019CMB, Chluba2020Thermalization, Bolliet:2020ofj, Acharya2021CMB}. Importantly, the cases where the DM decays into electrons or photons (i.e., with $f_{\rm em}=1$ and closely following $f_{\rm loss}$) are quite general since they can be recast to represent also many other decay channels simply by modifying $f_{\rm em}$ \cite{Slatyer2016General}. In fact, in most cases (that is, as long as the lifetime of the intermediate decay product is much shorter than its the interaction timescale with the plasma, which is usually the case) it cannot be distinguished whether the DM decays directly to, say, photons with a BR of 50\% or first to any other particle, which in turn decays to photons with the same BR. In other words, the constraints set on $f_\chi$ for the cases of $\chi\to\gamma,e^{\pm}$ can be interpreted as constraints on the parameter combination $f_\chi\,f_{\rm em}$, to be used for any model-dependent form of $f_{\rm em}$.

This approach has been used, for instance, in [\hyperlink{IX}{IX}] to constrain the decay of massive relics into neutrinos, i.e., processes of the type $\chi\to\nu\nu$ or $\chi\to\nu\bar{\nu}$, although the overall method is very general and can be applied to any other decaying massive relic scenario. Historically, these decay channels were often invoked in the context of more complex dark sectors to avoid cosmological constraints on the non-DM particles that had to disappear into SM particles. Yet, although these decay channels would \textit{per se} not lead to any energy deposition (because of the weak interaction of the neutrinos), as shown in the reference if the decaying relic is massive enough it could radiate on- and off-shell $W$ and $Z$ bosons that would in turn decay (dominantly via $\chi\to\nu f_1 f_2 f_3$, where $f_i$ are any SM fermions) and induce EM and hadronic showers of particles that can affect the aforementioned observables. 

\begin{figure}[t!]
	\centering
	\includegraphics[width=0.48\textwidth]{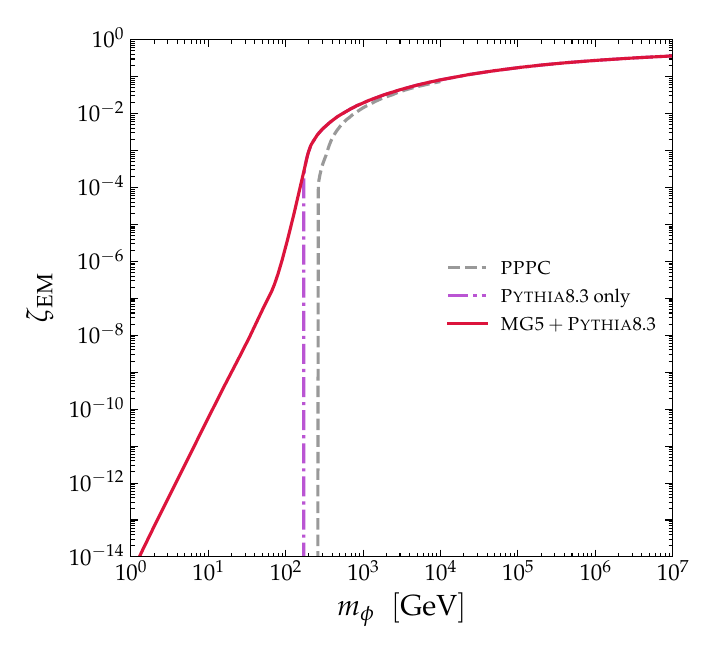}
	\includegraphics[width=0.48\textwidth]{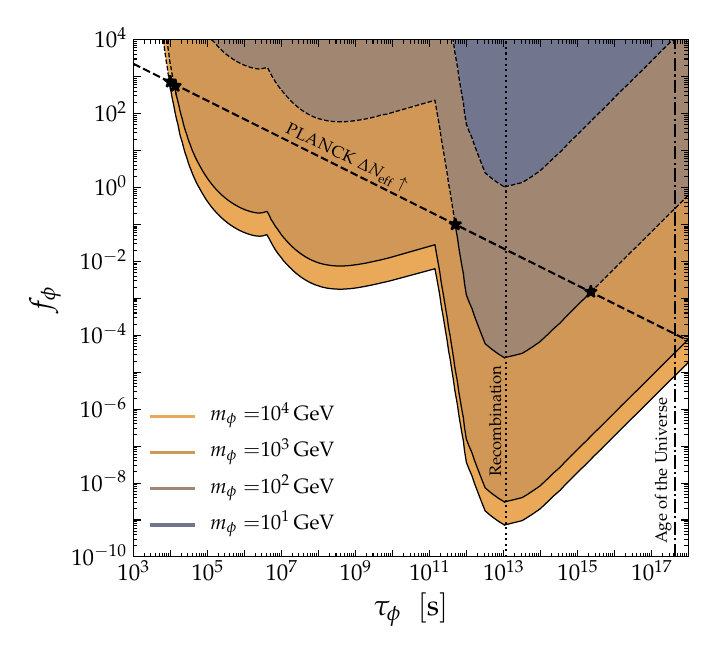} \\
	\hspace*{0.1 cm}
	\includegraphics[width=0.48\textwidth]{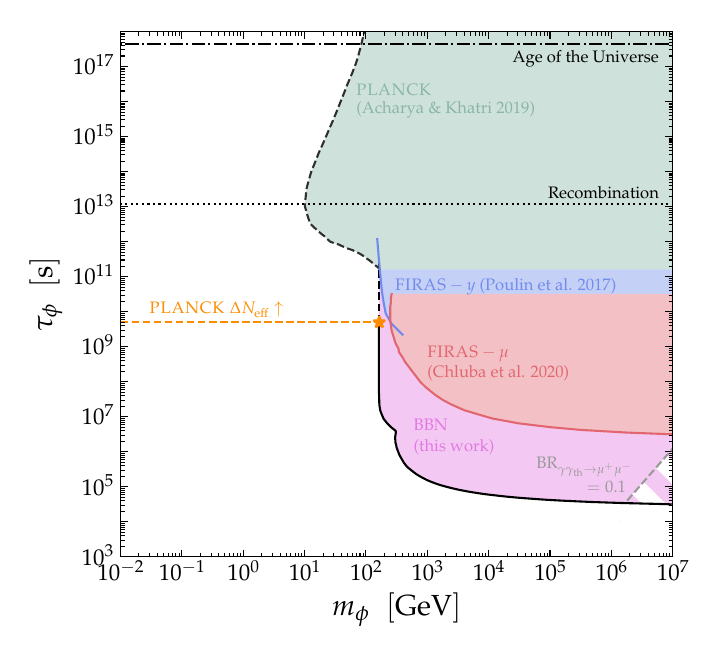}
	\includegraphics[width=0.48\textwidth]{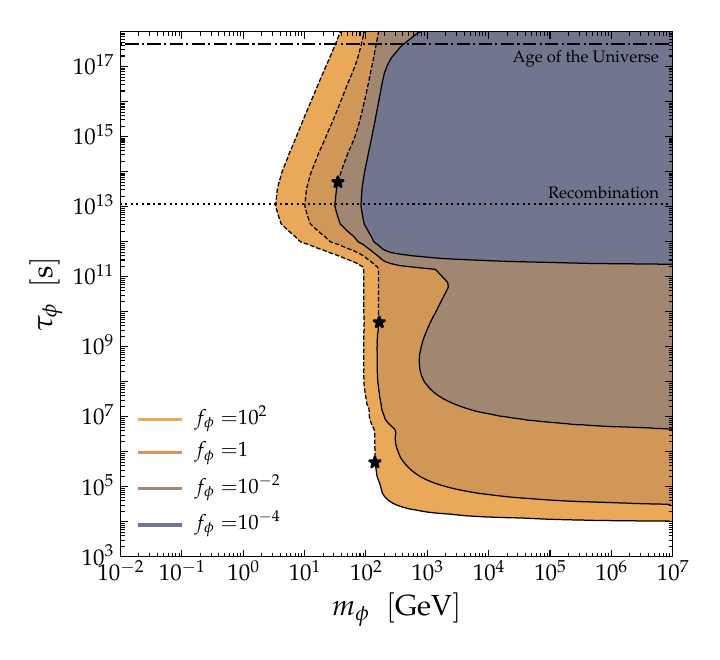}
	\caption{\textit{Top panels:} BR into EM radiation of the $\chi\to\nu\bar{\nu}$ decay (left) -- for consistency with the notation employed in the text we refer to $\phi$ in the figures with $\chi$ -- and corresponding cosmological constraints on the fractional $\chi$ abundance as a function of the particle lifetime and mass (right). \textit{Bottom panels:} Corresponding constrains on the $\tau_\chi - m_\chi$ plane for $f_\chi=1$ and with a graphical breakdown of the different probes (left) as well as comparing different $f_\chi$ values (right). Figures taken from [\protect\hyperlink{IX}{IX}].}
	\label{fig: DM_nu}
\end{figure}

The exact calculation of the resulting fraction of injected EM radiation leads to the relation between $f_{\rm em}$ ($\xi_{\rm em}$ in the figure) and $m_\chi$ ($m_\phi$ in the figure) displayed in the left top panel of Fig. \ref{fig: DM_nu}, taken from [\hyperlink{IX}{IX}] (red line, to be compared to previous works\footnote{An important result of the reference was to show that the \texttt{PPPC} \cite{Cirelli2011PPPC} and \texttt{PYTHIA} \cite{Sjostrand2014Introduction} (gray and purple lines in Fig. \ref{fig: DM_nu}, respectively) are unable to capture the full mass dependence of the EM production, strongly underestimating the latter at masses below $2\,m_{W/Z}$. In this regime other complementary numerical tools such as \texttt{FeynRules} \cite{Alloul2013FeynRules} and \texttt{MadGraph} \cite{Alwall:2011uj} need to be used. As pointed out in [\hyperlink{IX}{IX}], this finding could be used in the future to update late-time constraints on the same decay channel.}). For large masses ($m_\chi>2\,m_{W/Z}$) one finds that the production of EM radiation is rather high, of the order of $1-50\%$. Towards lower masses the production of the bosons becomes off-shell and the scaling of $f_\chi$ follows the mass dependence of the boson propagator\footnote{Proportional to $1/(q^2-m_{W/Z}^2)\sim1/m_{W/Z}^2$, where $q$ is the particle's momentum.}, i.e., $f_{\rm em}\propto(m_\chi/m_{W/Z})^4$. Given this relation one can then reformulate the constraints presented in Fig. \ref{fig: DM_decay_exclusion_region} (or equivalent) as a function of the mass of the decaying relic as reported in the right top panel of Fig. \ref{fig: DM_nu}, taken from [\hyperlink{IX}{IX}]. Alternatively, one could choose to fix the relic abundance (i.e., the value of $f_\chi$) and represent the same constraints in the $\tau_\chi - m_\chi$ plane. This has been done assuming $f_\chi=1$ in the bottom left panel of Fig.~\ref{fig: DM_nu}, taken from~[\hyperlink{IX}{IX}], with a breakdown of the various considered probes. The right bottom panel of the same figure, taken from~[\hyperlink{IX}{IX}], compares different $f_\chi$ values.

As a result, one finds that CMB anisotropy constraints can test $\chi$ masses as small as a few GeV for $f_\chi=1$, while current early-time limits only probe masses down to $100$ GeV. Of the latter, BBN constraints are currently more stringent than the SDs bounds (see left bottom panel of Fig.~\ref{fig: DM_nu}), although as argued in the reference on the basis of Fig.~\ref{fig: DM_decay_exclusion_region}
in the future also SDs will be able to test the $1-10$ GeV range also for lifetimes before recombination.

Another interesting application of SDs in the context of decaying DM revolves around ALPs $a$. Aside from the perspective of constraining their properties via their mixing with the photons (see below), also their decay can be constrained with CMB SDs \cite{Cadamuro:2010cz, Cadamuro:2011fd, Millea:2015qra, Balazs:2022tjl}. Assuming for instance an interaction term of the form
\begin{align}
	\mathcal{L}_{\rm int} = \frac{g_{\gamma a}}{4}\,a\,F^{\mu\nu}\tilde{F}_{\mu\nu}\,,
\end{align}	
where $g_{\gamma a}$ is the ALP-photon coupling, $F^{\mu\nu}$ is the EM field strength tensor and $\tilde{F}_{\mu\nu}$ its dual, the lifetime of the ALP can be determined to be \cite{Balazs:2022tjl}
\begin{align}
	\tau_{a} = 1.32\times10^8~\text{s}\left(\frac{g_{\gamma a}}{10^{-12}~\text{GeV}^{-1}}\right)^{-2}\left(\frac{m_a}{10~\text{MeV}}\right)^{-3}\,.
\end{align}
Focusing in particular on the most recent analysis of \cite{Balazs:2022tjl}, the use of the \texttt{GAMBIT} framework \cite{GAMBIT:2017yxo, Kvellestad:2019vxm} (and specifically of the \texttt{CosmoBit} module \cite{GAMBITCosmologyWorkgroup:2020htv}, whose SD calculation is based on the \texttt{CLASS}+\texttt{MontePython} implementation discussed in Secs. \ref{sec: num} and \ref{sec: num_2}) allowed to perform a combined analysis of BBN and CMB data as well as astrophysical information. The main advantage of this type of analyses is that it allows to fully capture the complementarity between the various probes.

\begin{figure}[t]
	\centering
	\includegraphics[width=0.48\textwidth]{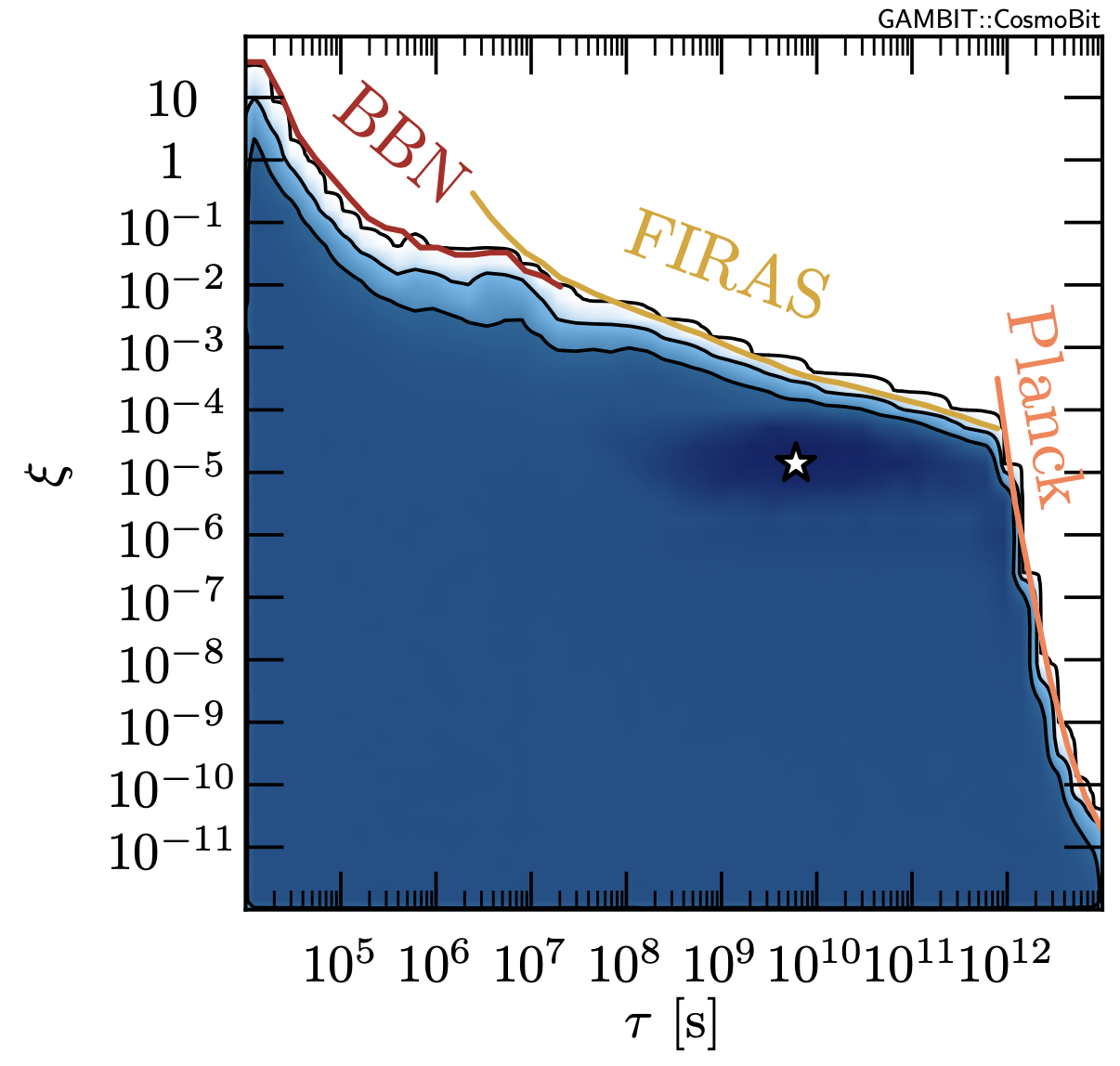}	\includegraphics[width=0.48\textwidth]{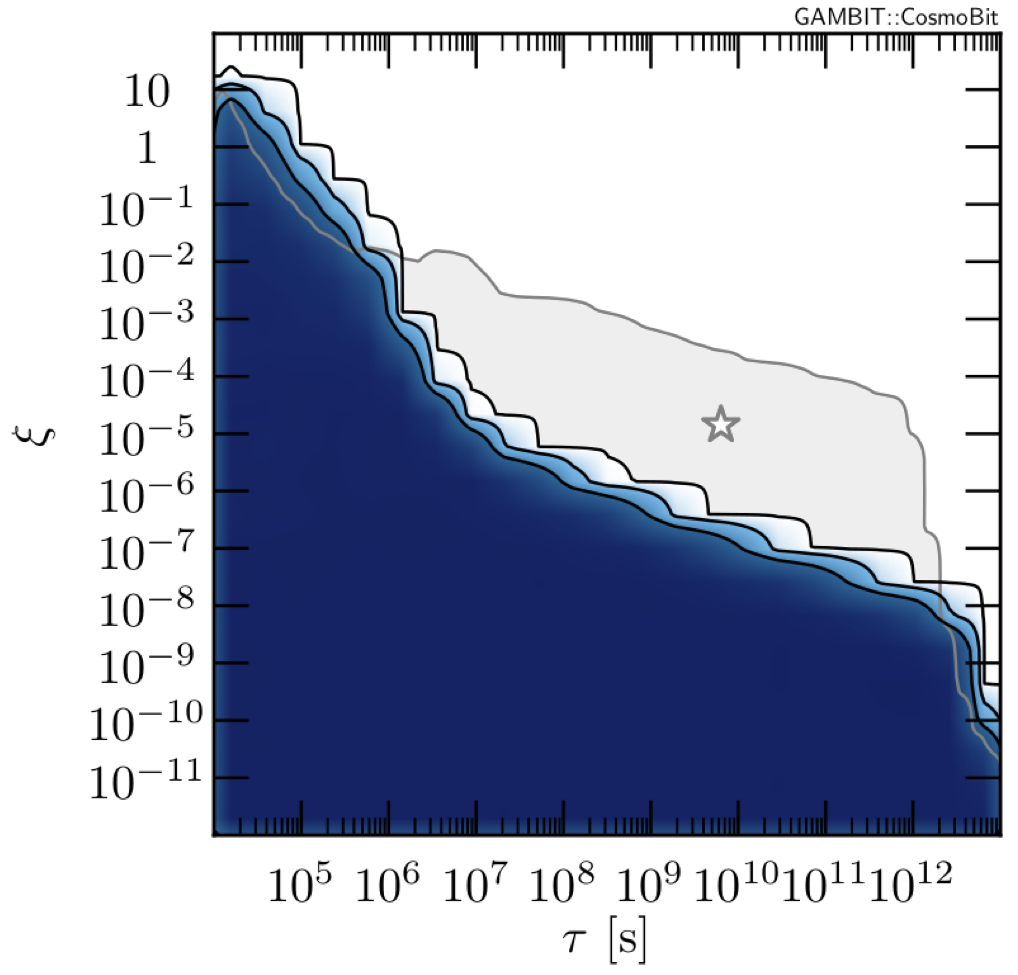}
	\caption{Exclusion contours on the fractional abundance of decaying ALPs ($\xi$ in the figures) as a function of their lifetime assuming FIRAS (left) and PIXIE (right). Figures taken from \cite{Balazs:2022tjl}.}
	\label{fig: DM_ALP_decay}
\end{figure}

The resulting constraints on the fractional ALPs abundance ($\xi$ in the figure) are shown in Fig.~\ref{fig: DM_ALP_decay}, taken from \cite{Balazs:2022tjl}, as a function of the lifetime for both the FIRAS (left) and PIXIE (right) missions. Very interesting, the global scan found a (mildly) preferred region of parameter space (white star) precisely at lifetimes that can be probed by future CMB missions (see right panel of the figure). The origin of the preference has been reconducted to the fact that the presence of the ALPs improves the agreement between the observed deuterium abundance (see Eq. \eqref{eq: BBN_obs}) and its theoretical prediction within \lcdm, which the reference reports to be D/H$\,\simeq2.454\pm0.046$ at 95\% CL (a value slightly lower than the one given in Eq. \eqref{eq: BBN_th}). In fact, as evident from Fig. 6 of the reference (from which the aforementioned value of D/H is inferred), the \lcdm determination of D/H slightly underestimates the observed one, while the decay of the ALPs can restore the concordance between the two estimates by photo-disintegrating $^4$He atoms and thereby producing more deuterium. Should this preference be real, the corresponding ALP model would lead to a production of SDs that could be probed already by a PIXIE-like mission, as graphically visible in the right panel of Fig.~\ref{fig: DM_ALP_decay}.

\vspace{0.3 cm}
\noindent\textbf{DM annihilation}

\noindent CMB SDs can constrain DM annihilation via the consequent energy injection it would involve. The related energy injection rate follows from Eq. \eqref{eq: E_inj} with the luminosity given by the product of the released energy per annihilation $E=2m_{\chi}$ and the annihilation probability $P=\langle\sigma v\rangle n_\chi$ (see e.g., \cite{Poulin2015Dark}), i.e., 
\begin{align}
	L_\chi= 2 \langle\sigma v\rangle \rho_\chi\,,
\end{align} 
where $\langle\sigma v\rangle$ is the thermal averaged cross section of the annihilation process. Assuming then that $\chi$ is the DM (i.e., $f_\chi=1$) and noting that $n_\chi$ in Eq. \eqref{eq: E_inj 1} needs to be divided by two because it takes two $\chi$ particles for each annihilation, one obtains
\begin{align}
	\dot{Q}_{\rm inj}=  \rho_{\rm cdm}^2 \, \frac{\langle\sigma v\rangle}{m_{\chi}} \,.
\end{align}
Depending on the DM model involved, the cross section can have different velocity (and hence redshift, since $v\propto T/m\propto (1+z)/m$) dependences. For instance, s-wave annihilation would predict $\langle\sigma v\rangle$ to be constant in time, while p-wave annihilation would have $\langle\sigma v\rangle\propto(1+z)$ or $\langle\sigma v\rangle\propto(1+z)^2$ (see e.g., \cite{mcdonald2000cosmic, Chluba2013Distinguishing}). A useful parameterization of the energy injection rate then becomes
\begin{align}
	\dot{Q}_{\rm inj}=  \rho_{\rm cdm}^2 (1+z)^\alpha\, p_{\rm ann} \quad \text{with} \quad p_{\rm ann}=\frac{\langle\sigma v\rangle_0}{m_{\chi}}\,,
\end{align}
where the index $\alpha$ depends on the chosen annihilation channel and $\langle\sigma v\rangle_0$ is the thermal averaged cross section today.

\begin{figure}
	\centering
	\includegraphics[width=0.65\textwidth]{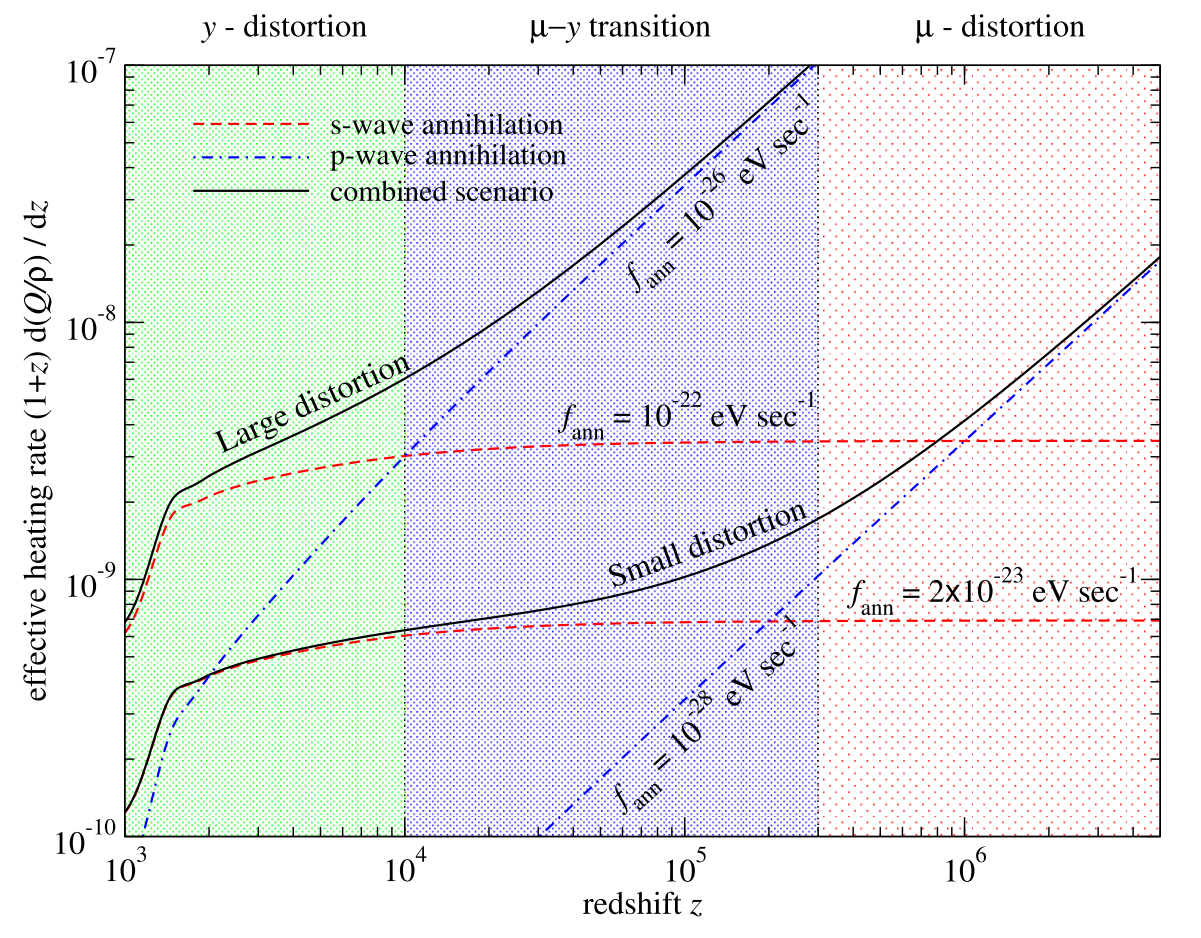}
	\caption{Heating rate following the annihilation of DM in the s-wave ($\alpha=0$) and p-wave ($\alpha=1$) case. Figure taken from \cite{Chluba2013Distinguishing}.}
	\label{fig: DM_heating_ann_pwave}
\end{figure}

The heating rate assuming s-wave annihilation (i.e., $\alpha=0$) and the maximum of $p_{\rm ann}$ allowed by data is reported in the left panel of Fig. \ref{fig: DM_heating_ann_dec}, with the corresponding SD signal in the right panel. From the figure it becomes clear that DM annihilation has only a minor impact on the SD spectrum. Nevertheless, as shown in Fig. \ref{fig: DM_heating_ann_pwave}, taken from \cite{Chluba2013Distinguishing}, assuming p-wave annihilation (blue lines assume $\alpha=1$ in the figure, to be compared with the red lines representing the $\alpha=0$ case) would significantly enhance the heating rate\footnote{Note that in the figure $f_{\rm ann}$ does not directly correspond to $p_{\rm ann}$, but the derived conclusions are independent of that.} and lead to a sizable SD signal.

For the case of s-wave annihilations (i.e., for $\alpha=0$), Planck 2018 data sets an upper bound on $p_{\rm ann}$ of the form $p_{\rm ann}<3.3\times10^{-28}$ cm$^3$/(s GeV) (at 95\% CL) \cite{Aghanim2018PlanckVI} and future high-precision SD missions are not expected to significantly improve upon it [\hyperlink{III}{III}] (in particular compared to the limits expected by up-coming CMB anisotropy missions). Nevertheless, for the case of p-wave annihilations with $\alpha=1$ the redshift dependence enhances the constraining power of SDs and a PIXIE-like mission would already be able to set the strongest constraints to date, of the order of $p_{\rm ann}<1\times10^{-32}$ cm$^3$/(s GeV) (at 95\% CL) \cite{Chluba2013Distinguishing}. Further increasing the value of $\alpha$ makes of BBN the most constraining early-time probe \cite{Depta2019BBN}.

\vspace{0.3 cm}
\noindent\textbf{Follow-up idea 11:} Updating systematically the SD constraints on annihilating DM for p-wave annihilation in the context of future missions would be very useful.

\vspace{0.3 cm}
\noindent\textbf{DM scattering}

\begin{figure}[t!]
	\centering
	\includegraphics[width=0.45\textwidth]{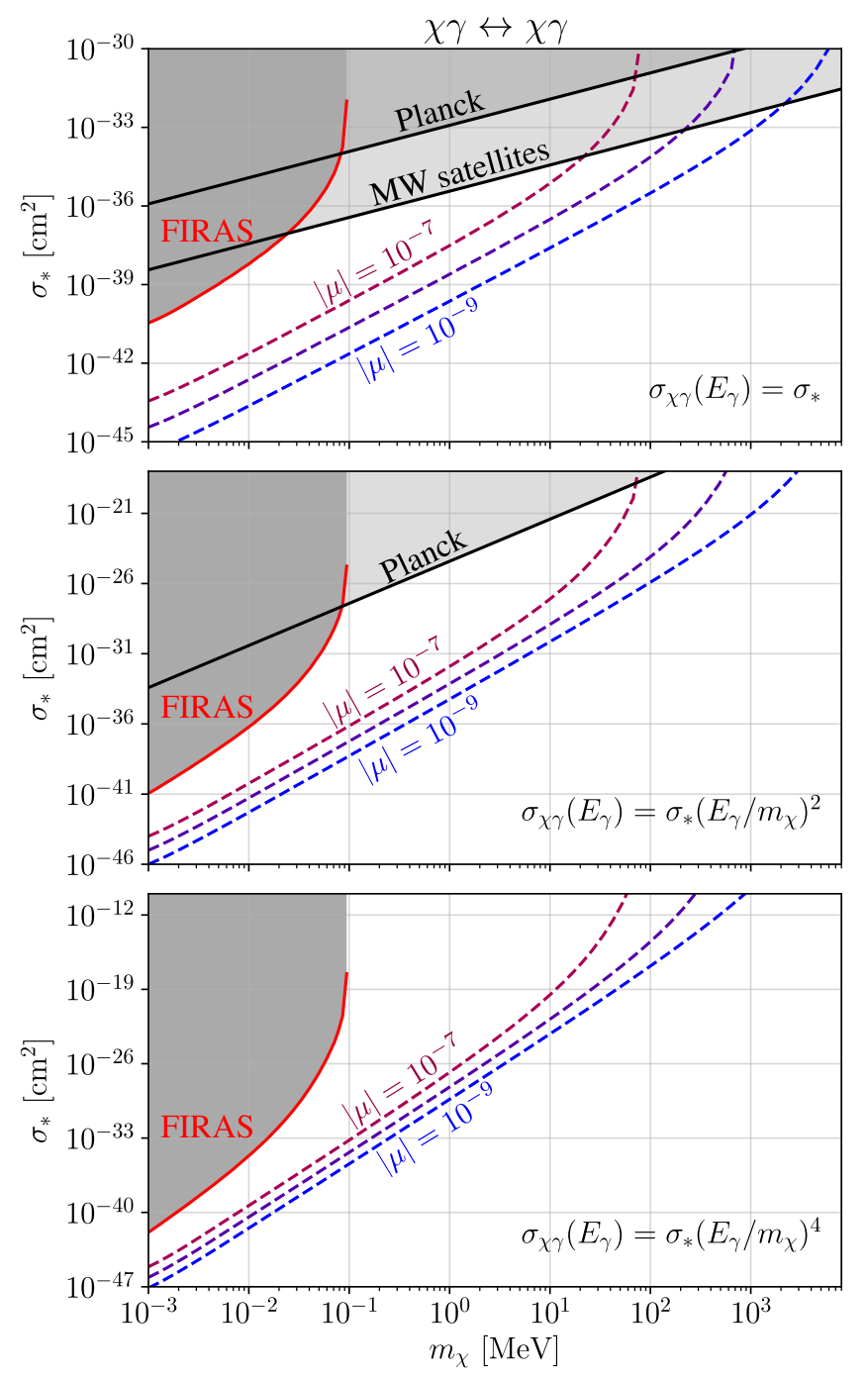}
	\caption{Exclusion contours of the DM-photon scattering cross section as a function of the DM mass for different velocity dependences of the cross section. Figure taken from \cite{Ali-Haimoud2021Testing}.}
	\label{fig: DM_scatterings}
\end{figure}

\noindent As discussed in Sec. \ref{sec: SD_non_lcdm}, another DM interaction type that can be probed with CMB SDs are elastic scatterings with photons, electrons and protons, since they would inevitably cool the photon bath, thereby effectively extracting energy from it and distorting the CMB energy spectrum \cite{AliHaimoud2015Constraints, Ali-Haimoud2021Testing}. Most recently, all of these interaction channels have been extensively considered in \cite{Ali-Haimoud2021Testing}. In Fig. \ref{fig: DM_scatterings}, taken from the reference, we report the case of DM-photon scatterings (the discussion is qualitatively similar also for the other cases, see Fig.~4 of the reference). As for the DM annihilation, also here a number of different models can be accounted for by considering different energy-scalings of the scattering cross section (see e.g., \cite{Dubovsky:2003yn, Sigurdson:2004zp, Bernabei:2007gr, Weiner:2012cb, Fitzpatrick:2012ix, Anand:2013yka, Boddy:2018kfv}). 

As it turns out, FIRAS data is already setting some of the strongest constraints (if not the only ones) for DM masses $m_\chi<100$ keV. A sensitivity to the $\mu$ signal of the order of $10^{-7}$ would guarantee a significant improvement of the current constraints on DM-photon and DM-electron scatterings for all considered cross sections over the keV$-$GeV mass range. The same would be true for DM-proton scatterings at $m_\chi<10$ keV, but it would require $\delta\mu\lesssim 10^{-9}$ above that mass.

\vspace{0.3 cm}
\noindent\textbf{DM-photon mixing}

\noindent In Sec. \ref{sec: SD_non_lcdm} we have introduced the idea that photon conversion to beyond-SM particles, such as ALPs $a$, might occur in the presence of an external magnetic field \cite{1983PhRvL..51.1415S, 1988PhRvD..37.1237R, 1988PhRvD..37.2001A}. Concretely, one could consider an ALP-photon interaction of the form  
\begin{align}
	\mathcal{L}_{\rm int} = g_{\gamma a} \textbf{E}_\gamma \textbf{B}_{\rm ext} \,a\,,
\end{align}
where $g_{\gamma a}$ is the corresponding coupling, 
$\textbf{E}_\gamma$ is the electric field of the photon and $\textbf{B}_{\rm ext} $ is the external magnetic field. The resulting oscillation probability has already been introduced in Sec.~\ref{sec: SD_non_lcdm} (see Eq.~\eqref{eq: P_ALP} there) and depends, among others, on the magnetic field strength. In a similar fashion, also light scalar particles $\phi$ could present couplings to the photon magnetic field $\textbf{B}_\gamma$, i.e., 
\begin{align}
	\mathcal{L}_{\rm int} = g_{\gamma \phi} \textbf{B}_\gamma \textbf{B}_{\rm ext} \,\phi\,.
\end{align}
Although not necessarily representing a DM candidate, we point out that a similar discussion would also apply to gravitons \cite{Dolgov:2013pwa}. 

As such, CMB photons traveling through voids, the intergalactic and inter-cluster medium and the Milky Way are susceptible to these frequency-dependent oscillations and CMB SDs might form \cite{Mukherjee2018Polarized}. Should PMFs exist, they would also lead to ALP-photon conversions at earlier times and to the consequent formation of primordial SDs \cite{Tashiro:2013yea, Ejlli:2013uda}. Because of the alignment with the external magnetic field, the resulting SD signal would be polarized on top of reflecting the aforementioned oscillatory behavior \cite{Mukherjee2018Polarized} (see in particular Fig.~1 therein). This would represent a unique signature for the existence of this mechanism to be looked for in current and future CMB \matteo{anisotropy and SD}\footnote{\matteo{A PIXIE-like spectrometer would in fact also be able to measure linear polarization \cite{Kogut2011Primordial}.}} experiments \cite{Mukherjee:2018zzg, Mukherjee:2019dsu}. As shown in particular in Fig. 11 of \cite{Mukherjee2018Polarized}, a Super-PIXIE-like mission would be able set the currently strongest bounds on the ALP-photon mixing for axion masses of the order of $m_a\lesssim 10^{-13}$ eV and would in general apply to $m_a\lesssim 10^{-11}$~eV \matteo{(assuming typical galactic parameters to characterize the magnetic field)}.

\vspace{0.3 cm}
\noindent\textbf{Summary}

\noindent In summary, CMB SDs would be able to constrain a wide range of interaction channels between the DM and the SM particles (photons and electrons in particular), like e.g., decay, annihilation, scattering and mixing. In almost all instances, the observation of SDs with a Super-PIXIE-like mission would deliver the strongest constrains to date in the relevant sections of parameter space. As such, CMB SDs could play a significant and unique role in our understanding of the particle physics properties of DM.

In the context of this section, the work carried out during the thesis is meant to open and highlight the synergy between CMB SDs and other complementary probes in constraining DM properties. This is particularly evident in the case of decaying DM, one of the representative examples considered in [\protect\hyperlink{I}{I}] and later also in [\protect\hyperlink{IX}{IX}], as shown in Figs.~\ref{fig: DM_decay_exclusion_region} and \ref{fig: DM_nu}. Furthermore, a systematic analysis of the interplay between CMB anisotropy and SD constrains has been performed in [\protect\hyperlink{III}{III}] in the context of (s-wave) DM annihilation. The very general numerical setup discussed in Sec. \ref{sec: num} also allows for the straightforward implementation of other sources of heating related to DM, making of the aforementioned studies possible guides for future analyses.

\subsection{Primordial black holes}\label{sec: res_PBH}
\begin{figure}
	\centering
	\includegraphics[width=0.85\textwidth]{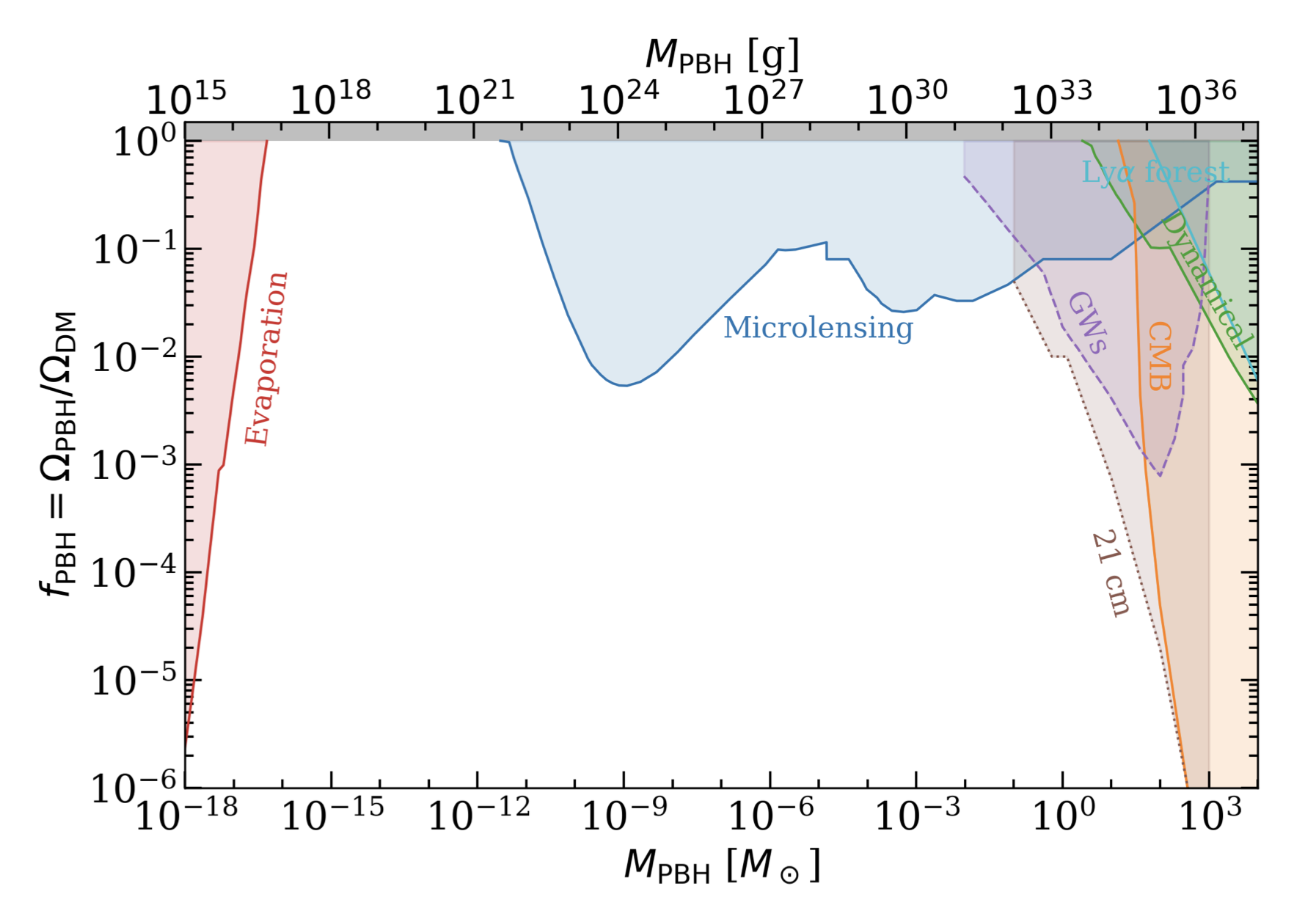}
	\caption{Summary plot of the constraints on the fractional PBH abundance as a function of the PBH mass. Figure taken from \cite{Villanueva-Domingo2021Brief}.}
	\label{fig: PBH_overview}
\end{figure}

Another very popular candidate to explain the presence of DM are PBHs (see e.g., \cite{Carr2010New, Carr2016Primordial, Carr2020Primordial, Villanueva-Domingo2021Brief} for reviews). Although their existence has been first postulated in the '60s and '70s \cite{Zeldovich1967Hypothesis, Hawking1971Gravitationally, carr1974black, chapline1975cosmological}, the topic saw a surge of popularity only in the mid '90s and much more recently with the first detection of the GW emission from a BH binary system by the Laser Interferometer Gravitational-Wave Observatory (LIGO) \cite{Abbott2016Observation}. Overall, based on a number of observational strategies the PBH abundance (parameterized via $f_{\rm PBH}=\Omega_{\rm PBH}/\Omega_{\rm cdm}$) has been severely constrained over a mass range that covers more than 25 orders of magnitude, as summarized in Fig.~\ref{fig: PBH_overview}, taken from \cite{Villanueva-Domingo2021Brief}. Of particular interest are the recent advances of the LIGO-Virgo-KAGRA (LVK) collaboration, which has now collected data on more than 100 events \cite{LIGOScientific:2021djp}. In fact, some of these systems seem to point to a PBH population with $f_{\rm PBH}\simeq10^{-2}-10^{-3}$ and $M_{\rm PBH}\simeq 10-100$ M$_\odot$ \cite{Clesse:2017bsw, Raidal:2017mfl, Raidal:2018bbj, Gow:2019pok, Carr:2019kxo, DeLuca:2020qqa, DeLuca2020Constraints, Hall:2020daa, Jedamzik:2020omx, Jedamzik:2020ypm, Clesse:2020ghq, Hutsi:2020sol, Franciolini:2022tfm, Escriva:2022bwe} and the question of whether this preference is compatible with current cosmological observations (especially of the CMB -- see Fig.~\ref{fig: PBH_overview}) has become matter of speculation as we will discuss more in depth below.

As it turns out, a particularly versatile probe of the existence of PBHs are CMB SDs. In fact, they can severely constrain their formation mechanism, their evaporation and their accretion of matter, placing significant limits on the PBH abundance on both ends of the mass spectrum.\footnote{As an initial remark, we point out here that all PBHs evaporate and accrete to a degree. Nevertheless, as we will see below and as clear from Fig.~\ref{fig: PBH_overview}, PBH evaporation is most efficient for very low-mass PBHs -- since the luminosity scales like $L_\chi\propto M^{-2}$, see Eq. \eqref{eq: L_PBH} -- while PBH accretion requires a lot of gravitational attraction to be efficient and is therefore most relevant for large PBH masses -- with $L_\chi\propto M^{3}$, see Eq. \eqref{eq: L_PBH_acc}.} Below we review all of these bounds in detail. The main contributions of the thesis are in the context of evaporation and accretion, while the discussion surrounding the formation constraint is mentioned for sake of completeness.

\subsubsection{PBH formation}\label{sec: res_PBH_form}
\begin{figure}
	\centering
	\includegraphics[width=\textwidth]{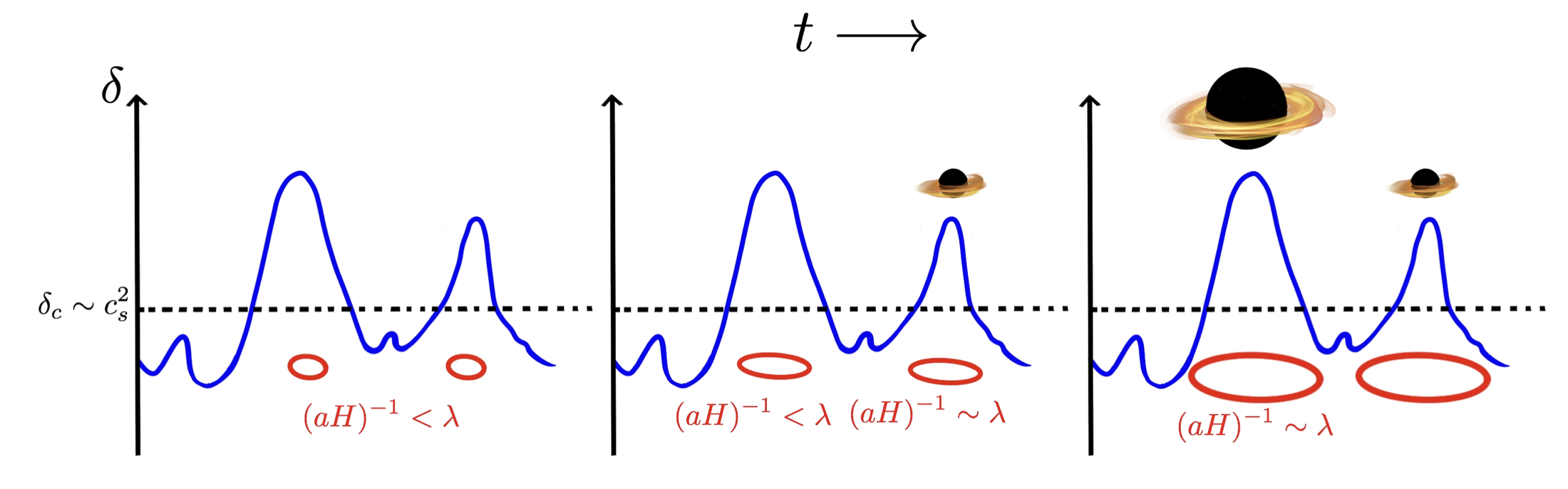}
	\caption{Schematic illustration of the standard PBH formation scenario. Figure taken from \cite{Villanueva-Domingo2021Brief}.}
	\label{fig: PBH_formation}
\end{figure}

In a nutshell, PBHs form whenever and wherever the energy density of the universe $\rho$ exceeds a critical value $\rho_c$.\footnote{Since the collapse is given by a balance between gravity, i.e., $\rho$, and pressure the value of $\rho_c$ is in general a function of the EOS parameter $\omega$ of the universe (see Sec. \ref{sec: cosmo_princ}).} Of course, if this was to happen at the background level, the whole universe would collapse to a PBH. Since we know this not to be the case, it is useful to consider the possibility that only fluctuations over the background $\delta=\delta \rho/\rho$ exceed the corresponding threshold $\delta_c$ and collapse. In this way, there is a direct proportionality between the mass of the PBH and the size $\lambda$ of the overdensity. As explained in Secs. \ref{sec: infl} and \ref{sec: CMB}, the size of a fluctuation determines when the fluctuation re-enters the Hubble horizon. Since only fluctuations that are on sub-Hubble scales can collapse to a BH, this introduces a further dependence of the PBH mass on the time $t$ (or scale $k$) of the Hubble-crossing of the relevant fluctuations (with $M_{\rm PBH}\propto\lambda\propto t$). This behavior is graphically represented in Fig.~\ref{fig: PBH_formation}, taken from \cite{Villanueva-Domingo2021Brief}. Following the aforementioned reviews, this relation can be quantitatively expressed as\footnote{In the last equality we neglect the role of the effective number of relativistic degrees of freedom $g_*$ for simplicity.}
\begin{align}\label{eq: Mpbh}
	M_{\rm PBH} = \gamma M_{\rm H} \simeq 1~\text{M}_\odot~\left(\frac{t}{2\times10^{-5}~\text{s}}\right)\simeq 1~\text{M}_\odot~\left(\frac{4\times10^{6}~\text{Mpc}^{-1}}{k}\right)^2\,.
\end{align}
The first equality exists because of the fact that not all of the mass contained in the collapsing Hubble patch ($M_{\rm H}$) participates in the gravitational collapse which ultimately leads to the PBH formation (depending, for instance, on the dynamics at horizon entry). The proportionality constant $\gamma$ takes this into account and has been determined to be of the order of $\gamma\simeq0.2-0.4$ \cite{carr1975primordial, Green:2004wb}. A similar relation to the ones expressed in Eq. \eqref{eq: Mpbh} can also be derived with respect to the temperature of the thermal bath during RD (see e.g., Eq. (2) of \cite{Carr:2019kxo}).

The question now is how probable it is for the primordial overdensities to exceed~$\delta_c$, as this determines the abundance of the PBHs. Intuitively, to have a sizable production of PBHs one \matteo{would expect $\delta$ not to be significantly lower than $\delta_c$}. However, numerical simulations suggest that $\delta_c\simeq0.4-0.6$ (see e.g., \cite{Escriva:2020tak, Musco:2020jjb} for recent treatments of the problem), which is significantly larger than what expected within the \lcdm model for $\delta$, since $\delta\propto\mathcal{P}(k)^{1/2}\sim 5\times 10^{-5}$. Therefore, although the formation of PBHs in the standard (single field) slow-roll inflation model is not \textit{per se} excluded, one often needs to resort to more complex scenarios able to enhance the PPS at the desired scales in order to produce a reasonable amount of PBHs. Out of the many consequences it would have, this means that the unquestioned observation of even one single PBH would radically change our understanding of the inflationary epoch. 

\begin{figure}
	\centering
	\includegraphics[width=0.8\textwidth]{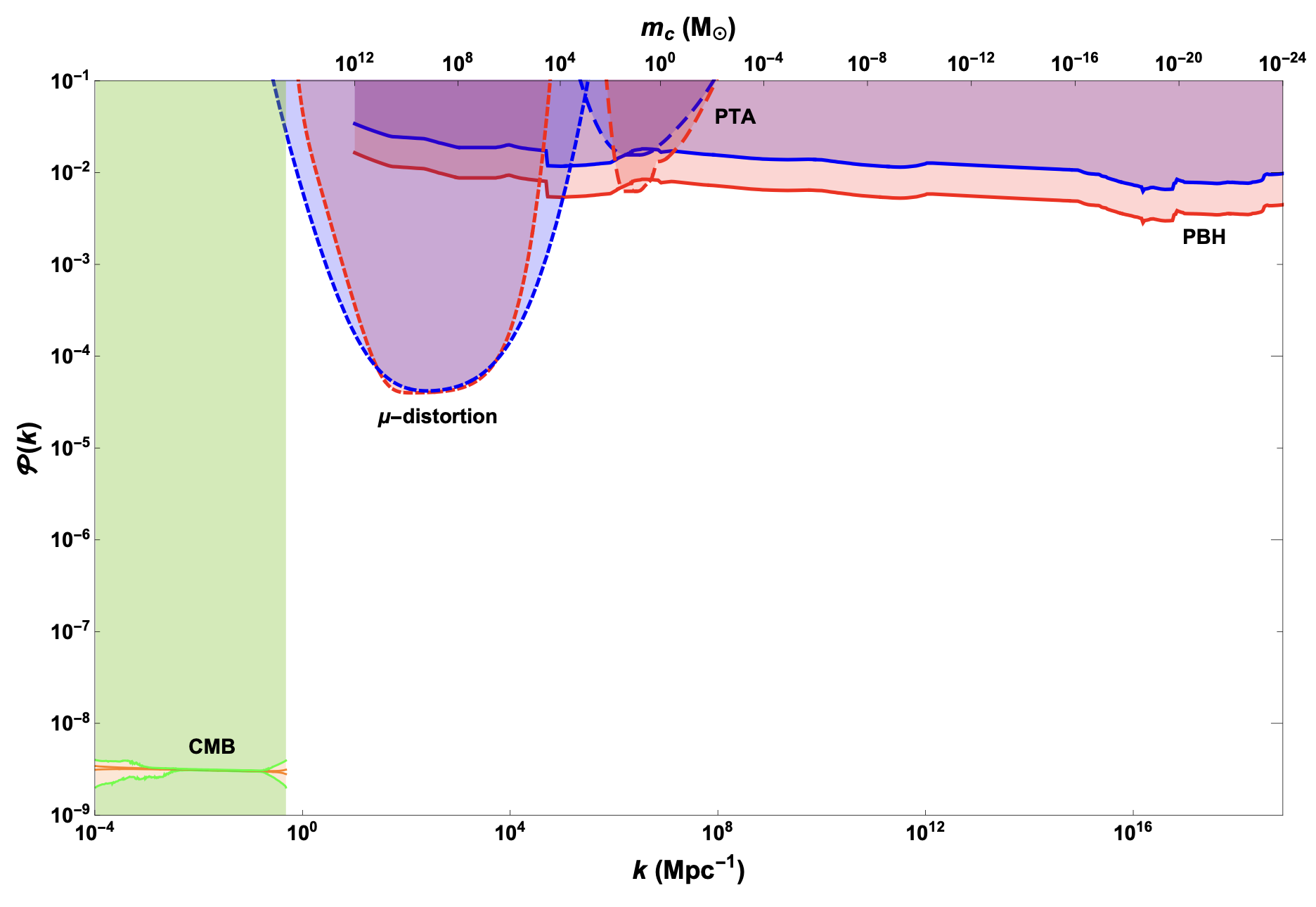}
	\caption{Cosmological constraints on the PPS for different PBH mass distributions (quasi-monochromatic in red and lognormal in blue). Figure taken from \cite{Gow:2020bzo}.}
	\label{fig: PBH_PPS}
\end{figure}

This relation between PBH abundance and PPS shape is very important as it allows to apply to both the constraints specifically derived on either one of them. For instance, as said above, if $\delta$ was too much larger than $\delta_c$ the gravitational pressure would overclose the universe. Therefore, by imposing that the PBH abundance is not too large one can infer upper bounds on the magnitude of the primordial perturbations and hence of the PPS. These are graphically represented in Fig. \ref{fig: PBH_PPS}, taken from \cite{Gow:2020bzo} (in the figure, the colors refer to different mass distributions -- quasi-monochromatic in red and lognormal in blue, see below). Since these limits do not depend on the direct observation of the PPS by a given cosmological probe, they can reach arbitrarily small scales.

\newpage
In turn, any constraint on the PPS severely limits the possibility to produce PBHs. As discussed in Secs. \ref{sec: CMB} and \ref{sec: res_infl}, both CMB anisotropies and SDs can be used in this regard\footnote{In the figure, also limits from Pulsar Timing Arrays (PTA) are shown. However, since they do not significantly constraint the PBH hypothesis we will not discuss them here.}, and their combined constraining power (assuming Planck and FIRAS) is shown in Fig. \ref{fig: PBH_PPS} at scales between $10^{-4}-10^4$ Mpc$^{-1}$ (see Fig.~6 of \cite{Gow:2020bzo} for the forecasted reach of future missions as well as Fig. 1 of \cite{FrancoAbellan:2023sby} for possible additional constraints in the same $k$ range). Since CMB SDs are probing the smallest scales, they are also the most constraining of the two. As immediately clear from the figure (red contours), the SD limits derived by FIRAS data can indeed exclude the existence of a significant population of PBHs with masses $M_{\rm PBH}\gtrsim5\times10^4$ M$_\odot$ (as confirmed by dedicated analyses \cite{Kohri:2014lza, Nakama:2017xvq, Juan2022QCD}). A Voyage 2050 mission would deliver an upper bound of the form $M_{\rm PBH}\lesssim10^3$ M$_\odot$ \cite{Nakama:2017xvq}.

Of course, these constraints depend on the assumed deviation of the PPS from its scale-invariant form. In fact, for simplicity one usually assumes a sharply peaked feature in the PPS that leads to a so-called monochromatic distribution of PBH masses and the aforementioned constraints are derived with this in mind. However, as justified by more realistic inflationary scenarios, also other (extended) mass distributions can be generated (see e.g., Sec. IIB of \cite{Carr2020Primordial} for a review), which would modify the value of the upper bound. 

For instance, assuming non-Gaussian curvature perturbations (see Sec. \ref{sec: res_infl}) enhances and sharpens the overdensities \cite{Byrnes:2012yx} (see in particular Fig.~1 therein), following in an increased PBH formation with respect to the Gaussian case. In turn, this relaxes the SD bounds~\cite{Nakama:2017xvq}, although only at most to $M_{\rm PBH}\lesssim5\times10^5$ M$_\odot$ for FIRAS \cite{Juan2022QCD}. On the other hand, however, other extended mass distributions (such as the lognormal one considered in e.g., \cite{Gow:2020bzo}) might involve a relatively soft (compared to the monochromatic case) transition of the PPS from Planck scales to the wished enhancement \cite{Byrnes2018Steepest, Cole:2022xqc} (see also Eq.~\eqref{eq: PPS_step} and related discussion). In this case, CMB SDs could constrain the transition itself, thereby limiting even more the maximum allowed mass of the PBHs, which would only start to form upon completion of the step in the PPS. An example of this behavior can be seen by comparing the red and blue contours in Fig.~\ref{fig: PBH_PPS}, which represent a narrow (quasi-monochromatic) and a less narrow mass distribution, respectively.

\vspace{0.3 cm}
\noindent\textbf{Follow-up idea 12:} To our knowledge, a systematic analysis of the CMB SD constraints on extended PBH mass distributions (such as the one considered in e.g., Fig.~\ref{fig: PBH_PPS}) has not been performed in the literature yet. Given the mathematical setup already developed in the aforementioned references and the analysis already carried out in [\hyperlink{V}{V}], such an exercise would be rather straightforward and of major interest. It would be in particular useful to know if future SD missions could reach the  $10-100$ M$_{\odot}$ range in any of these scenarios, and start to test the LVK observations. Based on Fig. 6 of \cite{Gow:2020bzo} this possibility should not be unrealistic.
\vspace{0.3 cm}

\matteo{As a final remark, it needs to be pointed out that also formation mechanisms other than the collapse of large density perturbations can lead to the formation of super-massive PBHs. Examples are for instance the collapse of cosmic string loops \cite{hawking1989black, Polnarev:1988dh, Charnock:2016nzm, Brandenberger:2021zvn, Cyr:2022urs} and bubble collisions \cite{hawking1982bubble, Jung:2021mku} in the early universe. In these cases, the requirements on the PPS and the consequent formation of SD would be different than what described above, introducing a degree of model-dependence in the aforementioned upper bounds.}

In summary, CMB SDs deliver the currently most stringent constraints imposed on PBHs via their formation mechanism \matteo{(assuming the collapse of large density perturbations)}.~In the monochromatic case and in absence of primordial non-Gaussianities SDs impose a strong upper bound on the PBH mass of the order of ${M\lesssim5\times10^4}$~M$_\odot$ for FIRAS, which  could become significantly more stringent with future SD missions. The limit relaxes to at most $M\lesssim10^5$ M$_\odot$ in the presence of non-Gaussianities but could become much stronger in the case of extended mass distributions. Although not considered in the course of the thesis, these classes of possibilities would be rather easy to study with the numerical pipeline introduced in Secs. \ref{sec: num} and~\ref{sec: num_2}.

\subsubsection{PBH evaporation}\label{sec: res_PBH_ev}
\textbf{Schwarzschild PBHs}

\noindent Once BHs are formed they are predicted to evaporate \cite{Hawking1974Black}. A simple interpretation of this phenomenon is that after the spontaneous production (from the vacuum) of a virtual particle pair in the immediate vicinity of the BH horizon, one particle (the one with ``negative'' energy) might fall into the BH while the other (the one with ``positive'' energy) can escape the gravitational pull. For an observer far away from the BH, this effectively looks like the BH is radiating particles. Nevertheless, because of energy conservation arguments the emitted energy must come a the cost of a reduction of the BH mass (due to the accretion of ``negative'' energy). As a consequence, the more the BH emits particles the more it loses mass and shrinks (due to the direct dependence of BH mass and size). Therefore, the BH is said to be evaporating.

Since the emitted radiation is thermal, one can associate a temperature to it that is found to be related to the BH mass via
\begin{align}\label{eq: T_BH}
	T=\frac{1}{8\pi M}\,.
\end{align}
Intuitively, this anti-proportionality can be understood by noting that the larger the BH mass the \matteo{weaker} the curvature close to its horizon is going to be. In turn, the \matteo{weaker the consequent tidal forces} the less particle \matteo{pairs can be stripped apart}, effectively reducing the evaporation (i.e., the radiation) and the associated temperature\footnote{The intensity of the BB spectrum is in fact directly related to its temperature, following the Stefan-Boltzmann law $L\propto T^4$. The same proportionality also applies to BHs.}.

The number and type of emitted particles $j$ depend on whether they are kinematically allowed, i.e., if their mass $m_j$ is smaller than $T$, and on their quantum numbers, such as their spin $s_j$. Mathematically, one can write
\begin{align}\label{eq: dN}
	\frac{\text{d}N_j}{\text{d}t\text{d}E} = \frac{1}{2\pi}\frac{\Gamma_j}{e^{(E-\mu_j)/T}-(-1)^{2s_j}}\,,
\end{align}
where $\Gamma_j=\Gamma_j(E,M)$ represent the dimensionless absorption probability of the given emitted species and the exponential factor proportional to $T$ suppresses the production of particles with $m_j>T$. The mass loss rate can then be computed by integrating $\text{d}N_j/\text{d}t\text{d}E$ over energy, i.e., 
\begin{align}\label{eq: dMdt 1}
	\frac{\text{d}M}{\text{d}t} = - \int \sum_j \frac{\text{d}N_j}{\text{d}t\text{d}E} E \text{d}E\,,
\end{align}
where the minus sign is due to the fact that the mass reduces in time. Accurate calculations of this quantity \cite{Macgibbon1990QuarkI, Macgibbon1990QuarkII} show that Eq. \eqref{eq: dMdt 1} can be expressed as
\begin{equation}
	\frac{\text{d}M}{\text{d}t} = -5.34\times10^{25} \frac{\mathcal{F}(M)}{M^2} \,\, \text{g/s}\,,
\end{equation}
where $\mathcal{F}(M)$ (or equivalently $\mathcal{F}(T)$) represents the effective number of emitted species and it is arbitrarily normalized to unity in the regime where only massless particles can be emitted, i.e., for $T\ll 0.1$ MeV or equivalently $M\gg10^{17}$ g. In practice, it is given by
\begin{align}\label{eq: f_M_PBH}
	\mathcal{F}(M)=\sum_j g_j f_{s,q} e^{-M/(\beta_s \tilde{M}_j)}\,.
\end{align}
Here, $g_j$ and $f_{s,q}$ encapsulate the number of degrees of freedom and the quantum numbers (like spin and charge) of the particle $j$. $\tilde{M}_j$ is the mass that a BH would have if its temperature was equal to the rest mass of the particle $j$ and $\beta_s$ is defined so as to ensure that the spectrum of a BH with $M=\beta_s \tilde{M}_j$ peaks at the mass of the particle $m_j$. The numerical coefficients can be found in e.g., Tab.~2 of [\hyperlink{I}{I}]. 

In its essence, this formulation of $\mathcal{F}(M)$ reflects the discussion surrounding Eq.~\eqref{eq: dN}, i.e., that only particles with masses smaller than the BH temperature (associated to its mass M) can be emitted. Nevertheless, this only applies to fundamental particles \cite{Macgibbon1990QuarkI, Macgibbon1990QuarkII} and in the case of quarks this definition depends on whether the BH temperature is above or below the QCD confinement scale $T_{\rm QCD}\sim 300$ MeV \cite{Stocker2018Exotic} (see Sec. \ref{sec: neu}). In fact, while above that threshold quarks can be radiated directly, below it the quarks can only be emitted in form of pions (since these are the only available mesons at these scales). This suppresses the naive estimate for the production of quarks below $T_{\rm QCD}$ (corresponding to roughly $3\times10^{13}$ g) to instead enhance the pion production by the same amount. To take this into account, \cite{Stocker2018Exotic} suggested the introduction of a tanh-like weighting factor
\begin{align}\label{eq: Q_QCD}
	Q_{\rm QCD}=\left[1+\exp\left(\pm\frac{\log_{10}(T/T_{\rm QCD})}{\sigma}\right)\right]^{-1}\,,
\end{align}
where its form with the plus applies to pions and that with the minus to gluons and quarks, and $\sigma$ (arbitrarily fixed to $\sigma=0.1$) sets the width of the cut-off (see Figs. 4 and 5 of \cite{Stocker2018Exotic} for a visual representation of the impact of $Q_{\rm QCD}$).

\begin{figure}
	\centering
	\includegraphics[width=0.63\textwidth]{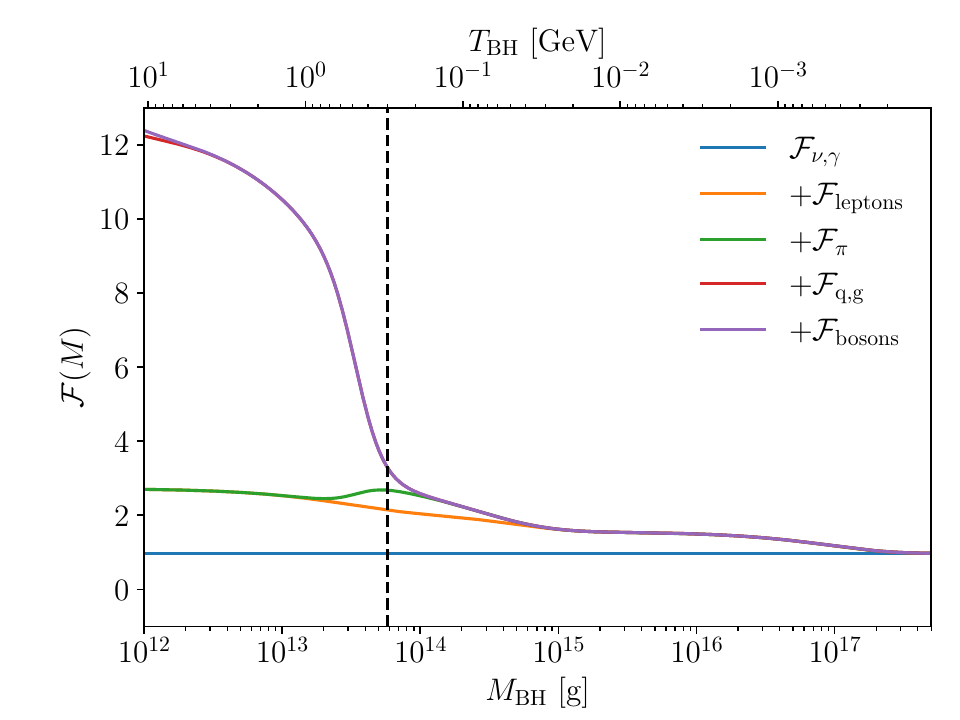}
	\caption{Effective number of emitted species due to BH evaporation. Figure adapted from [\protect\hyperlink{I}{I}].}
	\label{fig: PBH_f_m}
\end{figure} 

The final shape of $\mathcal{F}(M)$ is graphically reported in Fig. \ref{fig: PBH_f_m}, adapted from [\hyperlink{I}{I}], highlighting the different contributions of the various particles. For very large masses only photons and neutrinos are emitted, while in the approximate mass range between $10^{15}-10^{17}$ g the emission of leptons (in particular electrons) becomes as important. While pions play only a minor role between $10^{15}$ g and the QCD threshold, above the latter the production of quarks significantly dominates over all other particles and explains the raise of $\mathcal{F}(M)$. \matteo{To this last point, we note that, because of their confinement and the lack of direct knowledge of their nature at very high energies, the discussion surrounding the emission of quarks would likely be more complex than in the picture portrayed so far. Leaving a dedicated discussion for future work, this limitation needs to be kept in mind when discussing the results obtained for masses lower than about $5\times10^{13}$~g.}

Once the mass loss rate is known, also the lifetime (or evaporation time) of the BH can be computed by integration over time to find \cite{Macgibbon1990QuarkII}
\begin{equation}\label{eq: t_ev}
	t_{\rm ev}=6.24\times 10^{-27} \text{ s } \frac{M^3}{\mathcal{F}(M)} \simeq t_{\rm uni} \left(\frac{M}{5.2\times10^{14}~\text{g}}\right)^3 \left(\frac{2}{\mathcal{F}(M)}\right)\,,
\end{equation}
where $t_{\rm uni}=13.8~\text{Gyr}$ is the age of the universe. This equation implies on the one hand that the more massive a BH is the longer it takes for it to evaporate. This is consistent with the idea that a very massive BH would have a lower evaporation temperature (and efficiency). On the other hand, it also suggests that should a BH have a primordial origin, it would have been evaporated by now ($t_{\rm ev}<t_{\rm uni}$) if its mass was lower than $5\times10^{14}$ g.\footnote{As a note, for PBHs so massive that their lifetime is much larger than the age of the universe the mass loss rate quickly becomes negligible and the PBH mass can be considered constant \cite{Poulin2017Cosmological}.} This consideration already places a strong lower bound on the possibility for PBHs to be the DM.

Nevertheless, the relevance of PBHs goes beyond their ability to explain the DM abundance, as discussed for instance in the previous section in the context of the implications that their detection would have for inflation. Therefore, it is in general interesting to consider PBHs over the full mass spectrum and study how the impact of their evaporation would affect cosmological probes such as BBN, the CMB and the 21 cm signal, among others. To do so, one first of all needs to define the respective energy injection and hence the luminosity of the evaporation. Based on the discussion above, this simply reduces to
\begin{align}\label{eq: L_PBH}
	L_\chi = -\frac{\text{d}M}{\text{d}t}\,,
\end{align}
using the same notation as in Eq. \eqref{eq: E_inj}.

\begin{figure}
	\centering
	\includegraphics[width=0.63\textwidth]{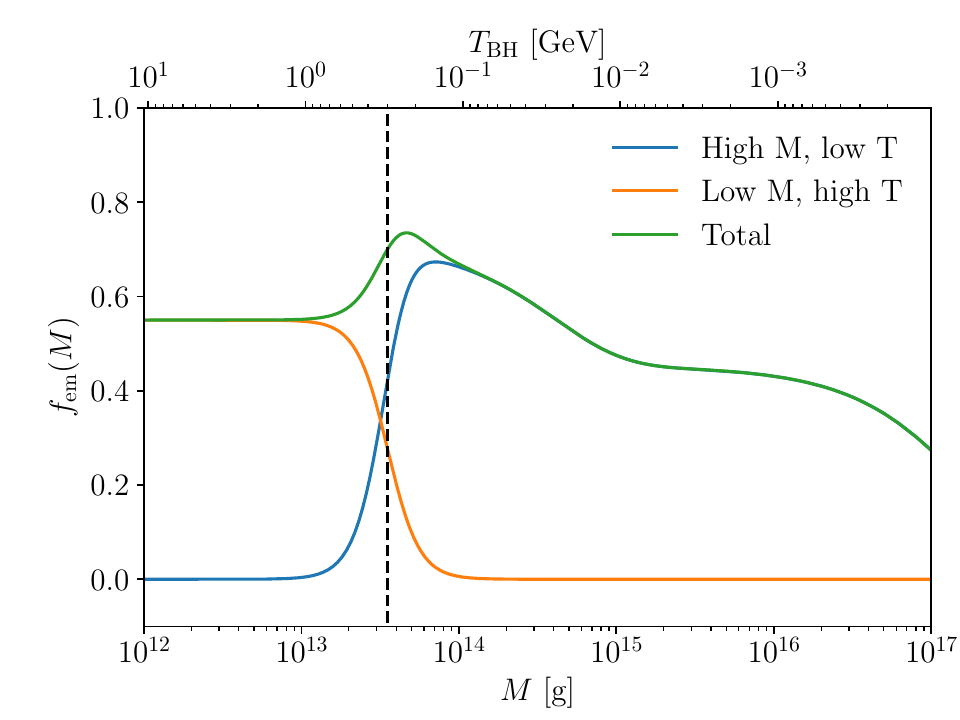}
	\caption{Fractional amount of the total emitted energy resulting in EM radiation following the evaporation of a BH. Figure taken from [\protect\hyperlink{I}{I}].}
	\label{fig: PBH_f_em_M}
\end{figure}

Differently than for the cases discussed in Sec.~\ref{sec: res_DM}, however, when describing the energy deposition from PBH evaporation one cannot just assume a constant value of the (fractional) amount of EM radiation $f_{\rm em}$ since the number and type of emitted particles depend on time. A first-order approximation of $f_{\rm em}$ was attempted in [\protect\hyperlink{I}{I}] (see App. C2 therein), where it was noted that for masses below the QCD threshold one can simply remove the neutrino component of $\mathcal{F}$, $\mathcal{F}_\nu$, from the total to find
\begin{align}
	f_{\rm em}^{\rm high\,M}=\frac{\mathcal{F}-\mathcal{F}_\nu}{\mathcal{F}}\,.
\end{align}
Although this implicitly assumes that all of the energy emitted in form of $\pi$, $\mu$ and $\tau$ particles eventually becomes EM radiation, thereby overestimating the value of $f_{\rm em}$, the corrections are expected to be small and in any case only relevant in the mass range between $10^{14}-10^{15}$ g. Moreover, for BH masses above the QCD limit [\protect\hyperlink{I}{I}] pointed out on the basis of the results of \cite{Macgibbon1990QuarkI} (see in particular Tab. III therein) that a relatively constant fraction of the emitted power is carried away by neutrinos, namely 45\%, up to temperatures as high as 100 GeV (corresponding to masses of the order of $10^{11}$ g). In this \linebreak \noindent regime, one can therefore express $f_{\rm em}$ as
\begin{align}
	f_{\rm em}^{\rm low\,M}=1-0.45\,.
\end{align}
Using Eq. \eqref{eq: Q_QCD} (with the minus) to smooth the transition between the two approximations, one obtains \vspace{-0.1 cm}
\begin{align}\label{eq: f_em_PBH}
	f_{\rm em} = 0.55 \, Q_{\rm QCD} + (1-Q_{\rm QCD}) \frac{\mathcal{F}-\mathcal{F}_\nu}{\mathcal{F}}\,,
\end{align}
which is graphically shown in Fig. \ref{fig: PBH_f_em_M}, taken from [\hyperlink{I}{I}].

\vspace{0.3 cm}
\noindent \textbf{Follow-up idea 13:} Analogously to the work performed in e.g., [\hyperlink{IX}{IX}] in the context of decaying DM into neutrinos, also this expression for $f_{\rm em}$ would need to be updated and extended to an even larger mass range \matteo{(also taking into account a better description of the quark emission as mentioned above)}.
\vspace{0.3 cm}

\begin{figure}
	\centering
	\includegraphics[width=0.48\textwidth]{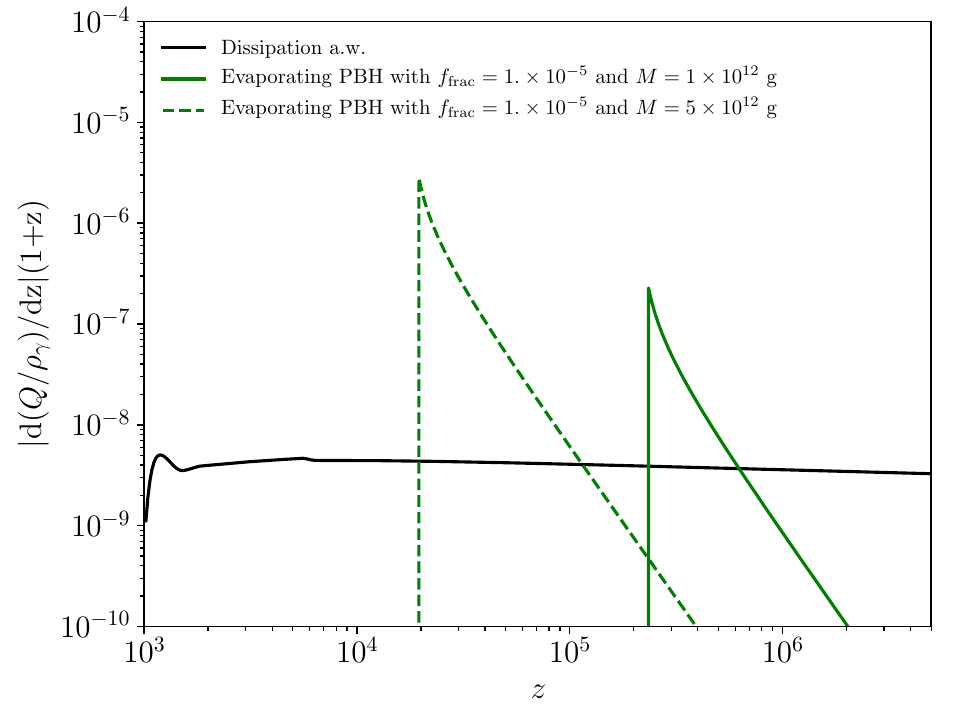}
	\includegraphics[width=0.48\textwidth]{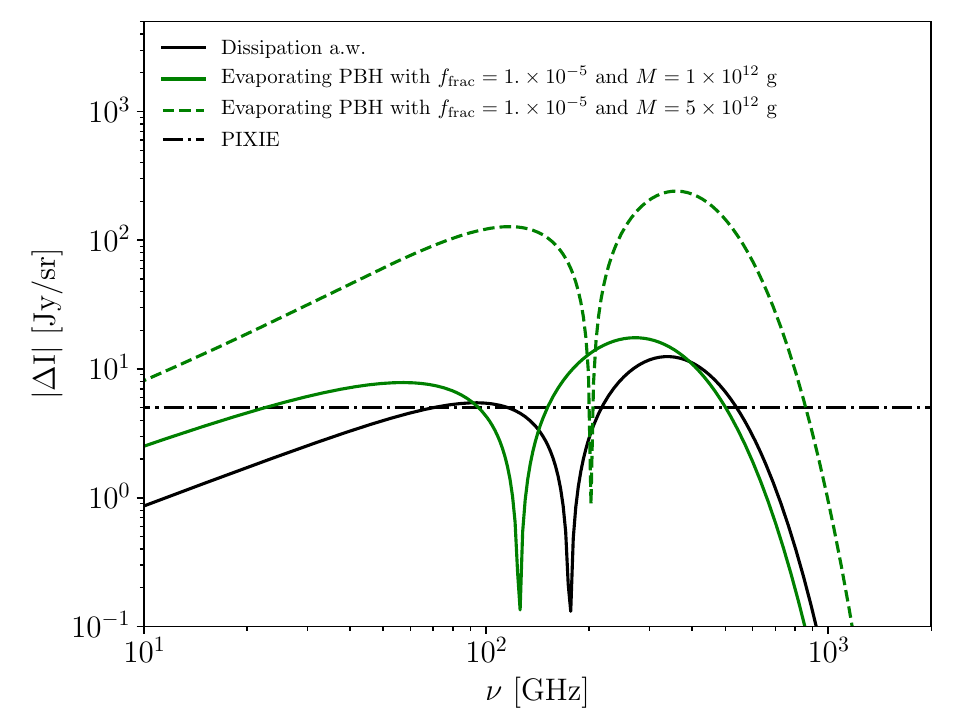}
	\caption{Same as in Fig.~\ref{fig: DM_heating_ann_dec}, but for the case of PBH evaporation. Figures taken from~[\protect\hyperlink{I}{I}].}
	\label{fig: PBH_heating_eva}
\end{figure}

By using the energy injection following from Eq. \eqref{eq: L_PBH} and the form of $f_{\rm em}$ introduced in Eq. \eqref{eq: f_em_PBH}, one can then calculate, among others, the heating rate in the pre-recombination era due to the evaporation of PBHs. Two representative examples thereof are shown in the left panel of Fig. \ref{fig: PBH_heating_eva}, taken from [\hyperlink{I}{I}]. From the figure, one can already draw the interesting conclusion that the majority of the heating rate due to PBH evaporation happens close to the evaporation time $t_{\rm ev}$ with a very spiked final energy injection. This means that the various considered observational strategies will be very sensitive to the mass dependence of the injection and therefore their complementarity becomes fundamental. In this sense, the discussion here is very similar to that carried out in the context of decaying DM (see Fig.~\ref{fig: DM_heating_ann_dec}). For completeness, the corresponding SD signals are shown in the right panel of the figure.

\begin{figure}[t]
	\centering
	\includegraphics[width=0.75\textwidth]{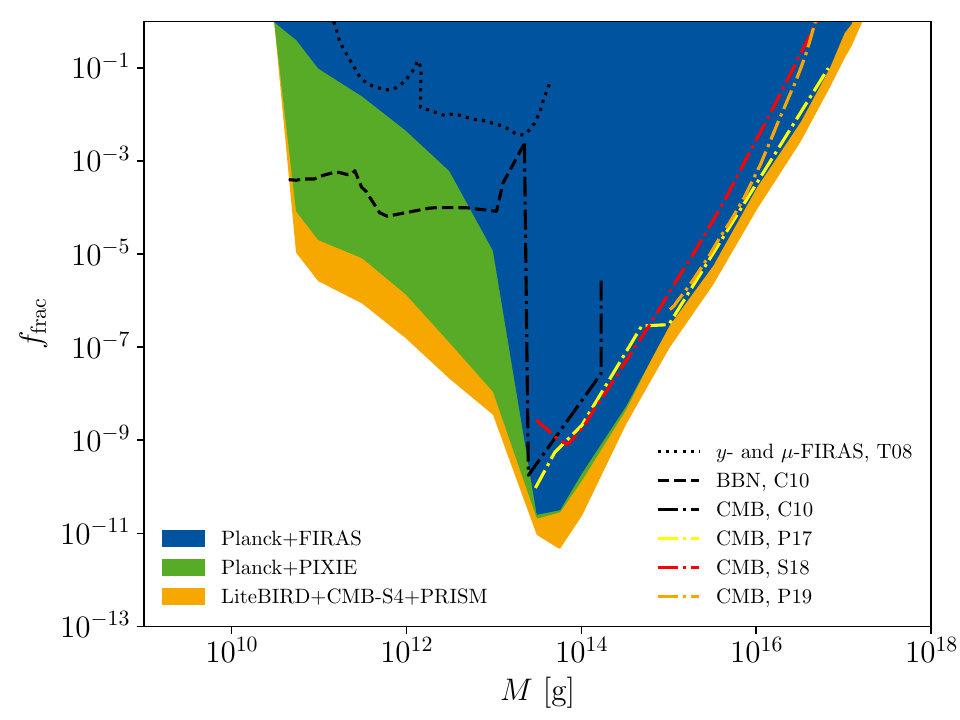}
	\caption{Exclusion contours on the fractional abundance of PBHs as a function of their mass. Figure taken from [\protect\hyperlink{I}{I}].}
	\label{fig: PBH_exclusion_region}
\end{figure}

Putting all the pieces together, one obtains the CMB and BBN constraints shown in Fig.~\ref{fig: PBH_exclusion_region}, taken from [\protect\hyperlink{I}{I}] (see e.g., \cite{Carr2010New, Stocker2018Exotic, Acharya2019CMB, Tashiro2008Constraints, Poulin2017Cosmological, Acharya:2019xla, Poulter2019CMB, Chluba2020Thermalization} for similar results).\footnote{The constraints for $M_{\rm PBH}>10^{15}$ g (i.e., for PBHs that have not already evaporated today) correspond to those labeled ``Evaporation'' in Fig. \ref{fig: PBH_overview}.} As above, in this case the same conclusions expressed for the case of decaying DM apply also to the evaporation of PBHs. One particular aspect to underline is the fact that the combination of CMB anisotropy and SD missions is able to tightly constrain about seven decades in PBH mass, with SD forecasted to eventually place the strongest bounds in the mass range between $3\times10^{10}-3\times10^{13}$~g. 

\vspace{0.3 cm}
\noindent\textbf{Quasi-extremal PBHs}

\noindent So far we have assumed the PBHs to be neutral and non-rotating. Based on the standard formation scenario discussed above, this is well justified given that the collapsing patch of the universe can be expected to have a zero net-charge and since matter would fall in radially also no angular momentum should build up. Moreover, even if the PBHs had a spin or a charge to begin with, they would be radiated away very efficiently \cite{carter1974charge, page1976particle}. However, alternative scenarios might lead to the formation of a non-vanishing charge $Q$ or spin parameter $a$, which is related to the angular momentum $J$ as $a=J/M$ (see e.g.,~[\hyperlink{XIII}{XIII}] for an overview).
PBHs with large values of these parameters are called quasi-extremal, and simply extremal if  $a/M=1$ or $Q/M=1$, i.e., their maximum allowed value (although arguments -- the so-called cosmic censorship hypotheses -- have been made that forbid BHs to become extremal \cite{penrose1999question}).

\vspace{0.3 cm}
\noindent \textbf{Follow-up idea 14:} Although [\hyperlink{XIII}{XIII}] does suggest some scenarios to justify the existence of quasi-extremal PBHs (qPBHs), the discussion \matteo{carried out} there is intentionally kept very general \matteo{and hypothetical, more focused on the cosmological implications of the possible existence of qPBHs and} meant to serve as a guide for future \matteo{dedicated} studies \matteo{in this direction}. A rigorous investigation of how to produce qPBHs in practice \matteo{(and whether it would be possible at all)} would be a natural follow-up.
\vspace{0.3 cm}

As shown in [\hyperlink{XIII}{XIII}], the quasi-extremality of a PBH can best be captured by the quasi-extremality parameter $\varepsilon$, which can be related to $Q$ and $a$ via 
\begin{align}
	\varepsilon^2=1-\frac{Q^2}{M^2}\quad\text{and}\quad\varepsilon^2=1-\frac{a^2}{M^2}\,,
\end{align}
but can also include the behavior of higher-dimensional PBHs (see App. A of the [\hyperlink{XIII}{XIII}]). When $\varepsilon=1$ one recovers the Schwarzschild case (i.e., the discussion above), when $\varepsilon\ll1$ the BH is quasi-extremal and when $\varepsilon=0$ it is extremal. In fact, as it turns out, the evaporation temperature of a qPBH becomes very generally (i.e., independently of the source of quasi-extremality)
\begin{equation}\label{eq: T_BH_2}
	T=\frac{1}{8 \pi M}\frac{2^2\, \varepsilon}{(1+\varepsilon)^2}\,,
\end{equation}
from which one recovers Eq. \eqref{eq: T_BH} in the limit $\varepsilon=1$, as expected. This result immediately shows that a qPBH evaporates at a lower temperature with respect to a PBH with the same mass.

This conclusion is reflected in the definition of the luminosity, which becomes 
\begin{equation}\label{eq: luminosity 2}
	L _\chi= -\frac{\text{d}M}{\text{d}t}\frac{2^6\, \varepsilon^4}{(1+\varepsilon)^6}\,,
\end{equation}
where in a first approximation the mass loss rate remains unchanged with respect to Eq.~\eqref{eq: f_M_PBH}, although now the mass needs to be corrected according to Eq. \eqref{eq: T_BH_2}.\footnote{Intuitively, at first order the amount and type of emitted particles should only depend on the temperature of the BH, regardless of the characteristics of the BH reaching that temperature. It therefore makes sense for Eq.~\eqref{eq: f_M_PBH} to also apply to qPBHs.} Because of the suppressed evaporation, also the lifetime of the PBHs changes when $\varepsilon\neq1$, and in particular it increases according to
\begin{equation}\label{eq: t_ev_2}
	t_{\rm ev}=6.24\times 10^{-27}\text{ s } \frac{M^3}{\mathcal{F}(T)}\frac{(1+\varepsilon)^6}{2^6\,\varepsilon^4}\,.
\end{equation}
Eqs. \eqref{eq: luminosity 2} and \eqref{eq: t_ev_2} have two important consequences for the energy injection. On the one hand, the former suppresses it. This can be seen explicitly by inserting Eq. \eqref{eq: luminosity 2} in Eq. \eqref{eq: E_inj} and Eq. \eqref{eq: E_inj} in Eq. \eqref{eq: Q_dep_1}, leading to
\begin{align}\label{eq: E_prop_PBH}
	\dot{Q}_{\text{dep}} \propto f_{\chi} \frac{\mathcal{F}(M)\,2^6\,\varepsilon^4}{M^3\,(1+\varepsilon)^6}\,.
\end{align}
On the other hand, the latter delays it. This implies that for a qPBH and a PBH to evaporate at the same time, one needs to assume a lower initial mass for the former. In terms of the constraints on the evaporation, this results in a vertical (towards less stringent $f_\chi$ values) and an horizontal (towards lower PBH masses) offset, respectively, depending on the value of $\varepsilon$. 

\begin{figure}
	\centering
	\includegraphics[width=0.48\textwidth]{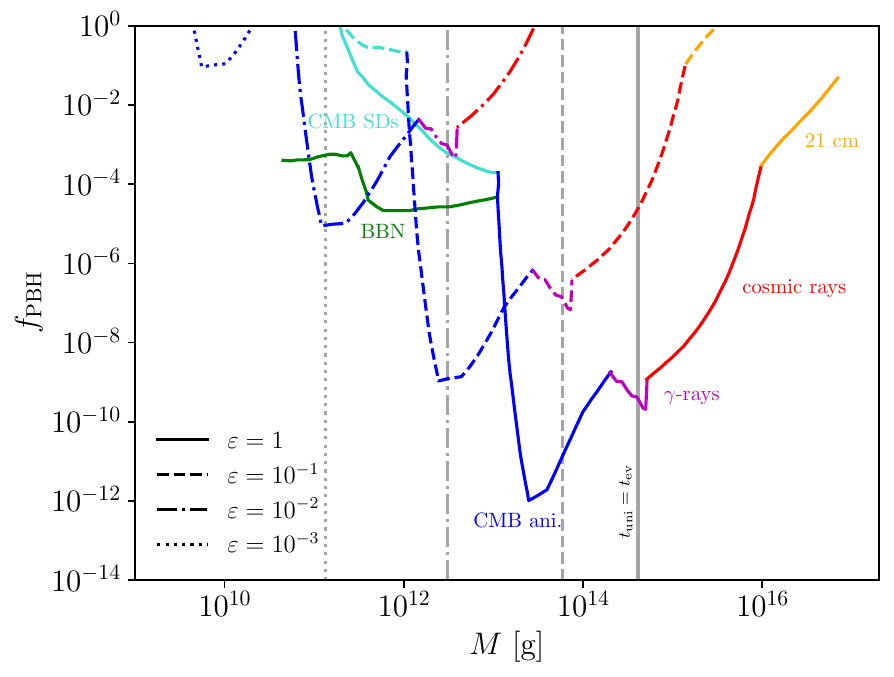}
	\includegraphics[width=0.48\textwidth]{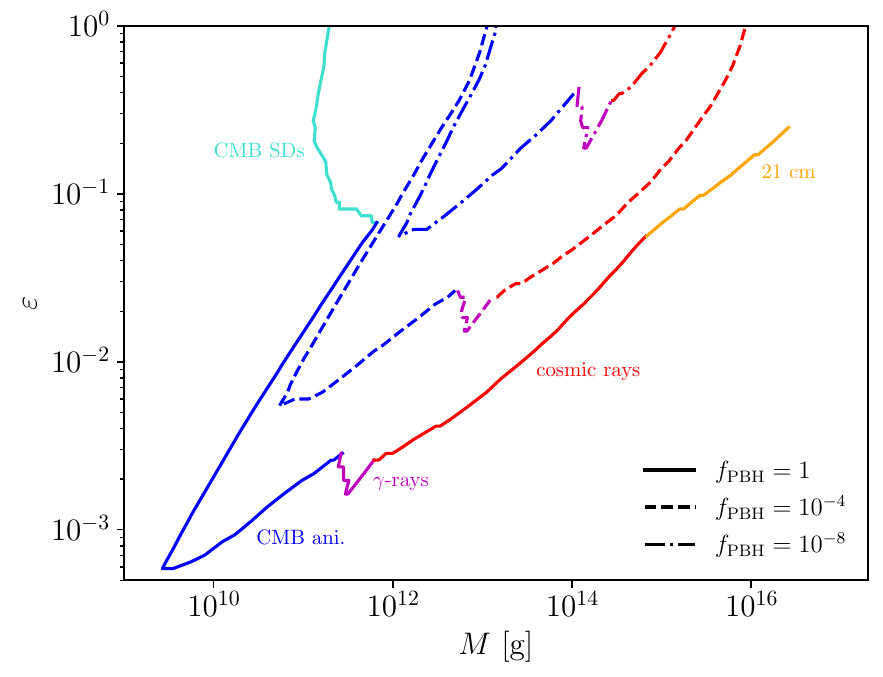}
	\caption{Constraints on the PBH abundance for different values of the  degree of quasi-extremality (left) and on the degree of quasi-extremality for different PBH abundances (right) as a function of the PBH mass. Figures taken from~[\protect\hyperlink{XIII}{XIII}].}
	\label{fig: PBH_bounds_qPBH}
\end{figure}

The left panel of Fig. \ref{fig: PBH_bounds_qPBH}, taken from [\protect\hyperlink{XIII}{XIII}], shows this behavior for several values of~$\varepsilon$. For reference, in the figure the gray vertical lines represent the PBH masses that would lead to $t_{\rm ev}=t_{\rm uni}$. As evident from the figure, for $\varepsilon\lesssim10^{-3}$ the constraints on the PBH abundance as a consequence of their evaporation completely vanish. This, combined with the fact that for the same values of $\varepsilon$ qPBHs as light as $10^{11}$ g (and below) would survive until today, completely reopens this section of parameter space for PBHs to be the DM.

In turn, one could also consider a model-building perspective where $\varepsilon$ is fixed by the formation mechanism and recast the constraints from the $f_\chi-M$ plane to the $\varepsilon-M$ plane for fixed values of $f_\chi$ (labeled $f_{\rm PBH}$ in the figure). The resulting bounds are shown in the right panel of Fig. \ref{fig: PBH_bounds_qPBH}, taken from [\protect\hyperlink{XIII}{XIII}]. Assuming one wanted to explain the totality of the DM abundance with qPBHs, one would need to justify values of $\varepsilon\lesssim10^{-3}$. The lower the expected abundance the looser the limit on the quasi-extremality parameter.

A particularly interesting, and currently open, question is nevertheless how to determine the degree of quasi-extremality of a qPBH, should we ever have evidence of its evaporation. In fact, if, say, a future CMB SD mission detected a signal compatible with PBH evaporation, the reconstruction of $\varepsilon$ and $M$ would be limited by the degeneracy between the two parameters given in Eq. \eqref{eq: E_prop_PBH}. Breaking this degeneracy would require complementary information from e.g., gravitational direct detection \cite{Carney:2019pza} or other direct detection \cite{Lehmann:2019zgt} techniques. Future GW missions have also been forecasted to be sensitive to the spin parameter to a very accurate degree \cite{Burke:2020vvk}.

This does not diminish, however, the important role that SDs could play in the context of qPBHs. First of all, the large discovery space that they could cover with future missions is unmatched by any other prospect (see Fig. \ref{fig: PBH_exclusion_region}), with the exception of 21 cm observations which are just as promising. Moreover, if a future SD mission did observe a signal, because of the limited mass range that could have generated it, one would already be able to set strong lower bounds on $\varepsilon$.

\vspace{0.3 cm}
\noindent \textbf{Follow-up idea 15:} Based on the foundation set by [\hyperlink{XIII}{XIII}], it would be very interesting to systematically investigate what would it take to observationally determine the value of $\varepsilon$ at a statistically significant level. 
\vspace{0.3 cm}

\noindent\textbf{Summary}

\noindent In summary, CMB SDs probe the evaporation of PBHs in the mass range between $10^{10}-10^{13}$~g and could deliver the by far strongest constraints in this region of parameter space with future measurements. Furthermore, should PBHs be quasi-extremal (i.e., have a charge, spin or live in a higher dimensional space, among others), SDs would also offer a very promising perspective for their discovery given the large discovery space they could cover in the future.

In the context of this section, the work carried out in the course of the thesis has been dedicated to the implementation of PBH evaporation in \texttt{CLASS} [\hyperlink{I}{I}] (see Sec. \ref{sec: num}). This allowed to derive state-of-the-art constraints on this process exploiting the combination of (current and future) CMB anisotropy and SD data. The development of these bounds has then enabled to explore their dependence on the assumption of quasi-extremality~[\hyperlink{XIII}{XIII}].

\subsubsection{PBH accretion}\label{sec: res_PBH_acc}
Another way via which PBHs could act as source of energy injection would be the accretion of matter in their surroundings. In a simplified manner, one could approximate the problem to a static sphere of radius $r_B$ that accretes with a cross section $\sigma=4\pi r_B^2$ all particles (with density $\rho_\infty$) moving towards it at a speed $v_\infty$, giving a flux of matter $\phi=\rho_\infty v_\infty$. The resulting accretion rate is the so-called Bondi accretion rate
\begin{align}
	\dot{M}=4\pi\rho_\infty v_\infty \lambda r_B^2 = 4\pi\rho_\infty \lambda \frac{M^2}{ v_\infty^3} \,.
\end{align}
where $\rho_\infty$ and $v_\infty$ are the density and velocity of the medium far away from the PBH, $\lambda$ is the dimensionless accretion rate (which accounts for eventual additional complexities of the problem, such as proper BH velocity, viscosity, outflows and more) and $r_B=M/v_\infty^2$ is the radius of the accretion sphere, also known as Bondi radius, which we used in the second equality. In a cosmological context one can usually assume $v_\infty=c_{s,\infty}$, where $c_{s,\infty}=\sqrt{P_\infty/\rho_\infty}$ is the sound speed far away from the accretion region. 

Once matter starts to fall in the accretion sphere, it compresses adiabatically (at least in some regimes, see below) and eventually emits radiation. The related luminosity (in analogy to the case of PBH evaporation and following again the notation of Eq. \eqref{eq: E_inj}) is
\begin{align}\label{eq: L_PBH_acc}
	L_\chi = \epsilon \dot{M}\,,
\end{align}
where $\epsilon$ is the emission efficiency which encapsulates the model dependence of the emission process (including the aspect of adiabaticity).

Given the physical complexity of the accretion process, its modeling is plagued by enormous uncertainties, which in turn translate into a wide range of possible values (and functional forms) of $\lambda$ and $\epsilon$. For instance, as reviewed in [\hyperlink{XI}{XI}], one of the most important unknowns is the geometry of the accretion. The two most popular proposals in this case involve spherical \matteo{\cite{Ricotti2007Effect, AliHaimoud2017Cosmic}} and disk \cite{Poulin2017CMB} accretion. Intuitively, while in the former case the accretion efficiency is larger than in the former case because of the much larger accretion surface, for the same reason the opposite is true for the emission efficiency (where the larger surface of the spherical accretion case impedes more strongly the emission of the radiation). As a result one has $\lambda\simeq 0.1-1$ and $\epsilon\simeq 10^{-11}-10^{-3}$ for the spherical accretion case (depending on the PBH mass), while for the disk accretion case one has $\lambda\simeq 0.01$ and $\epsilon\simeq 10^{-4}-10^{-1}$ (see e.g., Fig. 1 of [\hyperlink{XII}{XII}] and Fig. 2 of [\hyperlink{XI}{XI}] for a complete graphical comparison). Clearly, this already introduces up to five orders of magnitude uncertainty in the definition of the luminosity $L_\chi\propto \epsilon\lambda^2$. A similar discussion (with a similar impact on the luminosity) also applies to the ionization model (that follows the matter compression and leads to the radiation emission) \cite{AliHaimoud2017Cosmic}, the impact of structure formation \cite{Mack:2006gz, Inman:2019wvr, Adamek:2019gns, Serpico:2020ehh, DeLuca:2020fpg}, the potential presence of outflows \cite[\hyperlink{XI}{XI}]{Bosch-Ramon:2020pcz, Bosch-Ramon:2022eiy} and of ionization fronts that can build up ahead of the PBH along its direction of motion \cite[\hyperlink{XII}{XII}]{Park:2012cr, Sugimura:2020rdw, Scarcella:2020ssk} (see e.g., [\hyperlink{XI}{XI}] for a complete review).

\begin{figure}
	\centering
	\includegraphics[width=0.7\textwidth]{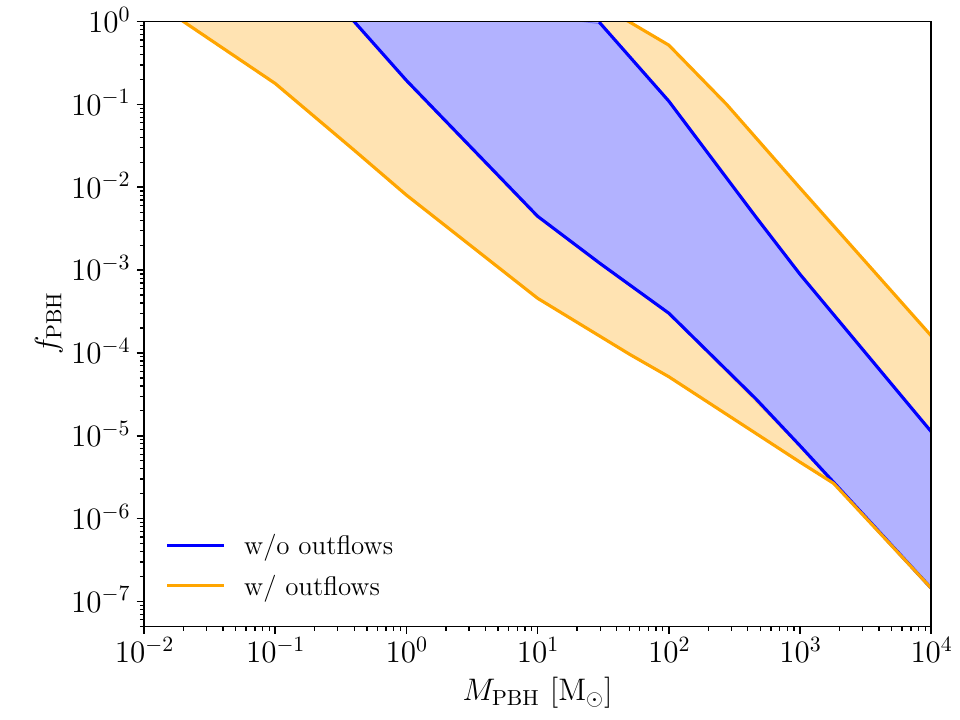}
	\caption{Uncertainty of the upper limit on the PBH abundance derived from CMB anisotropy constraints on PBH accretion. \matteo{The shaded area does \textit{not} represent the excluded region of parameter space, but rather where the upper bounds might lay depending on the assumed accretion model.} Figure taken from [\protect\hyperlink{XI}{XI}].}
	\label{fig: PBH_bounds_acc}
\end{figure}

This plethora of open questions is reflected in the possible range of cosmological constraints that can be imposed on the process. Fig. \ref{fig: PBH_bounds_acc}, taken from [\protect\hyperlink{XI}{XI}], focuses on the CMB anisotropy constraints derived from Planck data. In the figure, the shaded area does \matteo{\textit{not}} represent the excluded region of parameter space, but that where the upper bounds on the PBH abundance might lay depending on the assumed model.\footnote{\matteo{For a given choice of model, the bounds exclude the upper region of parameter space.}} Furthermore, the reference only accounted for geometry, ionization model and outflows, so the size of the actual uncertainty is even underestimated. As clear from the figure, the upper bounds on the PBH mass for $f_\chi=1$ ($f_{\rm PBH}$ in the figure) can vary by almost 4 orders of magnitude depending on the chosen modeling of the accretion. The potentially preferred LVK region lays exactly in the middle of the uncertainty band, meaning that CMB constraints on PBH accretion could potentially test the GW preference but only once the modeling uncertainties will be under control. The situation is inevitably similar also for the other cosmological probes constraining the same process, such as the 21 cm signal \cite{Mena:2019nhm}.

\begin{figure}
	\centering
	\includegraphics[width=0.62\textwidth]{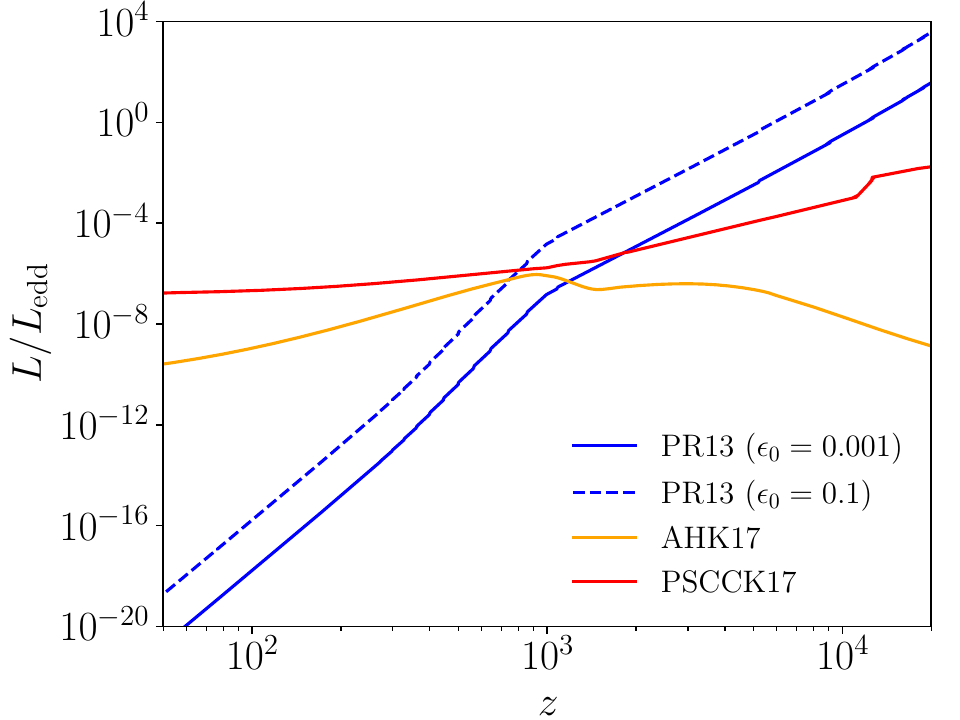}
	\caption{Comparison of the luminosity (normalized to the Eddington luminosity) for three different accretion models, namely with ionization fronts \cite[\protect\hyperlink{XII}{XII}]{Park:2012cr} (PR13), spherical accretion~\cite{AliHaimoud2017Cosmic} (AHK17) and disk accretion~\cite{Poulin2017CMB} (PSCCK17).~Figure taken from~[\protect\hyperlink{XII}{XII}].}
	\label{fig: PBH_L_over_Ledd}
\end{figure}

On the basis of the results obtained in \cite{AliHaimoud2017Cosmic} in the context of spherical accretion, it was argued that the production of CMB SDs can be neglected. This is indeed the case (as we confirm with the numerical setup discussed in Sec. \ref{sec: num} obtaining $y\simeq10^{-8}\,f_\chi$ and $\mu\simeq10^{-11}\,f_\chi$ for $M=10^4$ g -- the maximum mass where the numerical set up developed in the reference can be applied), but the same conclusion might not apply to other scenarios. In fact, as shown in Fig. \ref{fig: PBH_L_over_Ledd}, taken from [\protect\hyperlink{XII}{XII}], while the spherical accretion model (orange curve) predicts a relative flat luminosity, the disk accretion scenario (red curve) and the inclusion of ionization fronts (blue curve) might significantly increase the emission in the pre-recombination era.\footnote{In the latter case one should take the results of~[\protect\hyperlink{XII}{XII}] with a grain of salt in this regime because the model intrinsically relies on the assumption that the medium surrounding the PBH is neutral \cite{Park:2012cr}, which is not the case before recombination. Furthermore, as pointed out in the reference, at $z>5\times10^4$ the model brakes down, so that the calculation of the $\mu$ signal becomes even more unreliable. Yet, as shown in Fig. 1 of the reference the model naturally approximates the Bondi-Hoyle-Lyttleton limit for $z>10^3$ (see Eq. (3) of [\protect\hyperlink{XII}{XII}]), which does not suffer from the same limitations. Therefore, the qualitative conclusion made here remains the same, although the quantitative results most probably would not. \label{foot: PR13}} 

As a consequence, in the disk accretion scenario we find $y\simeq\mu\simeq 10^{-6}f_\chi$ for $M=10^4$ g  which might be within the reach of future SD missions (although not competitive with current CMB anisotropy measurements even with Voyage 2050-like sensitivities). Assuming the results of [\protect\hyperlink{XII}{XII}] hold in the relevant regimes (which is not the case, see however footnote \ref{foot: PR13}), or more in general an even stronger redshift dependence of the luminosity, might deliver exceptionally large SDs, in this particular case of the order of $y\simeq5\times 10^{4}\,f_\chi$ and $\mu\simeq5\times10^{8}\,f_\chi$ for $M=10^4$ g.\footnote{In the actual Bondi-Hoyle-Lyttleton limit these values would significantly reduce, to $y\simeq5\times 10^{-4}\,f_\chi$ and $\mu\simeq5\times10^{-2}\,f_\chi$. Nevertheless, although suppressed, these would still lead to constraints of the order of $f_\chi<10^{-3}$ for FIRAS and $f_\chi<10^{-6}$ for Voyage 2050 -- competitive with the most stringent bounds shown in Fig.~\ref{fig: PBH_bounds_acc}.}

Refraining from giving too much weight to these last values, the aforementioned argument mostly serves to point out that CMB SDs could constrain the emission following the accretion of matter onto PBHs and when they do the constraints can be very competitive.\footnote{The same discussion had been carried out in the context of DM annihilation, where SDs are most valuable when a redshift dependence of the emission is involved.} \matteo{Although this was already pointed out in \cite{Ricotti2007Effect}, the following updated calculations of \cite{AliHaimoud2017Cosmic} later suggested that the role of SDs could be neglected. The discussion carried out here re-opens this possibility.}

\vspace{0.3 cm}
\noindent\textbf{Follow-up idea 16:} This finding needs to be analyzed in greater detail, although its reliability might depend on the further development of the accretion model discussed in \cite[\protect\hyperlink{XII}{XII}]{Park:2012cr} in the pre-recombination era. Nevertheless, should the aforementioned predictions for $y$ and $\mu$ in this setup even remotely resemble reality (i.e., only vary by less than $6-7$ orders of magnitude), even FIRAS would completely exclude the LVK preference for PBHs. 
\vspace{0.3 cm}

In summary, CMB SDs can probe the accretion of matter onto PBHs. Contrary to previous findings, we argued here that this would indeed be the case should the redshift dependence of the accretion luminosity be enhanced in the pre-recombination era. Although the estimated constraining power of SDs seems promising, this preliminary result will have to be investigated in greater detail in future work.

\subsection{Inhomogeneous energy injections and recombination}\label{sec: res_PMFs}

One common aspect shared by many of the sources of CMB SDs discussed so far (both in Secs.~\ref{sec: app} and \ref{sec: SD_non_lcdm}) is the fact that they might lead to either inhomogeneous energy injections or to an anisotropic evolution of the recombination process. Below we dive into both of these possibilities, noting that while the latter has been treated in the course of the thesis, the former is mentioned for completeness.

\vspace{0.3 cm}
\noindent \textbf{Inhomogeneous energy injections}

\noindent When discussing the scenarios presented in Secs. \ref{sec: res_DM} and \ref{sec: res_PBH} we have implicitly assumed that both the heating (from e.g., energy injections or scatterings) and the thermalization/Comptonization processes happen homogeneously in space (see Sec.~\ref{sec: ped} for a definition of these concepts). Nevertheless, because of the spatial matter and temperature fluctuations that populate the universe (see Sec.~\ref{sec: CMB}) this is not necessarily accurate. For instance, if the energy density of a massive relic decaying in the early universe followed the same fluctuations of the DM energy density, it would induce an anisotropic energy injection. In turn, the efficiency of the thermalization of these energy injections will depend on the baryon and photon energy densities, which are themselves subject to fluctuations. Accounting for the inhomogeneity of these processes would lead to the production of an anisotropic SD signal \cite{Chluba:2022xsd, Chluba:2022efq,Kite:2022eye}. While its monopole (i.e., sky average) would correspond to the SD spectrum discussed in the previous sections, the higher multipoles would carry additional information on the perturbed universe as well as on the time dependence of the heating and thermalization process.

\begin{figure}[t!]
	\centering
	\includegraphics[width=0.75\textwidth]{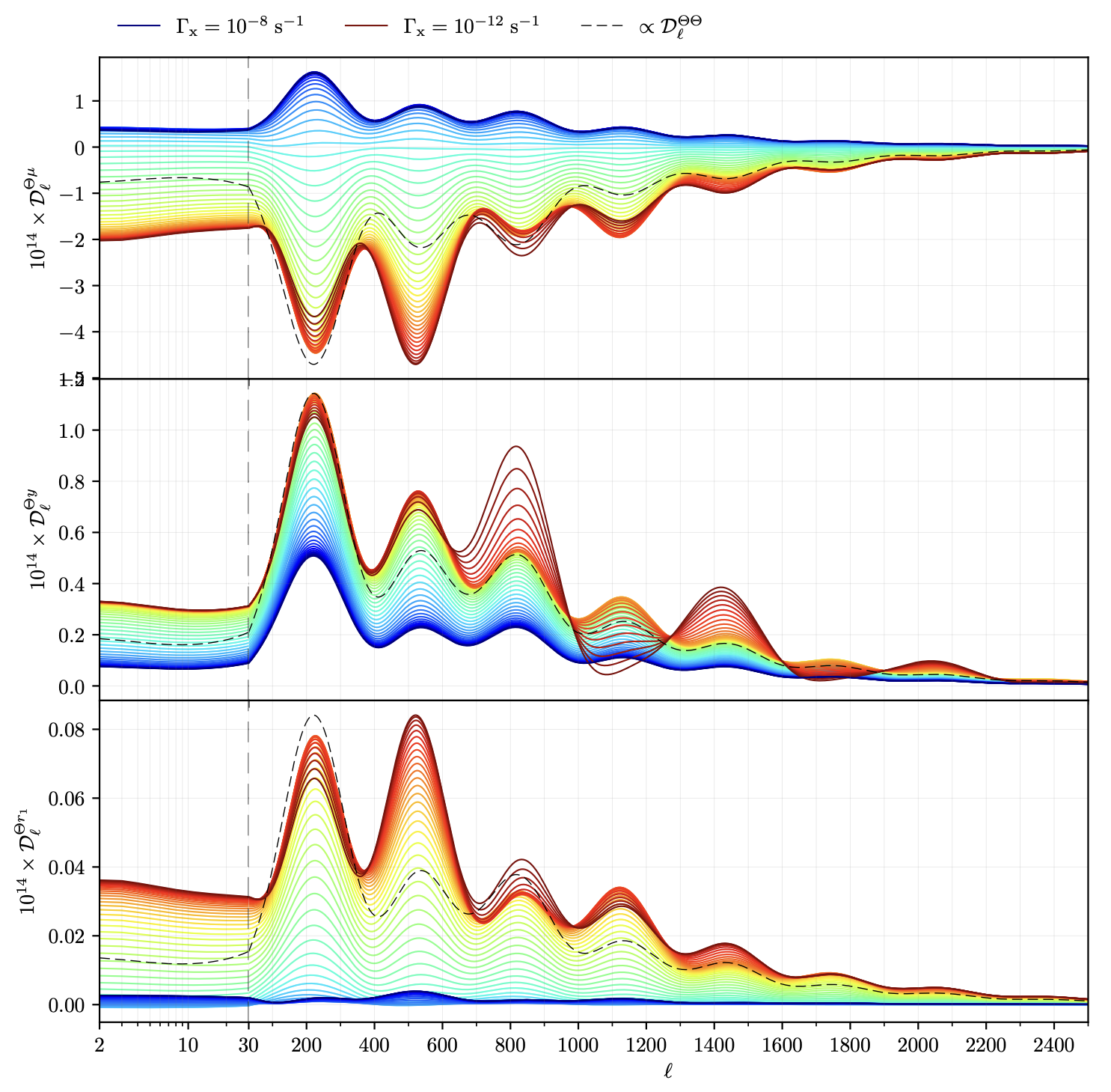}
	\caption{Temperature-SD anisotropy power spectra sourced by the energy injection following the decay of DM at different lifetimes. The color coding reflects the various lifetimes, ranging from $\tau_{\chi}=1/\Gamma_\chi=10^8$ s in blue (happening during the $\mu$ era) to $\tau_{\chi}=10^{12}$ s in red (happening during the $y$ era). The rescaled CMB temperature power spectrum (see Fig.~\ref{fig: CMB_spectra}) is reported as dashed black line for reference. Figure taken from \cite{Kite:2022eye}.}
	\label{fig: inhom_Cl}
\end{figure}

Similarly as for the discussion had in Sec.~\ref{sec: res_infl} in the context of non-Gaussianities, the resulting SD anisotropies can then be correlated with each other or with the CMB temperature and polarization anisotropies (which have been produced by the same energy density fluctuations) to produce the respective power spectra. The representative example of the temperature-SD cross-correlation power spectra\footnote{SD-SD correlations have been shown to be too faint to be observable in a realistic setup.} (from top to bottom assuming $\mu$, $y$ and residual distortions) sourced by the energy injection following the decay of DM\footnote{To a degree, SD anisotropies are expected to exist also within \lcdm (at least the aspect regarding the thermalization/Comptonization process), but in this case the heating rate is too low to produce any sizable signal.} is shown in Fig. \ref{fig: inhom_Cl}, taken from \cite{Kite:2022eye}, for different DM lifetimes. In the figure, the color coding reflects the various lifetimes, ranging from $\tau_{\chi}=1/\Gamma_\chi=10^8$ s in blue (happening during the $\mu$ era) to $\tau_{\chi}=10^{12}$ s in red (happening during the $y$ era). The rescaled CMB temperature power spectrum (see Fig.~\ref{fig: CMB_spectra}) is reported as dashed black line for reference.

\enlargethispage{\baselineskip}

Without dwelling too extensively on the details of these results (see \cite{Kite:2022eye} for more in-depth discussions), several interesting conclusions can be drawn from the figure. First of all, it is clear that the shape and amplitude of the power spectra carry a direct dependence on the injection time. In this sense, including the spatial information allows to go beyond the simpler $y/\mu$ decomposition (see Sec. \ref{sec: ped}), which is largely insensitive to exact epoch at which the injection takes place.\footnote{That is, because the $y$ and $\mu$ parameters are integrated variables. They can be distinguished from each other via the different spectral shapes, but many degenerate heating histories could lead to the same values of the amplitudes. The residuals (see Sec.~\ref{sec: num} and in particular Fig.~\ref{fig: SD_PCA} there) could help in this direction, but their signal would be faint and mostly localized to intermediate redshifts.} Furthermore, different heating processes (such as in the case for e.g., DM annihilation and scattering, PBH evaporation and accretion, and PMFs) would lead to different anisotropic signals and could therefore be distinguished, task that would be much more complicated (if possible at all) with the sole knowledge of the $y/\mu$ parameters. Finally, the overall behavior -- in particular the position of all peaks and the amplitude of the odd ones -- of the temperature-SD (colored lines) and CMB temperature power spectrum (dashed black line) is rather similar, reflecting the fact that the two quantities underwent a common perturbation history. As pointed out in \cite{Kite:2022eye}, the observation of the SD anisotropies would therefore allow to test a largely similar physics, but in a completely independent way.

As done in Sec.~\ref{sec: res_infl} in the context of primordial non-Gaussianities, once the cross-power spectra are known one can then compare them to the (non-)observations by imaging telescopes such as Planck. As a result, Planck would already be able to place competitive constraints with those inferred from FIRAS (see Fig. 30 of \cite{Kite:2022eye}), as derived in Sec.~\ref{sec: res_DM} under the assumption of homogeneity. Up-coming CMB anisotropy missions like LiteBIRD \cite{Matsumura2013Mission, Suzuki2018LiteBIRD} would even be able to improve upon the FIRAS constraints by a factor 2.5 for DM lifetimes $\tau_{\rm dec}\gtrsim 10^{10}$ s (see Fig. 28 of \cite{Kite:2022eye}).

\vspace{0.3 cm}
\noindent \textbf{Follow-up idea 17:} Including this formalism in \texttt{CLASS} would be very profitable in order to further exploit the complementarity between CMB anisotropies and SDs.
\vspace{0.3 cm}

\noindent \textbf{Inhomogeneous recombination}

\begin{figure}
	\centering
	\includegraphics[width=0.65\textwidth]{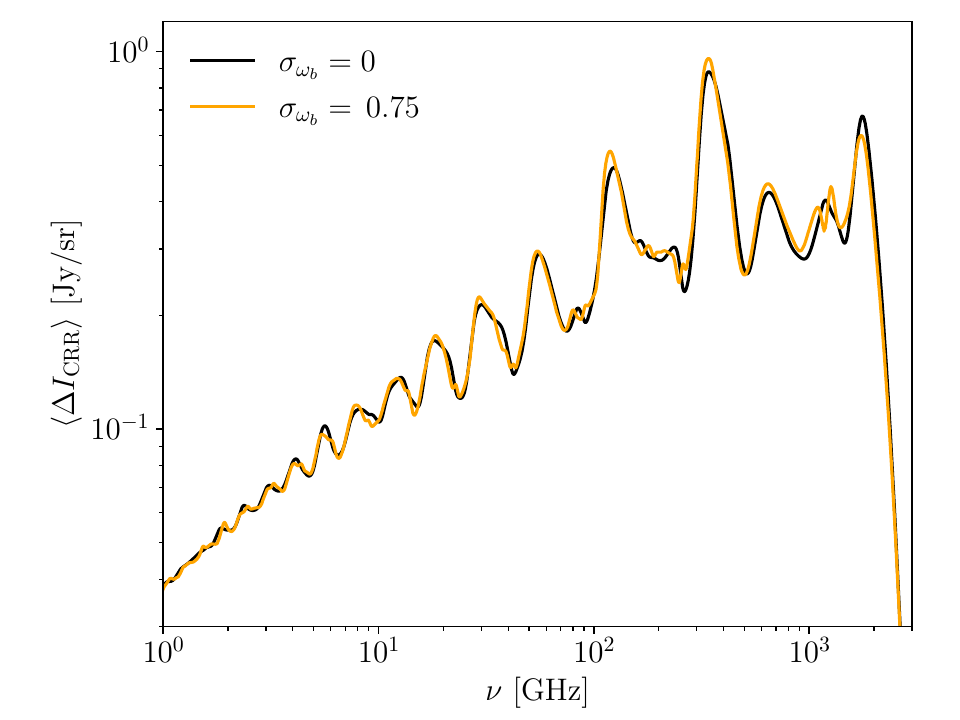}
	\caption{Sky-averaged CRR spectrum with (orange) and without (black) fluctuations of the baryon energy density. Figure taken from [\protect\hyperlink{XIV}{XIV}].}
	\label{fig: inhom_from_Taylor}
\end{figure}

\noindent The second way via which fluctuations of the matter and photon energy density might affect the SD signal is by perturbing the recombination process. In fact, as explained in Sec.~\ref{sec: SD_lcdm} (see Eq.~\eqref{eq: DI_CRR} there as well as the related footnote), when spatial fluctuations are accounted for the sky-averaged CRR spectrum gets a second-order contribution proportional to the size of the overdensities [\hyperlink{XIV}{XIV}]. A representation of the impact of a variation in~$\omega_b$, parameterized via $\sigma_{\omega_b} \equiv \langle \Delta\omega_b^2 \rangle^{1/2}/\bar{\omega}_b$ where $\bar{\omega}_b$ is the mean value and $\langle \Delta\omega_b^2 \rangle^{1/2}$ its variance, is shown in Fig.~\ref{fig: inhom_from_Taylor}, taken from [\protect\hyperlink{XIV}{XIV}] (orange curve -- to be compared to the black curve which represents the reference spectrum in absence of inhomogeneities). A similar exercise could be performed also for the variations $\sigma_{p}$ of the other parameters $p$ listed in Eq. \eqref{eq: params}, namely $Y_p$, $T_0$, $\omega_{\rm cdm}$ and $N_{\rm eff}$.

\begin{figure}
	\centering
	\includegraphics[width=0.48\textwidth]{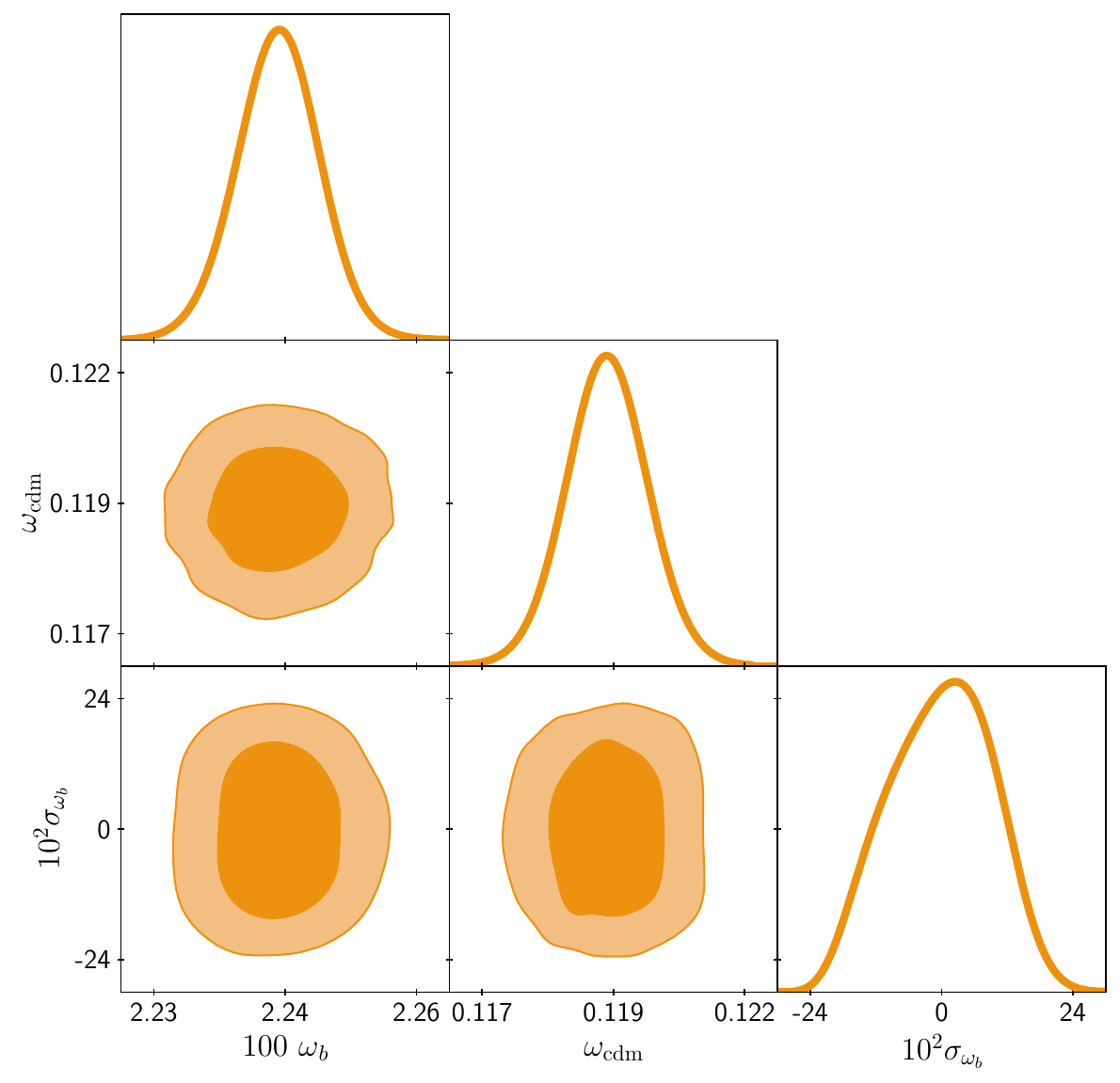}
	\includegraphics[width=0.48\textwidth]{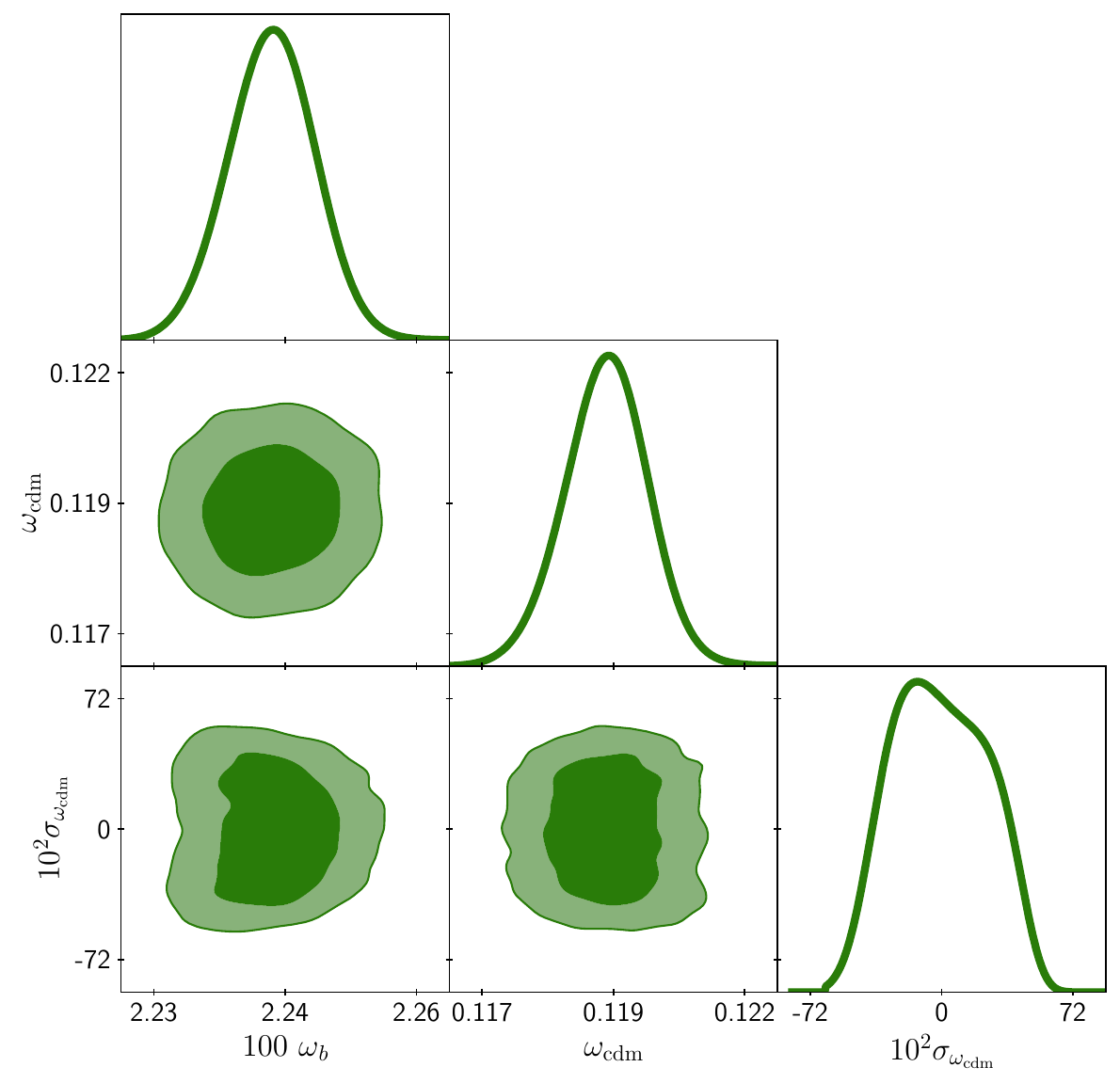}
	\caption{1D posteriors and 2D contours (at 68\% and 95\% CL) of the parameters describing scenarios with inhomogeneities in the baryon (left) and DM (right)  energy densities. A Voyage 2050+ mission is assumed. Figures taken from [\protect\hyperlink{XIV}{XIV}].}	
	\label{fig: inhom_MCMC_res}
\end{figure}

Clearly, the imprint of the fluctuations is very rich of features.\footnote{The spectral behavior of these second-order contributions can be understood by noting that they correspond to the second derivative of the spectrum with respect to the given parameter (see Eq.~\eqref{eq: DI_CRR}). Since in the case of $\omega_b$ one roughly has that $\partial\Delta I_{\rm CRR}/\partial \ln\omega_b\propto \Delta I_{\rm CRR}$ \cite{Hart2020Sensitivity}, the imprint of the inhomogeneities on the CRR spectrum approximately follows $(\Delta I_{\rm CRR}\,\sigma_{\omega_b})^2$.} Therefore, with sufficiently accurate SD missions these additional effects could be probed and the magnitude of the underlying fluctuations could be constrained. For a Voyage 2050+ mission, the sensitivity to the variations $\sigma_{p}$ of all parameters $p$ has been forecasted to be 13\%, 15\%, 0.7\%, 55\% and 165\%, respectively, when all $\sigma_{p}$ are varied at the same time. Restricting the parameter space to the single variations of e.g., $\omega_b$ and $\omega_{\rm cdm}$ improves the sensitivities to the corresponding $\sigma_{p}$ values to 10\% and 49\%, respectively. The posterior distributions of these last two cases are shown in Fig.~\ref{fig: inhom_MCMC_res}, taken from [\protect\hyperlink{XIV}{XIV}].

This level of sensitivity is very promising because, although it would not suffice to probe the primordial fluctuations expected within \lcdm (see Sec. 5.2 of [\protect\hyperlink{XIV}{XIV}]), it would allow to test a number of beyond-\lcdm models in relevant regions of parameter space. In~[\protect\hyperlink{XIV}{XIV}] the representative example of PMFs was considered. In brief, the idea behind PMFs is to justify with a primordial origin the presence of the observed magnetic fields that seem to populate the late-time universe at all scales, from voids and galaxy clusters to galaxies and stars (see e.g., \cite{Subramanian:2015lua} for a recent review). Although several formation mechanisms have been suggested (see Sec. 4 of \cite{Subramanian:2015lua} and references therein), no consensus around a particular scenario has been reached and there is therefore a large freedom in the initial conditions for the evolution of PMFs. Nevertheless, regardless of their exact generating effect, their existence leaves inevitable imprints on cosmological observables such as BBN and the CMB (see e.g., Sec. 7 of \cite{Subramanian:2015lua} for a detailed discussion). 

For instance, in analogy to the impact of primordial GWs discussed in Sec.~\ref{sec: res_further}, PMFs can strain (i.e., perturb) the energy momentum tensor. The following dissipation of energy\footnote{The energy spent to perturb the baryon-photon fluid is effectively taken away from the PMF. In this sense, the PMF dissipates its energy into the baryon-photon fluid.} results in the heating rate introduced in Eq. \eqref{eq: Q_PMFs} and affects the CMB anisotropies and SD spectrum. In this way, Planck and FIRAS data constrain the present-day value of the magnetic field $B_0$ to be $B_0\lesssim1$~nG \cite{Subramanian:2015lua, Kunze:2013uja}. This bound is relatively robust with respect to the scale-dependence of the PMF power spectrum, i.e., the value of $n_B$ (see Eq.~\eqref{eq: Q_PMFs} and below)\footnote{This is because for $n_B=-3$ (i.e., for a scale-invariant spectrum) the CMB anisotropy constraints are very stringent, while for $n_B>-3$ (i.e., for a spectrum tilted towards small scales) the role of SD is enhanced.}, and could even get significantly stronger in particular regimes and formation mechanisms. A future observation of the SD spectrum by PIXIE would further strengthen this upper bound by more than one order of magnitude \cite{Kunze:2013uja}.

Here, we focus instead on another way via which the presence of PMFs could leave an imprint on the CMB, and that is by producing a small-scale clumping of the baryons (see e.g., \cite{Jedamzik:2018itu, Jedamzik2020Relieving} for an overview). The magnitude of this effect is commonly parameterized by the so-called clumping factor
\begin{align}
	b=\frac{\langle n_b^2\rangle-\langle n_b\rangle^2}{\langle n_b\rangle^2}\,,
\end{align}
where $\langle n_b\rangle$ is the average baryon density and $\langle n_b^2\rangle$ its variance. For reference, a value of $b$ of the order of $b\simeq0.5$ could be generated by PMF values (today) of about 0.05 nG, well below the aforementioned limits. The corresponding size of the inhomogeneities would be of the order of $\sim1$ kpc. 

Interestingly, exactly these benchmark values have been shown to be able to significantly alleviate the Hubble tension \cite{Jedamzik2020Relieving} (see Sec. \ref{sec: tensions} -- noting in particular the follow-up work presented in \cite{Galli2021Consistency} debating this result). The mechanism that enables this partial resolution is the fact that the recombination process is accelerated in the inhomogeneous case and this reduces the sound horizon $r_s$ (see Eq.~\eqref{eq: sound_hor}) by increasing the value of $z_{\rm rec}$ which determines its integration limit. Since the actually observed quantity is $\theta_s$ (see Eq.~\eqref{eq: theta_s}), the ratio of $r_s$ and $D_a$ has to remain constant. If $r_s$ decreases, $D_a$ has to do so too, which can be achieved by increasing $H_0$ (see Eq. \eqref{eq: Da}), thereby moving the value of the latter towards the late-time measurements. In this regard, the potentially very stringent CMB anisotropy constraints are avoided by the fact that the clumping happens at much smaller scales than the ones observed by e.g., Planck (see e.g., Fig. 2 of \cite{Galli2021Consistency}).\footnote{This also explains why the inclusion of small-scale CMB anisotropy data from ACT and SPT reduces the available parameter space and the model does no longer solves the tension to the same degree \cite{Galli2021Consistency}.}

\begin{figure}
	\centering
	\includegraphics[width=0.63\textwidth]{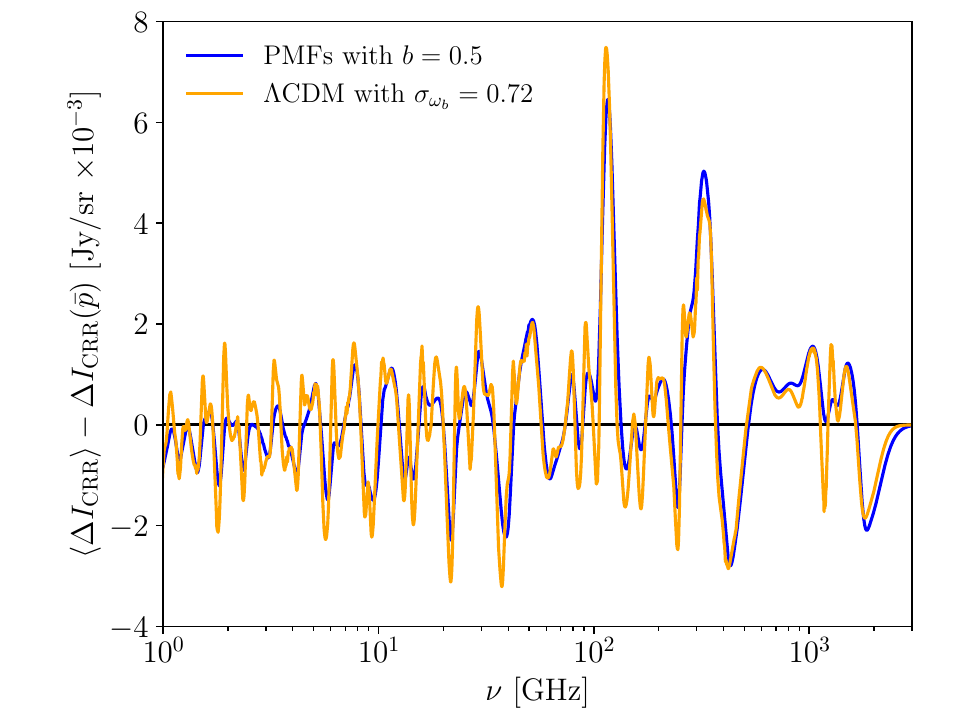}
	\caption{Comparison between the effects on the CRR spectrum of inhomogeneities in $\omega_b$ (orange) and those coming from PMFs (blue). Figure adapted from [\protect\hyperlink{XIV}{XIV}].}
	\label{fig: inhom_PMFs_vs_sigmab}
\end{figure}

At the same time, the presence of this baryon clumping and the consequent inhomogeneous evolution of recombination imprints second-order contributions to the sky-average CRR spectrum [\protect\hyperlink{XIV}{XIV}] (see Eq. \eqref{eq: DI_CRR}). As shown in Fig. \ref{fig: inhom_PMFs_vs_sigmab}, adapted from [\protect\hyperlink{XIV}{XIV}], indeed the effects of the PMFs on the CRR (blue curve) can be captured very closely by inhomogeneities in the value of $\omega_b$ (orange curve). Values of $\sigma_{\omega_b}$ of the order of 0.7 correspond to values of the clumping factor $b=0.5$ (as expected, since $\sigma_{\omega_b}\simeq b^{1/2}$). Since a Voyage 2050+ mission would be sensitive to $\sigma_{\omega_b}=0.1$, this implies that the same mission would be able to set bounds on the clumping factors of the order of 0.01, thereby fully exploring the region of parameter space relevant for the Hubble tension. Even beyond the Hubble tension, these would represent some of the most stringent constraints on PMFs to date, probing magnetic field strengths of the order of the pG in the pre-recombination era.

\vspace{0.3 cm}
\noindent\textbf{Follow-up idea 18:} Including PMFs in the numerical pipeline described in Sec. \ref{sec: num} would allow to fully explore these constrains, considering for instance the different possibilities to model their impact on the recombination history. Once consistently included in \texttt{CLASS}, the complementarity of all CMB anisotropy and SD constraints (from both the dissipation and clumping effect) could be investigated. Such a comprehensive study (which to our knowledge has never been performed before) would be of great interest.
\vspace{0.3 cm}

Of course, the validity of this type of constraints is not restricted to PMFs, but extends also to any model sourcing clumping of matter and more in general inhomogeneities in the recombination process. Examples thereof could be for instance inhomogeneous BBN (see e.g., \cite{Barrow:2018yyg, Scherrer:2021tbo} for recent treatments of the problem) and PBHs \cite{Jensen:2021mik} among others. In this sense the work carried out in [\protect\hyperlink{XIV}{XIV}] has to be considered as a proof of concept for future studies.

Finally, a particularly interesting aspect of this effect is that, even if the fluctuations had completely dissipated away by the time of photon-electron decoupling, they could in principle still be observed if they where sufficiently large during the previous stages of (helium) recombination. This memory effect would open the unique possibility to directly test the time dependence of cosmic perturbations around the recombination epoch.

\vspace{0.3 cm}
\noindent \textbf{Follow-up idea 19:} Studying which experimental sensitivity would be required to get access to this time-dependent information is an interesting avenue left for future work.
\vspace{0.3 cm}

\noindent \textbf{Summary}

\noindent In summary, CMB SDs are in principle able to probe the inhomogeneous universe in mainly two ways (on top of primordial non-Gaussianities explained in Sec.~\ref{sec: res_infl}). On the one hand, accounting for the energy density fluctuations of the cosmic fluids would lead to an inhomogeneous evolution of the heating history and of the thermalization/Comptonization processes, in turn generating SD anisotropies. Exploiting this additional spatial information in correlation with the CMB temperature/polarization anisotropies would already allow to improve upon current bounds with upcoming CMB anisotropy missions such LiteBIRD. On the other hand, the inhomogeneous evolution of the recombination process would lead to second-order contributions to the CRR spectrum. A future SD mission such as Voyage 2050+ able to observe the CRR very precisely could probe levels of inhomogeneities in (most notably) the baryon density of the order of 10\%. Such degree of precision would enable to test many cosmological models, such as PMFs, in relevant regions of parameter space.

In the context of this section, the work carried out during the thesis has focused on the relation between inhomogeneities in the recombination process and the CRR spectrum. The existence of this connection was suggested in [\protect\hyperlink{XIV}{XIV}] for the first time. Its implications have been discussed in the reference both in general and on a model-by-model basis, considering in particular \lcdm and PMFs. In this sense, the analysis performed in [\protect\hyperlink{XIV}{XIV}] presents several ramification, as highlighted in the course of the section and developed in more detail in the reference. 

\subsection{Hubble tension}\label{sec: res_tens}
In Sec. \ref{sec: tensions} we have discussed the emergence of a tension between the early-time inference and the late-time measurement of the expansion rate parameter $H_0$, known as the Hubble tension. In the same context, we have also presented evidence in favor of models that attempt to solve this discrepancy by modifying the expansion history of the universe prior to recombination, making of CMB SDs an ideal candidate to probe them.

In light of these considerations, CMB SDs have been shown to be able to play a role in the Hubble tension in several ways. First of all, a mission designed to observe them could also measure the value of $H_0$ \cite{Abitbol2019Measuring}, as already anticipated in Sec.~\ref{sec: res_lcdm}. As pointed out there, although a sub-percent measurement might be optimistic on Voyage 2050 timescales, it is not to be excluded. Secondly, in Sec. \ref{sec: res_PMFs} we have also highlighted the fact that models involving small-scale baryon clumping in order to address the Hubble tension, such as in the case of PMFs \cite{Jedamzik2020Relieving, Galli:2021mxk}, can be tightly constrained by SDs. Moreover, following the discussion carried out in Secs. \ref{sec: SD_lcdm} and~\ref{sec: res_lcdm}, SDs could also directly test the expansion history around the epoch of recombination, predominantly via the CRR. Finally, because of the fact that SDs can constrain several of the \lcdm parameters (see Sec.~\ref{sec: res_lcdm}), they can also be used in synergy with other cosmological probes to break eventual parameter degeneracies that might arise in some solutions to the $H_0$ tension to compensate for the new physics they introduce. 

Below, we focus on the last two approaches, the former  -- based on \cite{Hart:2022agu} -- for sake of completeness and the latter -- based on [\hyperlink{IV}{IV}] -- as a contribution of the thesis. In particular, we will consider them in context of the representative example of Early Dark Energy (EDE) \cite[\hyperlink{X}{X}]{Karwal2016Dark, Poulin2018Early, Smith2019Oscillating, Klypin2020Clustering, Hill:2021yec, Poulin:2021bjr, LaPosta:2021pgm, Poulin:2023lkg, Hill2020Early, Ivanov2020Constraining, Goldstein:2023gnw} as a proof of principle (see Sec. \ref{sec: tensions}). We remark, however, that the discussion is much more general and applies also to other models, as we will point out below. In this sense, the discussion is also independent of whether EDE can succesfully solve the $H_0$ or not (see e.g., \cite{Hill2020Early, Ivanov2020Constraining, Goldstein:2023gnw}).

In brief, the EDE model predicts the presence of an extremely light scalar field $\phi$ with mass $m\sim\mathcal{O}(10^{-27}$ eV) that, starting from an initial value $\phi_i$ (parameterized via $\theta_i=\phi_i/f$), slow rolls down an axion-like potential of the form $V(\phi)=V_0(1-\cos(\phi/f))^n$ until it becomes dynamical around the time of recombination (moment around which $m\sim H$). Once the field reaches the minimum of the potential it starts to oscillate around it, quickly diluting its energy density. In the definition of the potential, $f$ is the decay constant of the field, which takes values $f\sim\mathcal{O}(10^{27}$ eV). The index $n$ is commonly fixed to $n=3$. 

\begin{figure}
	\centering
	\includegraphics[width=0.6\textwidth]{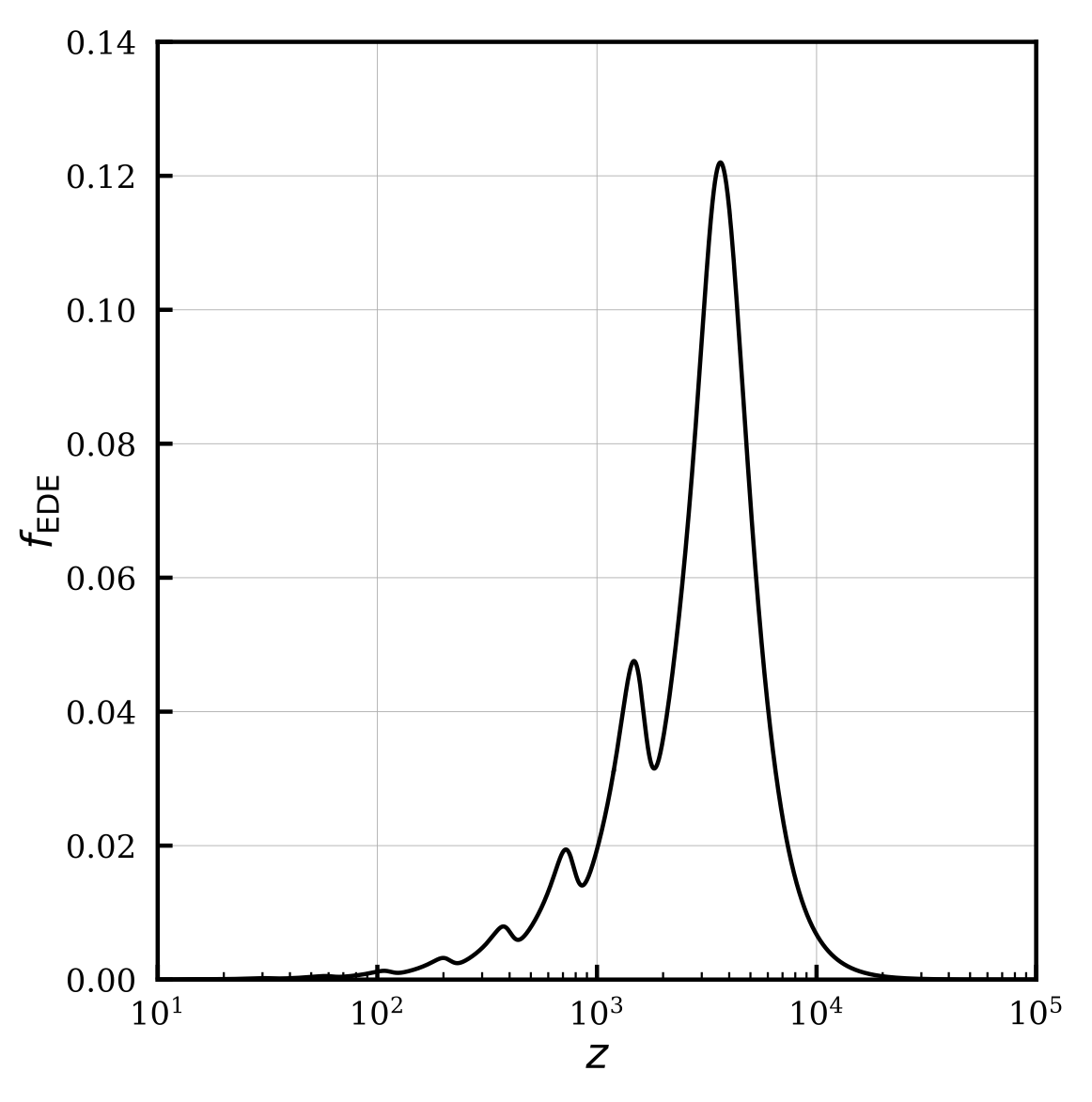}
	\caption{Representative example of the redshift evolution of the fractional energy density of EDE. Figure taken from \cite{Hill2020Early}.}
	\label{fig: tensions_fEDE}
\end{figure}

In a cosmological context, these phenomenological parameters ($m$, $\theta_i$ and $f$) can be recast in terms of the critical redshift $z_c$ at which $H<m$ and of the fractional energy density of the field (with respect to the total energy density) $f_{\rm EDE}$. As it turns out, values of the order of $f_{\rm EDE}(z_c)\simeq0.1-0.2$ and $\log(z_c)\simeq 3.5$ seem to be able to solve the Hubble tension with $H_0\simeq 72-77$ km/(s Mpc) for various combination of CMB, BAO, SNIa and late-time direct measurements of $H_0$ (see most recently \cite[\hyperlink{X}{X}]{Hill:2021yec, Poulin:2021bjr, LaPosta:2021pgm}), although large-scale structure data strongly disfavors this possibility due to the large increase of the amplitude of the matter power spectrum in EDE cosmologies \cite{Hill2020Early, Ivanov2020Constraining, Goldstein:2023gnw}. A graphical representation of the redshift evolution of $f_{\rm EDE}$ in one of these Hubble tension-solving parameter combinations is provided in Fig. \ref{fig: tensions_fEDE}, adapted from \cite{Hill2020Early} (see also Sec. VII therein for more details about the parameter conversion from \{$m$, $\theta_i$, $f$\} to \{$z_c$, $\theta_i$, $f_{\rm EDE}$\}).

\newpage
\noindent\textbf{Expansion history}

\begin{figure}
	\centering
	\includegraphics[width=0.48\textwidth]{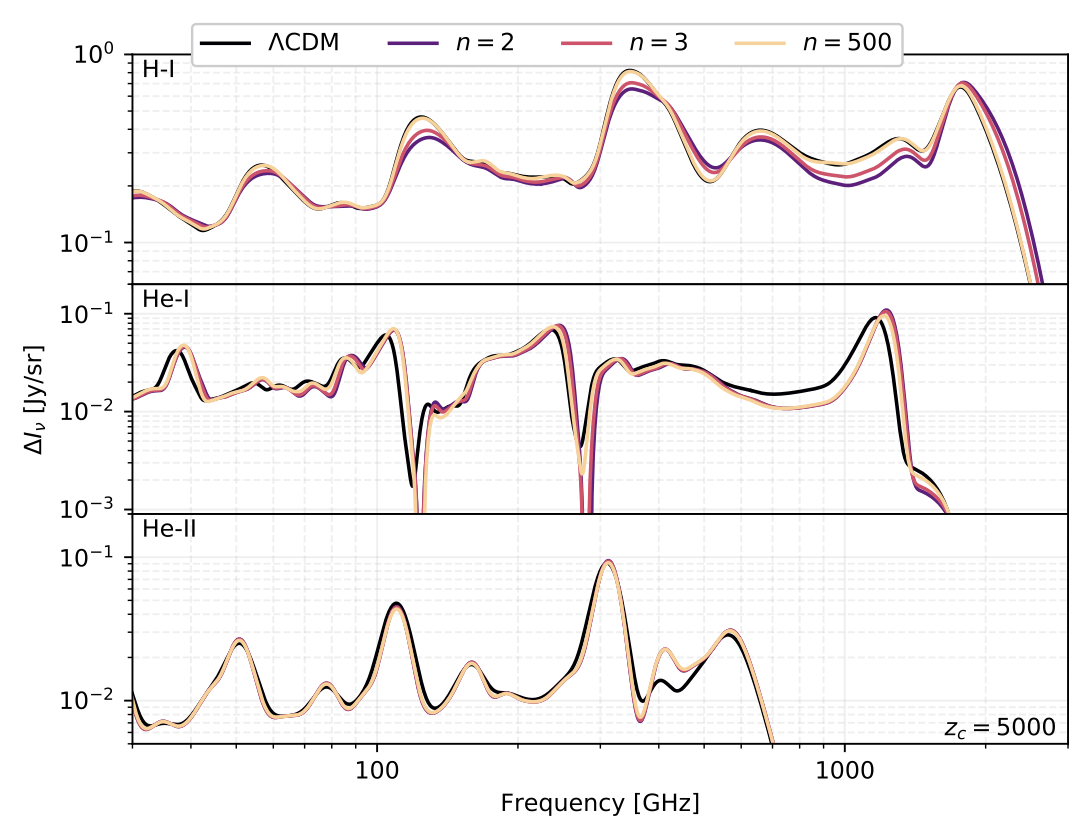}
	\includegraphics[width=0.48\textwidth]{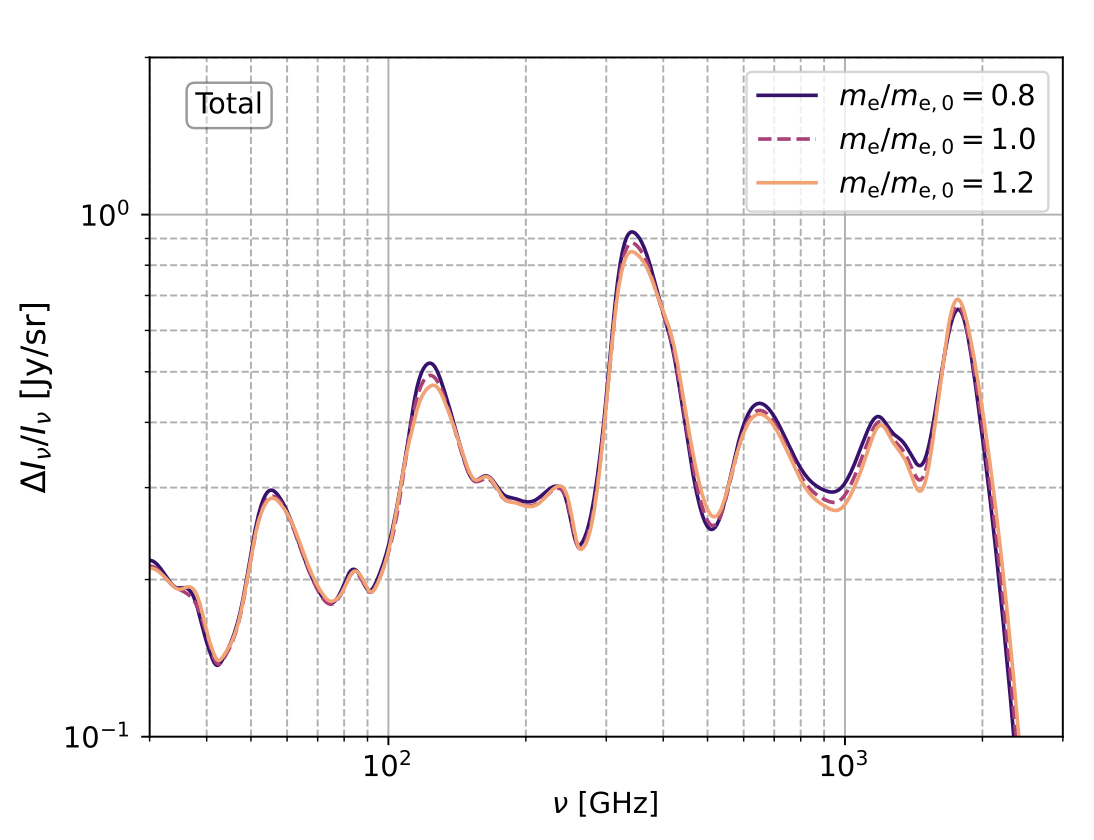}
	\caption{Impact of the presence of EDE (left) and variations of the electron mass $m_e$ (right) on the CRR. Figures taken from \cite{Hart:2022agu}.}
	\label{fig: tensions_CRR_EDE}
\end{figure}

\noindent As clear from the figure, EDE maximally modifies the expansion history of the universe just before the time of recombination, right when the CRR is building up. Therefore, according to the discussion had in Sec. \ref{sec: SD_lcdm}, it is to be expected that the CRR spectrum will be modified by the presence of EDE.\footnote{As found in Sec. \ref{sec: res_lcdm}, while in principle also the $y$ and $\mu$ signal is affected by modifications to the expansion history of the universe, the impact on (and hence the constraining power of) the CRR is much stronger. We will therefore only focus on this contribution here.} This possibility has been quantitatively considered in \cite{Hart:2022agu}, where it was found that this is indeed the case. A representation of how EDE affects the CRR in the three phases of recombination (assuming $f_{\rm EDE}(z_c)=0.8$ for graphical purposes) is reported in the left panel of Fig. \ref{fig: tensions_CRR_EDE}, taken from the reference.

However, despite showing that it would be in principle possible to constrain EDE with the CRR, the forecasts performed in \cite{Hart:2022agu} demonstrated that even Voyage 2050+ sensitivities (10-fold better than Voyage 2050) would not be able to set competitive bounds on EDE with SD information only. The expected sensitivity to $f_{\rm EDE}(z_c)$ would in fact be of the order of $0.5-1$ (depending on the value of $z_c$ and of the potential index $n$), slightly above the amount expected to (potentially) resolve the Hubble tension.  Instead, a Voyage 2050++ mission (50 times better than Voyage 2050) would be required to start probing the EDE physics on a level competitive with CMB anisotropy data.

\newpage
\noindent\textbf{Follow-up idea 20:} Whether SDs could improve upon current constraints on EDE once combined with CMB anisotropy data (among others) remains an open question to be investigated in future work. Including EDE in the numerical set up presented in Sec.~\ref{sec: num} to calculate the CRR would open this possibility.
\vspace{0.3 cm}

The authors further point out, nevertheless, that the same might not be true for other promising solutions to the Hubble tension. In the reference, the case of variations of fundamental constants, namely the fine structure constant $\alpha$ and the electron mass $m_e$, was analyzed. The impact of the latter on the CRR is shown in the right panel of Fig. \ref{fig: tensions_CRR_EDE}, taken from \cite{Hart:2022agu}. In particular for $m_e$, it was found that the addition of a Voyage 2050 mission to Planck would reduce the error bars on the determination of $H_0$ by a factor 3, thereby tightly testing the model ability to address the Hubble tension.

\vspace{0.3 cm}
\noindent\textbf{Braking parameter degeneracies}

\begin{figure}
	\centering
	\includegraphics[width=0.65\textwidth]{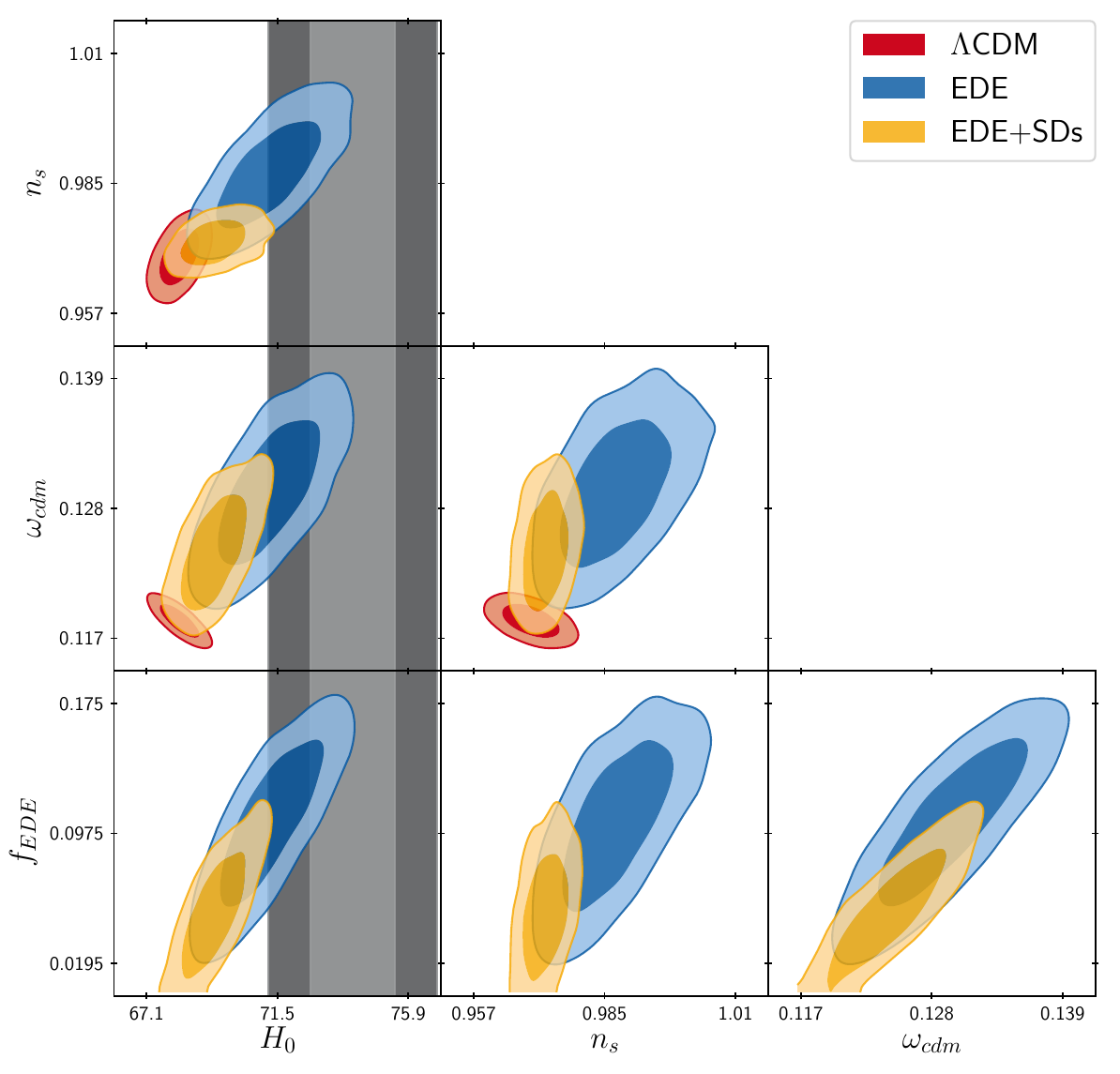}
	\caption{2D contours of the parameters most relevant for EDE as a solution to the Hubble tension. Figure taken from [\protect\hyperlink{IV}{IV}].}
	\label{fig: tensions_MCMC_res}
\end{figure}

\noindent As extensively explained in \cite{Hill2020Early}, the presence of EDE suppresses the growth of perturbations and this needs to be compensated by an increased DM energy density $\Omega_{\rm cdm}$ and scalar spectral index $n_s$. This leads strong degeneracies between these two parameters and the ones ruling the new physics introduced by EDE, most notably characterized by $f_{\rm EDE}(z_c)$. In turn, since the model's ability to solve the Hubble tension depends in large part on how significant the contribution of EDE is to the total energy budget, another degeneracy line can be expected between $f_{\rm EDE}(z_c)$ and $H_0$. All this expectations are confirmed on the basis on MCMC scans (see Sec. \ref{sec: num_2}) such as the one shown in Fig. \ref{fig: tensions_MCMC_res}, taken from [\protect\hyperlink{IV}{IV}], (blue contours) which isolates exactly these four quantities.

Based on these considerations, it was pointed out in [\protect\hyperlink{IV}{IV}] that, because of the fact that CMB SDs are highly sensitive to the value of $n_s$ via the dissipation of acoustic waves (see Sec. \ref{sec: SD_lcdm}), they can be used in combination with the other data sets to break the degeneracy this parameter shares with $f_{\rm EDE}(z_c)$ and indirectly limit the freedom that the model has to solve the Hubble tension. This logic can be graphically seen in Fig. \ref{fig: tensions_MCMC_res}. Focusing firstly on the distribution of $n_s$, one notices by comparing the \lcdm to EDE distributions (in red and blue, respectively) that the EDE case significantly elongates the contours. If one was assume that a future SD mission with sufficient sensitivity was to perfectly confirm the Planck determination of this quantity within \lcdm, i.e., $n_s\simeq 0.96$, SD would act as a strong prior on $n_s$ and further tighten the EDE distribution towards the \lcdm values. In turn, in this way EDE would not be able anymore to increase $n_s$ to compensate for non-zero values of $f_{\rm EDE}(z_c)$ (see the $f_{\rm EDE}(z_c)-n_s$ subplot in the bottom row), and therefore the model's ability to increase $H_0$ would be indirectly also hindered (see the $f_{\rm EDE}(z_c)-H_0$ subplot in the bottom row). Of course, one could just as easily assume SDs to observe an $n_s$ value much closer to unity and in that case SDs would be providing an even stronger evidence in favor of the presence of new physics around the epoch of recombination.

This type of increase in $n_s$ is very common in proposed solutions to the Hubble tension and the aforementioned approach is therefore widely applicable. Examples of such models are e.g., neutrino \cite{Kreisch2019Neutrino} and sterile neutrino \cite{Archidiacono:2020yey} self-interactions, interacting majoron-neutrinos \cite{Escudero2019CMB, Escudero2020Could}, modified theories of gravity (such as in the Brans–Dicke model) \cite{SolaPeracaula:2020vpg} as well as other rolling scalar field scenarios \cite{Agrawal2019Rock}.

\newpage
\noindent \textbf{Summary}

\noindent In summary, CMB SDs can play an important role in the context of the Hubble tension by directly constraining either the expansion history of the universe prior to recombination or the amount of baryon clumping at small scales. SDs can also be very valuable to break degeneracies that the beyond-\lcdm models parameters might share with the scalar spectral index $n_s$. Pointing out this last application of SDs to the Hubble tension has been the main contribution of the thesis in the context of this section.

\newpage
\section{Summary and closing remarks}\label{sec: conc}

\noindent This thesis broadly reviews the topic of Cosmic Microwave Background (CMB) Spectral Distortions (SDs) and places them in the general landscape of modern cosmology.

The theoretical aspects of CMB SDs are covered and reviewed in the first part of the thesis. Building on the general recap of the \lcdm model carried out in Sec.~\ref{sec: lcdm}, Sec.~\ref{sec: theory} provides a very pedagogical and illustrative introduction to the theory of SDs (Sec.~\ref{sec: ped}) and to the various effects that can source them (within \lcdm~-- Sec.~\ref{sec: SD_lcdm} -- and beyond -- Sec.~\ref{sec: SD_non_lcdm}). The content of the section is based on the reviewing works [\hyperlink{I}{I},\,\hyperlink{III}{III},\,\hyperlink{XIV}{XIV}]. In a nutshell, CMB SDs are any deviation from a pure Black Body (BB) shape of the CMB energy spectrum. They can be generated by any heating, number changing and photon mixing process taking place in the history of the universe as long as the resulting distortion cannot be thermalized (i.e., redistributed to obtain a BB spectrum). This is the case when the number changing (i.e., bremsstrahlung and double Compton scattering) and scattering (i.e., Compton scattering) processes start to become inefficient, which happens at redshifts $z\simeq10^6$ and $z\simeq10^4$, respectively. The SD signal can then be broken down into so-called $\mu$ and $y$ components (produced respectively at $z\simeq10^4-10^6$ and $z\lesssim10^4$), as well as into a residual component that captures the more complex spectral dependence of the signal. 
A graphical representation of the evolution of the SD spectrum as a function of the underlying processes is given in Fig.~\ref{fig: SD_th_example_box}.

The existence of CMB SDs is predicted even within the \lcdm model, where they can be sourced most notably by the dissipation of acoustic waves prior to recombination, by the Cosmological Recombination Radiation (CRR), i.e., the emission/absorption of photons during the recombination process, and by the Sunyaev-Zeldovich (SZ) effect, i.e., the up-scattering of CMB photons by the free hot electrons populating the late universe. The former two largely define the primordial SD signal, i.e., the pre-recombination contribution to the total SD signal, while the latter is a purely late-time effect. A summarizing collection of the effects sourcing SDs within the \lcdm (as explained in Sec.~\ref{sec: SD_lcdm}) is given in Fig.~\ref{fig: SD_LCDM_summary}, which combine to a guaranteed floor for the total SD signal. All together, these processes are sensitive to the inflationary, expansion and thermal history of the universe, and can therefore be used to constrain it. Of course, as discussed in Sec.~\ref{sec: SD_non_lcdm}, also a plethora of beyond-\lcdm models would lead to the production of SDs, ranging from beyond-slow roll inflationary scenarios to decaying Dark Matter (DM) and evaporating Primordial Black Holes (PBHs), among others (see below). This opens a wide region of parameter space for the discovery of new physics.

Nevertheless, so far no CMB SD has ever been observed, although very important steps forward are already in motion. To this point, Sec.~\ref{sec: exp} provides a full picture on the currently limited experimental status and on the promising perspectives for the observation of the SD signal in the future. The section is largely based on the knowledge gathered during the development of [\hyperlink{I}{I},\,\hyperlink{III}{III},\,\hyperlink{V}{V}]. In order to first of all illustrate why a SD detection cannot be as easily achieved, Sec.~\ref{sec: fore} reviews the main challenges behind the accurate observation of the CMB energy spectrum, i.e., atmospheric absorption (for ground-based and balloon experiments) and the unavoidable (extra-)galactic foregrounds. The latter are graphically collected in Fig. \ref{fig: SD_fore_gal}. Because of these contaminating effects, as discussed in Sec.~\ref{sec: firas} 
the measurements of the CMB energy spectrum 
have been severely limited so far. A collection of more than a hundred data points constraining the shape of the CMB spectrum is provided in Fig.~\ref{fig: SD_exp_spectrum}. Its most precise determination 
has been achieved
by the FIRAS mission in the early '90s,
although only accurate enough to place upper bounds on the $\mu$ and $y$ parameters of the order of $y<1.5\times 10^{-5}$ and $\mu<4.7\times10^{-5}$ (at 95\% CL). 

While the pioneering success of FIRAS towards the observation of the SD signal has 
not seen any follow-up to date,
as argued in Sec.~\ref{sec: current_exp}, the picture will be radically changing in the coming $1-3$ years with the deployment of a number of ground-based and balloon experiments with enough sensitivity to detect at high significance the late-time SD signal and, potentially, also the CRR at very low frequencies. A summary plot comparing the expected missions sensitivities with the SD signals is shown in Fig.~\ref{fig: SD_exp_upcoming}. This second generation of efforts (the first generation being FIRAS) will achieve two important results. On the one hand, it will deliver the first direct detection of the SD spectrum, with potentially enormous consequences for our understanding of the thermal history of the late universe and of structure formation (see below). On the other hand, it will significantly improve our characterization of the relevant foregrounds, which will in turn be fundamental to set the stage for the more sensitive missions targeting the primordial SD signal from space.

Such satellites will then represent the third generation of SD missions, and the state of the art of their development is reviewed in Sec.~\ref{sec: future_exp}. The expected sensitivities of many of the suggested experimental setups are compared in Fig. \ref{fig: SD_exp_future}. Although not itself under commissioning, the most relevant template for any future space mission is the PIXIE proposal, whose technical aspects have been carefully worked out. In fact, its design in an improved form has been selected by ESA in the context of the Voyage 2050 initiative as one of the priorities for its long-term science program. Such a mission would be able to measure all primordial effects with high significance, opening the way for the inference of cosmological information from CMB SDs.

In light of this exciting possibility, the first main contributions of the thesis to the field of SDs has been the development of a unified, comprehensive and easily extendable numerical pipeline for the cosmological analysis of CMB SDs. In fact, as presented in Sec.~\ref{sec: num}, the newly developed implementation of CMB SDs (performed in [\hyperlink{I}{I},\,\hyperlink{XIV}{XIV}]) in the publicly-available cosmological Boltzmann solver \texttt{CLASS} includes all \lcdm sources of SDs as well as many beyond it, and it allows for the consistent evaluation of any experimental setup. Furthermore, in order to perform sensitivity forecasts for the aforementioned missions, the \texttt{CLASS} implementation has been complemented in [\hyperlink{I}{I},\,\hyperlink{III}{III}] by the inclusion of a very general class of (mock) SD likelihoods in the parameter inference code \texttt{MontePython}. As overviewed in Sec.~\ref{sec: num_2}, the \texttt{MontePython} likelihoods share the same freedom as \texttt{CLASS} in the definition of the experimental design, thereby being applicable to all aforementioned SD missions. On top of this, the code includes a realistic treatment of galactic and extra-galactic foregrounds, implemented in [\hyperlink{V}{V}], so as to increase the degree of realism of the resulting forecasts.

By exploiting the variety of cosmological observables, models and data sets already included by default in the \texttt{CLASS}+\texttt{MontePython} pipeline, the new implementation of SDs allows to fully explore the constraining power of this observable alone and in combination with other complementary probes of the universe. The second main contribution of the thesis to the field of SDs has then been to highlight the versatility and competitiveness (in terms of constraining power) of CMB SDs for a number of different models, mission designs and data set combinations. Moreover, possibly even more importantly, due to the very general numerical setup described above the number of considered applications can (and hopefully will) be easily extended to include many more known and new effects. This will make the code increasingly more comprehensive and accurate, and at the same time it will further accentuate the fundamental contribution to cosmology that the future observation of CMB SDs would entail.

Sec.~\ref{sec: app} (together with Sec.~\ref{sec: sources}) reviews a number of these applications, synergetically integrating the results derived during the thesis and those obtained by other references (but whose incorporation in the aforementioned numerical pipeline would be both straightforward and very useful) for sake of completeness. The main conclusions from the section are succinctly summarized below, starting from the \lcdm model and subsequently following the chronology of the universe.
\begin{itemize}
	\item[--] Sec.~\ref{sec: res_lcdm}: CMB SDs can test the \lcdm model at both early and late times. With the observation of the primordial signal, a Voyage 2050-like mission could set competitive constraints on four (or, optimistically, five) of the six \lcdm parameters, with particular emphasis on the baryon energy density $\omega_b$ and the scalar spectral index $n_s$, thereby also improving present and future CMB anisotropy bounds. This has been particularly highlighted in [\hyperlink{I}{I},\,\protect\hyperlink{III}{III},\,\hyperlink{XIV}{XIV}] and can be graphically seen in Fig.~\ref{fig: SD_MCMC_res_1}. Moreover, the late time observation of the SZ effect would deliver strong constraints on the $\sigma_8$ parameter and the matter energy density $\Omega_m$ as well as on the galactic thermodynamics and astrophysics. This is graphically summarized in Figs.~\ref{fig: SD_SZ_signal} and~\ref{fig: SD_SZ_b_feed}.

	\item[--] Sec.~\ref{sec: res_infl}: Being sensitive to physics happening at very small scales (of the order of ${k\sim1-10^4}$~Mpc$^{-1}$), CMB SDs are an extremely powerful probe of the Primordial Power Spectrum (PPS) and hence of inflation. As such, they can be used to constrain the inflationary potential, both in the slow-roll limit and beyond, as well as specific features in the shape of the PPS such as kinks and steps (the state of the art in these regards being set by [\hyperlink{V}{V}]). A graphical overview of the constraining power of SDs in combination with Planck data can be found in Figs.~\ref{fig: infl_P18_V6}\,--\,\ref{fig: infl_P18_features}. As clear from the figures, CMB SDs severely constrain the shape of the PPS by placing an additional anchor at small scales, thereby significantly extending the lever arm in $k$ space in complementarity with the CMB anisotropy bounds at larger scales. Also the Gaussianity of primordial perturbations can be probed with CMB SDs, with wide-ranging consequences for a number of inflationary scenarios. 
	
	\item[--] Sec.~\ref{sec: res_further}: Should they be sourced in the very early universe, the existence of primordial Gravitational Waves (GWs) would lead (via their dissipation) to CMB SDs. The sensitivity window covers the $10^{-15}-10^{-9}$~Hz range and in this way CMB SDs nicely complement other known GW observational strategies, as illustrated by Fig.~\ref{fig: GWs_constr}. 
	
	\item[--] Sec.~\ref{sec: res_DM}: A wide range of interaction channels between the DM and the standard model particles, especially (but not limited to) photons and electrons, would generate potentially very large CMB SDs. The section focuses on decay (studied in [\hyperlink{I}{I},\,\hyperlink{IX}{IX}]), annihilation (studied in [\hyperlink{I}{I},\,\hyperlink{III}{III}]), scattering and photon mixing. In almost all instances, the observation of SDs with a Voyage 2050-like mission would deliver the strongest constrains to date in the relevant sections of parameter space. This is graphically confirmed on the basis of Figs.~\ref{fig: DM_decay_exclusion_region}\,--\,\ref{fig: DM_ALP_decay} and \ref{fig: DM_scatterings}. Of particular note, a future CMB SD measurement could play an important role in probing decaying massive relics with lifetimes $\tau_{\rm dec}\sim 10^{5}-10^{13}$~s and the sub-MeV mass range for DM candidates scattering with electromagnetically interacting particles.

	\item[--] Sec.~\ref{sec: res_PBH}: The existence of PBHs can be probed by CMB SDs in various ways, namely via their formation mechanism, their evaporation and their accretion of~matter. The former effect delivers a very strong upper bound on the PBH mass (conservatively) of the order of $M \lesssim 10^5$~M$_\odot$, as evident from Fig.~\ref{fig: PBH_PPS}. The evaporation constraints apply to the $10^{10}-10^{13}$~g mass range (as found in [\hyperlink{I}{I}] and as clear from Fig.~\ref{fig: PBH_exclusion_region}) and the future observation of CMB SDs will offer the exciting possibility to test the quasi-extremality of the PBHs (as found in [\hyperlink{XIII}{XIII}] and as clear from Fig.~\ref{fig: PBH_bounds_qPBH}). \matteo{It has finally been pointed out here that SDs could also probe the accretion of PBHs should the redshift dependence of the luminosity be enhanced in the pre-recombination era (see Fig.~\ref{fig: PBH_L_over_Ledd}), finding that} will need to be investigated in greater detail in the future.
	
	\item[--] Sec.~\ref{sec: res_PMFs}: The inhomogeneous evolution of the thermal history of the universe in the pre-recombination era would leave observable imprints on the CMB SD signal. For instance, should a source of heating happen inhomogeneously in the universe (like it would be the case if e.g., the energy density of a decaying massive relic followed the primordial DM energy density fluctuations), it would generate SD anisotropies. This additional spatial information can be used to extract useful information on the time dependence of the energy injection, as displayed in Fig.~\ref{fig: inhom_Cl}. Furthermore, if the recombination process was evolving inhomogeneously across the sky (like it would be the case in the presence of e.g., Primordial Magnetic Fields -- PMFs, a proposed solution to the Hubble tension), it would imprint second-order contributions to the CRR spectrum, as shown in Fig.~\ref{fig: inhom_PMFs_vs_sigmab}, which would be testable with a Voyage 2050-like mission.
	
	\item[--] Sec.~\ref{sec: res_tens}: CMB SDs could shed light on the Hubble tension by directly constraining either the expansion history of the universe prior to recombination (see Fig.~\ref{fig: tensions_CRR_EDE}) or the amount of baryon clumping at small scales (see previous point). On top of this, SDs could also be used to break degeneracies between the new physics introduced by the given model and the scalar spectral index $n_s$, as pointed out in [\hyperlink{IV}{IV}] and graphically represented in Fig.~\ref{fig: inhom_MCMC_res}.
\end{itemize}

The collection of these results and perspectives certainly makes a strong case for the relevance of CMB SDs in the future of cosmology and to prove this point is the ultimate goal of the thesis. Still, the richness of the topic extends even beyond the aforementioned examples and in the course of the manuscript several (20) possible future ramifications have been pointed out (see e.g., \cite{Chluba2019Spectral, Chluba2019Voyage} for additional interesting applications). The (in our opinion) most compelling ones are reported below.
\begin{itemize}
	\item[--] The current \texttt{CLASS} implementation of the CRR  (Sec.~\ref{sec: num}) could be extended in several directions. The follow-up ideas 4 and 5 (see same section) point out how this could be achieved to explore both a larger number of cosmological models and a more complex time dependence of the signal. Specific applications are highlighted in the follow-up ideas 18, 19 and 20 in the context of PMFs (Sec.~\ref{sec: res_PMFs}) and early dark energy (Sec.~\ref{sec: res_tens}), but could be applied more generally also to DM properties, cosmic strings and variations of fundamental constants, among many others. The inclusion of each of these scenarios in \texttt{CLASS} would be very valuable, in particular in synergy with the other observables already included in the code.
	
	\item[--] In terms of mission design, the follow-up ideas 7, 8 and 9 (Secs.~\ref{sec: firas}\,--\,\ref{sec: future_exp}) suggest possible applications of the \texttt{CLASS}+\texttt{MontePython} implementation to revive old experimental concepts (like sounding rockets following the COBRA experience) and to perform combined sensitivity forecasts for the up-coming SD missions as well as cost-reward analyses for the optimal setup of Voyage 2050-like missions. Exploring these possibilities would be very useful in order to progress towards the design of the ideal experimental setup, eventually able to extract the maximum amount of cosmological information from its observational cycle.
	
	\item[--] The aforementioned numerical pipeline could also be straightforwardly applied to the specific case of PBHs, as pointed out in the follow-up ideas 12 and 16 (Secs.~\ref{sec: res_PBH_form} and \ref{sec: res_PBH_acc}). The first suggests to consider popular extended PBH mass distributions and to systematically constrain them via the necessary enhancement of the PPS that would be required to produce them. The second proposes to further investigate the possibility to constrain PBH accretion with CMB SDs, an original idea put forward in the manuscript.
\end{itemize}

In conclusion, circling back to the hiking metaphor that opened the thesis, now is the time to persevere. The clouds have been thick, but are about to start clearing. The magnificent view that expects us all on the top of the mountain will soon be emerging. We hope this thesis could convince the reader of that.

\newpage

\section*{Bibliography}
\addcontentsline{toc}{section}{\protect\numberline{} Bibliography}
\vspace{-1 cm}
{\small
	\bibliography{bibliography}{}
	\bibliographystyle{JHEP}
}

\newpage

\section*{List of publications}
\addcontentsline{toc}{section}{\protect\numberline{} List of publications}
The content of thesis broadly relies on the following publications (in chronological order)

\begin{enumerate}[ {[}I{]} ]
	\item \hypertarget{I}{} \textbf{M. Lucca}, N. Sch\"oneberg, D. C. Hooper, J. Lesgourgues, and J. Chluba, \emph{The synergy between CMB spectral distortions and anisotropies}, 2019, \href{https://iopscience.iop.org/article/10.1088/1475-7516/2020/02/026}{\mbox{JCAP 02 (2020) 026}}, [\href{https://arxiv.org/abs/1910.04619}{arXiv:1910.04619}]
	
	\item \hypertarget{II}{} \textbf{M. Lucca}, and D. C. Hooper, \emph{Shedding light on Dark Matter-Dark Energy interactions}, 2020, \href{https://journals.aps.org/prd/abstract/10.1103/PhysRevD.102.123502}{\mbox{Phys. Rev. D 102 (2020) 12, 123502}}, [\href{https://arxiv.org/abs/2002.06127}{arXiv:2002.06127}]
	
	\item \hypertarget{III}{} H. Fu, \textbf{M. Lucca}, S. Galli, E. S. Battistelli, D. C. Hooper, J. Lesgourgues, and N. Sch\"oneberg, \emph{Unlocking the synergy between CMB spectral distortions and anisotropies}, 2020, \href{https://iopscience.iop.org/article/10.1088/1475-7516/2021/12/050}{\mbox{JCAP 12 (2021) 050}}, [\href{https://arxiv.org/abs/2006.12886}{arXiv:2006.12886}]
	
	\item \hypertarget{IV}{} \textbf{M. Lucca}, \emph{The role of CMB spectral distortions in the Hubble tension: a proof of principle}, 2020, \href{https://www.sciencedirect.com/science/article/pii/S0370269320305943}{Phys. Lett. B 810 (2020) 135791}, [\href{https://arxiv.org/abs/2008.01115}{arXiv:2008.01115}]
	
	\item \hypertarget{V}{} N. Sch\"oneberg, \textbf{M. Lucca}, and D. C. Hooper, \emph{Constraining the inflationary potential with spectral distortions}, 2020, \href{https://iopscience.iop.org/article/10.1088/1475-7516/2021/03/036}{\mbox{JCAP 03 (2021) 036}}, [\href{https://arxiv.org/abs/2010.07814}{arXiv:2010.07814}]
	
	\item \hypertarget{VI}{} \textbf{M. Lucca}, \emph{Dark energy-dark matter interactions as a solution to the $S_8$ tension}, 2021, \href{https://www.sciencedirect.com/science/article/abs/pii/S2212686421001291}{\mbox{Phys. Dark Univ. 34 (2021) 100899}}, [\href{https://arxiv.org/abs/2105.09249}{arXiv:2105.09249}]
	
	\item \hypertarget{VII}{} \textbf{M. Lucca}, \emph{Multi-interacting dark energy and its cosmological implications}, 2021, \href{https://journals.aps.org/prd/abstract/10.1103/PhysRevD.104.083510}{\mbox{Phys. Rev. D 104 (2021) 8, 083510}}, [\href{https://arxiv.org/abs/2106.15196}{arXiv:2106.15196}]
	
	\item \hypertarget{VIII}{} D. C. Hooper, and \textbf{M. Lucca}, \emph{Hints of dark matter-neutrino interactions in Lyman-$\alpha$ data}, 2021, \href{https://journals.aps.org/prd/abstract/10.1103/PhysRevD.105.103504}{\mbox{Phys. Rev. D 105 (2022) 10, 103504}}, [\href{https://arxiv.org/abs/2110.04024}{arXiv:2110.04024}]
	
	\item \hypertarget{IX}{} T. Hambye, M. Hufnagel, and \textbf{M. Lucca}, \emph{Cosmological constraints on the decay of heavy relics into neutrinos}, 2021, \href{https://iopscience.iop.org/article/10.1088/1475-7516/2022/05/033}{\mbox{JCAP 05 (2022) 033}}, [\href{https://arxiv.org/abs/2112.09137}{arXiv:2112.09137}]
	
	\item \hypertarget{X}{} T. Smith, \textbf{M. Lucca}, V. Poulin, G. F. Abellan, L. Balkenhol, K. Benabed, S. Galli, and R. Murgia,  \emph{Hints of Early Dark Energy in Planck, SPT, and ACT data: new physics or systematics?}, 2022, \href{https://journals.aps.org/prd/abstract/10.1103/PhysRevD.106.043526}{\mbox{Phys. Rev. D 106 (2022) 08, 043526}},  [\href{https://arxiv.org/abs/2202.09379}{arXiv:2202.09379}]
	
	\item \hypertarget{XI}{} L. Piga, \textbf{M. Lucca}, N. Bellomo, V. Bosch-Ramon, S. Matarrese, A. Racanelli, and L. Verde,  \emph{The effect of outflows on CMB bounds from Primordial Black Hole accretion}, 2022, \href{https://iopscience.iop.org/article/10.1088/1475-7516/2022/12/016}{\mbox{JCAP 12 (2022) 016}}, [\href{https://arxiv.org/abs/2210.14934}{arXiv:2210.14934}]
	
	\item \hypertarget{XII}{} G. Facchinetti, \textbf{M. Lucca}, and S. Clesse,   \emph{Relaxing CMB bounds on Primordial Black Holes: the role of radiative feedback}, 2022, \href{https://journals.aps.org/prd/abstract/10.1103/PhysRevD.107.043537}{\mbox{Phys.Rev.D 107 (2023) 4, 043537}}, [\href{https://arxiv.org/abs/2212.07969}{arXiv:2212.07969}]
	
	\item \hypertarget{XIII}{} Jose A. de Freitas Pacheco, E. Kiritsis, \textbf{M. Lucca}, and J. Silk, \emph{Quasi-extremal primordial black holes are a viable dark matter candidate}, 2023, \href{https://journals.aps.org/prd/abstract/10.1103/PhysRevD.107.123525}{Phys.Rev.D 107 (2023) 12, 123525}, [\href{https://arxiv.org/abs/2301.13215}{arXiv:2301.13215}]
	
	\item \hypertarget{XIV}{} \textbf{M. Lucca}, A. Rotti, and J. Chluba,  \emph{CRRfast: An emulator for the Cosmological Recombination Radiation with effects from inhomogeneous recombination}, 2023, [\href{https://arxiv.org/abs/2306.08085}{arXiv:2306.08085}]
\end{enumerate}
as well as on the following letters of interest I have contributed to as part of the Snowmass planning exercise
\begin{enumerate}[ {[}I{]} ]
	\setcounter{enumi}{14}
	\item \hypertarget{XV}{} E. Di Valentino et al., \emph{Cosmology Intertwined I: Perspectives for the Next Decade}, 2020, \href{https://www.sciencedirect.com/science/article/abs/pii/S0927650521000505}{Astropart. Phys. 131 (2021) 102606}, [\href{https://arxiv.org/abs/2008.11283}{arXiv:2008.11283}]
	
	\item \hypertarget{XVI}{} E. Di Valentino et al., \emph{Cosmology Intertwined II: The Hubble Constant Tension}, 2020, \href{https://www.sciencedirect.com/science/article/abs/pii/S0927650521000499}{Astropart. Phys. 131 (2021) 102605}, [\href{https://arxiv.org/abs/2008.11284}{arXiv:2008.11284}]
	
	\item \hypertarget{XVII}{} E. Di Valentino et al., \emph{Cosmology Intertwined III: $f\sigma_8$ and $S_8$}, 2020, \href{https://www.sciencedirect.com/science/article/abs/pii/S0927650521000487}{Astropart. Phys. 131 (2021) 102604}, [\href{https://arxiv.org/abs/2008.11285}{arXiv:2008.11285}]
	
	\item \hypertarget{XVIII}{} E. Di Valentino et al., \emph{Cosmology Intertwined IV: The Age of the Universe and its Curvature}, 2020, \href{https://www.sciencedirect.com/science/article/abs/pii/S0927650521000517}{Astropart. Phys. 131 (2021) 102607}, [\href{https://arxiv.org/abs/2008.11286}{arXiv:2008.11286}]
	
	\item \hypertarget{XIX}{} J. Chluba et al., \emph{CMB Spectral Distortions: A new window to fundamental physics}, 2020, [\href{https://www.snowmass21.org/docs/files/summaries/CF/SNOWMASS21-CF7_CF3-TF9_TF0_Jens_Chluba-069.pdf}{PDF}]
	
	\item \hypertarget{XX}{} E. Abdalla et al., \emph{Cosmology Intertwined: A Review of the Particle Physics, Astrophysics, and Cosmology Associated with the Cosmological Tensions and Anomalies}, 2022, \href{https://www.sciencedirect.com/science/article/pii/S2214404822000179}{JHEAp 34 (2022) 49-211},  [\href{https://arxiv.org/abs/2203.06142}{arXiv:2203.06142}]
\end{enumerate}
Also material adapted from various conference talks and seminars will be referred to. The original versions can be found in e.g.,
\begin{enumerate}[ {[}I{]} ]
	\setcounter{enumi}{20}
	\item \hypertarget{XXI}{} \textit{PONT 2020}, Avignon, France, [\href{https://indico.cern.ch/event/830633/contributions/3790286/}{Indico}]
	\item  \hypertarget{XXII}{}\textit{16th Marcel Grossmann meeting}, 2021, Rome, Italy, [\href{https://indico.icranet.org/event/1/contributions/174/}{Indico}]
\end{enumerate}
These references will be cited in the manuscript following this roman notation.

The following references have also been published during the same period, but their content is not related to the thesis and will therefore not be included:
\begin{itemize}
	\item[] \textbf{M. Lucca}, L. Sagunski, F. Guercilena, and C. Fromm, \emph{Shedding light on the angular momentum evolution of binary neutron star merger remnants: a semi-analytic model}, 2020, \href{https://www.sciencedirect.com/science/article/pii/S2214404820300598}{JHEAp 29 (2021) 19-30}, [\href{https://arxiv.org/abs/2010.11224}{arXiv:2010.11224}].
	
	\item[] T. Hambye, \textbf{M. Lucca}, and L. Vanderheyden, \emph{Dark matter as a heavy thermal hot relic}, 2020, \href{https://www.sciencedirect.com/science/article/pii/S0370269320303579}{Phys. Lett. B 807 (2020) 135553}, [\href{https://arxiv.org/abs/2003.04936}{arXiv:2003.04936}].
	
	\item[] \textbf{M. Lucca}, and L. Sagunski, \emph{The lifetime of binary neutron star merger remnants}, 2019, \href{https://www.sciencedirect.com/science/article/pii/S2214404820300240}{JHEAp 27 (2020) 33-37},  [\href{https://arxiv.org/abs/1909.08631}{arXiv:1909.08631}].
\end{itemize}

\end{document}